\documentclass[preprint,12pt]{elsarticle}

\usepackage{amssymb}
\usepackage{epsfig}
\usepackage{color}
\usepackage{bm}
\usepackage{mathrsfs}
\usepackage{geometry} 
\usepackage{times}
\usepackage{amsmath}
\usepackage{enumitem}

\biboptions{sort&compress}

\newcommand{\beq}[1]{\begin{equation}\label{#1}}
\newcommand{\eeq}{\end{equation}}
\newcommand{\bea}[1]{\begin{eqnarray} \label{#1}}
\newcommand{\eea}{\end{eqnarray}}
\newcommand{\ba}{\begin{array}}
\newcommand{\ea}{\end{array}}

\newcommand{\angstrom}{\textup{\AA}}

\usepackage{epsfig}
\usepackage{color}

\newcommand{\avg}[1]{\langle#1\rangle}

\def\be{\begin{equation}}
\def\ee{\end{equation}}

\def\bea{\begin{eqnarray}}
\def\eea{\end{eqnarray}}

\newcommand{\half}{\frac{1}{2}}

\newcommand{\Eps}{\mathcal{E}}

\newcommand{\numu}{\nu_{\mu}}

\usepackage[usenames,dvipsnames]{xcolor}
\definecolor{rossoCP3}{cmyk}{0,.88,.77,.40}
\usepackage{color}
\usepackage{amsmath}

\usepackage{epsfig}
\newcommand{\postscript}[2]{\setlength{\epsfxsize}{#2\hsize}
   \centerline{\epsfbox{#1}}}

\def\es{{\rm erg\ s}^{-1}}

\long\def\symbolfootnote[#1]#2{\begingroup%
\def\thefootnote{\fnsymbol{footnote}}\footnote[#1]{#2}\endgroup}

\def\lsim{\mathrel{\rlap{\lower4pt\hbox{\hskip1pt$\sim$}}
    \raise1pt\hbox{$<$}}}         
\def\gsim{\mathrel{\rlap{\lower4pt\hbox{\hskip1pt$\sim$}}
    \raise1pt\hbox{$>$}}}         

\begin{document}
\begin{frontmatter}



\thispagestyle{empty}
\title{ 
\color{rossoCP3} {\bf Ultra-High-Energy Cosmic Rays}} 
\author{\color{black} Luis A. Anchordoqui}
\address{\color{black} Department of Physics \& Astronomy, Lehman
  College, City University of New York, NY 10468, USA\\
Department of Physics, Graduate Center, City University of New York, NY 10016, USA\\
Department of Astrophysics, American Museum of Natural History, NY 10024, USA
}

\begin{abstract}
  In this report we review the important progress made in recent years
  towards understanding the experimental data on ultra-high-energy ($E
  \gtrsim 10^9~{\rm GeV}$) cosmic rays. We begin with a general survey
  of the available data, including a description of the energy
  spectrum, the nuclear composition, and the distribution of arrival
  directions. At this point we also give a synopsis of experimental
  techniques. After that, we introduce the fundamentals of cosmic ray
  acceleration and energy loss during propagation, with a view of
  discussing the conjectured nearby sources. Next, we survey the state
  of the art regarding the high- and ultra-high-energy cosmic
  neutrinos which may be produced in association with the observed
  cosmic rays. These neutrinos could constitute key messengers
  identifying currently unknown cosmic accelerators, possibly in the
  distant universe, because their propagation is not influenced by
  background photon or magnetic fields. Subsequently, we summarize the
  phenomenology of cosmic ray air showers. We describe the hadronic
  interaction models used to extrapolate results from collider data to
  ultra-high energies and the main electromagnetic processes that
  govern the longitudinal shower evolution. Armed with these two
  principal shower ingredients and motivation from the underlying
  physics, we describe the different methods proposed to distinguish
  the primary particle species. In the end, we explore how
  ultra-high-energy cosmic rays can be used as probes of beyond
  standard model physics models.
\end{abstract}

\end{frontmatter}
\newpage

\tableofcontents


\section{Introduction}

\label{sec:intro}

For biological reasons our perception of the Universe is based on the
observation of photons, most trivially by staring at the night-sky
with our bare eyes. Conventional astronomy covers many orders of
magnitude in photon wavelengths, from $10^5$~cm radio-waves to
$10^{-16}$~cm gamma-rays of TeV energy.  This 70 octave span in photon
frequency allows for a dramatic expansion of our observational
capacity beyond the approximately one octave perceivable by the human
eye. Photons are not, however, the only messengers of astrophysical
processes; we can also observe baryonic cosmic rays, neutrinos, and
gravitons (glaring as gravitational waves). On 2017 August 17, the
Advanced LIGO and Advanced Virgo gravitational-wave detectors made
their first observation of a binary neutron star merger, with
subsequent identification of transient counterparts across the entire
electromagnetic spectrum~\cite{GBM:2017lvd}. On 2017 September 22, the
blazar TXS 0506+056 was observed simultaneously in neutrinos and
photons~\cite{IceCube:2018dnn}. These unprecedented observations have been hailed
as the dawn of a new multi-frequency and multi-messenger era in
astronomy.

Baryon astronomy may be feasible for neutral particles or possibly
charged particles with energies high enough to render their
trajectories magnetically rigid. Indeed, on 2018 January 18, the
Pierre Auger Collaboration reported an indication of a possible
correlation between nearby starburst galaxies and cosmic rays of
energy $> 10^{10.6} {\rm GeV}$, with an a posteriori chance
probability in an isotropic cosmic ray sky of $4.2 \times 10^{-5}$,
corresponding to a 1-sided Gaussian significance of
$4\sigma$~\cite{Aab:2018chp}.  Should Mother Nature be so cooperative,
the emission of extremely high energetic particles by starburst
galaxies (predicted in the late '90s~\cite{Anchordoqui:1999cu}) would
become the first statistically ironclad observation of point sources
in cosmic rays. Although there may be some residual skepticism in the
broader community about the extreme-energy cosmic ray-starburst
connection, we expect that the very new data and arguments -- which we
will summarize and clarify in this article -- should soon dispel that skepticism.

The history of cosmic ray studies has witnessed many discoveries
central to the progress of high-energy physics, from the watershed
identification of new elementary particles in the early days, to the
confirmation of long-suspected neutrino oscillations. A major recent
achievement is establishing the suppression of the spectrum at the
highest energies~\cite{Abbasi:2007sv,Abraham:2008ru,Abraham:2010mj},
which may be the long-sought Greisen, Zatzepin, and Kuzmin (GZK)
cutoff~\cite{Greisen:1966jv,Zatsepin:1966jv}. The GZK effect is a
remarkable example of the profound links between different regimes of
physics, connecting as it does the behavior of the rarest,
highest-energy particles in the Cosmos to the existence of Nature's
most abundant particles -- the low energy photons in the relic
microwave radiation of the Big Bang -- while simultaneously demanding
the validity of Special Relativity over a mind-boggling range of
scales. Ultra-high-energy ($E \gtrsim 10^9~{\rm GeV}$) cosmic rays
(UHECRs) are the only particles with energies exceeding those
available at terrestrial accelerators. These UHECRs carry about seven
orders of magnitude more energy than the beam of the large hadron
collider (LHC). Hence, with UHECRs one can conduct particle physics
measurements in the center-of-mass (c.m.) frame up to about one order of
magnitude higher than the LHC energy reach.

In this review we concentrate on the physics of UHECRs focusing
tightly on the interface between experiment and phenomenology. The
layout is as follows. We begin in Sec.~\ref{sec:2} with a brief
summary of the most recent observations; guidance is given in the appendices to
statistical formulae and significance tests referred to in the main text.   In Sec.~\ref{sec:3} we
discuss the physics and astrophysics associated with the search for
the UHECR origin. We first summarize the main acceleration mechanisms
and then discuss the energy loss during propagation in the
intergalactic space and in the source environment. Subsequently, we
evaluate the multi-messenger relations connecting neutrino, gamma ray,
and UHECR observations.  In Sec.~\ref{sec:4} we focus attention on the
general properties and techniques for modeling extensive air showers
initiated when UHECRs interact in the atmosphere.  In Sec.~\ref{sec:5}
we examine how UHECRs can be used as probes of
new physics beyond the highly successful but conceptually incomplete
standard model (SM) of weak, electromagnetic, and strong
interactions. Finally, in Sec.~\ref{sec:6} we provide evidence-based
guidance to set strict criteria for research and development of future
UHECR experiments. 
 
Before proceeding, we pause to present our notation. Unless otherwise
stated, we work with natural (particle physicist's) Heaviside-Lorentz
(HL) units with
\begin{equation}
\hbar=c=k =\varepsilon_0=\mu_0=1\,.
\end{equation}
The fine structure constant is $\alpha=e^2/(4\pi\varepsilon_0\hbar c)\simeq1/137$.
All SI units can then be expressed in electron Volt (eV), namely
\begin{align}
1~{\rm m} &\simeq 5.1\times10^6~{\rm eV}^{-1}\,,&
1~{\rm s} & \simeq 1.5\times10^{15}~{\rm eV}^{-1}\,,&
1~{\rm kg} & \simeq 5.6\times10^{35}~{\rm eV}\,, \\
1~{\rm A} & \simeq 1244~{\rm eV}\,, &
1~{\rm G} &\simeq1.95\times10^{-2}{\rm eV}^2\,,&
1~{\rm K} &\simeq8.62\times10^{-5}~{\rm eV}\,.
\end{align}
Electromagnetic processes in astrophysical environments are often
described in terms of Gauss (G) units. For a comparison of formulas
used in the literature, we note some conversion factors: $(e^2)_{\rm
  HL} = 4\pi(e^2)_{\rm G}$, $(B^2)_{\rm HL} = (B^2)_{\rm G}/4\pi$, and
$(E^2)_{\rm HL} = (E^2)_{\rm G}/4\pi$. To avoid confusion we will
present most of the formulas in terms of {\it invariant quantities}
i.e., $eB$, $eE$ and the fine-structure constant $\alpha$. Distances
are generally measured in Mpc $\simeq 3.08 \times 10^{22}~{\rm
  m}$. Extreme energies are sometimes expressed in EeV $\equiv
10^9~{\rm GeV} \equiv 10^{18}~{\rm eV}$.  The following is a list of
additional useful conversion factors: $1~{\rm GeV} = 1.602 \times
10^{-3}~{\rm erg}$, $h = 6.626 \times 10^{-27}~{\rm erg \, Hz^{-1}}$,
$hc = 1.986 \times 10^{-16}~{\rm erg \ cm}= 1.986 \times 10^{-8}~{\rm
  erg} \, \angstrom$, $\hslash c = 1.973 \times 10^{-14}~{\rm GeV} \,
{\rm cm}$, $1~{\rm sr} = 3.283 \times 10^3~{\rm sq\, deg} = 4.255
\times 10^{10}~{\rm sq \, arcsec}$, $1~{\rm WB} = 10^{-8}~{\rm GeV} \,
{\rm cm}^{-2} {\rm s}^{-1} \, {\rm sr}^{-1}$, and one Jansky or
$1~{\rm Jy} = 10^{-23}~{\rm erg} \, {\rm cm}^{-2} \, {\rm s}^{-1} \,
{\rm Hz}^{-1}$. The Planck units of mass, length, and time are
$M_{\rm Pl} = \ell_{\rm Pl}^{-1} = t_{\rm Pl}^{-1} \sim 1.2 \times
10^{19}~{\rm GeV}$.  We adopt the usual concordance cosmology of a
flat universe dominated by a cosmological constant, with dark energy
density parameter $\Omega_\Lambda \approx 0.69$ and a cold dark matter
plus baryon component $\Omega_m \approx 0.31$~\cite{Ade:2015xua}. The
Hubble parameter as a function of redshift $z$ is given by
\mbox{$H^2(z) = H_0^2 [\Omega_m (1 + z)^3 + \Omega_\Lambda]$,}
normalized to its value today, $H_0 = 100 \, h~{\rm km} \ {\rm s}^{-1}
\, {\rm Mpc}^{-1}$, with $h = 0.678(9)$~\cite{Patrignani:2016xqp}. The
dependence of the cosmological time with redshift is $dz = - dt (1 +
z) H(z)$. This report will build upon the content of the lecture notes
from the 6th CERN-Latin-American School of High-Energy
Physics~\cite{Anchordoqui:2011gy}, the whitepaper contribution to the
US Snowmass planning process~\cite{Anchordoqui:2013eqa},
and~\cite{Anchordoqui:2002hs,Torres:2004hk,Anchordoqui:2004xb}.

\section{Experimental observations and searches}
\label{sec:2}
\subsection{Historical overview}

In 1912 Hess carried out a series of pioneering balloon flights during
which he measured the levels of ionizing radiation as high as 5~km
above the Earth's surface~\cite{Hess:1912srp}.  His discovery of
increased radiation at high altitude revealed that we are bombarded by
ionizing particles from above. These cosmic ray (CR) particles are now
known to consist primarily of protons, helium, carbon, nitrogen and
other heavy ions up to iron.  Measurements of energy and isotropy
showed conclusively that one obvious source, the Sun, is not the main
source. Only for CRs with kinetic energy $E_{\rm kin} \lesssim 100~{\rm MeV}$, where the solar wind
shields protons coming from outside the solar system, does the Sun
dominate the observed proton flux.\footnote{Cosmic rays entering the
  Solar System have to overcome the outward-flowing solar wind. The
  energy of incoming cosmic rays is reduced through interactions with
  this wind, preventing the lowest energy ones from reaching the
  Earth. This effect is known as solar modulation. The Sun has an
  11~yr activity cycle, which is echoed in the strength of the solar
  wind to modulate cosmic rays. Because of this effect the flux of
  cosmic rays reaching Earth is anti-correlated with the level of
  solar activity: when the solar activity is high and there are lots
  of sunspots, the flux of cosmic rays at Earth is low, and vice
  versa. The solar modulation is visible below about 1~GeV.}
Spacecraft missions far out into the solar system, well away from the
confusing effects of the Earth's atmosphere and magnetosphere, confirm
that the abundances around 1~GeV are strikingly similar to those found
in the ordinary material of the solar system. Exceptions are the
overabundance of elements like lithium, beryllium, and boron,
originating from the spallation of heavier nuclei in the interstellar
medium.

Above about $10^{5}~{\rm GeV},$ the rate of CR primaries is less than
one particle per square meter per year and direct observation in the
upper layers of the atmosphere (balloon or aircraft), or even above
(spacecraft) is inefficient. Only experiments with large apertures and
long exposure times can hope to acquire a significant number of
events. Such experiments exploit the atmosphere as a giant
calorimeter.  The incident cosmic radiation interacts with the atomic
nuclei of air molecules and produces air showers which spread out over
large areas.  Already in 1938, Auger concluded from the size of the
air showers that the spectrum extends up to and perhaps beyond
$10^{6}$~GeV~\cite{Auger:1938ef,Auger:1939sh}.  Data collected during
1954 and 1957  at the Agassiz Station of
the Harvard College Observatory provided evidence for primary
particles with $E \sim 10^9~{\rm GeV}$~\cite{Clark:1961mb}. In 1962,
the array of scintillation detectors at the MIT Volcano Ranch station
detected the first UHECR event with an estimated energy of ${\cal O}
(10^{11}~{\rm GeV})$~\cite{Linsley:1963km}.

In 1964, the cosmic microwave background (CMB) was
discovered~\cite{Penzias:1965wn}, and shortly thereafter it became
self-evident that the relic photons would make the universe opaque to
the propagation UHECRs~\cite{Greisen:1966jv,Zatsepin:1966jv}. The GZK
interactions occur, for example, for protons with energies beyond the
photopion production threshold,
\begin{equation}
E_{\rm th} = \frac{m_\pi \, (2m_p + m_\pi)}{4 \varepsilon} 
\approx 3.4 \times 10^{10}\,
\left(\frac{\varepsilon}{10^{-3}~{\rm eV}}\right)^{-1} \,\,{\rm GeV}\,,
\label{gzkEth}
\end{equation}
where $m_p$ ($m_\pi$) denotes the proton (pion) mass and
$\varepsilon \sim 10^{-3}$~eV is a typical CMB photon
energy. After pion production, the proton (or perhaps, instead, a
neutron) emerges with at least 50\% of the incoming energy. This
implies that the nucleon energy changes by an $e$-folding after a
propagation distance \mbox{$d^{\rm GZK}_{68} \lesssim
(\sigma_{p\gamma}\,n_\gamma\,\langle y \rangle)^{-1} \sim 15$~Mpc,}
where $n_\gamma \approx 410$~cm$^{-3}$ is the number density of the
CMB photons, $\sigma_{p \gamma} > 0.1$~mb is the photopion production
cross section, $\langle y \rangle$ is the average energy fraction (in
the laboratory system) lost by a nucleon per interaction, and where
the subscript of the GZK distance specifies the CR energy in EeV
units. For heavy nuclei, the giant dipole resonance (GDR) can be
excited at similar total energies and hence, for example, iron nuclei
do not survive fragmentation over comparable GZK
distances~\cite{Stecker:1969fw,Puget:1976nz}. Additionally, the
survival probability for extremely high energy ($\approx 10^{11}$~GeV)
$\gamma$-rays (propagating on magnetic fields~$\gg 10^{-11}$~G) to a
distance $d$, \mbox{$p(>d) \approx \exp[-d/6.6~{\rm Mpc}]$}, becomes
less than $10^{-4}$ after traversing a distance of 50~Mpc~\cite{Torres:2004hk}. All in all,
as our horizon shrinks dramatically for  $E \gtrsim
10^{11}$~GeV, one would expect a sudden suppression in the energy
spectrum if the CR sources follow a cosmological distribution.

Throughout the past few decades, continuously running monitoring using
sophisticated equipment on high altitude balloons and ingenious
installations on the Earth's surface uncovered a plummeting 
flux that goes down from $10^4$ m$^{-2}$ s$^{-1}$ at $E\sim 1$~GeV to
$10^{-2}$ km$^{-2}$ yr$^{-1}$ at $E \sim
10^{11}$~GeV~\cite{Patrignani:2016xqp}.  In recent years, substantial
progress has been made in measuring the extraordinarily low flux at
the high energy end of the spectrum. There are two primary detection
methods that have been successfully used in ground-based high exposure
experiments. In the following paragraph we provide a terse overview of
these approaches.  For an authoritative review on experimental
techniques and historical perspective see~\cite{Nagano:ve,Kampert:2012vi}.

\begin{figure}[tpb] \postscript{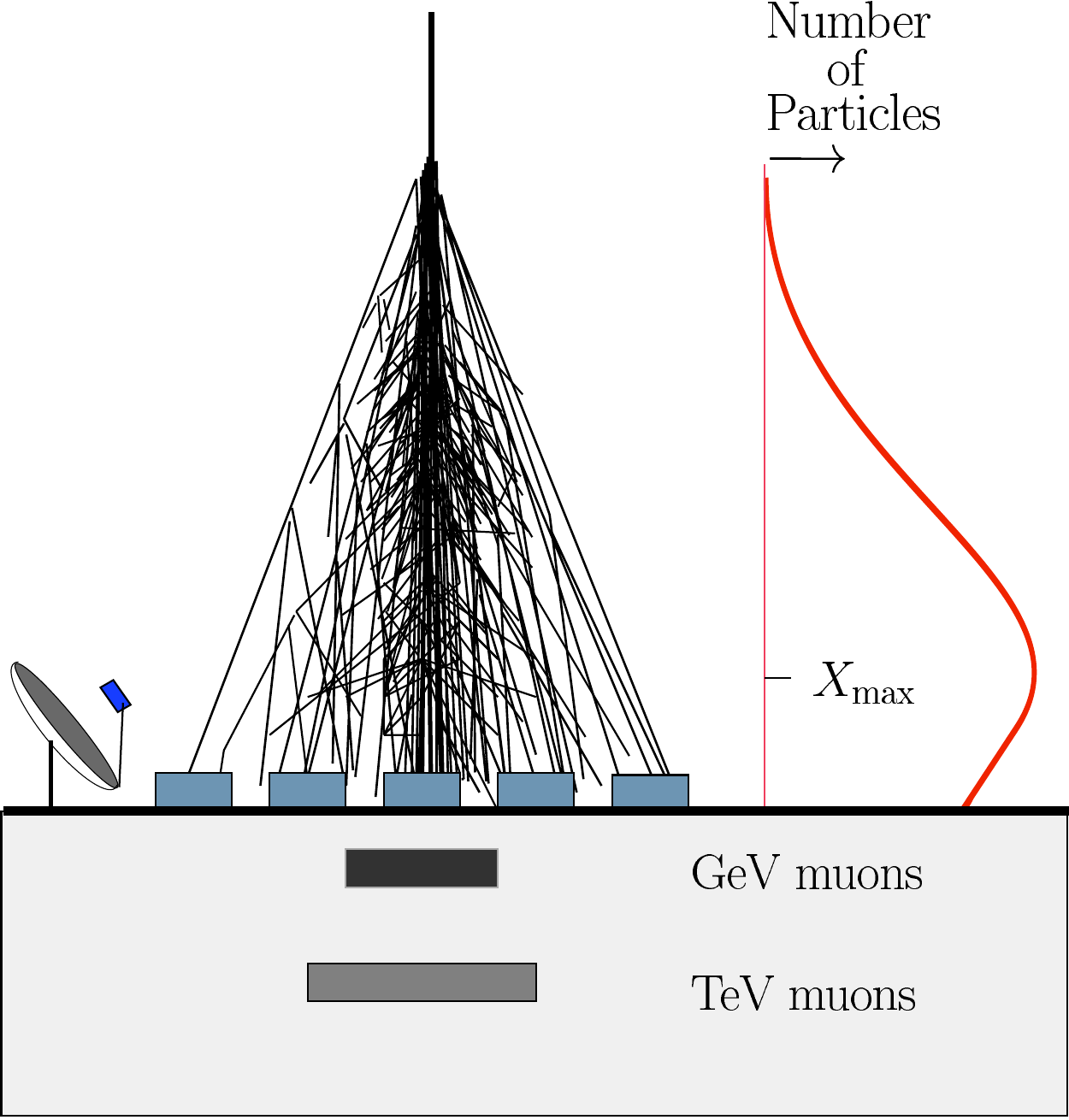}{0.4} \caption{Particles
    interacting near the top of the atmosphere initiate an
    electromagnetic and hadronic cascade. Its profile is shown on the
    right. The different detection methods are illustrated. Mirrors
    collect the Cherenkov and nitrogen fluorescent light, arrays of
    detectors sample the shower reaching the ground, and underground
    detectors identify the muon component of the shower. The number of
    particles as a function of the amount of atmosphere penetrated by
    the cascade ($X$ in ${\rm g \, cm}^{-2}$) is known as the
    longitudinal profile. The integrated longitudinal profile provides
    a calorimetric measurement of the energy of the primary CR, with a
    relatively small uncertainty due to the correction for energy lost
    to neutrinos and particles hitting the ground. From
    Ref.~\cite{Anchordoqui:2002hs}.}
\label{fig:1}
\end{figure}

When the incident cosmic radiation interacts with atomic nuclei of air
molecules, it produces fluxes of secondary, tertiary, and subsequent
generations of particles. All these particles together create a
cascade, called air shower. As the cascade develops longitudinally the
particles become less and less energetic since the energy of the
incoming cosmic ray is redistributed among more and more
participants. The transverse momenta acquired by the secondaries cause
the particles to spread laterally as they propagate through the
atmospheric target. Most of the air shower particles excite nitrogen
molecules in the atmosphere, which fluoresce in the ultraviolet
(UV). The size of an extensive air shower (EAS) at sea-level depends
on the primary energy and arrival direction. For showers of UHECRs,
the cascade is typically several hundreds of meters in diameter and
contains millions of secondary particles. Secondary electrons and
muons produced in the decay of pions may be observed in scintillation
counters or alternatively by the Cherenkov light emitted in water
tanks. The separation of these detectors may range from 10~m to 1~km
depending on the CR energy and the optimal cost-efficiency of the
detection array. The shower core and hence arrival direction can be
estimated by the relative arrival time and density of particles in the
grid of detectors. The lateral particle density of the showers can be
used to calibrate the primary energy.  Another well-established method
of detection (pioneered by the Fly's Eye
experiment~\cite{Baltrusaitis:1985mx} and its up-scaled version with
high-resolution dubbed HiRes~\cite{AbuZayyad:2000uu}) involves
measurement of the shower longitudinal development (number of
particles versus atmospheric depth, shown schematically in
Fig.~\ref{fig:1}) by sensing the fluorescence light produced via
interactions of the charged particles in the atmosphere.  The emitted
light is typically in the $300 - 400~{\rm nm}$ UV range to
which the atmosphere is quite transparent. Under favorable atmospheric
conditions, EASs can be detected at distances as large as 20~km, about
2 attenuation lengths in a standard desert atmosphere at ground
level. However, observations can only be done on clear moon{\it less}
nights, resulting in an average 10\% duty cycle.

In this review we concentrate on the latest results from the two UHECR
experiments currently in operation: the Telescope Array and the Pierre
Auger Observatory to which we will often refer as TA and
Auger.  TA has been collecting data since 2007
in Millard County, west central Utah. This facility is located at
$39.3^\circ$~N and $112.9^\circ$~W, 1.4~km above sea level (equivalent
to $880~{\rm g/cm^2}$ of atmospheric depth). The experiment consists
of 507 scintillation surface detectors sensitive to muons and
electrons~\cite{AbuZayyad:2012kk}, and 48 fluorescence telescopes
located in three fluorescence detector stations overlooking the
counters~\cite{Tokuno:2012mi}. The spacing of the counters in the
surface detector array is 1.2~km and they are placed over an area of
approximately $700~{\rm km^2}$.

The Pierre Auger Observatory is located on the vast plain of {\em
  Pampa Amarilla}, near the town of Malarg\"ue in Mendoza Province,
Argentina ($35.1^\circ - 35.5^\circ$~S, $69.6^\circ$~W, and
atmospheric depth of 875~g/cm$^2$)~\cite{Abraham:2004dt}. The
experiment began collecting data in 2004, with construction of the
baseline design completed by 2008.  From January 2004 until December 2016, Auger had
collected in excess of $6.7 \times 10^4~{\rm km}^2~\rm{sr}~\rm{yr}$ in
exposure, significantly more exposure than other cosmic ray
observatories combined~\cite{Aab:2017njo}.  Two types of instruments are employed.
Particle detectors on the ground sample air shower fronts as they
arrive at the Earth's surface, while fluorescence telescopes measure
the light produced as air shower particles excite atmospheric
nitrogen.

The surface array~\cite{Abraham:2010zz} comprises $1.6 \times 10^3$  
surface detector (SD) stations, each consisting of a tank 
filled with 12 tons of water and instrumented with 3 
photomultiplier tubes which detect the Cherenkov light 
produced as particles traverse the water.
The signals from the photomultipliers are read out with flash analog
to digital converters at 40~MHz and timestamped by a GPS unit, allowing
for detailed study of the arrival time profile of shower particles.
The tanks are arranged on a triangular grid with a 1.5~km spacing, 
covering about $3 \times 10^3~{\rm km}^2$. 
The surface array operates with close to a 100\% duty cycle,
and the acceptance for events above $10^{9.5}$~GeV
is nearly 100\%.

The fluorescence detector (FD) system~\cite{Abraham:2009pm} 
consists of 4 buildings, each housing 6
telescopes which overlook the surface array.
Each telescope employs an $11~{\rm m}^2$ segmented
mirror to focus the fluorescence light entering through a 2.2~m diaphragm
onto a camera which pixelizes the image using 440 photomultiplier tubes.
The photomultiplier signals are digitized at 10~MHz, providing
a time profile of the shower as it develops in the atmosphere.
The FD can be operated only when the sky is dark and clear,
and has a duty cycle of  10-15\%.  In contrast to the SD acceptance,
the acceptance of FD events depends strongly on 
energy~\cite{:2010zzl}, extending down to about $10^{9}$~GeV.

The two detector systems of Auger and TA provide complementary
information, as the SD measures the lateral distribution and time
structure of shower particles arriving at the ground, and the FD
measures the longitudinal development of the shower in the atmosphere.
A subset of showers is observed simultaneously by the SD and FD.
These ``hybrid'' events are very precisely measured and provide an
invaluable calibration tool.  In particular, the FD allows for a
roughly calorimetric measurement of the shower energy since the amount
of fluorescence light generated is proportional to the energy
deposited along the shower path; in contrast, extracting the shower
energy via analysis of particle densities at the ground relies on
predictions from hadronic interaction models describing physics at
energies beyond those accessible to existing collider experiments.  Hybrid
events can therefore be exploited to set a model-independent energy
scale for the SD array, which in turn has access to a greater data
sample than the FD due to the greater live time.

In the remainder of the section, we describe recent results from Auger
and TA, including measurements of energy spectrum and UHECR nuclear
composition, as well as searches for anisotropy in the CR arrival
directions.

\subsection{Measurements of the cosmic ray intensity and its nuclear composition}
\label{high-low}

The (differential) intensity of CRs is defined as the number of particles
crossing a unit area $dA$ per unit time $dt$ and unit solid angle
$d\Omega$, with energy within $E$ and $E + dE$~\cite{Rybicki},
\begin{equation}
 J (E) = \frac{dN}{dA \ dt \ d \Omega \ dE} \  .
\label{intensity}
\end{equation}
Throughout $J(>E) = \int_E^\infty dE' \ J(E')$ symbolizes the number of
particles with energy $> E$ crossing an unit area per unit time and
unit solid angle.  For an isotropic intensity, the flux of particles
from one hemisphere traversing a planar detector is found to be
\begin{equation}
{\cal F}(E) = \int d\Omega \ J(E) \ \cos \theta = J(E) \int_0^{2 \pi} d \phi
\int_0^{\pi/2} d \theta \sin \theta \cos \theta = \pi J(E) \, ,
\end{equation}
where $\theta$ and $\phi$ are the zenith and azimuthal angles,
respectively. The (differential) number density of
cosmic rays with velocity $v$ is given by
\begin{equation}
n(E) = \frac{4 \pi}{v} J(E) \, .
\label{nCR}
\end{equation}
On the whole, the intensity could depend both on the position $\mathbf
x$ of the detector and on its orientation $(\theta,\phi)$. The intensity
can be related to the CR phase space distribution, $f(\mathbf x, \mathbf p)$, 
by substituting 
\begin{equation}
 dN = f(\mathbf x, \mathbf p) \  d^3x \ d^3p 
\end{equation}
into (\ref{intensity}), with $d^3x = dA \ v \ dt$ and $d^3p = p^2 \ dp \ d\Omega$. This leads to
\begin{equation}
J(\mathbf x,p,\theta,\phi) = v p^2 \frac{dp}{dE} \ f(\mathbf x, \mathbf
p)  = p^2 \ f(\mathbf x,\mathbf p) \  .
\label{phase-space}
\end{equation}

The kinetic energy density of CRs can be expressed as
\begin{equation}
\epsilon_{\rm CR} = \int dE \ E_{\rm kin} \ n(E) = 4 \pi \int dE
\ \frac{E_{\rm kin}}{v} \  J(E)
\, .
\label{rho-cr}
\end{equation}
The energy density of CRs in the local interstellar medium
(LIS), i.e. with the intensity in (\ref{rho-cr})  extrapolated outside the
reach of the solar wind, is found to be $\epsilon_{\rm CR} \approx 0.8~{\rm
  eV/cm^3}$~\cite{Kachelriess:2008ze}. It is of interest to compare this value with the
average energy density: {\it (i)}~of baryons in the Universe $\epsilon_b
\approx 100~{\rm eV/cm^3}$, {\it (ii)}~of starlight in the Galactic
disc $\epsilon_{\rm light} \approx 5~{\rm eV/cm^3}$,  {\it (iii)}~of
the Galactic magnetic field $\epsilon_B = B^2/2 \approx 0.6~{\rm eV/cm^3}$ (assuming $B \approx
5~\mu{\rm G}$). Note that if the LIS energy density of cosmic rays would be
taken as a fiducial value for the entire Universe, then roughly 1\% of
the energy of all baryons would be in the form of relativistic
particles. This is highly unlikely. Therefore, the preceding estimates seem to indicate that most
CRs are accumulated inside the Milky Way.

Above about 1~GeV, the energy spectrum is observed to fall roughly as
a power law; namely, the CR intensity decreases nearly three
orders of magnitude per energy decade, until eventually suffering a
strong suppression around $E_{\rm supp} \sim 10^{10.6}~{\rm
  GeV}$~\cite{Patrignani:2016xqp}. Individual experiments cover only a
part of these 11 decades in energy, which means that a common picture
needs to be pieced together from many data sets. The database of
charged cosmic rays (CRDB)  is a meritorious endeavor which maintains
central machine-readable access to CR data collected by satellite and
balloon experiments~\cite{Maurin:2013lwa}. Because of their direct
detection method, these experiments provide characterization of the
separate elements in terms of their charge $Ze$, and hence they report
the CR intensity per element. Above about $10^{6}~{\rm GeV}$ direct
detection methods become inefficient and ground-based experiment take
over. Unfortunately, because of the highly indirect method of
measurement, extracting precise information from EASs has proved to be
exceedingly difficult, and hence the ability to discriminate the
charge $Ze$ of each CR is lost. The most fundamental problem is that
the first generations of particles in the cascade are subject to large
inherent fluctuations and consequently this limits the event-by-event
energy resolution of the experiments. In addition, the 
c.m. energy of the first few cascade steps is well beyond any
reached in collider experiments. Therefore, one needs to rely on
models that attempt to extrapolate, using different mixtures of theory
and phenomenology, our understanding of particle physics.

If the primary cosmic ray is a baryon, the air-shower development is
driven by soft QCD interactions, which are computed from
phenomenological models tuned to collider data. Hundreds to thousands
of secondary particles are usually produced at the interaction vertex,
many of which also have energies above the highest accelerator
energies~\cite{Anchordoqui:1998nq}. These secondary products are of
course intrinsically hadrons. Generally speaking, by extrapolating
final states observed at collider experiments, we can infer that, for
$pp$ collisions at c.m. energy $\sqrt{s} \gtrsim 100~{\rm
  TeV}$, the jet of hadrons contains about 75\% pions (including 25\%
$\pi^0$'s, in accord with isospin invariance), 15\% kaons, and 10\%
nucleons~\cite{GarciaCanal:2009xq}. During the shower evolution, the
hadrons propagate through a medium with an increasing density as the
altitude decreases and the hadron-air cross section rises slowly with
energy. Therefore, the probability for interacting with air before
decay increases with rising energy. Moreover, the relativistic time
dilation increases the decay length by a factor $E_h/m_h$, where $E_h$
and $m_h$ are the energy and mass of the produced hadron. When the
$\pi^0$'s (with a lifetime of $\simeq 8.4 \times10^{-17}~{\rm s}$) do
decay promptly to two photons, they feed the electromagnetic component
of the shower. For other longer-lived mesons, it is instructive to
estimate the critical energy at which the chances for interaction and
decay are equal. For a vertical transversal of the atmosphere, such a
critical energy is found to be: $\xi_c^{\pi^\pm} \sim 115~{\rm GeV}$,
$\xi_c^{K^\pm} \sim 850~{\rm GeV}$, $\xi_c^{K^0_L} \sim 210~{\rm
  GeV}$, $\xi_c^{K^0_S} \sim 30~{\rm TeV}$~\cite{Gondolo:1995fq}.  The
dominant $K^+$ branching ratios are to $\mu^+ \nu_\mu\ (64\%)$, to
$\pi^+ \pi^0\ (21\%)$, to $\pi^+\pi^+\pi^-\ (6\%)$, and to $\pi^+
\pi^0\pi^0\ (2\%)$, whereas those of the $K^0_S$ are to $\pi^+\pi^- \
(60\%)$, to $\pi^0 \pi^0 \ (30\%)$, and for $K^0_L$ we have $\pi^\pm
e^\mp \nu_e \ (40\%)$, $\pi^\pm \mu^\mp \nu_\mu \ (27\%)$, $\pi^0
\pi^0 \pi^0 \ (19\%)$, $\pi^+ \pi^- \pi^0 \
(12\%)$~\cite{Patrignani:2016xqp}.  With these figures in mind, to a first
approximation it seems reasonable to assume that in each generation of
particles about 25\% of the energy is transferred to the
electromagnetic shower, and all hadrons with
energy~$\gtrsim \xi_c^{\pi^\pm}$ interact rather than decay, continuing
to produce the hadronic shower.\footnote{The electromagnetic
shower fraction from pions only is less than 25\%, but simulations show that inclusion of
  other hadronic resonances brings the electromagnetic shower fraction
  up to about 25\%~\cite{Aab:2016hkv}. Thus, 25\% is a reasonable
  estimate of the energy transfer to the electromagnetic shower
  in each generation of particles.}
Eventually, the electromagnetic cascade dissipates around 90\% of the
primary particle's energy and the remaining 10\% is carried by muons
and neutrinos. 

As the cascade process develops in the atmosphere, the number of
particles in the shower increases until the energy of the secondary
particles is degraded to the level where ionization losses dominate.
At this point the density of particles starts to decline. A
well-defined peak in the longitudinal development, $X_{\rm max}$,
occurs where the number of $e^\pm$ in the electromagnetic shower is at
its maximum; see Fig.~\ref{fig:1}. $X_{\rm max}$ increases with
primary energy, as more cascade generations are required to degrade
the secondary particle energies. Evaluating $X_{\rm max}$ is a
fundamental part of many of the composition analyses done when
studying air showers. For showers of a given total energy $E$, heavier
nuclei have smaller $X_{\rm max}$ because the shower is already
subdivided into $A$ nucleons when it enters the atmosphere. The
average depth of maximum $\langle X_{\rm max} \rangle$ scales
approximately as $\ln(E/A)$~\cite{Linsley:1981gh}. Therefore, since
$\langle X_{\rm max} \rangle$ can be determined directly from the
longitudinal shower profiles measured with a fluorescence detector,
the composition can be extracted after estimating $E$ from the total
fluorescence yield. Indeed, the parameter often measured is $D_{10}$,
the rate of change of $\langle X_{\rm max} \rangle$ per {\it decade}
of energy.

Photons penetrate quite deeply into the atmosphere due to decreased
secondary multiplicities and suppression of cross sections by the
Landau-Pomeranchuk-Migdal (LPM) effect~\cite{Landau:um,Migdal:1956tc}.
Indeed, it is rather easier to distinguish photons from protons and
iron than protons and iron are to distinguish from one another. For
example, at $10^{10}~{\rm GeV}$, the $\langle X_{\rm max} \rangle$ for
a photon is about $1,000~{\rm g/cm}^2$, while for protons and iron the
numbers are 800~g/cm$^2$ and 700~g/cm$^2$, respectively. Searches for
photon primaries have been conducted using both the surface and
fluorescence instruments of Auger. While analysis of the fluorescence
data exploits the direct view of shower development, analysis of data
from the surface detector relies on measurement of quantities which
are indirectly related to the $X_{\rm max}$, such as the signal
risetime at 1000~m from the shower core and the curvature of the
shower front. Presently, the 95\% CL upper limits on the integrated
photon intensity, $\Phi_\gamma (>E_0)$, are given in Table~\ref{Apr1}. Further details on the analysis procedures can be found
in~\cite{Abraham:2006ar,Aglietta:2007yx,Abraham:2009qb}.

\begin{table}
\caption{Upper limits on the integral photon intensity with a spectrum
  $\propto E_\gamma^{-2}$ and no background subtraction~\cite{Aab:2016agp}. \label{Apr1}}
\begin{center}
\begin{tabular}{cc}
\hline
\hline
~~~~~~~~~~~~~~~~~~~~$\log_{10} (E_0/{\rm GeV})$ ~~~~~~~~~~~~~~~~~~~~ &
~~~~~~~~~~~~~~~~~~~~$\log_{10}[ \Phi_\gamma (> E_0)/({\rm km^{-2} \ sr^{-1} \ yr^{-1}})]$~~~~~~~~~~~~~~~~~~\\
\hline
$9.0$ & $-1.60$ \\
$9.3$ & $-2.05$ \\
 $9.5$  & $-2.10$ \\
 $9.7$  & $-2.10$ \\
$10.0$ &  $-2.65$ \\
$10.3$ & $-3.00$ \\
$10.7$ & $-3.40$ \\
\hline
\hline
\end{tabular}
\end{center}
\end{table}

The Auger Collaboration reported a measurement of the average shower
maximum as a function of energy using FD data ~\cite{Aab:2017njo}. The
study is based on $42,466$ events with $E > 10^{8.2}~{\rm
  GeV}$, out of which 62 have been detected at $E > 10^{10.5}~{\rm
  GeV}$. The $\langle X_{\rm max} \rangle$ evolves with energy at a
rate of $D_{10} = 79~{\rm g/cm^{2}}$ for $E< 10^{9.33\pm0.02}~{\rm
  GeV}$, and $D_{10} = 26~{\rm g/cm^2}$ for $E> 10^{9.33\pm0.02}~{\rm
  GeV}$, in agreement with previous
studies~\cite{Abraham:2010yv,Aab:2014kda}.  Predictions of the energy
evolution of $\langle X_{\rm max} \rangle$ from air shower simulations
are around $D_{10} = 60~{\rm g/cm^2}$, irrespective of the baryon
number $A$ of the primary and the model used to simulate hadronic
interactions~\cite{ObservatoryMichaelUngerforthePierreAuger:2017fhr}. This
implies that Auger high-quality, high-statistics data, when
interpreted with existing hadronic event generators, exhibit a strong
likelihood for a composition that becomes gradually lighter in the
range $10^{8.2} \lesssim E/{\rm GeV} \lesssim 10^{9.33}$,
qualitatively consistent with a transition from a heavy Galactic
composition, to a light extragalactic composition. For $E>
10^{9.33\pm0.02}~{\rm GeV}$, the trend is reversed and the average
baryon number increases with energy. Within uncertainties, the data
from TA are consistent with these findings~\cite{Abbasi:2014sfa,
  Abbasi:2015xga,Abbasi:2018nun,Hanlon:2018dhz}. Studies aim to identify the primary particle
species using Auger SD data yield results in agreement with FD
observations~\cite{Aab:2016htd,Aab:2017cgk}. An interpretation of the
full $X_{\rm max}$ distribution in each energy bin is achieved by
fitting a superposition of $X_{\rm max}$-templates obtained from
simulations of $p$-, He-, N- and Fe-induced air showers to the data.

\begin{figure}[tbp] 
\begin{minipage}[t]{0.49\textwidth}
    \postscript{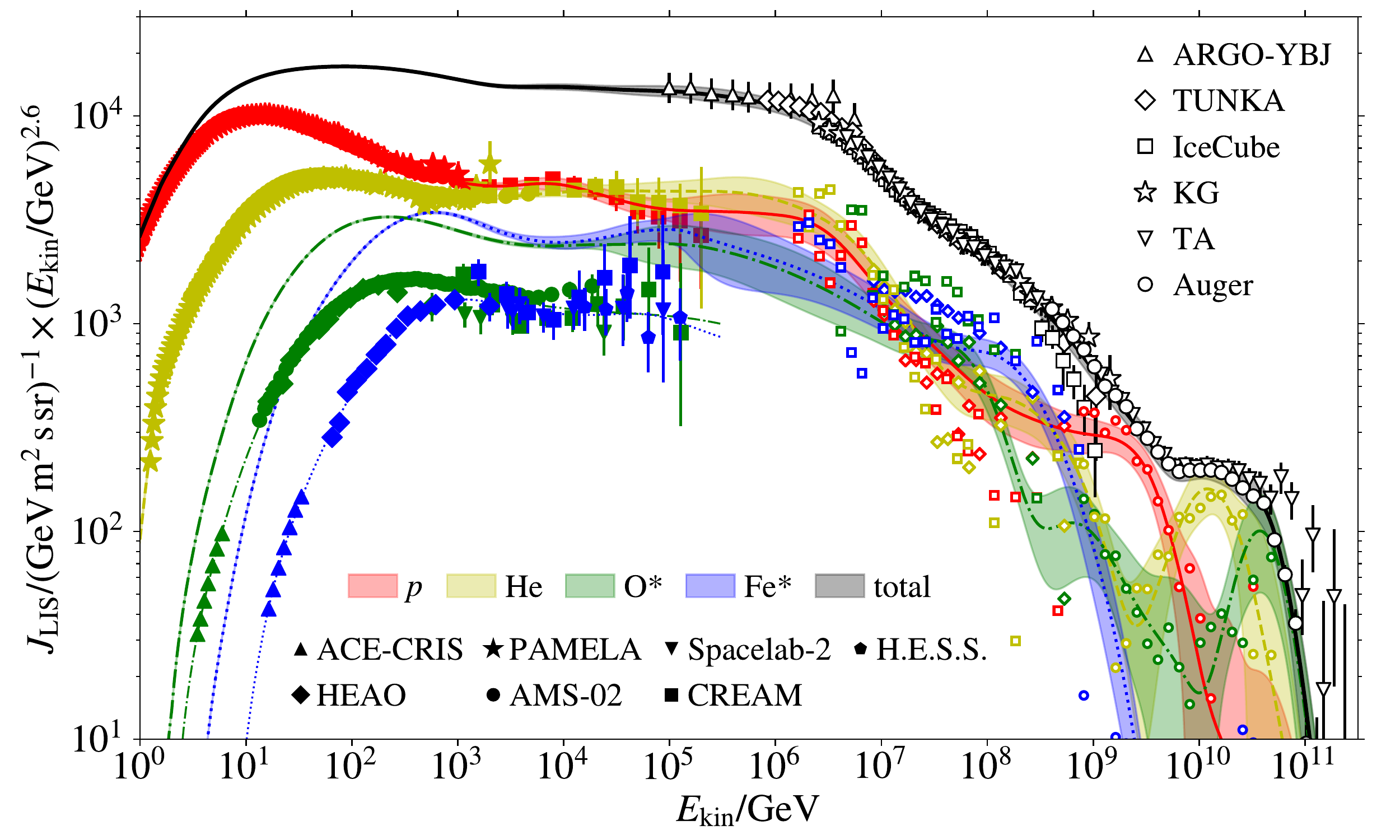}{0.99} 
\end{minipage} 
\hfill \begin{minipage}[t]{0.49\textwidth}
  \postscript{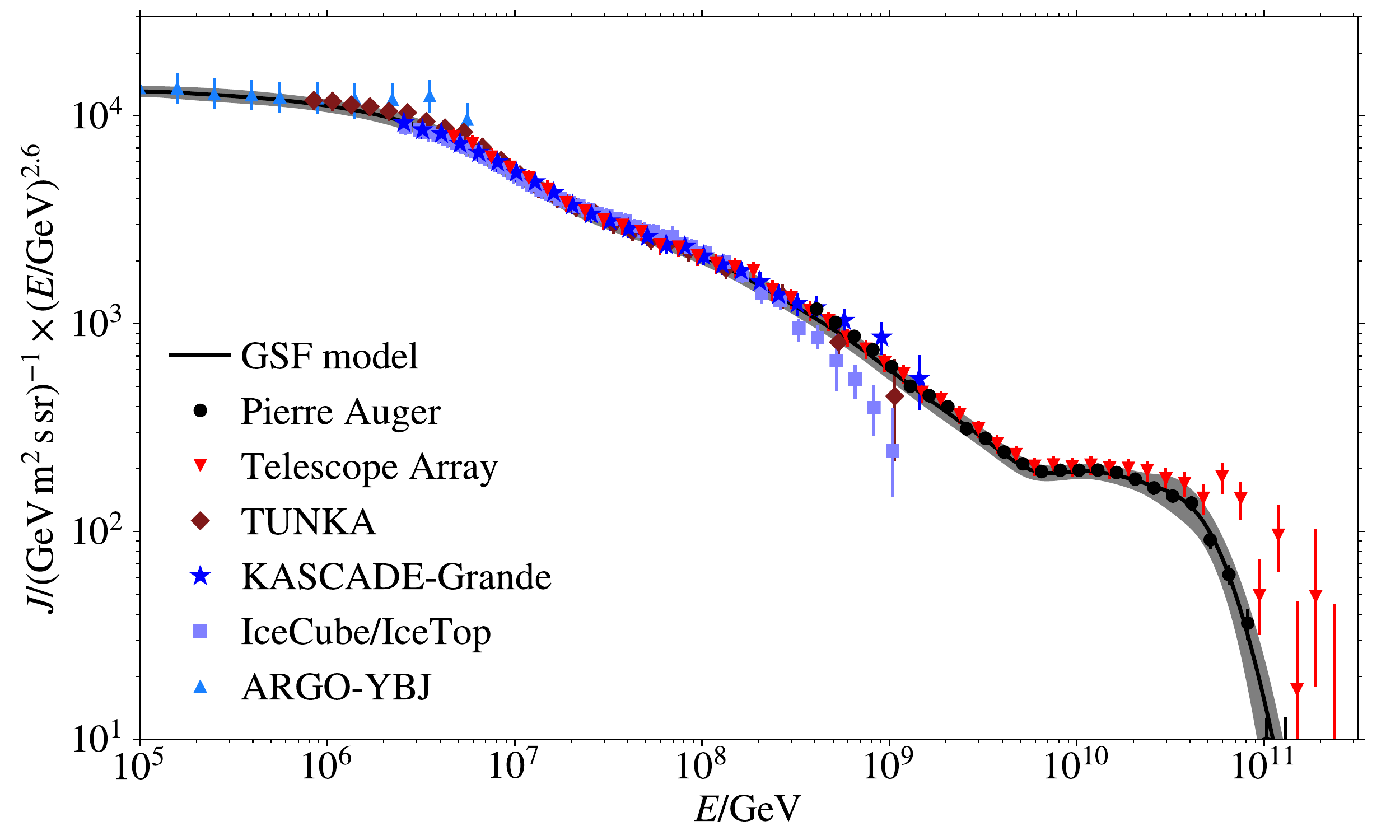}{0.99} 
\end{minipage}  
\caption{All-particle flux (black thick solid line), the flux
  contributed by protons (red line solid line), helium (yellow dashed
  line), the oxygen group (green dash-dotted line), and the iron group
  (blue dotted line). Bands around the model lines show a variation of
  one standard deviation. Data points show measurements which were
  energy-scale adjusted as described in~\cite{Dembinski:2017zsh}. Error bars represent
  combined statistical and systematic uncertainties. Data points of
  composition measurements from air-showers are not shown without
  error bars for clarity. In case of oxygen and iron, both the
  elemental flux and the group flux are shown; the smaller intensity
  without error band is the elemental intensity in each case. The
  intensity has been corrected for solar modulation as a function of
  kinetic energy of the nucleus using the force-field approximation~\cite{Gleeson:1968zza}
  and modulation parameters from the CRDB~\cite{Maurin:2013lwa,Usoskin}.  The
  distinction of kinetic energy and total energy makes no difference
  for high-energy cosmic rays, but the low energy results are easier
  to interpret when kinetic energy is used instead of total
  energy. Updated from Ref.~\cite{Dembinski:2017zsh}; courtesy of Hans Dembinski, Ralph Engel,
  Anatoli Fedynitch, Tom Gaisser, Felix Riehn, and Todor Stanev.}
\label{fig:2} 
\end{figure}

Combining TA~\cite{Ivanov:2015pqx} and
Auger~\cite{Aab:2014aea,Aab:2015bza} data with direct measurements
from the HEAO satellite~\cite{Engelmann:1990zz}, HEN~\cite{Juliusson},
PAMELA~\cite{Adriani:2011cu,Adriani:2013as},
AMS-02~\cite{Aguilar:2015ooa,Aguilar:2015ctt,Aguilar:2017hno},
CREAM-I, II and
III~\cite{Ahn:2009tb,Yoon:2011aa,Maestro:2009zz,Yoon:2017qjx},
ACE-CRIS~\cite{deNolfo:2006qj,Lave}, Spacelab2~\cite{Swordy,Mueller},
as well as air-shower measurements of high-energy CRs from
H.E.S.S.~\cite{Aharonian:2007zja}, ARGO-YBJ~\cite{Montini:2016fvq},
TUNKA~\cite{Prosin:2014dxa}, IceCube~\cite{Aartsen:2015awa}, and
KASCADE-Grande (KG)~\cite{Apel:2012rm,Schoo:2015oxd} a global spline
fit (GSF) to the CR spectrum was implemented in~\cite{Dembinski:2017zsh}.
The fitted GSF model is shown in Fig.~\ref{fig:2}; details are given
in~\ref{app1}. The global fit together with the adjusted energy-scales
reveals detailed structure in the all-particle flux. The most salient
features are: {\it (i)}~a steepening of the spectrum, which is known
in the cosmic vernacular as the ``knee'', occurring at $E_{\rm knee}
\sim 10^{6.6}~{\rm GeV}$; {\it (ii)}~a less prominent ``second knee'',
corresponding to a further softening, appears at $E_{\rm 2knee} \sim
10^{8.0}~{\rm GeV}$; {\it (iii)}~a pronounced hardening of the
spectrum, generating the so-called ``ankle'' feature, becomes evident
at $E_{\rm ankle} \sim 10^{9.7}~{\rm GeV}$. The results of the GSF are
consistent with those obtained by the Auger Collaboration through a 5
parameter fit of the UHECR spectrum,
\begin{equation}
J (E) = \left\{\begin{array}{ll}
J_0 \left(\dfrac{E}{E_{\rm ankle}} \right)^{-\gamma_1}  & \quad ;~E\leq E_{\rm
  ankle} \\
J_0 \left(\dfrac{E}{E_{\rm ankle}} \right)^{-\gamma_2} \left[ 1 +
  \left(\dfrac{E_{\rm ankle}}{E_{\rm supp}}\right)^{\Delta
      \gamma} \right] \left[ 1 + \left(\dfrac{E}{E_{\rm supp}}
      \right)^{\Delta \gamma} \right]^{-1} & \quad ;~E > E_{\rm
      ankle} \end{array} \right. \,, 
\end{equation}
with $E_{\rm ankle} = [5.08 \pm 0.06 ({\rm stat.}) \pm 0.8 ({\rm
  syst.})] \times 10^9~{\rm GeV}$, $E_{\rm supp} = [3.9 \pm 0.2 ({\rm
  stat.}) \pm 0.8 ({\rm syst.})] \times 10^{10}~{\rm GeV}$, $\gamma_1
= 3.293 \pm 0.002 ({\rm stat.}) \pm 0.05 ({\rm syst.})$, $\gamma_2 =
2.53 \pm 0.02 ({\rm stat.}) \pm 0.1 ({\rm syst.)}$, and $\Delta \gamma
= 2.5 \pm 0.1 ({\rm stat}) \pm 0.4 ({\rm
  syst.})$~\cite{Aab:2017njo}. Note that $E_{\rm ankle}$ is about a
factor 2.5 higher in energy than the observed break in $D_{10}$.  The
energy $E_{1/2}$ at which the integral spectrum drops by a factor of
two below what would be the expected with no steepening is $E_{1/2} =
[2.26 \pm 0.08 ({\rm stat.}) \pm 0.4 ({\rm syst.} )] \times
10^{10}~{\rm GeV}$.  TA and Auger observations show remarkable
agreement, except for a notable discrepancy at $E \gtrsim E_{\rm
  supp}$. On January 2018, the TA Collaboration reported evidence for
a declination dependence on the high energy end of the
spectrum~\cite{Abbasi:2018ygn}, suggesting that the shape of the
  spectrum for $E \gtrsim E_{\rm supp}$ could carry an imprint of the source
  density distribution along the line of sight (which would be
  different in different directions of the sky)~\cite{Abbasi:2017vru}. The GSF model,
however, is built on the assumption that the CR intensity is isotropic
and therefore pulled towards the Auger data points, which have much
smaller uncertainties.

  The variations of the spectral index $\gamma$ in the energy spectrum
  reflect various aspects of cosmic ray production, source
  distribution, and propagation.  The first and second knee seem to
  reflect the maximum energy of Galactic magnetic confinement, which
  grows linearly in the charge $Ze$ of the nucleus. A comparison of
  $E_{\rm knee}$ and $E_{\rm 2knee}$ shows that the spectral
  structures are separated in energy by a factor of $(E_{\rm 2knee} -
  E_{\rm knee})/E_{\rm knee} \sim 25$. Then, if $E_{\rm knee}$ is
  interpreted as the energy at which protons escape the Galaxy (or
  ``proton knee''), the higher energy break $E_{\rm 2knee}$ occurs
  where we would expect the ``iron knee'' according to the idea of
  rigidity dependent cutoffs in the spectra of individual
  nuclei~\cite{Peters}. This would imply that the second knee
  indicates where the highest-$Z$ CRs escape the Galaxy.  It is
  worthwhile to stress again, that this picture is in qualitative
  agreement with Auger measurements of $D_{10}$, and it is also
  supported by measurements of the Telescope Array low-energy
  extension (TALE)~\cite{Abbasi:2018xsn}.

  There are three different models that can explain the ankle in terms
  of source characteristics and propagation effects.  It has been
  advocated that the ankle feature could be well reproduced by a
  proton-dominated power-law spectrum, where the ankle is formed as a
  {\it dip} in the spectrum from the energy loss of protons via
  Bethe-Heitler pair
  production~\cite{Hillas:1967,Berezinsky:2002nc}. However, the
  apparent dominance of heavy nuclei in the vicinity of $10^{10}~{\rm
    GeV}$ is in contradiction with this interpretation of the ankle. A
  second model explains the ankle as the superposition of an
  extragalactic  population of UHECR nuclei with hard source spectra $\propto E^{-1.6}$ and
  maximum energy $E \sim  Z \times 10^{9.7}~{\rm GeV}$,  and an {\it ad hoc}
  light ($p$ + He) extragalactic component with steep emission spectra $\propto
  E^{-2.7}$~\cite{Aloisio:2013hya}.  A more {\it natural} explanation of the
  entire spectrum and nuclear composition emerges while accounting for
  the ``post-processing'' of UHECRs through photodisintegration in the
  environment surrounding the
  source~\cite{Unger:2015laa,Anchordoqui:2014pca,Farrar:2015ikt}.  In this model
  relativistic nuclei accelerated by a central engine to extremely
  high energies remain trapped in the turbulent magnetic field of the
  source environment. Their escape time decreases faster than the
  interaction time with increasing energy, so that only the highest
  energy nuclei can escape the source unscathed. In effect, the source
  environment acts as a {\it high-pass filter} on the spectrum of CRs. All
  nuclei below the energy filter interact, scattering off the
  far-infrared photons in the source environment.  These photonuclear
  interactions produce a steep spectrum of secondary nucleons, which
  is overtaken by the harder spectrum of the surviving nucleus
  fragments above about $10^{9.6}~{\rm GeV}$. These overlapping
  spectra could then carve an ankle-like feature into the source
  emission spectrum. The spectrum above the ankle exhibits a
  progressive transition to heavy nuclei, as the escape of
  non-interacting nuclei becomes efficient. Reproducing the data with
  such a model requires  hard spectra $\propto E^{-1}$ at the sources. In
  fact, as shown by the Auger Collaboration, simultaneously
  reproducing just the component of the spectrum above the ankle
  together with the observed nuclear composition also requires hard
  spectra at the sources~\cite{Aab:2016zth}.

\subsection{Anisotropy searches}
\label{DoAD}

There exists ``lore'' that convinces us that the highest energy CRs
observed should exhibit trajectories which are {\em relatively}
unperturbed by Galactic and intergalactic magnetic fields. Hence, it
is natural to wonder whether anisotropy begins to emerge at these high
energies.  Furthermore, if the observed flux suppression is the GZK
effect, there is necessarily some distance, ${\cal O} (100~{\rm
  Mpc)}$, beyond which cosmic rays with energies near $10^{11}$~GeV
will not be seen.  Since the matter density within about 100~Mpc is
not isotropic, this compounds the potential for anisotropy to emerge
in the UHECR sample. On the one hand, if the distribution of arrival
directions exhibits a large-scale anisotropy, this could indicate
whether or not certain classes of sources are associated with
large-scale structures (such as the Galactic plane, the Galactic halo,
or the super-Galactic plane).  On the other hand, if cosmic rays
cluster within a small  angular region or show directional
alignment with powerful  objects, one might be able to
associate them with isolated sources in the sky.

The directional exposure $\omega (\mathbf {\hat n})$ provides the effective
time-integrated collecting area for a flux from each direction of the
sky $\mathbf {\hat n} (\alpha, \delta)$, characterized by the right ascension
$\alpha$ and the declination $\delta$. For an experiment at latitude
$\lambda$, which is fully efficient for particles arriving with zenith
angle $< \theta_{\rm max}$ and
that experiences stable operation, $\omega (\mathbf{\hat n})$ actually becomes
independent of $\alpha$ when integrating the local-angle-detection
efficiency over full periods of sidereal revolution of the Earth. Full
efficiency means that the acceptance depends on $\theta$ only through
the reduction in the perpendicular area given by $\cos \theta$. The
$\omega$ dependence on $\delta$ relies on geometrical acceptance
terms and is given by
\begin{equation}
\label{eqn:omegageom}
\omega (\delta)= \frac{{\cal S} \ \Delta t}{2 \pi} \left(\cos \lambda \ \cos \delta \
  \sin \zeta +\zeta \ \sin \lambda \ \sin \delta \right),
\end{equation}
where ${\cal S}$ is the surface of the detector array, $\Delta t$ is
the time of data collection, and
\begin{equation}
\label{eqn:alpham}
\zeta = \begin{cases}
0 & ;\xi>1,\\
\pi & ;\xi<-1,\\
\arccos\xi & ;\text{otherwise,}
\end{cases}
\end{equation}
with $\xi
\equiv(\cos\theta_\text{max}-\sin\lambda\,\sin\delta)/(\cos\lambda\,\cos\delta)$~\cite{Sommers:2000us}.

Before proceeding, we pause to note that there are differences as to
how the primary energies are derived at Auger and TA, with systematic
uncertainties in the energy scale of the experiments amounting to
about $14\%$ and $21\%$ respectively, corresponding to about $70\%$
uncertainty in the flux above a fixed energy
threshold~\cite{diMatteo:2018vmr}. Therefore, when combining Auger and
TA data
to obtain full sky coverage it is necessary to cross-calibrate the
energy scales of the two datasets to avoid introducing a
spurious North/South asymmetry due to an energy scale mismatch. This
is accomplished by exploiting the wide declination band ($- 16^\circ
\lesssim \delta \lesssim +45^\circ$) where the two datasets
overlap. Regardless of the true arrival direction distribution, within
a region of the sky $\Delta \Omega$ fully contained in the field of
view (FoV) of both observatories, the sum over observed events $\sum_i
1/\omega(\mathbf{\hat n}_i )$ (with $\omega$ in ${\rm km \, yr}$ units) is
an unbiased estimator of $\int_{\Delta \Omega} J (> E_{\rm th},
\mathbf{\hat n}) \, d \mathbf{\hat n}$ and should be the same for both
experiments except for statistical fluctuations. Here, $J (> E_{\rm
  th}, \mathbf{\hat n})$ is the directional UHECR integrated intensity in ${\rm
  km^{-2} \, yr^{-1} \, sr^{-1}}$ units. This yardstick is generally
adopted to cross-calibrate the energy scales and to determine $E_{\rm
  th, Auger}$ and $E_{\rm th, TA}$ such that the Auger flux above
$E_{\rm th, Auger}$ matches the TA flux above $E_{\rm th,
  TA}$~\cite{diMatteo:2018vmr}.\footnote{Actually, the region of the
  sky which is mostly used spans the declination band $-12^\circ
  \lesssim \delta \lesssim +42^\circ$. This is because including
  directions too close to the edge of the FoV of one of the
  observatories would result in larger statistical fluctuations due to
  very large values of $1/\omega(\mathbf{\hat n}_i)$ near the edge.}

\begin{figure}[tbp] 
    \postscript{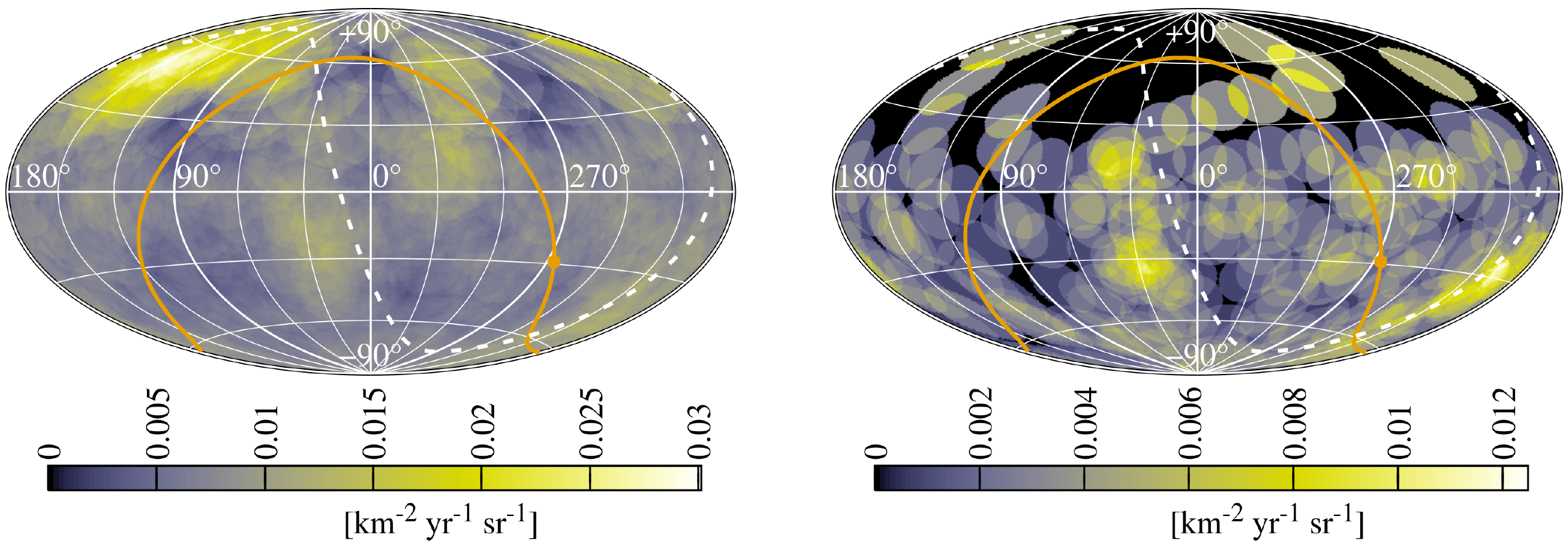}{0.99} 
\caption{Estimated intensity of UHECRs above the indicated energy
  thresholds, averaged over circular moving windows, in equatorial
  coordinates. The galactic plane (orange solid), galactic center
  (orange dot) and super-galactic plane (white dashed) are also
  shown. From Ref.~\cite{diMatteo:2018vmr}.
}  
\label{fig:3} 
\end{figure}

The observed arrival direction distribution is obtained by convoluting
the flux per steradian with the detector exposure, giving
\begin{equation}
\frac{dN}{d\Omega} (\mathbf{ \hat n}) = J (>E_{\rm th}, \mathbf{ \hat
  n}) \ \omega (\mathbf{ \hat n}) \, .
\end{equation}
The  distribution of arrival directions can be exposed
through the average directional (integral) intensity smoothed out at an angular scale $\Theta$,
\begin{equation}
\langle J (> E_{\rm th}, \mathbf{\hat n}) \rangle_\Theta = \frac{1}{\int_\Theta d \mathbf{\hat n}}
\ \int_\Theta d \mathbf{\hat n}' \ \ f(\mathbf{\hat n}, \mathbf{\hat n'})  \ \frac{1}{\omega(\mathbf{\hat
  n}')} \ \
  \frac{dN}{d\mathbf{ \hat n}'} \,,
\end{equation}
where $f(\mathbf{\hat n}, \mathbf{\hat n}')$ is the top-hat filter function of radius
$\Theta$ and $E_{\rm th}$ the threshold energy. For an illustration of
this technique,
we consider 602 events collected by Auger from January 2004 to March
2014, with a total exposure of approximately $6.6 \times 10^4~{\rm
  km^2 \, sr \, yr}$, $\theta \leq 80^\circ$, and $E \geq   40~{\rm
  EeV}$, as well as 83 events collected by TA from May 2008 to May
2015 with a total exposure of about $8.7 \times 10^3~{\rm km^2 \, sr
  \, yr}$, $\theta \leq 55^\circ$, and $E \geq 57~{\rm EeV}$. The left panel of Fig.~\ref{fig:3} shows the directional intensity of UHECRs
recorded by  TA and Auger, with energy above $E_{\rm th,TA} = 57~{\rm
  EeV}$ and $E_{\rm th,Auger} = 42~{\rm EeV}$ averaged out at $\Theta
= 20^\circ$. The fiducial values for the energy threshold and the
smoothing angle were extracted from searches of over-densities
anywhere in the sky by the TA Collaboration~\cite{Abbasi:2014lda}.
Actually, the selected energy threshold of 57~EeV follows from a prior
analysis by the Auger Collaboration~\cite{Abraham:2007bb}, which had
initially led to capture an anisotropy with 99\% confidence level that
was not confirmed by subsequent data~\cite{PierreAuger:2014yba}. A hot
spot is visible in the direction $(\alpha, \delta) =
(147^\circ,43^\circ)$~\cite{Abbasi:2014lda}, which is found to be the
region with the largest intensity. The right panel of Fig.~\ref{fig:3} displays the
intensity sky map for higher energy thresholds $E_{\rm th,TA} = 89~{\rm
  EeV}$ and $E_{\rm th,Auger} = 54~{\rm EeV}$ averaged out at $\Theta
= 12^\circ$. This energy threshold and smoothing angle correspond to
the search for local over-densities in Auger
data~\cite{PierreAuger:2014yba}. A clustering of events is visible in
the direction $(\alpha, \delta) = (198^\circ,
-25^\circ)$~\cite{PierreAuger:2014yba}, which is found to be the
region with the largest intensity. 

The existence of a signal in a region {\it on-source} is judged by the
event count number $N_{\rm on}$ originating from that region. The
counts in it are due to the potential source and the background. To
estimate the statistical significance of a measured signal one can
compare the number of events $N_{\rm on}$ that are detected in the
{\it on-source} region to the number of background events $N_{\rm bg}
= \eta N_{\rm off}$ that are expected in the {\it on-source}
region. The number of {\it off-source} events, $N_{\rm off}$, is
estimated through numerical simulations of virtual events sampled
randomly and uniformly in $(\alpha, \sin \delta)$, with $0\leq \alpha
\leq 2\pi $ and $-\pi/2 \leq \delta \leq \pi/2$. The normalization
factor is given by $\eta = N/N_{\rm sim}$, where $N$
is the total number of observed events in the data-sample and $N_{\rm
  sim}$ is the number of virtual events. A widely accepted method for
statistically inferring the existence of a source is the {\it
  hypothesis test} based on maximum likelihood principles. This test statistic (TS)
requires the use of two hypotheses and a statistical description of
each of them. The usual approach is to confront the null hypothesis
(events coming only from background; no source present), with its
negation (events coming from background and a source). This method
provides the {\it statistical Li-Ma significance} for the
rejection of the null (background only) hypothesis, 
\begin{equation}
S_1^{\rm LM} = \sqrt{2} \ \left\{ N_{\rm on} \ \ln \left[ \frac{(1+ \eta)
      N_{\rm on}}{\eta (N_{\rm on} + N_{\rm off})} \right] + N_{\rm
  off} \ \ln \left[ \frac{(1 + \eta) N_{\rm off} }{ N_{\rm on} +
    N_{\rm  off}} \right] \right\}^{1/2} \,,
\label{Li-Ma}
\end{equation}
which is valid only in the limit of large
statistics~\cite{Li:1983fv};  see~\ref{app2} for details.

\begin{figure}[tbp] 
    \postscript{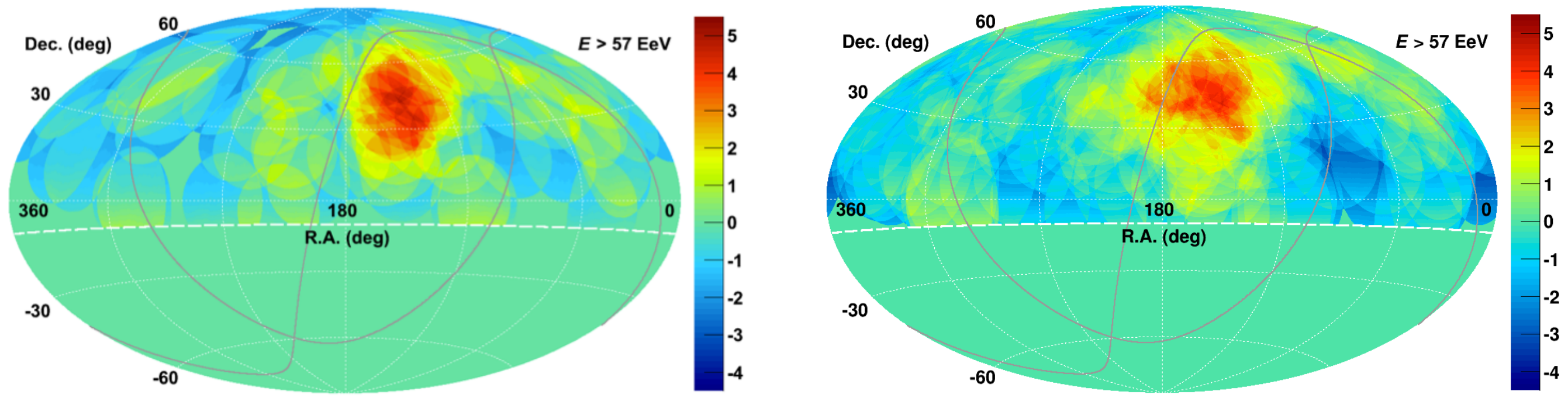}{0.99} 
\caption{The Li-Ma significance  skymap representing
  direction-dependent excesses and deficit with respect to the
  isotropic background in equatorial coordinates. The left panel
  corresponds to 5~yr of TA data with $E > 57~{\rm EeV}$ and with the original
  $20^\circ$ oversampling (from Ref.~\cite{Abbasi:2014lda}). The right panel corresponds to  9 yr of TA data 
with $25^\circ$ oversampling (from Ref.~\cite{Matthews}).}  
\label{fig:4} 
\end{figure}

The largest evidence for intermediate-scale ($\Theta = 20^\circ$)
clustering above the statistical expectation was actually unmasked
with $N_{\rm ob} = 72$ events above $E_{\rm th, TA} = 57~{\rm EeV}$
recorded between 2008 May 11 and 2013 May 4~\cite{Abbasi:2014lda}. The
Li-Ma significance map of this data-sample is shown in the left panel
of Fig.~\ref{fig:4}.  The maximum excess, characterized by $N_{\rm on}
= 19$, $N_{\rm sim} = 10^5$, $N_{\rm bg} = 4.49$, is centered at
$(\alpha, \delta) = (147^\circ, 43^\circ$) and has a Li-Ma
significance $S^{\rm LM}_{1, \rm max} = 5.1\sigma$.  Note that this is not the
hot-spot chance probability as the Li-Ma significance does not take
into account random clustering, i.e. the probability of such a hotspot
appearing by chance anywhere in an isotropic sky. To estimate the
correction (a.k.a. penalty) factor, the TA Collaboration generated
$10^6$ Monte Carlo data sets each having 72 spatially random events
within the experiment field of view (i.e., reproducing the statistics
of the experimental data-sample), assuming a uniform distribution over
the TA surface detector exposure. The maximum $S_{1, \rm max}^{\rm LM}$
significances were calculated for each Monte Carlo dataset following
the same prescription adopted for the observed data.  For $10^6$
simulations, in 365 instances $S^{\rm LM}_{1,\rm max} > 5.1 \sigma$, yielding
a chance probability to observe the hotspot in an isotropic cosmic-ray
sky of $p_{\rm TA} \simeq 365/10^6 \simeq 3.7 \times 10^{-4}$, equivalent to a one-sided
Gaussian significance of $3.4 \sigma$.

On 2017, the significance
and shape of the hot-spot were updated using $N_{\rm ob} = 143$ events above
$E_{\rm th, TA} = 57~{\rm EeV}$ collected through Spring
2017~\cite{Matthews}. Initially the excess search was performed following the original
$\Theta = 20^\circ$ oversampling, but it was apparent from the sky map
that the hot-spot became bigger in radius. The TA Collaboration then
made a scan over $\Theta$, limited to $15^\circ$, $20^\circ$,
$25^\circ$, $30^\circ$, and $35^\circ$, to determine the best fit to
the data. The maximum significance shows up at $\Theta = 25^\circ$,
with $N_{\rm on} = 34$ and $N_{\rm bg} = 13.5$.  The Li-Ma
significance map of this new data-sample is shown in the right panel
of Fig.~\ref{fig:4}. The hot-spot is now re-centered at $(\alpha,
\delta) = (148^\circ, 45^\circ)$, about $1.5^\circ$ away from the
center position found in the original search. The local (pre-trial)
Li-Ma significance is $5.1\sigma$ and the global (post-trial)
significance is $3.0\sigma$. A Kolmagorov-Smirnov test shows that the
rate of arrival of events to the hot-spot is consistent with the
fluctuations expected from a Poisson distribution with a mean of 2.8
events per year. However, the rate of event arrivals is also
inconsistent with chance excess from an isotropic distribution, a
Poisson average of 0.9 events per year, at about $2.6\sigma$. 

A new interesting development in this direction is that the TA
Collaboration has also reported evidence of an energy dependent
intermediate scale anisotropy appearing as a deficit of low energy
$(10^{10.20} \leq E/{\rm GeV} \leq 10^{10.75}$) events followed by an
excess of high energy ($E \geq 10^{10.75}~{\rm GeV}$) events in the
same region of the sky, with a maximum local pre-trial significance of
$6.2\sigma$~\cite{Abbasi:2018qlh}.  The anisotropy is contained inside a spherical bin of
radius $28^\circ$ centered at $(\alpha, \delta) = (139^\circ,45^\circ)$, which is
roughly $7^\circ$ away from the center of the hot-spot.  The
post-trial probability of this coincidence appearing by chance
anywhere on an isotropic sky is found to be $9 \times 10^{-5}$,
equivalent to a $3.7\sigma$ global significance. This new feature is
very suggestive of energy dependent magnetic deflection of UHECRs.

Spherical harmonic moments are well-suited for capturing anisotropies
at any angular scale in the intensity of
UHECRs~\cite{Sommers:2000us}. However, an unambiguous measurement of
the full set of spherical harmonic coefficients requires a full-sky
coverage~\cite{Anchordoqui:2003bx,Denton:2014hfa,Aab:2014ila,Denton:2015bga}. Because
of the partial-sky coverage of ground-based experiments, extraction of
the multipolar moments turns out to be nearly impossible without
explicit assumptions on the shape of the underlying angular
distribution.  Under the circumstances, the search for anisotropies in
the distribution of arrival directions is customarily performed
through harmonic analyses in right ascension of the event counting
rate, within the declination band defined by the field of view of the
experiment~\cite{Linsley:1975kp}. Note that ground-based facilities
which experience stable operation over a period of a year or more can
have a relatively uniform exposure in $\alpha$. The relevant
declination-integrated directional intensity, which embraces genuine
anisotropies $\delta J(\alpha)$, can be written as
\begin{equation}
J(>E_{\rm th}, \alpha) = \frac{{\cal F}_0}{2 \pi} [ 1 + \delta J(\alpha)] \, ,
\label{jofalpha}
\end{equation}
where ${\cal F}_0/(2\pi)$ is the isotropic component. Of course, because of changes
in the experimental conditions small variations of the exposure in
sidereal time would be expected. These variations translate into small
variations of the directional exposure in right ascension $\delta
\omega (\alpha)$, and therefore
\begin{equation}
\omega (\alpha) = \int d \delta \  \cos \delta \ \omega (\alpha,
\delta) = \omega_0 [1 + \delta \omega (\alpha)] \, ,
\end{equation}
where $\omega_0$ indicates the uniform exposure. The 1-dimensional
distribution of arrival directions which is driven by the right
ascension of events, $dN/d\alpha$, also embraces authentic
anisotropies $\delta J(\alpha)$, but coupled to the directional
exposure variations, i.e.,
\begin{eqnarray}
\frac{dN}{d \alpha} &  = & \frac{{\cal F}_0 \ \omega_0}{2 \pi} [1 + \delta \omega
(\alpha)] [1 + \delta J(\alpha)] \nonumber \\
& = & \frac{{\cal F}_0  \ \omega_0}{2 \pi} [1 + \delta \omega (\alpha)] \left[1
  + \sum_{m=1}^\infty a_m  \, \cos (m \alpha) + \sum_{m=1}^\infty b_m \,
  \sin(m \alpha)  \right] \, ,
\label{Fourier}
\end{eqnarray}
where $\delta J (\alpha)$ has been expanded in a Fourier series with
coefficients $a_m$ and $b_m$. Note that the Fourier coefficients
defined by the harmonic
decomposition in (\ref{Fourier}) give a direct measure of the anisotropy in right
ascension relative to the monopole, isotropic flux. Using the basic
orthogonality relation of trigonometric functions it follows that
\begin{subequations}
\label{orthogonality}
\begin{eqnarray}
a_m & = & \frac{1}{\pi a_0} \int \frac{d\alpha}{1 + \delta \omega
  (\alpha)} \ \frac{dN}{d\alpha} \cos (m \, \alpha) \,, \\
b_m & = & \frac{1}{\pi a_0} \int \frac{d\alpha}{1 + \delta \omega
  (\alpha)} \ \frac{dN}{d\alpha} \sin (m \, \alpha)  \, ,
\end{eqnarray}
with
\begin{equation}
a_0  =  \frac{1}{2 \pi} \int \frac{d \alpha}{1 + \delta \omega(\alpha)}
\frac{dN}{d \alpha} \, .
\end{equation}
\end{subequations}
The event counting rate $dN /d\alpha$ is the intrinsic
estimator of the distribution of arrival directions. The
coefficients of the Fourier expansion can be estimated through a
representation of the counting rate by a sum of Dirac functions over
the circle, i.e., $d N/d\alpha =\delta (\alpha, \alpha_i)$. Namely,  the estimators $\hat a_m$ and
$\hat b_m$ can be calculated  using discrete sums
running over the $N$ observed events: 
\begin{subequations}
\label{Rayleigh-estimators}
\begin{equation}
\hat a_m = \frac{2}{{\cal N}_\alpha} \sum_{i = 1}^{N} \frac{\cos (m \,
  \alpha_i)}{1 + \delta \omega(\alpha_i)} \quad {\rm and} \quad
\hat b_m = \frac{2}{{\cal N}_\alpha} \sum_{i = 1}^{N} \frac{\sin (m \,
  \alpha_i) }{1 + \delta \omega(\alpha_i)} \, ,
\end{equation}
where the coefficient $a_0$ has been estimated as $\hat a_0 = {\cal
  N}_\alpha/ (2 \pi)$, and 
\begin{equation}
{\cal N}_\alpha = \sum_{i=1}^N \frac{1}{ 1 + \delta
\omega (\alpha_i)}
\end{equation}
\end{subequations}
indicates the numbers of events that would
have been detected should the directional exposure been completely
uniform in $\alpha$. 

The results of the harmonic analysis are generally given using more
intuitive geometrical parameters, such as the amplitude of the harmonic
modulation
\begin{equation}
\hat r_m = \sqrt{\hat a_m^2 + \hat b_m^2} \,,
\end{equation}
and the phase associated to the right ascension of the maximum flux
\begin{equation}
\hat \varphi_m = \arctan (\hat b_m/\hat a_m) \,,  
\end{equation}
defined modulo $2\pi/m$. The probability that an amplitude equal to or
larger than $\hat r_m$ arises from an isotropic distribution can be
safely approximated by the cumulative distribution function of the
Rayleigh distribution 
\begin{equation}
p (\geq \hat r_m) = \exp \left( - \frac{{\cal N}_\alpha \hat r_m^2}{4} \right)\,;
\end{equation}
see~\ref{app3} for details. To give a specific example, a vector of
length ${\cal N}_\alpha \hat r_m^2/4 \geq~15$ would be required to
claim an observation whose probability of arising from random
fluctuation was $3 \times 10^{-7}$, equivalent to a two-sided Gaussian
significance $5\sigma$. 

Results of numerical simulations, which take into account all three
spatial degrees of freedom and the cosmological time-evolution of the
universe (assuming the spatial distribution of sources follows the
local mass distribution) suggest that the arrival directions of UHECRs
are expected to have a pronounced dipolar anisotropy and rather weak
higher-order contributions~\cite{Wittkowski:2017nfd}. With this in mind, it is
reasonable to assume that the UHECR flux per steradian is of the form: 
\begin{equation}
J(>E_{\rm th},\alpha,\delta) = \frac{{\cal F}_0}{4 \pi} (1+ \Bbbk \ {\bm{\hat \Bbbk}} \cdot
\mathbf {\hat n}) \,,
\end{equation}
where $\mathbf {\hat n}$ and $\bm{\hat {\Bbbk}}$ respectively denote
the unit vector in the direction of an arrival direction and in the
direction of the dipole.  In other words, the (differential) intensity
in the direction $\mathbf {\hat n}$ consists of an isotropic part,
$J_0= {\cal F}_0/(4\pi)$, modulated by a dipolar component in $\cos
(\bm{\hat{\Bbbk}, \mathbf {\hat n}})$, and higher order terms
are negligible. Here, $\Bbbk \in [0,1]$ is the amplitude of the
dipole, relative to the monopole. This means one has to derive from
the data three parameters: one amplitude, $\Bbbk$, and two angles,
$\bm{\hat \Bbbk}$).

The first harmonic amplitude of the UHECR right ascension distribution
can be directly related to $\Bbbk$. Namely, setting $m =1$, we can
rewrite $\hat a_1$, $\hat b_1$ and $\mathcal{N}_\alpha$ as:
\begin{subequations}
\label{corolo}
\begin{eqnarray}
\hat a_1&=& \frac{2}{\mathcal{N}_\alpha} \int_{\delta_{min}}^{\delta_{max}} d\delta \int_0^{2\pi}
d\alpha \cos \delta \ J(>E_{\rm th},\alpha,\delta) \ \omega(\delta) \cos \alpha,  \\  
\hat b_1&=& \frac{2}{\mathcal{N}_\alpha} \int_{\delta_{min}}^{\delta_{max}} d\delta \int_0^{2\pi}
d\alpha \cos \delta \ J(>E_{\rm th},\alpha,\delta) \ \omega(\delta) \sin \alpha, \\ 
\mathcal{N}_\alpha&=& \int_{\delta_{min}}^{\delta_{max}} d\delta \int_0^{2\pi}
d\alpha \cos \delta \ J(>E_{\rm th},\alpha,\delta) \ \omega(\delta) \, ,
\end{eqnarray}
\end{subequations}
because the effects of the small modulation in $\alpha$ are already
accounted for in the  Rayleigh analysis~\cite{Aublin:2005nv}.  Next, we write the angular dependence in 
$J(>E_{\rm th}, \alpha,\delta)$ as $ \bm{\hat \Bbbk} \cdot \mathbf {\hat n} = \cos \delta
\cos \delta_{\Bbbk} 
\cos (\alpha-\alpha_{\Bbbk}) + \sin \delta \sin
\delta_{\Bbbk}$, where $\alpha_{\Bbbk}$ and
$\delta_{\Bbbk}$
are the right ascension and declination of the apparent origin of the dipole, respectively. Performing the $\alpha$
integration  in (\ref{corolo}) it follows that
\begin{subequations}
\begin{equation}
\label{eqn:amplitudes}
\hat r_1=\left| \frac{A \Bbbk_\perp}{1+B \Bbbk_\parallel} \right|
\end{equation}
where
\begin{equation}
A =\frac{\int d\delta\,\omega(\delta) \cos^2 \delta}
{\int d\delta\,\omega(\delta) \cos \delta} \,, \hspace{1cm} B
=\frac{\int d\delta\,\omega(\delta) \cos \delta \sin \delta} {\int
  d\delta\,\omega(\delta) \cos \delta} \,,
\end{equation}
\end{subequations}
 $\Bbbk_\parallel=\Bbbk \sin{\delta_{\Bbbk}}$ is the component
of the dipole along the Earth rotation axis, and $\Bbbk_\perp=\Bbbk\cos \delta_{\Bbbk}$ is the component in the
equatorial plane. The coefficients $A$ and $B$
can be estimated from the data as the mean values of the cosine and
the sine of the event declinations: $\langle \sin \delta \rangle$ and
$\langle \cos \delta \rangle$. For a dipole amplitude $\Bbbk$, the measured amplitude of the first harmonic in right ascension
$\hat r_1$ thus depends on the region of the sky observed, which is
essentially a function of the latitude of the observatory $\lambda$,
and the range of zenith angles considered.  In the case of a small $B
\Bbbk_\parallel$ factor, the dipole component in the equatorial
plane can be obtained to linear order and is given by
$\Bbbk_\perp\simeq \hat r_1/\langle\cos{\delta}\rangle$, whereas  the phase
 corresponds to the right ascension of the dipole direction
$\hat \varphi_1 = \alpha_{\Bbbk}$~\cite{Abreu:2011ve}.

Note that the Rayleigh analysis in right ascension is sensitive only
to $\Bbbk_\perp$. A dipole component in the direction of the rotation
axis of Earth induces no modulation of the flux in right ascension,
but does so in the azimuthal distribution of arrival directions. A
non-vanishing value of $\Bbbk_\parallel$ leads to a sinusoidal
modulation in azimuth with a maximum toward the northern or the
southern direction.  Thus, to reconstruct both dipole components, it
is plausible to combine the first-harmonic analysis in $\alpha$ with a similar one
in the azimuthal angle $\phi$ (say, measured counterclockwise from the
east). The relevant estimators $\hat c_1$ and $\hat d_1$ are given by
an expression analogous to (\ref{Rayleigh-estimators}), but in terms
of the azimuth of the arrival direction of the shower rather than in
terms of the right ascension,
\begin{equation}
\hat c_1 = \frac{2}{{\cal N}_\phi} \sum_{i = 1}^{N} \frac{\cos 
  \phi_i}{1 + \delta \omega(\alpha^0_i,\phi_i)} \quad {\rm and} \quad
\hat d_1 = \frac{2}{{\cal N}_\phi} \sum_{i = 1}^{N} \frac{\sin 
  \phi_i }{1 + \delta \omega(\alpha^0_i, \phi_i)} \,,
\end{equation}
where $\alpha^0_i$ is the local sidereal time (for practical reasons
chosen so that it is always equal to the right ascension of the zenith
at the center of the experiment), and $\mathcal{N}_\phi = \sum_{i=1}^N
[1 + \delta \omega(\alpha^0_i, \phi_i)]^{-1}$.  Writing $\bm{\hat{\Bbbk}} \cdot \mathbf{\hat n}$
as a function of the local coordinates ($\theta$,$\phi$,$\alpha^0$),
\begin{eqnarray}
\bm{\hat{\Bbbk}} \cdot \mathbf{\hat n} & = &\sin\delta_{\Bbbk}(\cos\theta
\sin\lambda +\sin\theta \cos \lambda
\sin\phi)+\cos\delta_{\Bbbk}\cos\alpha_{\Bbbk}(-\sin\theta \cos\phi
\sin\alpha^0  +  \cos\theta  \nonumber \\
& \times &\cos\lambda  \cos\alpha^0-\sin\theta
\sin\lambda \sin\phi \cos\alpha^0)+\cos\delta_{\Bbbk}
\sin\alpha_{\Bbbk}(\sin\theta \cos\phi \cos\alpha_0+\cos\theta \nonumber
\\ 
& \times & \cos\lambda   \sin\alpha^0-\sin\theta \sin\lambda
\sin\phi \sin\alpha^0) \,,
\end{eqnarray}
the angular dependence of the intensity in the first harmonic amplitudes in
$\phi$ can be expressed as
\begin{eqnarray}
\hat c_1 &=&\frac{2}{\mathscr{N}_\phi}\int_0^{2\pi}
d\alpha^0\int_0^{2\pi} d \phi
\int_{\theta_{\rm{min}}}^{\theta_{\rm{max}}} d\theta \sin\theta
\cos\phi\, J(>E_{\rm th},\theta,\phi,\alpha^0)=0,\label{a1phid}\\
\hat d_1 &=&\frac{2}{\mathscr{N}_\phi}\int_0^{2\pi}
d\alpha^0\int_0^{2\pi} d \phi
\int_{\theta_{\rm{min}}}^{\theta_{\rm{max}}}  d\theta \sin\theta
\sin\phi\, J(>E_{\rm th},
\theta,\phi,\alpha_0)=\frac{\pi}{\mathscr{N}_\phi} {\cal F}_0
\Bbbk_\parallel \, \cos\lambda {\overline {\sin\theta}},\label{b1phid}\\
{\mathscr{N}_\phi}&=&\int_0^{2\pi} d\alpha^0\int_0^{2\pi} d\phi
\int_{\theta_{\rm{min}}}^{\theta_{\rm{max}}} d\theta \sin\theta\,
J(>E_{\rm th},\theta,\phi,\alpha_0)=\pi {\cal F}_0 ({\overline 1}+ \Bbbk_\parallel
\sin\lambda \, {\overline {\cos\theta}}),
\end{eqnarray}
where ${\overline
  {f(\theta)}}\equiv\int_{\theta_{\rm{min}}}^{\theta_{\rm{max}}}
d\theta f(\theta) \sin\theta $~\cite{ThePierreAuger:2014nja}. The
coefficient $\hat c_1$ vanishes as anticipated, while $\hat d_1$ is
related to $\Bbbk_\parallel$ by
\begin{equation}
\hat d_1 =\frac{\Bbbk_\parallel \cos\lambda \langle
  \sin\theta\rangle}{1+\Bbbk_\parallel \sin\lambda \langle \cos\theta\rangle},
\end{equation}
where ${\overline {\sin\theta}}/{\overline 1}$ has been estimated as
the mean value of $\sin \theta$ of the events themselves and likewise
${\overline {\cos\theta}}/{\overline 1} \simeq \langle
\cos\theta\rangle$. Note that for $\Bbbk_\parallel \sin\lambda
\langle \cos\theta\rangle \ll 1$, the dipole component along the
Earth's rotation axis can be obtained to linear order as 
\begin{equation}
\Bbbk_\parallel = \frac{\hat d _1}{\cos\lambda \langle
  \sin\theta\rangle} \, ,
\end{equation}
with $\tan \delta_{\Bbbk} = \Bbbk_\parallel/\Bbbk_\perp$. It is
noteworthy that a pure dipole distribution is not possible, because the
CR intensity cannot be negative in half of the sky.  A {\it
  pure dipole deviation from isotropy} means a superposition of
monopole and dipole, with the intensity everywhere $\geq 0$, as
expressed by the Fourier expansion (\ref{Fourier}).

\begin{figure}[tbp] 
\begin{minipage}[t]{0.49\textwidth}
    \postscript{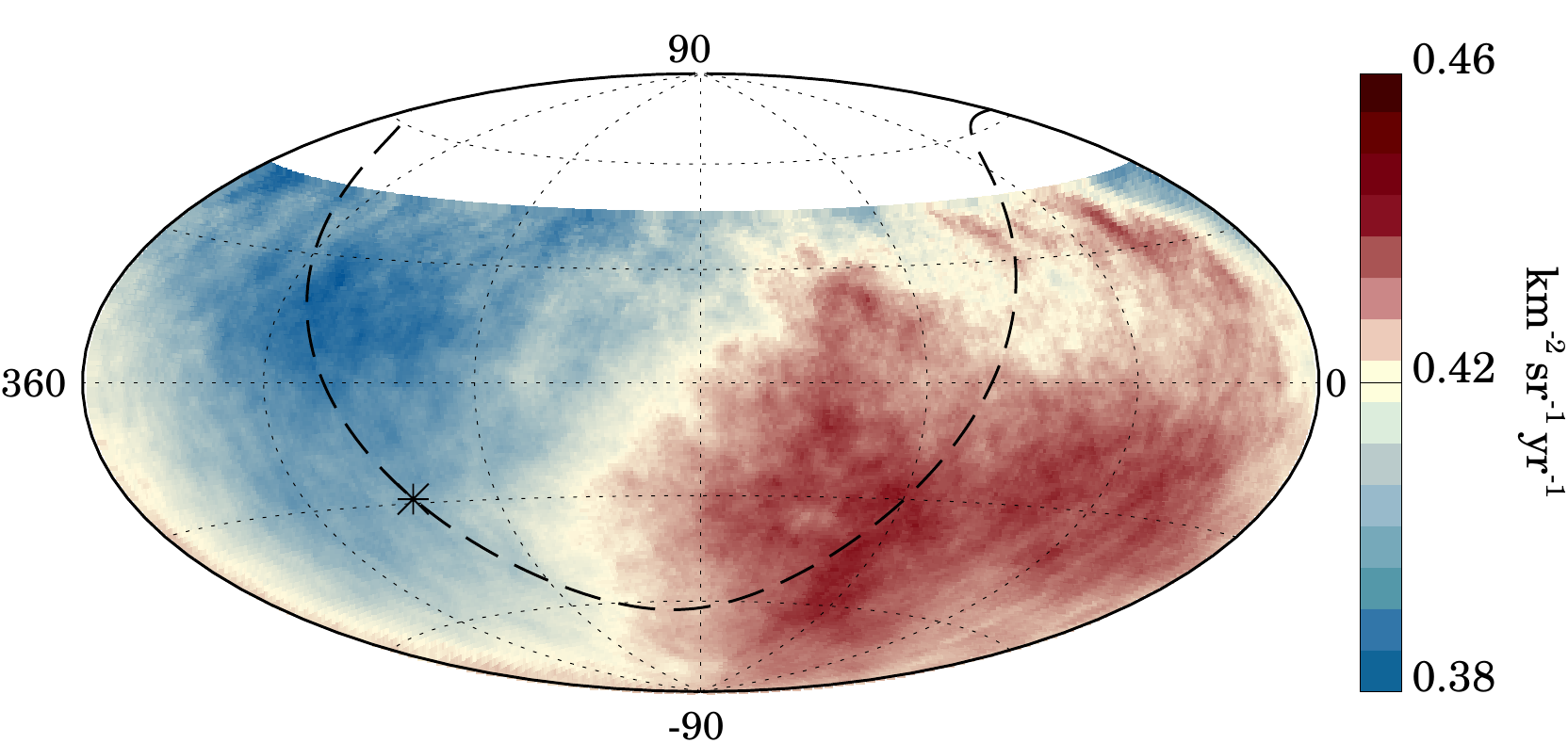}{0.99} 
\end{minipage} 
\hfill \begin{minipage}[t]{0.49\textwidth}
  \postscript{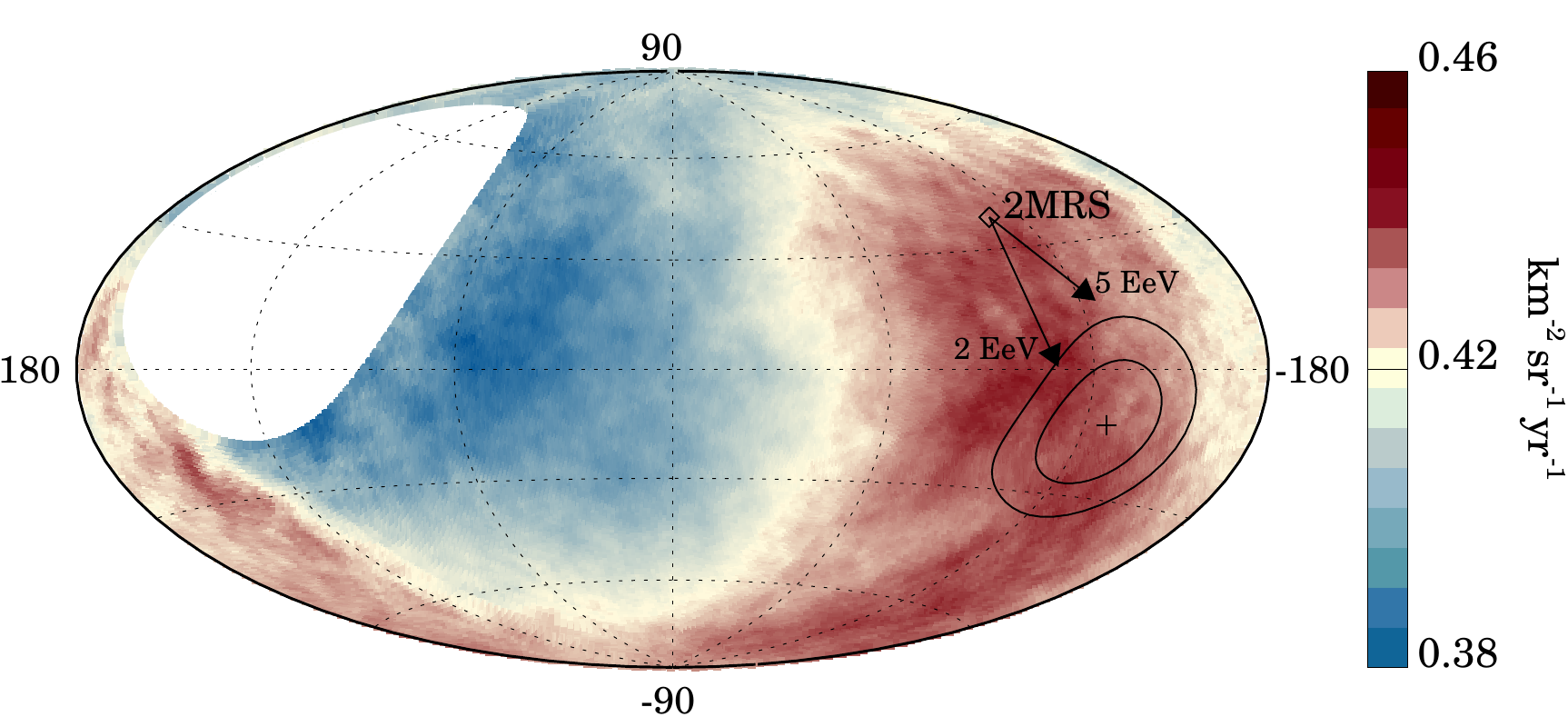}{0.99} 
\end{minipage} 
\caption{Sky map in equatorial coordinates using a Hammer projection
  (left) and Galactic coordinates (right) showing the CR intensity
  above 8~EeV smoothed with a $45^\circ$ top-hat function. On the left
  panel, the Galactic center is marked with an asterisk and the
  Galactic plane is shown by a dashed line. On the right panel, the
  Galactic center is at the origin. The cross indicates the measured
  dipole direction; the contours denote the 68\% and 95\%
  confidence-level regions. The dipole in the 2MRS galaxy distribution
  is indicated. Arrows show the deflections expected for a particular
  model of the Galactic magnetic
  field~\cite{Jansson:2012pc,Jansson:2012rt,Unger:2017kfh} on particles with $E/Z =
  5~{\rm EeV}$ and 2~EeV.  From Ref.~\cite{Aab:2017tyv}}
\label{fig:5} 
\end{figure}

On November 2017, the Pierre Auger Collaboration reported a
significant ($>5 \sigma$) large scale hemispherical asymmetry in the
arrival directions of UHECR recorded between 1 January 2004 and 31
August 2016, from a total exposure of about $7.68 \times 10^4~{\rm
  km^2} \ {\rm sr} \ {\rm yr}$~\cite{Aab:2017tyv}.  The mean cosine of
the declinations of the events in the sample is found to be $\langle
\cos \delta \rangle = 0.78$, whereas the mean sine of the zenith
angles of the events is $\langle \sin \theta \rangle = 0.65$. The
Rayleigh analysis of the first harmonic in right ascension of 32,187
events with $E \geq 8~{\rm EeV}$ yields $\hat r_1 =
0.047^{+0.008}_{-0.007}$ and $\hat \varphi_1 = (100 \pm 10)^\circ$,
with a random chance probability $p (\geq \hat r_1) = 2.6 \times
10^{-8}$. However, for the 81,701 events recorded with $4 < E/{\rm
  EeV} < 8$, the amplitude of the first harmonic is significantly
smaller: $\hat r_1 = 0.005^{+0.006}_{-0.002}$. In this energy bin
$\hat \varphi_1 = (80 \pm 60)^\circ$ and the probability by chance is
$p (\geq \hat r_1) = 0.6$. The data analysis in azimuthal angle leads
to $\hat d_1 = -0.013 \pm 0.005$ in the $4 <E/{\rm EeV} < 8$ energy
bin and $\hat d_1 = - 0.014 \pm 0.008$ in the $E \geq 8~{\rm EeV}$
bin. The probabilities that larger or equal absolute values for $\hat
d_1$ arise from an isotropic distribution are 0.8\% and 8\%,
respectively. The $E \geq 8~{\rm EeV}$ data are well-represented by a
dipole with amplitude $\Bbbk = 0.065^{+ 0.013}_{-0.009}$ pointing in
the direction $(\alpha_{\Bbbk}, \delta_{\Bbbk}) =
({100^{+10}_{-10}}^\circ, {-24^{+12}_{-13}}^\circ)$.  In a second
study, the Auger Collaboration further scrutinized the highest-energy
bin by splitting it into three energy ranges~\cite{Aab:2018mmi}.  They
found that the amplitude of the dipole increases with energy above
$4~{\rm EeV}$. The growth can be fitted to a power law with
index $= 0.79 \pm 0.19$. The Auger Collaboration also estimated the
quadrupolar components of the anisotropy, which are shown to be not
statistically significant. The dipolar pattern is clearly seen in the
flux map in Fig.~\ref{fig:5}. In Galactic coordinates, the direction
of this dipole is $(l_\Bbbk , b_\Bbbk) = (233^\circ, -13^\circ )$.
This direction is about $125^\circ$ from the Galactic center,
suggesting that the UHECRs creating the anisotropy have an
extragalactic origin.

Because of Liouville's theorem, the distribution of cosmic rays must
be anisotropic outside of the Galaxy for an anisotropy to be observed
at Earth~\cite{Lemaitre,Swann:1933zz}. An anisotropy cannot arise
through deflections of an originally isotropic flux by a magnetic
field. This is because the intensity is constant along any possible CR
trajectory. To get the picture, consider the time evolution of the
phase space distribution within $t$ and $t + dt$. The number of
particles in an open ball around $(\mathbf x', \mathbf p')$ at $t' = t
+ dt$ is given by $dN = f (\mathbf x ', \mathbf p') \ d^3x' \ d^3
p'$. Now, $f (\mathbf x , \mathbf p)$ would remain constant if the
Jacobian of the $(\mathbf x, \mathbf p) \to (\mathbf x', \mathbf p')$
transformation satisfies
\begin{equation}
\mathcal J = \frac{\partial (\mathbf x', \mathbf p')}{\partial
  (\mathbf x, \mathbf p)} = 1 \, .
\label{jacobian}
\end{equation}
For (\ref{jacobian}) to stay staunch, it is sufficient to prove that the time-derivative $d{\cal J}/dt
= 0$.  In other words, the expansion of ${\cal J}$ to first-order
terms in $dt$ must vanish. The evolution from $t$ to $t + dt$ entails
$\mathbf x' = \mathbf x + \mathbf v \ dt$ and $\mathbf p' = \mathbf p +
\mathbf F \ dt$, and so
\begin{equation}
{\rm diag} \, {\cal J} = \left(1,\ 1,\ 1,\  1+ \frac{\partial F_x}{\partial
    p_x} \ dt, \  1+ \frac{\partial F_y}{\partial p_y} \ dt , \ 1+
  \frac{\partial F_z}{\partial p_z} \  dt  \right) \,,
\end{equation}
whereas the off-diagonal elements are ${\cal O} (dt)$. All in all, the expansion
of the Jacobian to first order in $dt$ is given by
\begin{equation}
{\cal J} = 1 + \left(\frac{\partial F_x}{\partial p_x} + \frac{\partial
    F_y}{\partial p_y} + \frac{\partial F_z}{\partial p_z} \right) +
\cdots \, ,
\end{equation}  
and therefore since the electromagnetic force satisfies
$\boldsymbol{\nabla}_p\cdot \mathbf F = 0$ the phase space distribution is
constant along the CR trajectory~\cite{Kachelriess:2008ze}. In addition, the magnetic field does
not change the magnitude of the cosmic ray momentum $p$, and so $p^2
f(\mathbf x, \mathbf p)$ is preserved too. Now, it is straightforward
to see from (\ref{phase-space}) that the intensity is constant along
any possible CR trajectory.

A dipole anisotropy is expected due to the net motion of Earth with
respect to the rest frame of UHECR sources, the so-called
Compton-Getting effect~\cite{Compton:1935wde}. Note that any observer
moving relative to the coordinate system in which the distribution of
UHECRs is isotropic will observe an anisotropic flux. Let $f(\mathbf
x, \mathbf p)$ be the distribution function of UHECRs in the frame $S$
for which the intensity is isotropic, and $f(\mathbf x', \mathbf p')$
the one in $S'$ associated to the observer frame $S$ moving with
velocity $\mathbf u$ with respect to $S$. The differentials $d^3x$ and
$d^3p$ transform opposite under Lorentz transformations and the
particle number $dN$ is (of course) a scalar. Thus, Lorentz invariance
implies $f(\mathbf x, \mathbf p) = f(\mathbf x', \mathbf p')$. The
particle momentum in the frame $S'$ is related to that in $S$ by a
Lorentz transformation
\begin{equation}
\mathbf p'  = \gamma_u \left( \mathbf p - \frac{p}{v} \mathbf u \right) \,,
\end{equation}
where $v \sim 1$ is the UHECR velocity in $S$.  For a non-relativistic
motion of the observer, $\mathbf u \ll 1$ and so the anisotropy
induced by the Compton-Getting effect is dominated by the lowest
moment: its dipole moment. In addition, $\gamma_u \sim
1$, and so an expansion in the small parameter $\mathbf p -
\mathbf p' = - p \mathbf u$, leads to
\begin{equation}
f'(\mathbf p')  =  f(\mathbf p') - p \mathbf u \cdot
\left. \frac{\partial  f}{\partial \mathbf p} \right|_{\mathbf p =
  \mathbf p'} + {\cal O} (u^2) 
 =  f (\mathbf p') \left( 1 - \frac{\mathbf u \cdot \mathbf p}{p}
 \  \frac{\partial \ln f}{\partial\ln p} \right) \cdots \, .
\label{CG-expansion}
\end{equation}
From (\ref{phase-space}) it follows that $\ln J = 2 \ln p + \ln f$ and  so
\begin{equation}
\frac{\partial \ln f}{\partial \ln p} = \frac{\partial \ln J}{\partial \ln
  p} - 2  =  \frac{\partial \ln
  J}{\partial \ln E} \frac{\partial \ln E}{\partial \ln p} - 2 \, .
\label{CG-relation}
\end{equation}
Substituting (\ref{CG-relation}) into  (\ref{CG-expansion}) and
multiplying by $p^2$ one arrives at
\begin{equation}
J'(E') \simeq J(E) \left[ 1 + \left(2 - \frac{\partial \ln J}{\partial
      \ln E} \right)  \frac{\mathbf u \cdot \mathbf p}{p}   \right] \, .
\end{equation}
For particles with spectrum $J(E) \propto E^{-\gamma}$,  the intensity
observed in $S'$ is given by
\begin{equation}
J'(E') \simeq J(E) \left[ 1 + (2 + \gamma ) \ u \ \cos(\mathbf{\hat u} ,
  \mathbf {\hat p}) \right]
  \, .
\end{equation}
Assuming that the sources of UHECRs are on average at rest with
respect to the cosmological frame, the magnitude and direction of the
velocity of the solar system $\mathbf u$ can be inferred from the detection of the
dipole anisotropy in the CMB. This gives $u = 369.0 \pm 0.9~{\rm km/s}$
in the direction $(l_{\rm CMB}, b_{\rm CMB}) = (264.00^\circ \pm 0.03^\circ, 48.24^\circ
\pm 0.02^\circ)$~\cite{Kogut:1993ag,Hinshaw:2008kr,Adam:2015rua}. For
$\gamma \simeq 2.53$, the predicted amplitude of the Compton-Getting 
effect, $\Bbbk = (2 + \gamma) u$, 
is only 0.56\%~\cite{Kachelriess:2006aq}. This is about an order of
magnitude smaller than the amplitude unmasked by the Pierre Auger Collaboration~\cite{Aab:2017tyv}.

We have seen that at energies beyond a certain threshold, the GZK
interactions between UHECRs and the universal photon backgrounds
limits the distances that the particles would travel.  For He, $d^{\rm
  GZK}_{10} \sim 10^3~{\rm Mpc}$, whereas for $p$ and CNO $d^{\rm
  GZK}_{10} \sim 10^{3.3}~{\rm Mpc}$~\cite{Globus:2017fym}. At lower energies the GZK distance
increases, e.g., for $p$, $d^{\rm GZK}_{4} \sim 10^{3.3}~{\rm Mpc}$ and
for He and CNO is of the order of the Hubble distance. Note that if
the intervening magnetic fields are negligible, then the UHECR horizon
would be characterized by the GZK distance. Hence, since the GZK
distance for CR energies at which the dipole has been observed is
almost the Hubble distance, the UHECR dipole axis must be aligned with
the CMB dipole (that follows the large-scale structure dipole matter
distribution). Note, however, that both $\Bbbk$ and $(l_\Bbbk,
b_\Bbbk)$ are not compatible with the CMB dipole~\cite{Globus:2017fym}.

The lifetime of a CR with velocity $v \sim 1$ is limited to $\tau^{\rm
  GZK}_{E_{\rm EeV}} \sim d^{GZK}_{E_{\rm EeV}}/v$, where $E_{\rm EeV}
\equiv E/{\rm EeV}$. The diffusion of this CR in the extragalactic
magnetic field could limit its magnetic horizon to much less than
$d^{GZK}_{E_{\rm EeV}}$. For a diffusive propagation in a magnetic
field $\mathbf B$, the horizon scale is the diffusion distance
\begin{equation}
d^{\mathbf B}_{E_{\rm EeV}} \sim \sqrt{6 \ D \ {\rm min} \left\{\tau^{\rm GZK}_{E_{\rm EeV}},
  t_{\rm age} \right\}} \,,
\end{equation}
where $D$ is the diffusion
coefficient and $t_{\rm age}$ is the age of the source ($t_{\rm age} \sim$
Hubble time if the source has always been active)~\cite{Globus:2017fym}. The diffusion
coefficient depends on the rigidity of the particles and on the
strength and coherence length of the magnetic field. For a Kolmogorov
turbulence, the diffusion coefficient can be approximated via a fitting
function that 
accounts for both the resonant and non-resonant diffusion regimes
\begin{equation}
D \approx \left[0.03 \left( \frac{\lambda^2_{\rm Mpc} E_{\rm EeV}}{Z B_{\rm
      nG}} \right)^{1/3} + 0.5 \left( \frac{E_{\rm EeV}}{Z B_{\rm nG}
      \lambda_{\rm Mpc}^{0.5}} \right)^2 \right]~{\rm Mpc^2 \ Myr^{-1}} \,,
\end{equation}
where $Ze$ is the CR charge, $B_{\rm nG}$ is the extragalactic
magnetic field strength in nG and $\lambda_{\rm Mpc}$ its coherence
length in Mpc~\cite{Globus:2007bi}.\footnote{Measurements of the
  Faraday rotation in the linearly polarized radio emission from
  distant quasars yield upper limits on the
  extragalactic magnetic field strength as a function of the reversal
  scale~\cite{Kronberg:1993vk}. If electron densities follow that of the Lyman-$\alpha$
  forest~\cite{Blasi:1999hu}, the average magnitude of the magnetic
  field receives an upper limit of $B \sim 0.65$~nG for reversals on
  the scale of the horizon, and $B \sim 1.7$~nG for reversal scales on
  the order of 1~Mpc at the $2\sigma$ level~\cite{Pshirkov:2015tua}.} The diffusive approximation is valid for $6D < v \, d^{\rm
  GZK}_{E_{\rm EeV}} $, or equivalently when $d^B_{E_{\rm EeV}} <
d^{\rm GZK}_{E_{\rm EeV}}$. If it takes more than the age of the
Universe to enter the diffusion regime, then the
propagation of UHECR becomes quasi-rectilinear. Altogether, the size of the region 
contributing to the observed CR intensity (and consequently to the
observed anisotropy) is 
set by the CR horizon~\cite{Parizot:2004wh} 
\begin{equation}
H_{E_{\rm EeV}} = {\rm min}  \left\{\sqrt{6 D \tau^{\rm GZK}_{E_{\rm EeV}}},
d^{\rm GZK}_{E_{\rm EeV}} \right\} \, .
\end{equation}
For a homogeneous extragalactic magnetic field, characterized by
$B_{\rm nG} \approx 10$ and $\lambda_{\rm Mpc} \approx 0.2$, the CR
horizon shrinks dramatically; e.g. $H_{10} \sim 100~{\rm Mpc}$ for He
and $H_{10} \sim 200~{\rm Mpc}$ for CNO and $p$~\cite{Globus:2017fym}.

Of particular interest here, the distribution of nearby -- distance
${\cal O} (100~{\rm Mpc}$) -- galaxies, as mapped by the 2 Micron
All-Sky Redshift Survey (2MRS)~\cite{Huchra:2011ii}, exhibits a
dipolar structure in the direction $(l_{\rm 2MRS}, b_{\rm 2MRS}) =
(251^\circ \pm 12^\circ, 37^\circ \pm
10^\circ)$~\cite{Erdogdu:2005wi}. If the sources of UHECRs are a
subset of these galaxies, then the arrival direction of CRs at Earth
would follow the same structure. Numerical simulations indicate that a
mixed-composition of CRs with $E \sim 10~{\rm EeV}$, propagating in a
$B_{\rm nG} \sim 1$ field, would create a dipole anisotropy with an
amplitude of about 10\% if the source distribution follows that of the
2MRS catalog up to about 100~Mpc~\cite{Harari:2015hba}.\footnote{Note,
  however, that for $B_{\rm nG} \sim 1$, the horizon is larger than
  the extend of the 2MRS catalog. Sources at larger distances would
  somewhat lower the anisotropy.} The dipole of the flux-weighted
distribution of infrared-detected galaxies in the 2MRS catalogue is
shown as an open diamond in Fig.~\ref{fig:5}. The direction of the
2MRS dipole is $55^\circ$ away from the central direction of the
dipole discovered by the Auger Collaboration. To illustrate how the
Galactic magnetic field could influence the observed direction of the
2MRS dipole, the deflected positions of this dipole as predicted by
the Jansson-Farrar (JF) model of the Galactic magnetic
field~\cite{Jansson:2012pc,Jansson:2012rt,Unger:2017kfh} are indicated by arrows in
Fig.~\ref{fig:5}, for two different CR rigidities that are compatible
with the composition fractions shown in Fig.~\ref{fig:2}. The
agreement between the directions of the UHECR and 2MRS dipoles is
improved by adopting these assumptions about the nuclear composition
and the deflections in the Galactic magnetic field. Note that if the
UHECR sources are within 100~{\rm Mpc} and the UHECR dipole is
He-dominated, this would imply an extragalactic magnetic filed
strength of ${\cal O} (10~{\rm nG})$~\cite{Globus:2017fym}.

One way to increase the chance of success in finding out the sources
of UHECRs is to check for correlations between CR arrival directions
and known candidate astrophysical objects. This is because even if the
distribution of UHECRs is quasi-isotropic, the arrival directions
could get stacked around some pre-defined directions. To calculate a
meaningful statistical significance in such an analysis, it is
important to define the search procedure {\it a priori} in order to
ensure it is not inadvertently devised especially to suit the
particular data set after having studied it. With the aim of avoiding
accidental bias on the number of trials performed in selecting the
cuts, the anisotropy analysis scheme must follow a pre-defined
process. First an exploratory data sample should be employed for
comparison with various source catalogs and for tests of various cut
choices.  The results of this exploratory period should then be used
to design prescriptions to be applied to subsequently gathered data.
The Auger Collaboration began the anisotropy searches with a
prescription protocol~\cite{Clay:2003pv}.  However, collecting the rare ultra-high energy
events is very expensive in terms of time. Given that the nominal 
lifetime of the experiment extends to 2025, the formality of a
prescription at this stage becomes unpractical.

Following the latest report of the Auger
Collaboration~\cite{Aab:2018chp}, the ensuing discussion is focussed
on the search for intermediate-scale anisotropies in UHECR arrival
directions associated with two prominent groups of extragalactic
sources detected by the Large Area Telescope on board the Fermi Gamma
Ray Space Telescope spacecraft ({\it Fermi-LAT}): {\it (i)}~active galactic nuclei (AGNs) that emit $\gamma$-rays,
 so-called  ``$\gamma$AGNs''  and {\it (ii)} starburst galaxies
 (SBGs). The $\gamma$AGN source population is constructed using the 2FHL catalog, which
includes 360 sources detected by {\it Fermi}-LAT above
50~GeV~\cite{Ackermann:2015uya}. A selection of radio-loud objects
within a 250~Mpc radius reduces the sample to 17 blazars and radio
galaxies. Their $0.05 \leq E/{\rm TeV} \leq 2$ integral flux
${\cal F}_\gamma$ is used as a proxy for the UHECR flux. Given the distance
of these objects, the $\gamma$-ray absorption by the infrared 
background light can be safely neglected; see Sec.~\ref{sec:DEBRA}. The
detections of seven SBGs have been reported using {\it Fermi}-LAT
data: NGC 253, M82, NGC 4945, NGC 1068~\cite{Ackermann:2012vca}, NGC
2146~\cite{Tang:2014dia}, Arp 220~\cite{Peng:2016nsx}, and
Circinus~\cite{Hayashida:2013wha}. Their gamma-ray luminosity has been
shown to scale almost linearly with their continuum radio flux
${\cal F}_\gamma$~\cite{Ackermann:2012vca}, and therefore the continuum
emission of SBGs at 1.4 GHz (for which a larger census exists) is
adopted as a proxy for the UHECR flux.\footnote{TeV
  $\gamma$-ray emission has been observed from M82~\cite{Acciari:2009wq} and NGC 253~\cite{Abdalla:2018nlz}.} Among the 63 objects within
250~Mpc searched for gamma-ray emission in~\cite{Ackermann:2012vca}, 23
SBGs with a flux larger than 0.3~Jy are selected to define the working
sample.

The UHECR sky is modelled as the sum of an isotropic component plus
the anisotropic contribution (with signal fraction $f_{\rm sig}$) from
the sources.  For the anisotropic component, each source is modeled as
a Fisher-Von Mises distribution $\mathfrak{F} (\mathbf{\hat n},
\mathbf {\hat s}_i; \Theta)$ centered on the coordinates of the source
location $\mathbf{\hat s}_i$, with the angular width (or search
radius $\Theta$) being a free parameter common to all sources.\footnote{The
  Fisher-Von Mises distribution, $\mathfrak{F} (\mathbf{\hat n },
  \mathbf {\hat s}; \kappa) = \kappa \exp(\kappa \
  \mathbf{\hat n} \cdot \mathbf{\hat s})/(4 \pi \sinh \kappa)$,
  is the equivalent of a Gaussian on the sphere $\mathbb S^2$, where
  $\mathbf{\hat s} \in \mathbb S^2$ is the mean direction,
  $\kappa \geq 0$ is the concentration parameter, $\mathbf{\hat n}
  \in \mathbb S^2$ is a random unit vector on the sphere, and the
  remaining terms serve to normalize the distribution~\cite{Fisher}.  The parameter
  $\kappa$ controls the concentration of data points $\mathbf{\hat
    n}_j$ around the mean direction $\mathbf{\hat s}$, with $j = 1,
  \cdots N$. In particular, for $\kappa = 0$,  $\mathfrak{F}
  (\mathbf{\hat n }, \mathbf {\hat s}; \varkappa)$ reduces to the
  uniform density on $\mathbb S^2$, whereas as $\kappa \to \infty$,
  $\mathfrak{F} (\mathbf{\hat n }, \mathbf {\hat s}; \kappa)$
  tends to a point density. The parameters $\mathbf{\hat s}$  and
  $\kappa^{-1}$ are analogous to the mean and variance in the Gaussian distribution.
Thus, the search radius is defined as the inverse
  square root of Fisher's concentration parameter, {\it viz.} $\kappa = \Theta^{-2}$.} Smoothed density maps are constructed from a
superposition of catalog sources, weighted by the electromagnetic flux
of the source ${\cal F}_\gamma$. The smoothed density maps are
described by a function $F(\mathbf{\hat n})$, such that its value in a
given direction $\mathbf{\hat n}$ is proportional to the probability
of detecting a CR in that direction, according to the
model. Collectively, the probability density map function is given by
\begin{equation}
F(\mathbf{\hat n}; f_{\rm sig}, \Theta) = \frac{\omega(\mathbf{\hat
    n})}     {\mathcal{C}} \left[( 1- f_{\rm sig}) + f_{\rm sig}
  \sum_{i=1}^{N_{\rm cat}} {\cal F}_{\gamma,i} \ w(z_i) \  \mathfrak{F}
  (\mathbf{\hat n}, \mathbf {\hat s}_i; \Theta) \right] \, ,
\end{equation}
where $\mathcal{C}$ is an overall normalization constant that guarantees
$\int F \ d \mathbf{\hat n} = 1$, $w(z_i)$ is the weight attributed to
the $i$th source located at $z_i$ to account for the attenuation
factor because of GZK interactions, and the sum extends over all
sources in the catalog $N_{\rm cat}$~\cite{Abreu:2010ab}. Then, the model map depends on
two free parameters aimed at maximizing the degree of correlation with
UHECR events: the fraction of all events due to the sources
(anisotropic fraction) and the root-mean-square angular separation
between an event and its source (search radius) in the anisotropic
fraction. The search signal fraction $f_{\rm sig}$ controls to what
extent a contribution from the considered astrophysical sources is
preferred to over a purely isotropic distribution. The search radius
$\Theta$ provides an effective description of CR deflections in the
intervening magnetic fields. The fraction $(1 - f_{\rm sig})$
parametrizes the isotropic component. Note that this isotropic
contribution could originate in faint unresolved objects absent from
the considered catalog, or else account for highly deflected nuclei in
the Galactic $\mathbf B$-field. The GZK-weights are evaluated as the
fraction of the events produced above a given energy threshold, which
are able to reach the Earth from a source at a redshift $z$ with an
energy still above that same threshold~\cite{Harari:2008zp}. The
weights depend on both the shape of the emission spectra and the nuclear
composition at the sources. They are determined through a fit that
simultaneously reproduces Auger data on the spectrum and composition~\cite{Aab:2016zth}.

The most probable values of the free parameters ($f_{\rm sig}$ and $\Theta$)
are estimated using a maximum-likelihood ratio test, which also
quantifies the strength of each model by contrast with isotropy. The
likelihood is defined as the product over the UHECR events of the
model density in every UHECR direction
\begin{equation}
\mathscr{L} (f_{\rm sig}, \Theta; \mathbf{\hat n}_j) = \prod_{j=1}^{N}  F(\mathbf{\hat n_j}; f_{\rm sig}, \Theta)  \,,
\end{equation}
where $\mathbf{\hat n}_j$ is the direction of the $j$th event and $N$
the number of events in the UHECR data sample~\cite{Abreu:2010ab}.
As aforestated, the TS for deviation from isotropy is the likelihood
ratio test, $-2 \ln \lambda_0$,
between two nested hypotheses: the UHECR sky model and an isotropic
model (null hypothesis); see~\ref{app2} for details.  Note that the
condition $f_{\rm sig} = 0$ yields the density map of isotropy, and
consequently defines the null hypothesis. The TS is maximized as a
function of two parameters: the search radius and the anisotropic
fraction. The analysis is repeated for a sequence of energy
thresholds varying in the range $10^{10.3} \lesssim E_{\rm th}/{\rm
  GeV} \lesssim 10^{10.9}$.

For SBGs, the maximum TS $\equiv -2 \ln \lambda_0 = 24.9$ is obtained
with 894 events of $E > 39~{\rm EeV}$. This corresponds to a local
$p$-value of $3 \times 10^{-6}$, see Fig.~\ref{fig:25}. The smearing angle and the
anisotropic fraction corresponding to the best-fit parameters are
${13^{+4}_{-3}}^\circ$ and $(10 \pm 4)\%$, respectively. Remarkably,
the energy threshold of largest statistical significance coincides
with the observed suppression in the
spectrum~\cite{Aab:2017njo}, implying
that when we properly account for the barriers to UHECR propagation in
the form of energy loss
mechanisms~\cite{Greisen:1966jv,Zatsepin:1966jv} we obtain a self
consistent picture for the observed UHECR horizon. The scan in energy
thresholds comes out with a penalty factor, which was estimated
through Monte-Carlo simulations.  The post-trial chance probability in
an isotropic cosmic ray sky is $4.2 \times 10^{-5}$, corresponding to
a 1-sided Gaussian significance of $4\sigma$~\cite{Aab:2018chp}.
For $\gamma$AGNs, the maximum
TS $\equiv -2 \ln \lambda_0 = 15.2$ is obtained with 177 events of $E
> 60~{\rm EeV}$.  The maximum deviation for $\gamma$AGNs is found at an
intermediate angular scale of $\Theta = {7^{+4}_{-2}}^\circ$ with an
anisotropic fraction $f_{\rm sig} = (7 \pm 4) \%$. Penalizing for the
energy scan, the maximum TS obtained for $\gamma$AGNs corresponds to a
$2.7\sigma$ deviation from isotropy.

Because of possible incompleteness of the source-list
in~\cite{Ackermann:2012vca} near the Galactic plane ($|b| < 10^\circ$)
and in the southern sky $(\delta < 35^\circ$), relevant SBGs could be
missing from the selected sample. However, the Auger Collaboration
verified that the conclusions remain unchanged if: {\it (i)}~one uses
all 63 objects listed in~~\cite{Ackermann:2012vca}; {\it (ii)}~one
uses the catalog given in~\cite{Becker:2009hw} with 32 SBGs above
0.3~Jy, {\it (iii)}~one adds the Circinus SBG absent from {\it (i)}
and {\it (ii)}; {\it (iv)} one uses only the six SBGs (NGC 253, M82,
NGC 4945, NGC 1068, Circinus, NGC 2146) reported in the third {\it
  Fermi}-LAT source catalog (3FGL)~\cite{Acero:2015hja}, and their 1
to 100~GeV integral flux as a UHECR proxy.

\begin{figure}
\postscript{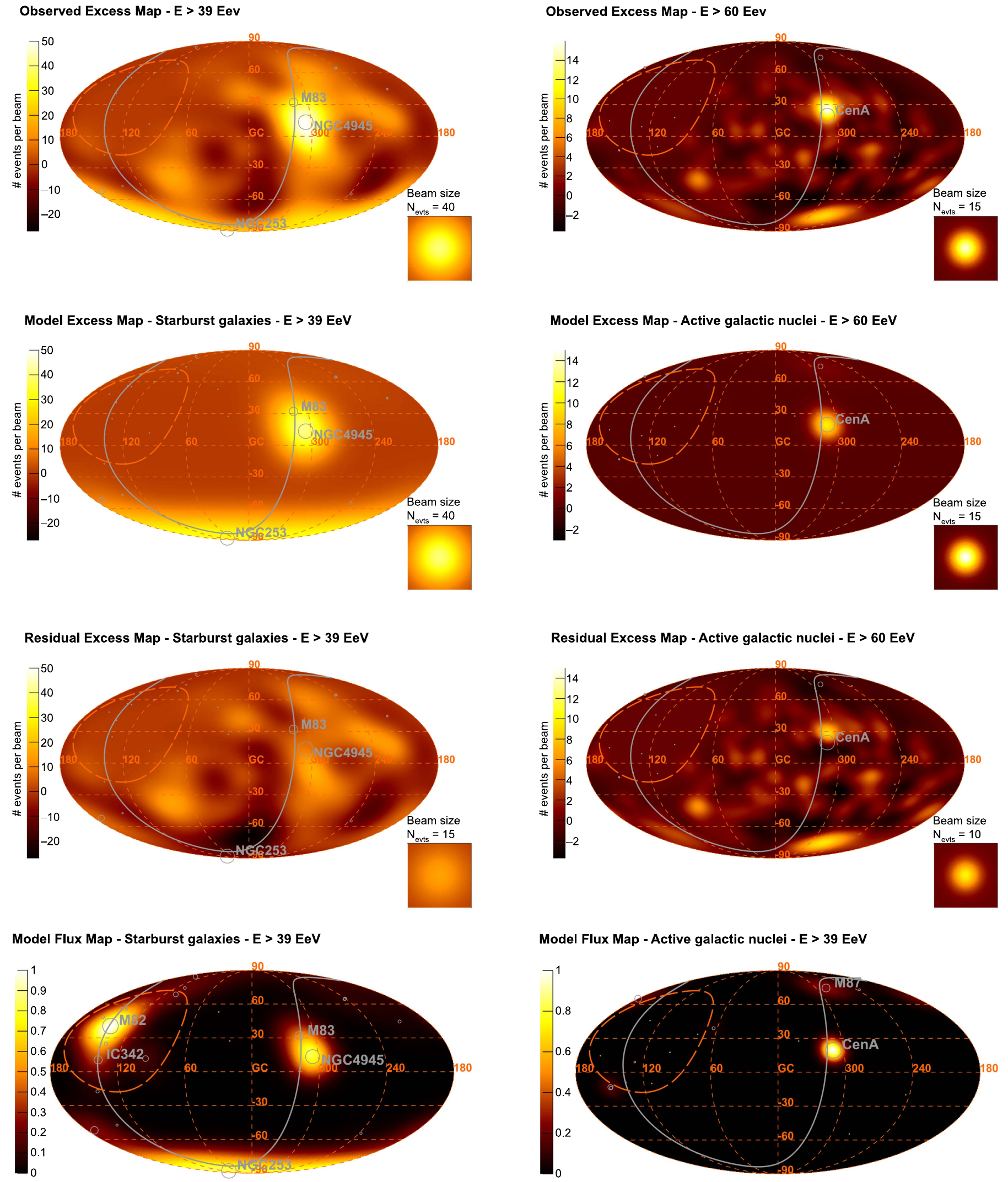}{0.9}
\caption{From top to bottom: {\it (i)}~observed excess map, {\it (ii)}~model
  excess map, {\it (iii)}~residual map, and {\it (iv)}~model flux map, for the best-fit
  parameters obtained with starburst galaxies above $39~{\rm EeV}$ (left)
  and $\gamma$AGNs above $60~{\rm EeV}$ (right). The excess
  maps (best-fit isotropic component subtracted) and residual maps
  (observed minus model) are smeared at the best-fit angular
  scale. The color scale indicates the number of events per smearing
  beam (see inset). The model flux map corresponds to a uniform
  full-sky exposure. The supergalactic plane is shown as a solid gray
  line. An orange dashed line delimits the field of view of the
  array. From Ref.~\cite{Aab:2018chp}}
\label{fig:6}
\end{figure}

Starburst galaxies provide the most significant indication that UHECRs
are not distributed isotropically on an intermediate angular scale,
with an a posteriori chance probability $p_{\rm Auger} \simeq 4.2
\times 10^{-5}$. As shown in Fig,~\ref{fig:6} the Auger signal is
dominated by three nearby starbursts NGC 4945, M83, and NGC 253. In
the bottom panel of Fig.~\ref{fig:6} it can be seen that M82 is
expected to be one of the dominant starbursts in the full-sky. Its
declination of $\delta \approx 70^\circ$~N is outside the exposure of
the Auger Observatory, but is covered in the northern hemisphere by
the TA. Thought-provoking, the starburst galaxy M82 is close to the
best-fit source position of the TA hot
spot~\cite{Anchordoqui:2014yva,Fang:2014uja,He:2014mqa,Pfeffer:2015idq,Attallah:2018euc}. The multiplicative
$p$-value for the two non-correlated observations is
\begin{equation}
p = p_{\rm TA} \otimes p_{\rm Auger} = 1.5 \times 10^{-8}, 
\label{p-value}
\end{equation}
yielding a statistical significance $\gtrsim 5
\sigma$~\cite{Anchordoqui:2017abg}.  However, caution must be
exercised in all-sky comparisons~\cite{Globus:2016gvy}.  Moreover,
(\ref{p-value}) combines a catalog-based cross-correlated search
(Auger) with a blind search (TA). Therefore, (\ref{p-value}) provides
a rough estimate of the statistical significance under the strong
assumption that M82 (which is at the border of the excess of TA
events) is the only source contributing to the TA hot-spot. It is
clear that new data are needed to confirm the suggested correlation.\footnote{First generation of UHECR
  observatories also pointed to a starburst origin for the highest
  energy events~\cite{Anchordoqui:2002dj}.} 

For $\gamma$AGNs, a compelling concentration of events is observed in
the region around the direction of the nearest active galaxy,
Centaurus A (Cen A).  A separate analysis~\cite{Aab:2017njo} shows
that the maximum departure from isotropy occurs for a ring of
$15^\circ$ around the object, in which 19 (out of a total of 203)
events with $E \geq 58~{\rm EeV}$ are observed compared to an
expectation of 6.0 from isotropy. The significance of this excess can
be obtained by penalizing for the scan in energy and angular
scale. Performing such a process one obtains a statistical
significance of $\sim 3.1~\sigma$~\cite{Aab:2017njo}. There are no
events coming from less than $15^\circ$ around M87, which is almost 5
times more distant than Cen A and lies at the core of the Virgo
cluster. The Auger exposure is 3 times smaller for M87 than for Cen
A~\cite{LetessierSelvon:2011dy}. Using these two rough numbers and
assuming equal luminosity, one expects 75 times fewer events from M87
than from Cen A. Hence, the lack of events in this region is not
completely unexpected.

The Centaurus cluster lies 45~Mpc behind Cen A. An interesting
question then is whether some of the events in the $15^\circ$ circle
could come from the Centaurus cluster rather than from Cen A. This
does not appear likely because the Centaurus cluster is farther away
than the Virgo cluster and for comparable CR luminosities one would
expect a small fraction of events coming from Virgo~\cite{Gorbunov:2008ef}.

It has been proposed that Fornax A could be the source of the apparent
excess above 60~EeV right to the Galactic South pole in
Fig.~\ref{fig:6}~\cite{Matthews:2018laz}. However, if the Galactic magnetic field is
approximated well by the JF 
model~\cite{Jansson:2012pc,Jansson:2012rt,Unger:2017kfh}, then Fornax A is unlikely
to be the source of the excess because the magnetic field would deflect
UHECRs into a different location near the Galactic equator~\cite{Smida:2015kga,Anjos:2018mgr}.

\begin{figure}[tpb]
\postscript{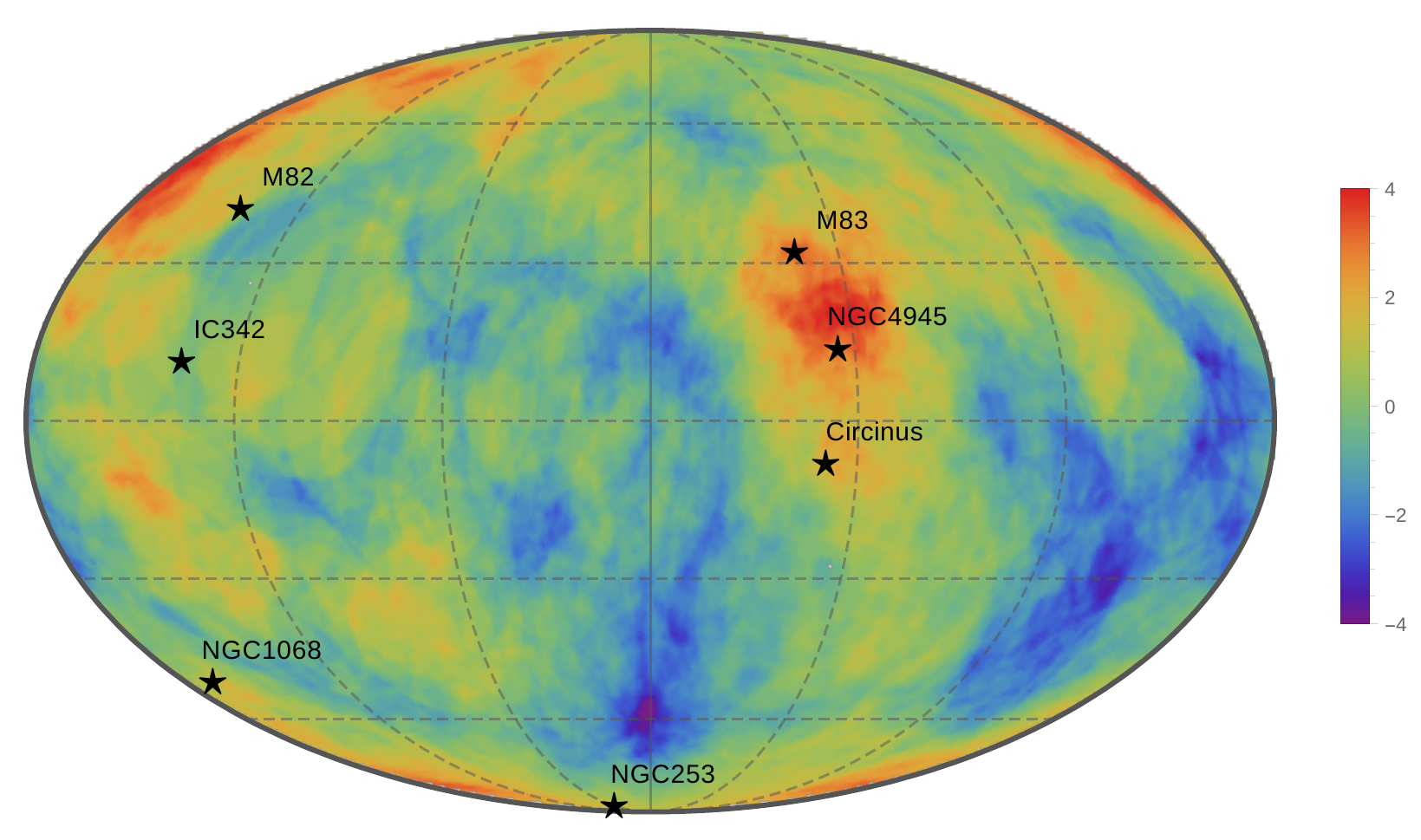}{0.7}
\begin{center}
\caption{Skymap in Galactic coordinates of the Li-Ma significances of
  over-densities in $20^\circ$ radius windows for  840 events recorded by
Auger with $E> E_{\rm th, Auger}$ and 130 events
recorded by TA with $E> E_{\rm th, TA}$. The color scale
indicates the significance in units of standard deviations;
negative values follow the convention of indicating the (positive)
significance of deficits. Nearby SBGs providing a significant contribution
to the UHECR correlation signal of Auger~\cite{Aab:2018chp} and TA~\cite{Abbasi:2018tqo} are indicated by stars. From Ref.~\cite{Anjos:2018mgr}.
\label{fig:extra}}
\end{center}
\end{figure}

The latest search for hot spot anisotropies is a joint effort by
the two collaborations considering 840 events recorded by
Auger with $E> E_{\rm th, Auger} = 40~{\rm EeV}$ and 130 events
recorded by TA with $E > E_{\rm th, TA} = 53.2~{\rm
  EeV}$~\cite{Biteau:2018paris}. The most significant
excesses observed in a $20^\circ$ search are at Galactic longitude and
latitude: $(l, b) \approx (303.0^\circ, 12.9^\circ)$ and
($l, b) \approx (162.5^\circ,44.4^\circ)$, with local Li-Ma statistical significance for the rejection of
the null (background only) hypothesis of $4.7\sigma$ and $4.2\sigma$,
respectively. The Li-Ma significance map of
this data-sample is shown in Fig.~\ref{fig:extra}. 

In addition, the TA
Collaboration carried out an independent test of the reported
correlation between the arrival directions of UHECRs and
SBGs. The data sample used for this analysis includes
CRs with $E >E'_{\rm th,TA} = 43~{\rm EeV}$ detected by TA in a
nine year period from May 2008 to May 2017. These data are compatible
with isotropy to within $1.1\sigma$ and with Auger result to within
$1.4\sigma$, and so the TA Collaboration concluded that with their
current statistics they cannot make a statistically significant
corroboration or refutation of the reported possible correlation
between UHECRs and SBGs~\cite{Abbasi:2018tqo}. It is important to
stress, however, that $E'_{\rm th, TA} < E_{\rm th, TA}$. Most importantly,
 $E_{\rm th, TA}$ is above the energy at which TA observes the
suppression in the spectrum~\cite{AbuZayyad:2012ru}, but
$E'_{\rm th, TA}$ is below. This implies that
the data sample of the test carried out by the TA Collaboration is
most likely contaminated from the isotropic background of UHECRs emitted by far
away sources, and consequently this would tend to reduce the
significance of any possible correlation with nearby sources.

We end with two observations:
\begin{itemize}[noitemsep,topsep=0pt]
\item It is important to keep in mind that if a source produces an anisotropy signal at energy $E$ with
  cosmic ray nuclei of charge $Ze$, it should also produce a similar
  anisotropy pattern at energies $E/Z$ via the proton component that
  is emitted along with the nuclei, given that the trajectory of
  cosmic rays within a magnetic field is only
  rigidity-dependent~\cite{Lemoine:2009pw}.  Moreover, 
  secondary protons produced during propagation could also create an
  anisotropy pattern in the ``low'' energy regime~\cite{Liu:2013ppa}. This sets a
  constraint on the maximum distance to nucleus-emitting-sources.
  Making the extreme assumption that these sources do not emit any
  protons, the hypothetical source(s) responsible for anisotropies
  should lie closer than $\sim 20$ to 30, 80 to 100, and 180 to
  200~Mpc, if the anisotropy signal is mainly composed of oxygen,
  silicon and iron nuclei, respectively~\cite{Liu:2013ppa}. This sets
  an interesting constraint on source models of UHECR nuclei and provides a distinctive
  signal to be tested by future data.
\item  It is also important to keep in mind that the anisotropies searches
  discussed above are all {\em a
  posteriori} studies, so {\em one cannot use them to determine a
  completely unbiased confidence level for anisotropy as the number of trials is unknown}. 
\end{itemize}

In summary, the inaugural years of data taking at TA and the Pierre
Auger Observatory have yielded a large, high-quality data sample.  The
enormous area covered by the Auger surface array together with an
excellent fluorescence system and hybrid detection techniques have
provided us with large statistics, good energy resolution, and solid
control of systematic uncertainties. Presently, Auger is collecting
some $7,000~{\rm km}^2~{\rm sr}~{\rm yr}$ of exposure each year, and
is expected to run until 2025.  New detector systems are being
deployed, which will lower the energy detection threshold down to
$10^{8}$~GeV~\cite{Aab:2016vlz}.  In particular, the addition of
planar plastic scintillator of $4~{\rm m}^2$ area to each Cherenkov
detector will provide baryonic-sensitive observables for each shower
enabling charge-discriminated studies with a duty cycle of nearly
100\%. An experimental radio detection
program is also co-located with the observatory and shows promising
results~\cite{Aab:2016eeq,Aab:2018ytv}. In addition,  new surface and fluorescence detectors are planned to be
constructed for the TA$\times$4 experiment to cover 4 times larger
area than TA to observe cosmic rays, especially with the highest
energies using high statistics~\cite{Kido}. As
always, the development of new analysis techniques is ongoing, and
interesting new results can be expected.

\section{Quest for the origin(s) of UHECRs}
\label{sec:3}

\subsection{Acceleration processes}

\subsubsection{Phenomenological considerations}
\label{sec:pheno}

It is most likely that the bulk of the cosmic radiation is a result of
some very general magneto-hydrodynamic (MHD) phenomenon in space which
transfers kinetic or magnetic energy into CR energy. The
details of the acceleration process and the maximum attainable energy
depend on the particular physical situation under consideration. There
are basically two types of processes that one might invoke. The first
type assumes the particles are accelerated directly to very high
energy by an extended electric field~\cite{Hillas:1985is}. This
idea can be traced back to the early '30s when Swann~\cite{Swann}
pointed out that betatron acceleration may take place in the
increasing magnetic field of a sunspot. These so-called ``one-shot''
mechanisms have been worked out in greatest detail, and the electric
field in question is now generally associated with the rapid rotation
of small, highly magnetized objects such as white
dwarfs~\cite{deJager,Ikhsanov:2005qf}, neutron stars
(pulsars)~\cite{Gunn:1969ej,Blasi:2000xm,Arons:2002yj,Fang:2012rx,Fang:2013cba},
or black holes~\cite{Blandford:1977ds,Znajek,Lovelace}. Electric field
acceleration has the advantage of being fast, but suffers from the
circumstance that the acceleration occurs in astrophysical sites of
very high energy density, where many opportunities for energy loss
exist. The second type assumes particles gain energy gradually through
multiple stochastic encounters with moving magnetized plasmas. This
idea was pioneered by Fermi~\cite{Fermi:1949ee,Fermi:1954ofk}. A
variety of astrophysical environments have been suggested as sites of
stochastic acceleration, including the interplanetary
medium~\cite{Jokipii:1971,Wenzel:1989}, supernova remnants
(SNRs)~\cite{Scott:1975,Chevalier:1976,Chevalier:1978qk,Cowsik:1984yya,Torres:2002af,Blasi:2010gr},
the Galactic disk and
halo~\cite{Jokipii:1985,Jokipii:1987,Bustard:2016swa,Merten:2018qoa},
AGNs~\cite{Protheroe:1983,Kazanas:1985ud,Protheroe:1992qs},
large-scale jets and lobes of giant radio-galaxies
(RG)~\cite{Biermann:1987ep,Rachen:1992pg,Romero:1995tn},
blazars~\cite{Blandford:1979za,Mannheim:1993jg,Dermer:2008cy,Caprioli:2015zka},
gamma-ray bursts (GRBs)~\cite{Waxman:1995vg,Vietri:1995hs}, starburst
superwinds~\cite{Anchordoqui:1999cu,Anchordoqui:2018vji}, Galactic
microquasar systems~\cite{Levinson:2001as,Aharonian:2005cx}, and
clusters of
galaxies~\cite{Norman:1995,Kang:1996rp,Ryu:2003cd}. Stochastic
acceleration has the disadvantage of being slow, and it is also hard
to keep the relativistic particles confined within the Fermi engine.

\begin{figure}[tpb]
\postscript{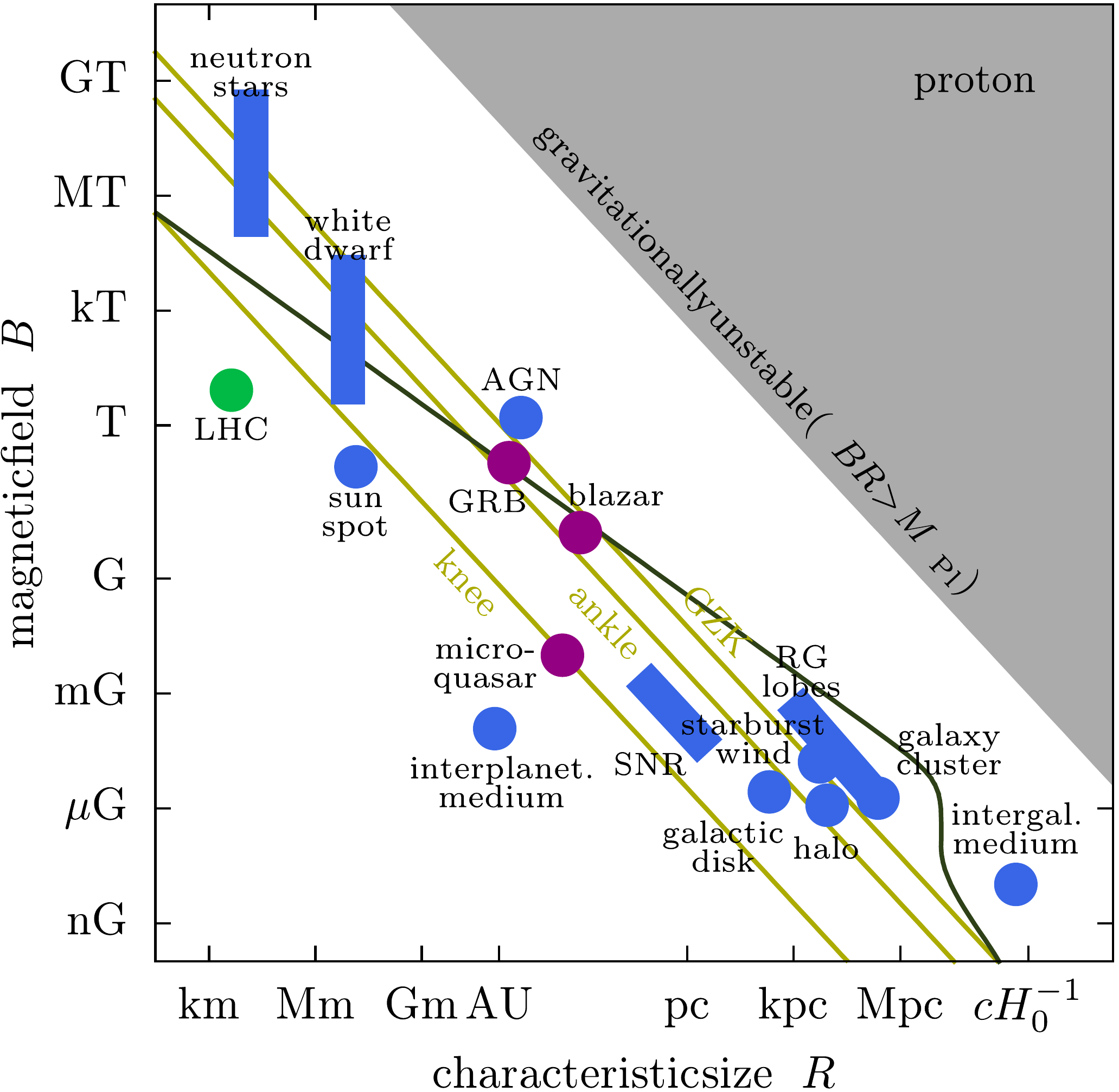}{0.9}
\caption{The ``Hillas plot'' for various CR source candidates, (blue
  areas). Also shown are jet-frame parameters for blazars, GRBs, and
  microquasars (purple areas). The corresponding point for the LHC
  beam is also shown. The straight lines show the {\it lower limit}
  for accelerators of protons at the {\it knee} ($E_{\rm knee} \sim
  10^{6.6}$~GeV), the {\it ankle} ($E_{\rm ankle} \sim 10^{9.7}$~GeV),
  and the GZK suppression ($E_{\rm supp} \sim 10^{10.6}$~GeV). The
  curve is the {\it upper limit} from synchrotron losses and proton
  interactions in the cosmic photon background ($R\gg1$~Mpc). The grey
  area corresponds to astrophysical environments with extremely large
  magnetic field energy that would be gravitationally unstable. Here,
  AGN indicates the unipolar inductor which originates in the rotating
  accretion disk around a $10^8 M_\odot$ black hole; such a disk could
  draw in magnetic flux with the gas to give a magnetic field $B \sim
  10^4~{\rm G}$ parallel to the rotation axis~\cite{Lovelace}. Blazars
  and RGs are AGNs that are oriented at different angles~\cite{Torres:2004hk}.
  When the galaxy is oriented so that the jets of energy exiting the
  black hole are pointed toward Earth the objects is called
  a blazar. When the Earth is oriented perpendicular to the jet axis,
  the full extent of the jets may be seen particularly at low
  frequencies, giving rise to a morphology typical of RGs. For blazars
  and RG lobes, particle
  acceleration proceeds via Fermi mechanism~\cite{Rachen:1992pg,Romero:1995tn,Dermer:2008cy}.  From Ref.~\cite{yellowbook}. }
\label{fig:7}
\end{figure}

The length scale characterizing the propagation of an UHECR of
energy $E$ and charge $Ze$ in a magnetic field $B$ is the Larmor radius
\begin{equation}
\label{LARMOR}
r_L   = \frac{1}{\sqrt{4\pi\alpha}} \ \frac{E}{ ZB} 
= 1.1 \ \frac{1}{Z} \ \left(\frac{E}{10^9~\text{GeV}}\right)\left(\frac{B}{\mu\text{G}}\right)^{-1}\,{\rm
  kpc}\, ;
\end{equation} 
a greater Larmor radius implies a less curved trajectory. If the CR energy
originates via an acceleration process, a general estimate of the
maximal energy can be obtained by requiring the Larmor radius of the
UHECR to be no larger than the linear size $R$ of the
accelerator. This constraint provides a qualitative criterion to identify
potential sources of UHECRs by simply looking at the largest values of
the product $BR$; namely,
\begin{equation}
E \lesssim Z \left(\frac{R}{\rm kpc}\right) \left(\frac{B}{\mu {\rm
    G}}\right) \times 10^9~{\rm GeV} \, .
\label{Hillas}
\end{equation}
The limitation in energy is conveniently visualized in the `` Hillas
plot''~\cite{Hillas:1985is} shown in Fig.~\ref{fig:7}, where the
characteristic magnetic field $B$ of candidate cosmic accelerators is
plotted against their characteristic size $R$. It is striking that the
potential accelerators range from neutron stars (for which $R \sim
10~ {\rm km}$), up to clusters of galaxies (for which $R \sim 1~{\rm
  Mpc}$).  Exceptions to the limit (\ref{Hillas}) may occur for
astrophysical systems containing jets which move relativistically in
the host-galaxy frame. Such relativistic jets are ubiquitous in
astrophysical systems that contain compact objects, such as blazars,
GRBs, and microquasars.  The Hillas criterion is a necessary
condition, but not sufficient. An important caveat is that
(\ref{Hillas}) neglects the finite lifetime of the acceleration region
and energy loss due to interactions with the environment, such as
synchrotron radiation in the magnetic field and the production of
secondary particles. For example, Gpc scale shocks from structure
formation with ${\cal O}$ (nG) magnetic fields would satisfy the Hillas
criterion, but the acceleration at such shocks could be much too slow
and consequently subject to large energy loss; see in Fig.~\ref{fig:7}
intergalactic medium.

Along these lines, in this section we will scrutinize some general
constraints on UHECR accelerators. The essence of these constraints
are briefly summarized in the following points~\cite{Ptitsyna:2008zs}:
\begin{itemize}[noitemsep,topsep=0pt]
\item geometry -- the accelerated particle should need be kept inside the source while being accelerated;
\item power -- the source should posses the required amount of energy to give it to accelerated particles;
\item radiation losses -- the energy lost by a particle for radiation in the accelerating field should not exceed the energy gain;
\item interaction losses -- the energy lost by a particle in interactions with other particles should not exceed the energy gain;
\item emissivity -- the total number (density) and power of sources should be able to provide the observed UHECR flux;
\item accompanying radiation of photons, neutrinos, and low-energy cosmic rays should not exceed the observed fluxes, both for a given source and for the diffuse background.
\end{itemize}

\subsubsection{Unipolar induction}
\label{sec:unipolar}

A neutron star is a compact object of radius $R_\ast \sim 10~{\rm
  km}$, which evolved from the gravitational collapse of an ordinary
massive star of radius $R_{\star} \sim 10^{6}~{\rm km}$, with a
magnetic field $B_{\star} \sim1~{\rm G}$, and a rotation period
$P_\star \sim 10 - 100~{\rm yr}$, or else from white dwarfs that
because of accretion exceeded the Chandrasekhar mass of $1.4 M_\odot$,
where $M_\odot$ is the solar mass. The conservation of angular
momentum, $M R_\star^2 \Omega_\star = M R_\ast^2 \Omega_\ast$, and
magnetic flux, $R_\star^2 B_\star = R_\ast^2 B_\ast$, entail that when
compressed to the size $R_\ast$ the rotation period and the magnetic
field of the neutron star are of order $P_\ast \sim (R_\ast/R_\star)^2
P_\star \sim (0.01 - 1)~{\rm s}$ and $B_\ast\sim (R_\ast/R_\star)^2
B_\star \sim 10^{12}~{\rm G} = 100~{\rm MT}$, where $\Omega_i =
2\pi/P_i$ is the angular velocity. Therefore, according to the Hillas
criterion neutron stars can accelerate CRs to the maximum observed
energies.

Neutron-star surfaces are thought to be composed of anisotropic,
tightly-bound condensed matter. The crust of neutron stars extends
down to about 1~km below the surface, with densities ranging from a
few ${\rm g/cm^3}$ on the exterior surface up to nuclear density
$10^{14}~{\rm g/cm^3}$ in the interior~\cite{Chamel:2008ca}.  The
outermost layers of the star are composed of long molecular chains of
$^{56}$Fe, with axes parallel to the magnetic field.  $^{56}$Fe ions
can thus be stripped off the surface and be accelerated to extremely-high
energies.

The acceleration process is Faraday's unipolar induction, in which the
rotational energy of the highly conducting plasma surrounding the
homogeneously magnetized star is converted into electromagnetic energy.
The electromagnetic force $F_{\rm em} \sim Ze E$ acting on a CR
of charge $Ze$ and mass $m$ near the neutron star surface turns out
to be many orders of magnitude greater than the gravitational force $F_{\rm g} =
GM m/R_\ast^2$. This condition allows us to disregard the electromagnetic
field distortion connected with the space curvature in the vicinity of
the neutron star. Therefore,  the  charges co-rotating with the star only
experience the magnetic force, which is orthogonal to both the
magnetic field and velocity. If unlimited plasma particles can be
supplied into the system the electric charges move freely until 
the electric force and the magnetic force are equal and opposite,
\begin{equation}
\mathbf{E} + \mathbf{v} \times \mathbf{B}  = \bm{0} \, .
\label{frozenin}
\end{equation}
The condition (\ref{frozenin}) simply implies that $\mathbf{E} \cdot
\mathbf{B} =0$. In other words, if the plasma surrounding the star is a
perfect conductor, the electric field in the prime coordinate system
co-rotating with the star is zero, $\mathbf{E'} = \bm{0}$.

Because of the presence of plasma in the pulsar magnetosphere
condition (\ref{frozenin}) is, with adequate accuracy, satisfied not
only in the interior of the neutron star but also in the whole
magnetosphere. It is evident that the rigid co-rotation becomes
impossible at large distances from the rotation axis. Indeed, the
plasma rigidly co-rotates with the star within a zone $r < R_L$
limited by the finiteness of the speed of light. The so-called {\em
  light cylinder} defines the distance from the rotational axis at
which a co-rotating particle would reach the speed of light; namely
\begin{equation}
R_L=\frac{c}{\Omega_\ast} \,,
\label{Rele}
\end{equation}
where for clarity, we have written explicitly the
speed of light while defining $R_L$;
hereafter we continue using natural units and we will drop $c$ from
our formulae.  Note that $R_L$ provides a natural scale to define the magnetosphere
boundary. For ordinary pulsars, $R_L \sim 10^3 - 10^4~{\rm km}$. This
implies that  the light cylinder is at distances several thousand times larger
than the neutron star radius.

To discuss the generalities of the acceleration mechanism following~\cite{Ruderman:1975ju} we
consider the simplest case of a perfectly conducting neutron star rotating
with angular velocity ${\boldsymbol \Omega_\ast}$. The star is endowed with a magnetic dipole
moment aligned with the  rotation axis, taken here in the vertical 
$\mathbf{\hat{z}}$ direction. The magnetic moment tends to be antiparallel (as
opposed to parallel) to its spin angular momentum and has a magnitude $\mu =
B_p R_\ast^3 /2$, where $B_p$ is the magnetic field
strength at the pole. In spherical coordinates the components of the
magnetic field are given by
\begin{subequations}
\begin{equation}
B (r, \theta) = B_p \ \left(\frac{R_\ast}{r} \right)^3\cos \theta
\ \mathbf{\hat{r}} 
\end{equation}
and
\begin{equation}
B (r, \theta) = \frac{B_p}{2} \ \left(\frac{R_\ast}{r} \right)^3 \
\sin \theta \ \mathbf{\hat{{\boldsymbol \theta}}} \, .
\label{vossosdelab}
\end{equation}
\end{subequations}
To first order approximation we can neglect the magnetic field
contribution from magnetospheric currents, such that the magnetic
field far from the neutron star surface is dominated by the star's own
dipole field.

\begin{figure}[tpb]
\postscript{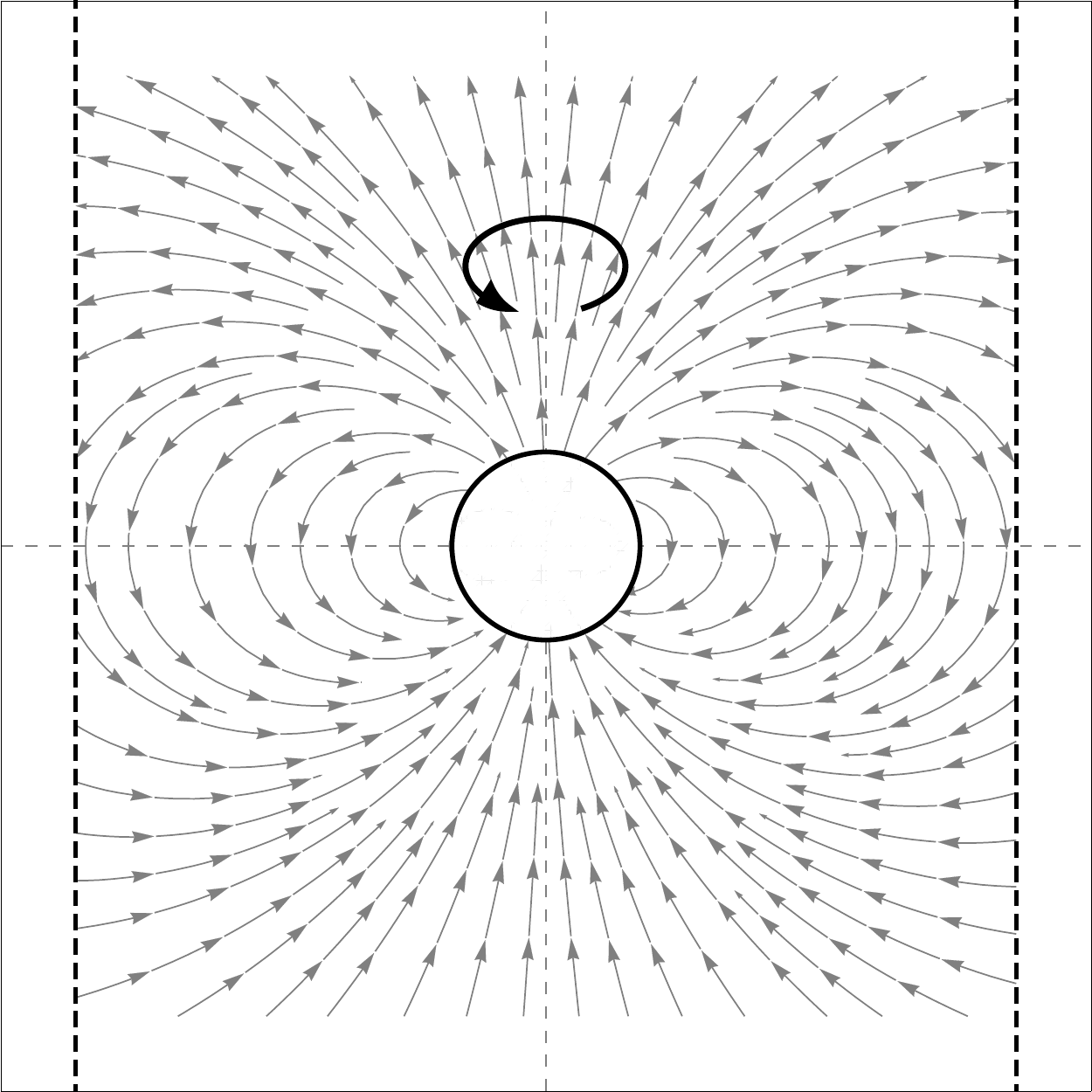}{0.5}
\caption{The axisymmetric magnetosphere structure of a neutron
  star. The rotating homogeneously magnetized star generates a dipole
  magnetic field. The open field lines coming out from the magnetic
  poles cross the light cylinder (dashed line). }
\label{fig:8}
\end{figure}

As one can see in Fig.~\ref{fig:8}, some magnetic field lines close
inside the light cylinder, while those connected to the polar region
cross it, and the particles moving along them cannot co-rotate.  These
{\it open field lines} define two {\it polar caps} on the stellar surface
from which charged particles leaving the star can move along field
lines and escape from the co-rotating magnetosphere by passing through
the light cylinder. At the light cylinder the co-rotating
magnetosphere carries one sign of net charge along an equatorial belt
and the opposite sign above and below it. A
(centrifugally induced) loss of charged particles out through the
light cylinder would cause $\mathbf E \cdot \mathbf B \neq 0$. 
The separatrix is the line dividing the {\em co-rotating
  magnetosphere} from the open field lines region. The {\it polar cap}
is defined as the portion of the star surface connected with the open
field lines. The semi-opening angle
for a dipolar magnetic field can be approximated by
\begin{equation}
  \sin\theta_0=\sqrt{\frac{R_\ast}{R_L}}=\sqrt{R_\ast \ \Omega_\ast} \, ,
\label{semian}
\end{equation}
and so for $R_\ast = 10~{\rm km}$, (\ref{semian}) leads  to  $\theta_0 \sim 0.8^\circ/\sqrt{P/{\rm s}}$~\cite{Vigano:2013uia}.

A polar magnetospheric gap is formed that spans the open field lines
from the stellar surface. In the gap $\mathbf E \cdot \mathbf B \neq
0$, although as we have seen the scalar product vanishes essentially everywhere else
in the near magnetosphere. Co-rotation requires a Goldreich-Julian
charge density, 
\begin{eqnarray}
\rho_{\rm GJ} & = & {\boldsymbol \nabla} \cdot \mathbf E = - {\boldsymbol
  \nabla} \cdot ({\boldsymbol \Omega} \times \mathbf r \times \mathbf B) = -
2 {\boldsymbol \Omega} \cdot \mathbf B \ + \ {\rm relativistic \
  corrections}  \sim
B_p R_\ast^3 \Omega_\ast^3\,,
\end{eqnarray}
screening $\mathbf E \cdot \mathbf B$~\cite{Goldreich:1969sb}. Here, we have used ${\boldsymbol
  \nabla} \times ({\boldsymbol \Omega} \times \mathbf r) = 2
{\boldsymbol \Omega}$, (\ref{Rele}), and (\ref{vossosdelab}). The potential difference between the pole
($p$) and the point ($c$) at the edge of the polar cap (where the
first open force line begins) is given by
\begin{eqnarray}
\Delta \phi  & = &   \int_p^c \mathbf{E} \cdot d \mathbf s= -
\int_p^c (\mathbf{v} \times \mathbf{B}) \cdot d\bm{l} =
\Omega_\ast \ R_\ast \ B_p \int_p^c
\left[(\mathbf{\hat z} \times \mathbf{\hat r}) \times \mathbf{\hat r} \right] \cdot d \bm{l}  
= R_\ast^2 \ \Omega_\ast \ B_p \int_0^{\theta_0} \sin \theta \ d
\theta  \nonumber \\
& = &  R_\ast^2 \ \Omega_\ast \ B_p \ (1-\cos \theta_0 ) \, ,
\label{lunipo}
\end{eqnarray}
where $\mathbf{\hat z} = \cos \theta \ \mathbf{\hat r} - \sin \theta \
\bm{\hat} {\boldsymbol \theta}$, 
$\mathbf{\hat z} \times
\mathbf{\hat{r}} = \sin \theta \, \bm{\hat}{\boldsymbol \phi}$,
$(\mathbf {\hat z} \times \mathbf{\hat{r}}) \times \mathbf{\hat{r}} = \sin \theta \
\bm{\hat} {\boldsymbol \theta}$,  $d \mathbf{s} = R_\ast \ d\theta \
\bm{\hat} {\boldsymbol \theta}$, and $\mathbf v = R_\ast \ {\boldsymbol
  \Omega_\ast} \times \mathbf{\hat r}$ is the velocity on the surface~\cite{Berezinsky:1983}. The
integration goes over the contour on the pulsar surface where to a good approximation the
magnetic field is constant and in the radial direction, i.e., $\mathbf
B \approx  B_p \mathbf{\hat r}$. The gap continually breaks down
(sparking) by forming $e^+e^-$ pairs on a time scale of a few
microseconds. The gap positrons move out along the open field lines,
and the electrons flow to the stellar surface to close the pulsar's
unipolar generator circuit. 

For $\theta_0 \ll 1$, we have $\sin \theta_0 \approx \theta_0$ and because $(1- \cos \theta_0 ) = 2 \sin^2 (\theta_0/2)$  we can rewrite (\ref{lunipo}) as
\begin{equation}
\Delta \phi = \frac{1}{2} \ \Omega_\ast \ B_p \ R^2_\ast \ \theta_0^2 \sim
\frac{1}{2} \ B_p  \  R_\ast^3  \ \Omega_\ast^2 \sim 6.6 \times
10^{19} \left(\frac{B_p}{10^{13}~{\rm G}} \right) \left(\frac{R_\ast}{10~{\rm
        km}}\right)^3 \left(\frac{P_\ast}{1~{\rm ms}}\right)^{-2}~{\rm V} \, .
\label{sunipo}
\end{equation}
Provided that particles of charge $Ze$ can experience the total potential
drop  (\ref{sunipo}), they will be accelerated to the
energy 
\begin{equation}
E  =  Ze \,  \Delta \phi \,,
\label{ocho}
\end{equation}
which corresponds to a maximum achievable particle Lorentz factor of
\begin{eqnarray}
 \gamma_{\rm max}^{\rm acc}=\frac{Ze}{Am_N}\,\Delta \phi\,=7\times
 10^{10}\,\,\frac{Z}{A}\, \left(\frac{B_p}{10^{13}~{\rm G}} \right) \left(\frac{R_\ast}{10~{\rm
        km}}\right)^3 \left(\frac{P_\ast}{1~{\rm ms}}\right)^{-2} \, ,
\end{eqnarray}
where $m_N \sim 1~{\rm GeV}$ is the nucleon mass and $A$ the nucleus baryon
number~\cite{Blasi:2000xm}.  The fiducial value of $P_\ast$ adopted in (\ref{sunipo})
corresponds to the exceptionally fast spinning young pulsars. The
majority of pulsars are born spinning slower. Indeed, the distribution
of pulsar-birth spin periods is Gaussian, centered at
300~ms, with standard deviation of
150~ms~\cite{FaucherGiguere:2005ny}.  Note that most of the pulsars
would accelerate heavy nuclei up a few $10^{7}~{\rm GeV}$. However,
the period of a uniformly rotating neutron star could be as low as
\mbox{$P_\ast \approx 0.288~{\rm ms}$~\cite{Haensel:1999mi}.} Hence,
proto-pulsars spinning initially with $P_\ast \approx 40~{\rm ms}$
would already reach $E \sim 10^{9}~{\rm GeV}$, which is roughly the maximum
energy of Galactic cosmic rays~\cite{Fang:2013cba}.

In the real world, the maximum CR energy within the corotating region
will be limited by energy loss. For example, within the potential drop
the charged particles follow the curved magnetic field lines and so emit
curvature-radiation photons. The energy loss rate or total power
radiated away by a single cosmic ray is~\cite{Ochelkov}
\begin{equation}
- \frac{d E}{dt}  = \frac{2}{3} \frac{Z^2 e^2 }{r_c^2}
\gamma^4 \,,
\end{equation}
where $r_c$ is the curvature radius of the magnetic field
lines. Acceleration gains are balanced by radiative losses,
\begin{equation}
\dot{\gamma} = \frac{Ze \ \Delta \phi}{Am_N} \frac{2 \pi}{\xi P_\ast}-\frac{8\pi^2}{3 P_\ast^2}\frac{Z^2e^2}{Am_N}\ \gamma^4\ ,
\label{eq:dgdt}
\end{equation}
where we have assumed that the total potential drop is available for
particle acceleration over a gap of length $\xi R_{\rm L}$ and that
$r_c \sim R_L$. In the absence of other damping mechanisms, the
radiation reaction limit turns out to be,
\begin{equation}
\label{eq:gamma_curv}
\gamma_{\rm max}^{\rm rad} = \left( \frac{3\pi B_pR_\ast^3}{2 Z e
    P_\ast \xi}\right)^{1/4}\sim 1.1\times 10^8  \left(\frac{Z}{26}
\right)^{-1/4} \xi^{-1/4}    \left(\frac{B_p}{10^{13}~{\rm G}}
    \right)^{1/4}  \left(\frac{P_\ast}{1~{\rm ms}}\right)^{-1/4}
    \left(\frac{R_\ast}{10~{\rm km}}\right)^{3/4}\ .
 \end{equation}
The actual maximum energy that particles can reach at any time within
the corotating magnetosphere is found to be
\begin{equation}
\gamma_{\rm max} = {\rm min} \, \left\{\gamma_{\rm max}^{\rm acc}, \
\gamma_{\rm max}^{\rm rad} \right\}\, .
\end{equation}
Before proceeding we note that $\gamma_{\rm max}^{\rm rad}$ has a very
weak dependence on the fraction $\xi$ of $R_{\rm L}$ over which the
gap extends, with $R_\ast/R_L < \xi <1$. However, the acceleration time to a given energy is strongly
dependent on the unknown $\xi$; namely,
\begin{equation}
\dot{\gamma}^{-1}=\frac{A m_N \gamma}{Z e \ \Delta \phi}\frac
{\xi P_\ast}{2 \pi} =5\times 10^{-6}~{\rm s} \
\left(\frac{\gamma}{10^9} \right) \
\left(\frac{A}{56} \right) \   \left(\frac{Z}{26}\right)^{-1} \
\left(\frac{B_p}{10^{13}~{\rm G} }\right)^{-1}\
  \left(\frac{R_{\ast}}{10~{\rm km} } \right)^{-3}\
    \left(\frac{P_\ast}{1{\rm ms}}\right)^3\ \xi \, .
\end{equation}
Note that for newly-born fast spinning pulsars, 
\begin{equation}
\frac{R_\ast}{R_L} \sim 0.2 \left(\frac{R_\ast}{10~{\rm km}} \right)
  \left(\frac{P_\ast}{1~{\rm ms}} \right)^{-1} \,,
\end{equation}
suggesting that $\xi = {\cal O} (1)$ and consequently that the gap cannot
be far from the star surface.

UHECR nuclei could also suffer photodisintegration in the thermal radiation
fields generated by the star and on the source environment. We will
discuss these phenomena in detail in Sec.~\ref{UFA}. We advance that for the most reasonable
range of neutron star surface temperatures ($T < 10^7~{\rm K}$), a
large fraction of nuclei survive {\it complete} photo-disintegration in the
hostile environment sustained by the thermal radiation field from the
star~\cite{Kotera:2015pya}. However, the  apparently inconsequential photo-disintegration
losses could still be enough to produce a mixed nuclear
composition at the source, with a non-negligible CNO component.  

The spectrum of accelerated UHECRs is determined by the evolution of
the rotational frequency. As the star spins down, the energy of the particles decreases. The total
fluence of UHECRs between energy $E$ and $E + dE$ is found to be
\begin{equation}
\frac{dN}{dE} = \frac{dN}{dt} \ \frac{dt}{d\Omega_\ast} \ \frac{d\Omega_\ast}{dE} \ dE
\, ,
\label{spectrumpulsar}
\end{equation}
where $\dot N = n_{\rm GJ} \pi R_L c$ and  $n_{\rm GJ} = \rho_{\rm
  GJ}/(Ze)$~\cite{Blasi:2000xm}. It is easily seen that when the effect of the
magnetosphere is taken into account the
spin-down luminosity of the pulsar can be well approximated by $I \Omega_\ast
\dot{\Omega}_\ast = \mu^2 \Omega_\ast^4$, where $I \sim \tfrac{2}{5} MR_\ast^2
\sim 10^{45}~{\rm g/cm^2}$ is the moment of
inertia~\cite{Spitkovsky:2006np}. Finally, by differentiating
(\ref{ocho}) it follows that 
$dE/d\Omega_\ast = 2 E/\Omega_\ast$. Therefore,  (\ref{spectrumpulsar}) can be rewritten as
\begin{equation}
\frac{dN}{dE} \sim \frac{\pi I}{Ze \mu E} \, ,
\end{equation}
which explicitly shows that the total fluence of UHECRs accelerated in the neutron star
magnetosphere yields a very hard  spectrum $\propto
E^{-1}$~\cite{Blasi:2000xm}. Recall that simultaneously
reproducing Auger data on the spectrum together with the observed
nuclear composition requires  hard source
spectra~\cite{Aab:2016zth}.

Faraday's dynamo may also operate in the vicinity of a spinning
super-massive black hole. The Blandford-Znajek
model~\cite{Blandford:1977ds,Znajek} permits a direct calculation of
the potential drop in polar cap regions of black hole magnetospheres,
and can explain acceleration of UHECR nuclei with choice of
parameters~\cite{Boldt:1999ge,Boldt:2000dx,Neronov:2007mh}.  However,
photo-nuclear interactions on the ambient photon fields surrounding
the accelerator present formidable challenges for the
model~\cite{Moncada:2017hvq}.

\subsubsection{Fermi acceleration at shock waves}
\label{fermi_acceleration}

\begin{figure}[tpb]
\postscript{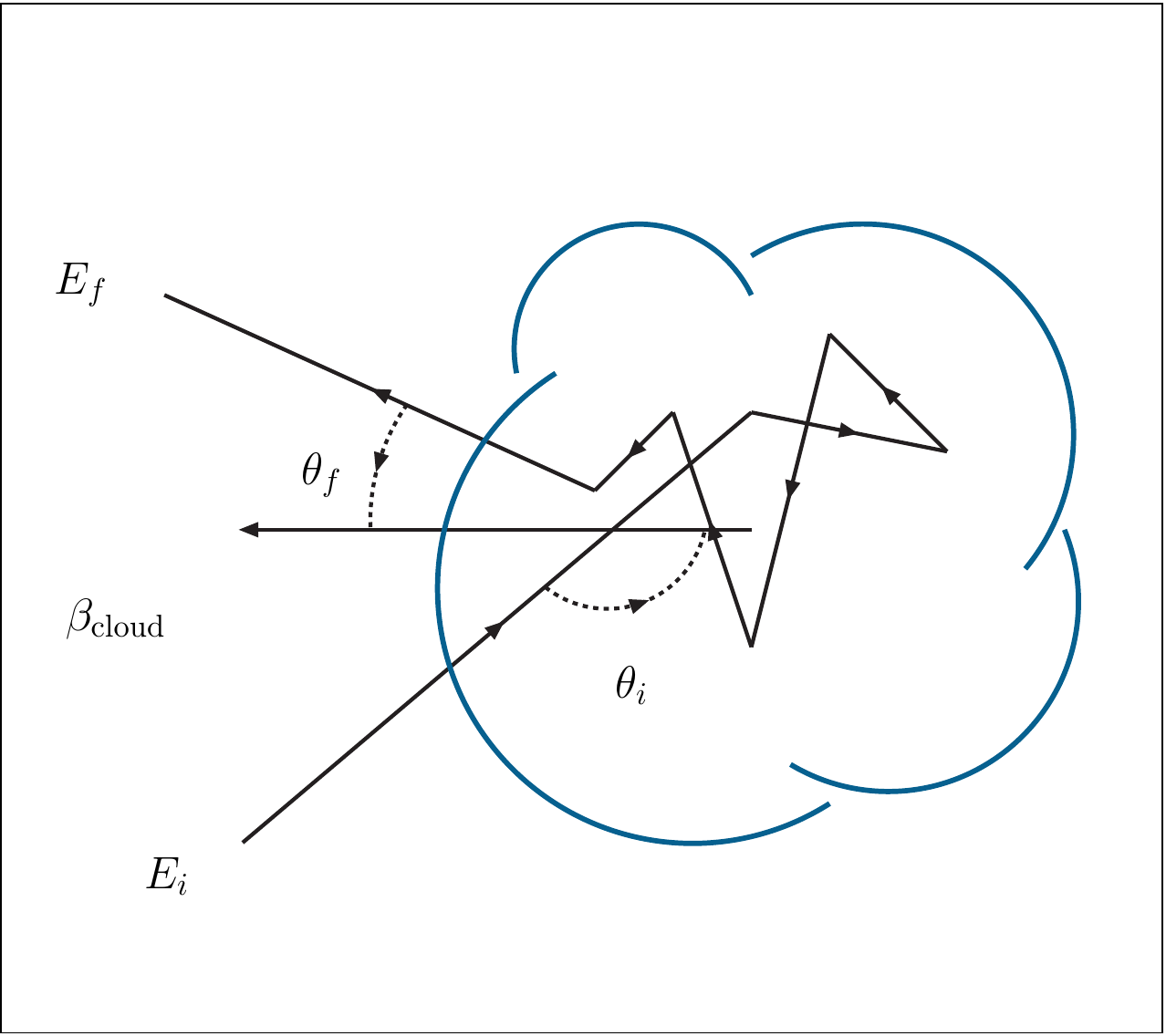}{0.5}
\caption{A sketch of 2nd order Fermi acceleration by scattering off magnetic clouds.}
\label{fig:9}
\end{figure}

In his original analysis of the late '40s, Fermi~\cite{Fermi:1949ee} considered the
scattering of CRs on moving magnetized clouds. A sketch of one of
these encounters is shown in Fig.~\ref{fig:9}. Consider a CR entering
into a single cloud with energy $E_i$ and incident angle $\theta_i$
with the cloud's direction undergoing diffuse scattering on the
irregularities in the magnetic field.  After diffusing inside the
cloud, the particle's average motion coincides with that of the gas
cloud. The energy gain by the particle, which emerges at an angle
$\theta_f$ with energy $E_f$, can be obtained by applying Lorentz
transformations between the laboratory frame (unprimed) and the cloud
frame (primed).  In the rest frame of the moving cloud, the CR
particle has a total initial energy
\begin{equation}
  E_i' = \gamma_{\rm cloud} \,E_i\, (1 - u_{\rm cloud}\, \cos \theta_i)\,, 
\end{equation}
where $\gamma_{\rm cloud}$ and $u_{\rm cloud}$ are the Lorentz factor
and velocity of the cloud, respectively.  In the frame of the cloud we
expect no change in energy ($E_i' = E_f'$), because all the
scatterings inside the cloud are due only to motion in the magnetic
field (so-called collisionless scattering).\footnote{In
  collision-dominated plasmas, particle-particle collisions drive the
  plasma to thermal equilibrium. If an individual particle gets more
  energy than average, it will immediately transfer energy via
  collisions to slower particles. Scatterings are inelastic. In
  collisionless plasmas, individual charged particles interact with
  the background magnetic field. Scatterings are nearly elastic.} There is elastic
scattering between the CR and the cloud as a whole, which is much more
massive than the CR. Transforming to the laboratory frame we find that
the energy of the particle after its encounter with the cloud is
\begin{equation}
E_f = \gamma_{\rm cloud} \,E_f'\, (1 + u_{\rm cloud} \cos \theta_f)\,.
\end{equation}
The fractional energy change in the 
laboratory frame is then
\begin{equation}
\frac{\Delta E}{E} = \frac{E_f-E_i}{E_i} = \frac{1 - u_{\rm cloud} \cos \theta_i + u_{\rm cloud} \cos \theta_f - u_{\rm cloud}^2 \cos \theta_i \cos \theta_f}{1 - u_{\rm cloud}^2} - 1 \,.
\label{flash0}
\end{equation}
To determine the average energy gain one must obtain the average
values of $\cos \theta_i$ and $\cos \theta_f$.  Inside the cloud, the
CR scatters off magnetic irregularities many times and its direction
becomes randomized, so that $\langle \cos \theta_f\rangle = 0.$ The
average value of $\cos \theta_i$ depends on the rate at which CRs
collide with clouds at different angles. The collisionless scattered
particle will gain energy in a head-on collision ($\theta_i>\pi/2$)
and lose energy by tail-end ($\theta_i<\pi/2$) scattering. The net
increase of its energy is a statistical effect. The average value of
$\cos \theta_i$ depends on the relative velocity between the cloud and
the particle. The probability $P$ per unit solid angle $\Omega$ of
having a collision at angle $\theta_i$ is proportional to $(v - u_{\rm
  cloud} \cos \theta_i)$, where $v$ is the CR speed. In the
ultrarelativistic limit, i.e., $v \sim 1$ (as seen in the laboratory
frame),
\begin{equation}
\frac{dP}{d \Omega_i} \propto (1 - u_{\rm cloud} \cos\theta_i)\,,
\end{equation}
so
\begin{equation}
\langle \cos \theta_i \rangle  = \left\{\int_0^\pi  \cos \theta_i \
  \frac{dP}{d\Omega_i} \right\} {\large /} \left\{\int_0^\pi \frac{dP}{d\Omega_i}
    d\Omega_i \right\} = 
-\frac{u_{\rm cloud}}{3}\,.
\label{flash}
\end{equation}
Now, inserting~(\ref{flash}) into (\ref{flash0}), one obtains for 
$u_{\rm cloud} \ll 1$, 
\begin{equation}
\frac{\langle\Delta E\rangle}{E} = \frac{1 +u_{\rm cloud}^2/3}{1-u_{\rm cloud}^2} - 1 \approx \frac{4}{3} \,u_{\rm cloud}^2\,.
\label{ja}
\end{equation}
Note that $\langle \Delta E \rangle/E \propto u_{\rm cloud}^2$, so even though
the average magnetic field may vanish, there can still be a net
transfer of the macroscopic kinetic energy from the moving cloud to
the particle. However, the average energy gain is very small, because
$u_{\rm cloud}^2 \ll 1$.  This acceleration process is very similar to a
thermodynamical system of two gases, which tries to come into thermal
equilibrium~\cite{Drury:1994fg}. Correspondingly, the spectrum of CRs
should follow a thermal spectrum which might be in conflict with the
observed power-law.

By the mid '50s, Fermi realized that a more efficient acceleration
process may occur when particles scatter back and forth between two
ends of a contracting magnetic bottle, as the particles would gain
energy at every scattering~\cite{Fermi:1954ofk}. Ever since the late '70s it became clear that
a simple version of this acceleration process may occur in the
vicinity of magnetized collisionless plasma shocks occurring in
astrophysical
environments~\cite{Krymskii:1977,Axford:1977,Bell:1978zc,Bell:1978fj,Blandford:1978ky,Lagage:1983zz,Drury:1983zz,Blandford:1987pw}. These
shocks originate in the sudden change of density $\rho$,
temperature $T$, and pressure $P$ that decelerate a supersonic
flow. Before proceeding, we pause to introduce some notation.  The
state of a system in (local) thermodynamic equilibrium may be defined
by any two intensive variables, such as $T$, $P$, the specific volume
$v \equiv \rho^{-1}$, the internal energy per unit mass $\varepsilon$,
or the entropy per unit mass $s$. The first law of thermodynamics may
be stated that in going from state 1 to state 2, the change in
internal energy per unit mass must equal the sum of the heat added per
unit mass $q$ to the work done per unit mass on the system, $d
\varepsilon = T ds - P dv$. This relation is valid for both reversible
and irreversible processes. Any process in an open system which does
not exchange heat with the environment is said to be adiabatic. If the
process is furthermore reversible, it follows that $\delta q = 0$ in
each infinitesimal step, so that the $\delta s = \delta q/T = 0$. The
specific entropy must in other words stay constant in any reversible,
adiabatic process. Such a process is for this reason called
isentropic. The sound speed is defined as
\begin{equation}
c_s^2 = \left. \left(\frac{\partial P}{\partial \rho} \right)\right|_s   = \frac{\gamma
  P}{\rho} \, ,
\label{sound}
\end{equation}
where the derivative is taken isentropically so that $P \rho^{-\gamma} =$
constant, with $\gamma$ the adiabatic index. The value of the
adiabatic index is $\gamma = 5/3$ for monatomic gases, $\gamma = 7/5$ for diatomic
gases, $\gamma = 9/7$ for three-atomic gases with non-static bindings,
and $\gamma = 4/3$ for a relativistic gas.

Hereafter we assume that: {\it (i)}~the gas
density is low enough so particle-particle collisions are rare, and
{\it (ii)}~the gas passing through the shock is not dissociated, i.e.,
both the mass of a gas molecule and the adiabatic index remain
unchanged. Then changes occur in $\rho$, $T$, $P$, and velocity. 
 For simplicity, throughout we consider magnetized plasmas near a
perpendicular shock, i.e. one in which the propagation direction of
the shock is aligned with the incident magnetic field.  The
description of the plasma and fields near a parallel shock is
considerably more complicated than that of a perpendicular shock and
can be found in e.g.,~\cite{Protheroe:1998hp,Bell:2013vxa}.  At this stage, it is worthwhile to point out that despite the fact we
cannot describe the way that the fluid behaves inside the shock (say,
in a few mean free paths of the shock), we can portray how the fluid
conditions differ from side to side of the shock, i.e., as a result of
shock passing. We will show that this characterization is sufficient
for the purpose of modeling UHECR acceleration.

\begin{figure}[tpb]
\postscript{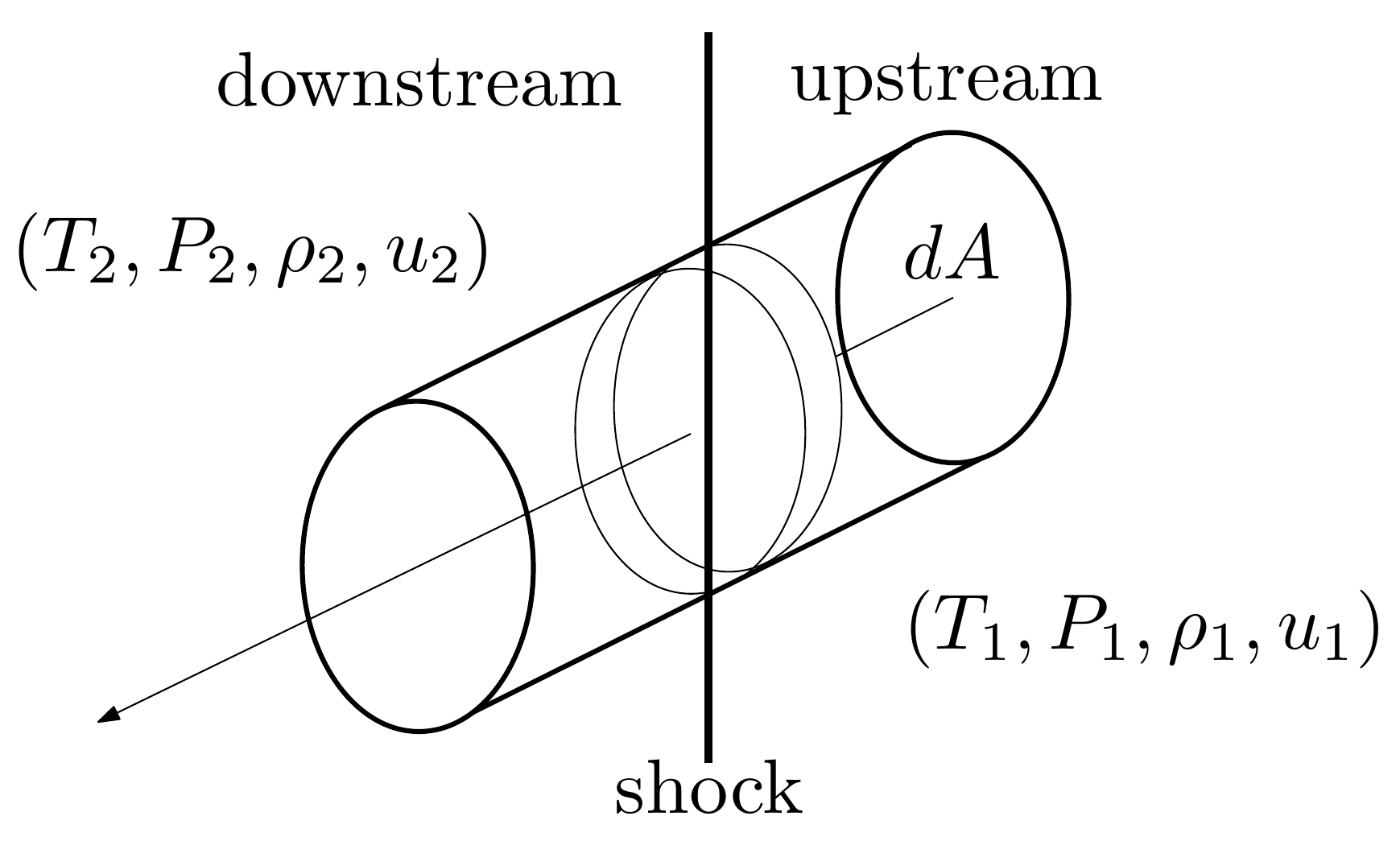}{0.5}
\caption{A sketch of a perpendicular shock.}
\label{fig:10}
\end{figure}

Consider a propagating supersonic flow. In the rest frame of the
shock, unshocked gas moving faster than its sound speed approaches the
shock from the right direction.  The preshock conditions are
characterized by $T_1,\ P_1,\ \rho_1, \, u_1$, whereas the postshock
conditions are described by $T_2,\ P_2,\ \rho_2$, $u_2$.  For a steady
shock, the mass contained in the cylinder shown in Fig.~\ref{fig:10}
is constant. This implies that the mass flux into the shock is the
same as the mass flux out,
\begin{equation}
\rho_1 \ u_1 = \rho_2 \ u_2 \, .
\label{RH1}
\end{equation}
The momentum density contained in the cylinder is also constant. Thus,
the net pressure force on the cylinder plus the net momentum flux
into the cylinder must be zero, yielding
\begin{equation}
(P_1 \ dA - P_2 \ dA ) + (\rho_1 \ u_1^2 \
dA - \rho_2 \ u_2^2 \ dA) = 0 \,, 
\end{equation}
or equivalently
\begin{equation}
P_1 + \rho_1 \ u_1^2 = P_2 + \rho_2 \  u_2^2 \, .
\label{RH2}
\end{equation}
Finally, the energy contained in the cylinder is a
constant. Therefore, the net flow of energy into the cylinder plus the
work done on a gas in the cylinder is equal to zero. This leads to
\begin{equation}
\left[u_1 \left(\rho_1 \ \varepsilon_1 +
  \frac{1}{2} \ \rho_1 \ u_1^2 \right) \ dA - u_2
\left(\rho_2 \ \varepsilon_2 + \frac{1}{2} \ \rho_2 \ u_2^2 \right) \ dA
\right] + ( P_1 \ u_1
- P_2 \ u_2 ) \ dA =  0 \,,
\end{equation}
or equivalently,
\begin{equation}
u_1 \left[P_1 + \rho_1 \left(\varepsilon_1 +
  \frac{1}{2}  \ \rho_1   \ u_1^2 \right) \right] = u_2
\left[ P_2 + \rho_2 \left( \varepsilon_2 + \frac{1}{2} \ \rho_2 \  u_2^2  \right) \right] \, .
\label{RHenergy}
\end{equation}
Substituting (\ref{RH1}) into (\ref{RHenergy}) it follows that 
\begin{equation}
\varepsilon_1 + \frac{1}{2} u_1^2 + \frac{P_1}{\rho_1} = \varepsilon_2
+ \frac{1}{2}  u_2^2 + \frac{P_2}{\rho_2} \, .
\label{RH3}
\end{equation}
Equations (\ref{RH1}), (\ref{RH2}), and (\ref{RH3}) are the well-known
Rankine-Hugoniot (RH) relations. Substituting (\ref{sound}) in (\ref{RH2}) and (\ref{RH3}) we obtain.
\begin{subequations}
\label{memd1}
\begin{equation}
u_1 \left( 1 + \frac{c_{s,1}^2}{\gamma \ u_1^2} \right) = u_2 \left( 1 +
  \frac{c_{s,2}^2}{\gamma \ u_2^2} \right) \, .
\end{equation}
and
\begin{equation}
\frac{1}{2} u_1^2 + \frac{c_{s,1}^2}{\gamma -1} = \frac{1}{2} u_2^2  +
\frac{c_{s,2}^2}{\gamma -1} \, .
\end{equation}
\end{subequations}
It is easily seen that the entropy per unit mass of the
matter entering the shock is lower than that of the gas leaving the
shock: the shock produces a great disorganization of the gas.

The dimensionless number that characterizes the strength of a shock is
the Mach number, the ratio of the unshocked gas speed to the upstream sound
speed,
\begin{equation}
{\cal M} = \frac{u_1}{c_{s,1}} = \left(\frac{\rho_1 u_1^2}{\gamma P_1}
\right)^{1/2} \, .
\end{equation}
Given that ${\cal M} > 1$, it is straightforward to see that $\rho_2 >
\rho_1$ (shocks compress), $u_2 < u_1$ (shocks decelerate), $P_2 >
P_1$ (shocks increase pressure), and $T_2 > T_1$ (shocks heat). The
latter may seem surprising, given that the shock is considered to be
adiabatic: although the process has been adiabatic, in that $\delta
q/dt = 0$, the gas has changed its adiabat; its entropy has increased
as a consequence of the shock converting kinetic energy into thermal,
internal energy.\footnote{A shock converts supersonic gas into denser,
  slower moving, higher pressure, subsonic gas. It increases the
  specific entropy of the gas. In another terminology, a shock shifts
  gas to a higher adiabat.  An adiabat is a locus of constant entropy ($T\propto \rho^{\gamma -1}$) 
in the density-temperature plane. Gas
    can move adiabatically along an adiabat, while changes in entropy
    move it from one adiabat to another.} The total energy of the
  post-shock gas is lower (in the shock rest frame) because of the
  work done on the gas by {\it viscosity} and pressure in the shock.
  It is this aspect of the shock that causes irreversibility, thus
  defining an ``arrow of time.''

The compression ratio achieved by the shock is defined by 
\begin{equation}
\zeta =\frac{\rho_2}{\rho_1} = \frac{u_1}{u_2}  ,
\end{equation}
so that $\psi = \zeta^{-1}$ is the inverse compression ratio, and
therefore $u_1 = c_{s,1}  {\cal M}$ and $u_2 = c_{s,1} \psi 
{\cal M}$.  Substituting these expressions in (\ref{memd1}) leads to
\begin{subequations} 
\begin{equation}
\left(1 + \frac{1}{\gamma \ \psi \ {\cal M}^2} \right) = \psi \left(1 +
    \frac{c_{s,2}^2}{c_{s,1}^2} \frac{1}{\gamma  \ \psi \ {\cal M}^2}  \right)
\end{equation}
and
\begin{equation}
\frac{1}{2} {\cal M}^2 + \frac{1}{\gamma -1} = \frac{1}{2} \psi^2
{\cal M}^2 + \frac{1}{\gamma -1} \frac{c_{s,2}^2}{c_{s,1}^2} \,,
\end{equation}
\end{subequations}
which can be rewritten as two expressions for $c_{s,2}^2/c_{s,1}^2$;
namely,
\begin{equation}
\frac{c_{s,2}^2}{c_{s,1}^2} = \psi + \gamma \psi {\cal M}^2 ( 1 -
\psi) = 1 + \frac{\gamma -1}{2} {\cal M}^2 ( 1 -\psi^2) \, .
\end{equation}
Combining these two expressions yields an equation for $\psi$
\begin{equation}
(\psi -1 ) + \gamma \psi {\cal M}^2 (1 -\psi) - \frac{\gamma -1 }{2}
{\cal M}^2 (\psi +1) = 0 \, .
\end{equation}
Note that if $\psi = 1$ there is no shock as $\rho_1 = \rho_2$, $u_1 =
u_2$, and $P_1 = P_2$. Thus,
\begin{equation}
1 - \gamma \psi {\cal M}^2 + \frac{\gamma -1}{2} {\cal M}^2 (\psi +1)
= 0
\end{equation}
yielding
\begin{equation}
\psi = \frac{\gamma -1}{\gamma +1} + \frac{2}{\gamma +1} \frac{1}{{\cal
  M}^2} \  .
\label{memd2}
\end{equation}
We can now substitute (\ref{memd2}) into (\ref{RH1}), (\ref{RH2}), and
(\ref{RH3}) to rewrite the RH jump conditions in a more useful form
\begin{subequations}
\begin{equation}
\frac{\rho_1}{\rho_2} = \frac{u_2}{u_1} = \frac{\gamma -1}{\gamma +1} +
\frac{2}{\gamma +1 } \frac{1}{{\cal M}^2} \,,
  \end{equation}
\begin{equation}
\frac{P_2}{P_1} = \frac{2 \gamma}{\gamma +1} {\cal M}^2 - \frac{\gamma
  -1}{\gamma +1} \,,
\end{equation}
and
\begin{equation}
\frac{T_2}{T_1} = \frac{c_{s,2}^2}{c_{s,1}^2} = \frac{P_2}{P_1}
  \frac{\rho_1}{\rho_2} = \left(\frac{2 \gamma}{\gamma +1} {\cal M}^2
    - \frac{\gamma -1}{\gamma +1} \right) \left( \frac{\gamma
      -1}{\gamma +1} + \frac{2}{1 + \gamma} \frac{1}{{\cal M}^2}
  \right) \, .
\end{equation}
\end{subequations}
Note that the thermodynamics variables depend on the Mach number of
the upstream gas ${\cal M}$ and the adiabatic index 
$\gamma$, which specify the way that energy is shared between the
internal energy and the kinetic flow. For a very strong shock, ${\cal
  M} \to \infty$, yielding $\zeta \to (\gamma +1)/(\gamma -1)$ and $P_2/P_1
\to \infty$. Note also that $\gamma \to 1$ gives the maximum compression
ratio, for which $\rho_2/\rho_1 \to {\cal M}^2$ and $T_1 \to T_2$, but $P_2/P_1$ can become arbitrarily large.

\begin{figure}[tpb]
\postscript{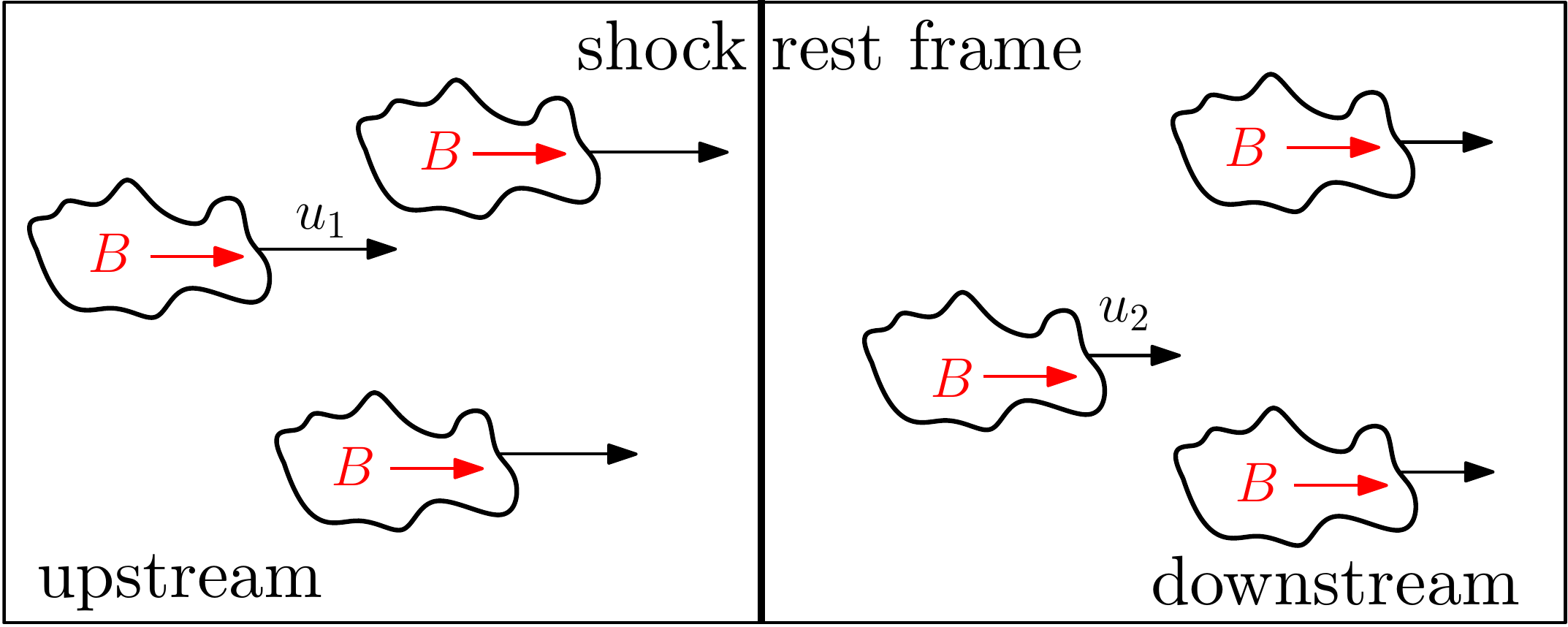}{0.65}
\postscript{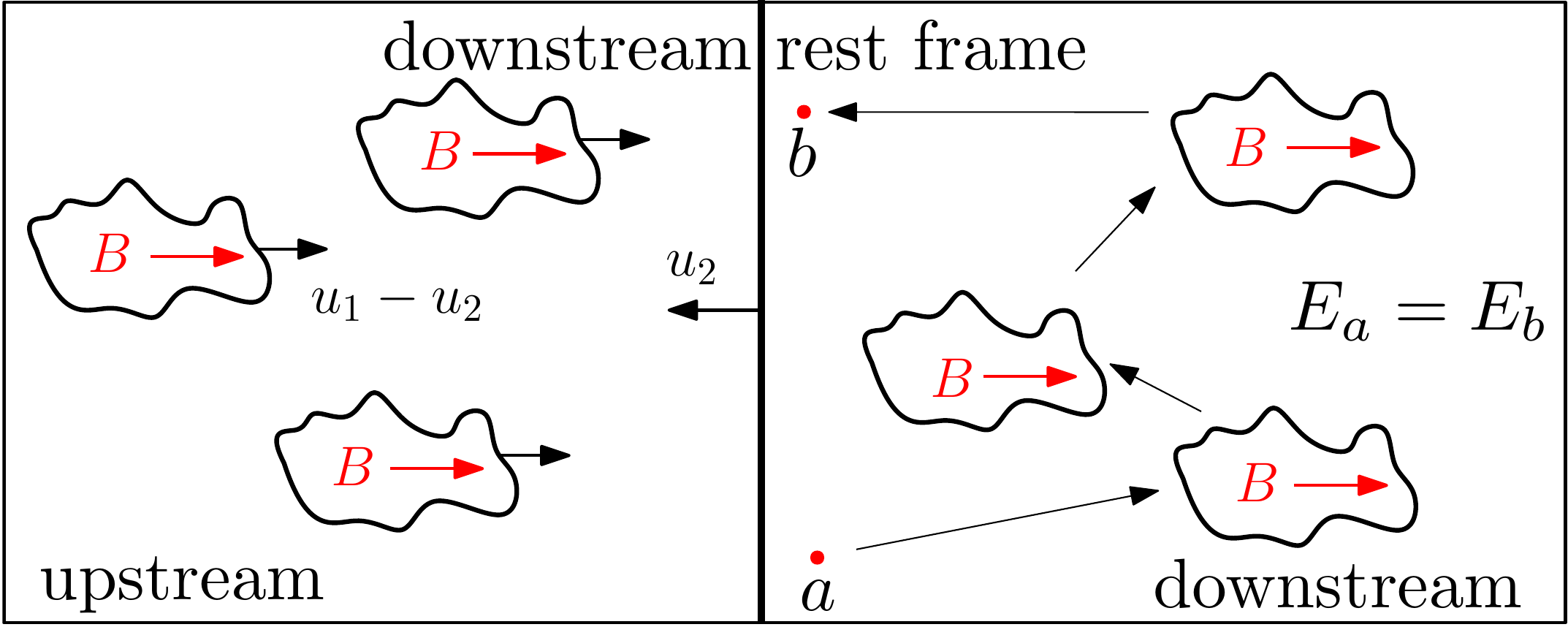}{0.65}
\postscript{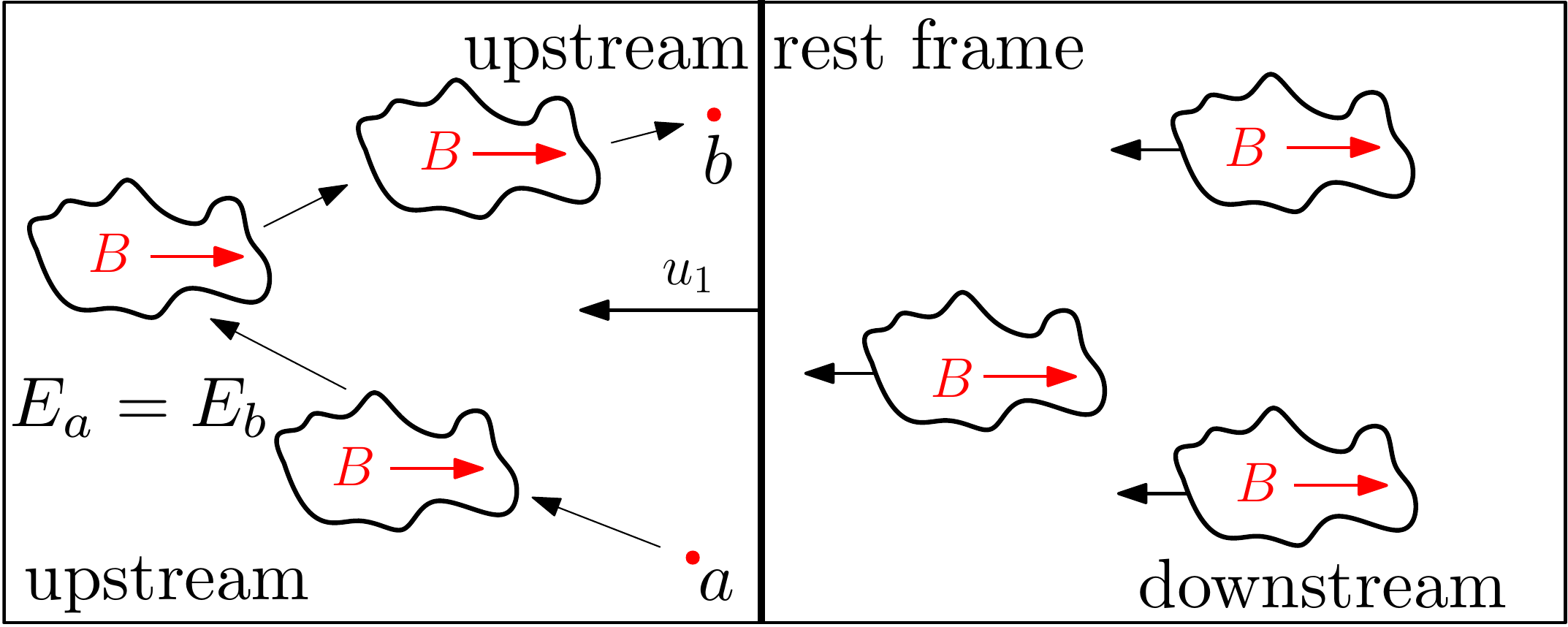}{0.65}
\caption{Different perspectives of 1st order Fermi acceleration by scattering off
  a plasma shock.}
\label{fig:11}
\end{figure}

Suppose that a strong (nonrelativistic) shock wave propagates through
the plasma as sketched in the left panel of Fig.~\ref{fig:11}.  In the
frame stationary with respect to the shock, the upstream flow
approaches with speed $u_1$ and the downstream flow recedes with speed
$u_2$. The RH conservation relations imply that the upstream velocity
$u_1$ (ahead of the shock) is much higher than the downstream velocity
$u_2$ (behind the shock).  Therefore, when measured in the upstream
rest frame, the quantity $u = u_1 - u_2$ is the speed of the shocked
fluid and $u_1 = u_{\rm shock}$ is the speed of the shock.  In the
primed frame stationary with respect to the downstream fluid, $u_2 =
u_{\rm shock}$ and $u = u_1-u_2$ is the speed of the upstream
fluid. Hence, because of the converging flow -- whichever side of the
shock you are on, if you are moving with the plasma, the plasma on the
other side of the shock is approaching you with velocity $u$ -- to
first order there are only head-on collisions for particles crossing
the shock front; see Fig.~\ref{fig:11}. The acceleration process,
although stochastic, always leads to a gain in energy. In order to
work out the energy gain per shock crossing, we can visualize magnetic
irregularities on either side of the shock as clouds of magnetized
plasma of Fermi's original theory. As one can see in Fig.~\ref{fig:11}
there is an asymmetry because upstream particles always return to the
shock, whereas downstream particles may be advected and never come
back to the shock.

\begin{figure}[tpb]
\postscript{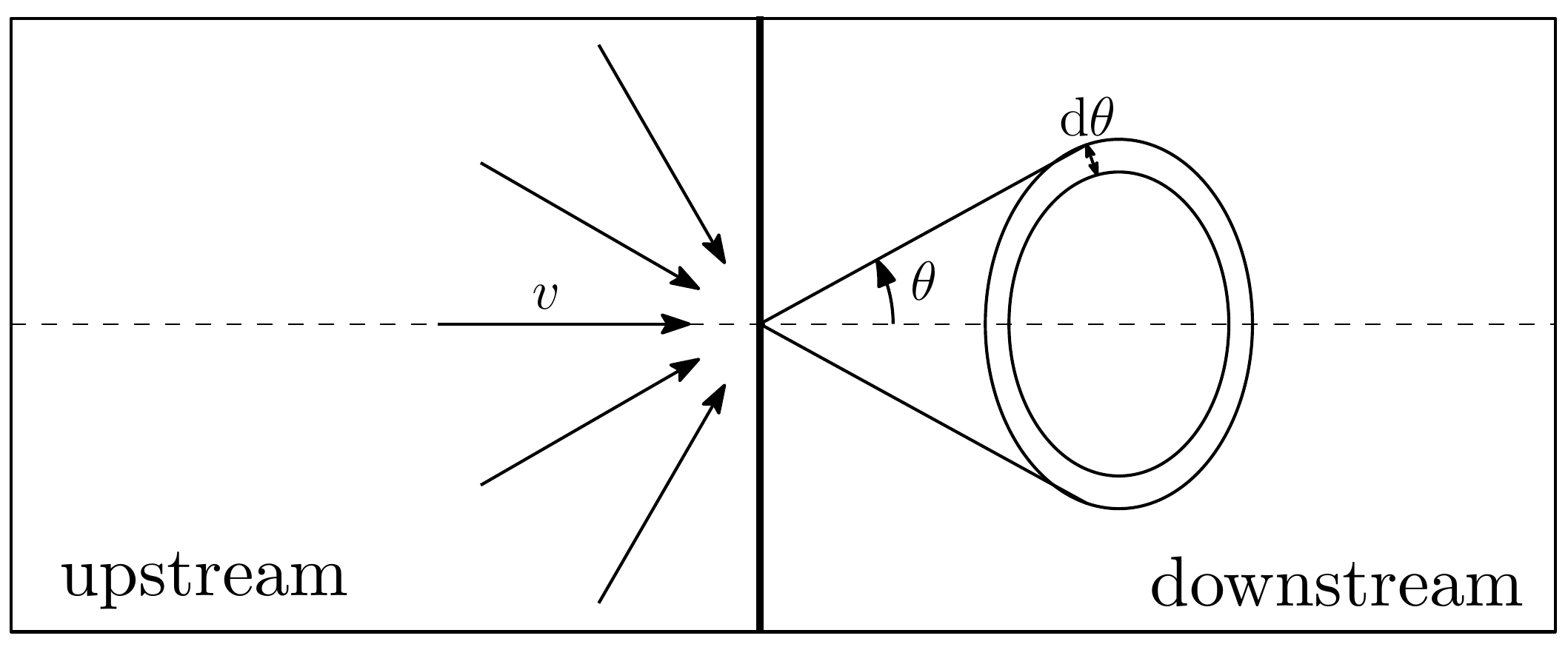}{0.65}
\caption{Properties of CRs undergoing diffusive shock
  acceleration. The number density of accelerated particles close to
  the shock is isotropic, $dn_{1,2} = n_{1,2} \ d\Omega/ (4\pi)$. The
  rate at which particles cross the shock is proportional to $ v \cos
  \theta$, with $0 < \theta < \pi/2$. The number of particles between
  $\theta$ and $\theta + d \theta$ is proportional to $\sin \theta \,
  d\theta$. The differential probability for an ultra-relativistic ($v
  \simeq c$) particle to cross the shock is $dP \propto \sin \theta \,
  \cos \theta \, d\theta$.}
\label{fig:12}
\end{figure}

Consider an upstream particle that has initial energy $E$ and initial
momentum $p$. The particle ``sees'' the downstream flow with a
velocity $u = u_1 -u_2$ and a Lorentz factor $\gamma_u$. To determine
the energy of the particle in the downstream rest frame  we perform a
Lorentz transformation
\begin{equation}
E' = \gamma_u (E + u \ p \cos \theta) \, ,
\end{equation}
where $\theta$ is the incident angle; see Fig.~\ref{fig:12}.
Now, assume that the shock is non-relativistic (i.e., $\gamma_u \sim 1$)
and that the particle is relativistic (i.e., $E = p$).  Under these conditions the
energy gain per half-cycle (say, upstream $\to$ downstream) is
\begin{equation}
\frac{\delta E}{E} = u \cos \theta \, .
\label{deltaEsobreE}
\end{equation}
Assuming that the particles upstream and downstream of the shock are
rapidly isotropized by magnetic field irregularities, the distribution
function for particles crossing the shock is $f(\theta) \propto \sin
\theta \ \cos \theta$; see Fig.~\ref{fig:12}. The total probability must equal
unity, $\int_0^{\pi/2} f(\theta) d\theta = 1$, and so the normalized
differential probability for a particle to cross the shock is found to
be, 
\begin{equation}
dP = 2 \sin \theta \ \cos \theta \ d\theta \, .
\label{P-dist-shock}
\end{equation}
The average gain per half-cycle $\langle \delta E \rangle/E$ is (\ref{deltaEsobreE})
averaged over the differential probability distribution (\ref{P-dist-shock})
\begin{equation}
\frac{\langle \delta E \rangle}{E}   = 2 u \int_0^{\pi/2} d
\theta \ \cos^2 \theta \ \sin \theta = \frac{2}{3} u \, .
\end{equation}
Now, it is straightforward to see that $\langle \delta E \rangle/E |_{\rm up \to down} =
\langle \delta E \rangle/E |_{\rm down \to up}$, and therefore for a full cycle
up $\to$ down and down $\to$ up the energy gain is
\begin{equation}
\frac{\langle \Delta E \rangle}{E}  = \frac{4}{3} u = \frac{4}{3} (u_1
- u_2) \, .
\label{deltaEfermi2nd}
\end{equation} 
Note this is first order in $u$, and is therefore more efficient than Fermi's original mechanism.

An attractive feature of Fermi acceleration is its prediction of a
power-law flux of CRs. Consider a test-particle with momentum $p$ in
the rest frame of the upstream fluid. The
particle's momentum distribution is isotropic in the fluid rest
frame. For pitch angles $\pi/2<\theta_i<\pi$ relative to the shock
velocity vector  the particle enters the
downstream region and has {\it on average} the relative momentum
$p[1+2(u_1-u_2)/3]$. Subsequent diffusion in
the downstream region `re-isotropizes' the particle's momentum
distribution in the fluid rest frame. As the particle diffuses back
into the upstream region (for pitch angles $0<\theta_f<\pi/2$) it has
{\it gained} an average momentum of $\langle\Delta p\rangle/p \simeq
4(u_1-u_2)/3$. This means that the momentum
gain of a particle per time is proportional to its momentum,
\begin{equation}
\dot{p} = p/t_\text{gain} \, .
\label{ganancia}
\end{equation}
On the other hand, the loss of particles from the acceleration region
is proportional to the CR distribution function
\begin{equation}
\dot{f} = - f/t_\text{loss} \, . 
\label{perdida}
\end{equation}
Therefore, taking the ratio (\ref{ganancia})/(\ref{perdida}) we first obtain 
\begin{equation}
df/dp= -\alpha f/p \,,   
\end{equation}
and after integration $f(p) \propto p^{-\alpha},$ with
$\alpha=t_\text{gain}/t_\text{loss}$.  If the acceleration cycle
across the shock takes the time $\Delta t$ we have already identified
$\Delta t/t_\text{gain} = \langle\Delta p\rangle/p \simeq
4(u_1 - u_2)/3$. Now, $\Delta
t/t_\text{loss}$ represents the probability that the
particle leaves the accelerator after each cycle. This is given by
\begin{equation}
\frac{\Delta t}{t_{\rm loss}}  \simeq \frac{R_{\rm out}}{R_{\rm in}} \,,
\end{equation}
where $R_{\rm in}$ is the number of particles per unit time (rate)
that begins a cycle and $R_{\rm out}$ is the number of particles per
unit time that leaves the system. The rate at which CRs cross from upstream to downstream is given by the
projection of the isotropic CR flux onto the plane shock front
\begin{equation}
R_{\rm in} ({\rm up \to down})= \int_{\rm up \to down} dn_1 \ v \ \cos \theta = \frac{v \, n_1 }{4
  \pi} \int_0^{2\pi} d\phi \int_0^{1} d\cos \theta = \frac{1}{4} v \, n_1  \, ,
\end{equation}
see Fig.~\ref{fig:12}. The particle lost rate (advected) downstream is $R_{\rm
  out} = n_2 u_2$, and so taking $v \simeq  c$
\begin{equation}
\frac{R_{\rm iout}}{R_{\rm in}} = \frac{4 n_1 u_2}{n_2} = 4 u_1 \ll 1 \,,
\end{equation}
showing that most of the particles perform many cycles. Putting all
this together, 
\begin{equation}
\alpha \simeq
 \frac{3 \, u_1}{u_1- u_2} = \frac{3 \, \zeta}{\zeta-1} \ .
\end{equation}
The differential energy spectrum $dN/dE \propto E^{-\gamma}$ is related to the
phase space distribution in momentum space by $dN = 4\pi p^2
f(p) dp$ and hence $\gamma = \alpha-2$. 

All in all, the steepness of the power-law spectrum at the sources
depends on the shock compression ratio via the adiabatic index. The
different regions of the parameter space can be easily visualized with
a simple phenomenological
argument~\cite{Baerwald:2013pu}.  Consider an expanding shell that
magnetically confines UHECRs. Assuming that the CRs are isotropically
distributed in the shell, the number of escaping particles is
proportional to the volume. The shell width expands as $\delta r
\propto r$. This implies that the volume of the plasma increases as $V
\propto r^3$ and the total energy scales as $U \propto V^{-(\gamma
  -1)} \propto r^{-3 (\gamma -1)}$. Now, using the scaling of the
volume and the total energy we can derive the scaling of the magnetic
field inside the plasma $B \propto \sqrt{U/V} \propto r^{-3
  \gamma/2}$. If we further assume that the energy of a single
particle in the plasma scales in the same way as the total energy of
the plasma, then the Larmor radius of the particle changes with time
(or radius) as $r_L \propto E/B \propto r^{-3 (\gamma -2)/2}$. For a
relativistic gas, $\gamma = 4/3$ yielding $r_L \propto r$, and so the
ratio $r_L/\delta r$ is constant.  This means that a relativistic gas
provides a critical balance for stability between losses and
escape. For $\gamma > 4/3$, the adiabatic energy loss is faster than
the escape, and the particles are more strongly confined for larger
radii.  For $\gamma < 4/3$, the Larmor radius increases faster than
the particles lose energy, and the particles are getting less confined
at larger radii.  For a monoatomic gas, $\zeta = 4$ and the spectral
index becomes $\gamma =2$. For a three-atomic gas with non-static
bindings $\zeta = 8$, yielding a hard source spectrum with spectral
index $\gamma = 1.4$. 

The acceleration time scale is given by
\begin{equation}
\left(\frac{1}{E} \frac{dE}{dt} \right)^{-1} = \frac{t_{\rm
      cycle}}{ \langle \Delta E \rangle/E} \,,
\label{Erate}
\end{equation}
where $t_{\rm cycle}$ is the cycle time for one back-and-forth
encounter. Diffusion takes place in the presence of advection at speed $u_2$
in the downstream direction.  The characteristic distance a CR 
diffuses in time $t_D$ is $l_D = \sqrt{D_2 t_D}$ where $D_2$ is the diffusion
coefficient in the downstream region.  The distance advected in
this time is simply $u_2 t_D$.  If $\sqrt{D_2 t_D} \gg u_2 t_D$ the
CR has a very high probability of returning to the shock, whereas
 if $\sqrt{D_2 t_D} \ll u_2 t_D$ the CR has a very high
probability of never returning to the shock (i.e. it has
effectively escaped downstream).  So, we set  $\sqrt{D_2 t_D} = u_2
t_D$
to characterized the diffusion time $t_D = D_2/u_2^2$ and define a
distance $l_D = D_2/u_2$ downstream of the shock which is
effectively a boundary between the region closer to the shock
where the particles will usually return to the shock and the
region farther from the shock in which the particles will usually
be advected downstream never to return.  There are $n_2
D_2/u_2$ particles per unit area of shock between the shock and
this boundary.  Dividing this by $R_{\rm in} ({\rm down \to up}) $ we obtain the
average time spent downstream before returning to the shock
\begin{equation}
t_2 \approx \frac{4}{v} \, \frac{D_2}{u_2}.
\end{equation}
Now, we must consider the other half of the cycle, after the CR has
crossed the shock from downstream to upstream until it returns to
the shock.  In this case we can define a boundary at a distance
$D_1/u_1$ upstream of the shock such that nearly all particles
upstream of this boundary have never encountered the shock, and
nearly all the particles between this boundary and the shock have
diffused there from the shock.  Then dividing the number of
particles per unit area of shock between the shock and this
boundary, $n_1 D_1/u_1$, by $R_{\rm in}({\rm up \to down})$ we obtain the
average time spent upstream before returning to the shock
\begin{equation}
t_1 \approx {4 \over v} {D_1 \over u_1},
\end{equation}
and hence the time for a full  cycle is
\begin{equation}
t_{\rm cycle} \approx {4 \over v} \left( {D_1 \over u_1} + {D_2
\over u_2} \right).
\label{tcycle}
\end{equation}
To proceed we must estimate the diffusion coefficient, which can be interpreted as
\begin{equation}
D = \frac{1}{3} \lambda_D \ v \,, 
\label{Dgen}
\end{equation}
where $v$ is the particle velocity and $\lambda_D$ the diffusion mean
free path.  The diffusion length cannot be smaller than the Larmor
radius because energetic particles cannot respond to irregularities in
the magnetic field smaller that the particle
gyroradius~\cite{Lagage:1983zz}.  The minimum diffusion
coefficient, which gives the maximum possible acceleration rate,
corresponds to the Bohm diffusion limit.  Substituting (\ref{LARMOR}) into
(\ref{Dgen}) gives
\begin{equation}
D_{\rm min} = \frac{r_L v}{3} \sim \frac{1}{3} \ \frac{E \ v}{Z
  \ \sqrt{4 \pi \, \alpha}  B} \, .
\end{equation}
Taking $D_1 = D_2 = D_{\rm min}$ in (\ref{tcycle}), and
inserting the output into (\ref{Erate}) yields  an
expression for the acceleration rate,  which does not depend on the
CR energy  because $t_{\rm cycle} \propto E$.  
Assuming that the acceleration is continuous, the constraint due to the finite lifetime $\tau$ of the shock yields,
\begin{equation}
E  \lesssim  \varkappa (\zeta) \  Z  \sqrt{4 \pi \, \alpha} \ B  \
u_1^2 \ \tau \,,
\label{ColoEngla}
\end{equation}
where $\varkappa (\zeta) = 3/20$ for a monoatomic classic
gas~\cite{Gaisser:1990vg}, and $\varkappa (\zeta) = 1/12$ for a three
atomic gas with non-static bindings~\cite{Anchordoqui:2018vji}.  At
this stage, it is worthwhile to remind the reader that in the preceding discussion it was
implicitly assumed that the magnetic field is parallel to the shock
normal.  Injecting additional hypotheses into the model may
reduce~\cite{Bustard:2016swa,Romero:2018mnb} or
increase~\cite{Jokipii:1986,Ferrand:2014laa} the maximum achievable
energy.

For a relativistic outflow, with $\gamma_u \equiv(1-u_1^2)^{-1/2} \gg 1$,
the calculation is somewhat more complicated, because the CR is allowed to be accelerated only over a
fraction of the characteristic length of the accelerator, which is comparable to $R/\gamma_u$~\cite{Waxman:1995vg,Waxman:2005id}. To visualize this, one
must realize that as the plasma expands, its magnetic field decreases,
and consequently the time available for acceleration corresponds to the time of
expansion from $R$ to, say, $2R$. In the observer frame this time is
$R/u_1$, while in the plasma rest frame it is $R/(\gamma_u u_1)$. Therefore, a
CR moving with the magnetized plasma can only be accelerated over a
transverse distance $\sim R/\gamma_u$. This shows that the maximal
energy is also inversely proportional to the Lorentz factor,
\begin{equation}
E \lesssim  \varkappa (\zeta) \  \ Z \sqrt{4 \, \pi
  \, \alpha} \ B  \
\frac{u_1}{\gamma_u} \ R \, .
\label{Hillas3}
\end{equation}
When a GRB erupts, the internal plasma is accelerated to
ultra-relativistic velocities, making GRBs premium astrophysical sites
to explore how relativistic collisionless shocks can accelerate
UHECRs~\cite{Waxman:1995vg,Vietri:1995hs}. For typical source
parameters, the plasma is opaque to the propagation of UHECR
nuclei~\cite{Anchordoqui:2007tn}, and so it appears that these
powerful compact objects would only accelerate protons up to
ultra-high energies; see,
however,~\cite{Wang:2007xj,Murase:2008mr,Globus:2014fka,Biehl:2017zlw,Zhang:2017moz}.

For the generic case of acceleration in an outflow, one can compare (\ref{Hillas3})
with the magnetic luminosity $L_B$ to set a lower bound on the
luminosity that a source must posses to accelerate UHECRs. Namely,  the magnetic field carries with it an
energy density $B^2/2$ and the flow carries with it an energy flux $>
v B^2/2$ so
 (\ref{Hillas3})  sets a lower limit on the rate
\begin{equation}
  L_B>\frac{1}{8 \pi \alpha } \ \frac{1}{ \varkappa^2 (\zeta)}  \
  \frac{\gamma^2_u}{u_1} \ E_p^2  \sim 10^{42} \ \frac{1}{\varkappa^2
    (\zeta)} \ 
  \frac{\gamma_u^2}{ u_1 } \left(\frac{E_p}{10^{10}\rm GeV}\right)^2{\rm erg/s}
\label{LP-rigidity}
\end{equation}
at which the energy must carried by the outflowing plasma to
accelerate a nucleus to a given rigidity
$E_p$~\cite{Matthews:2018laz}. The Poynting luminosity of the
accelerator~\cite{Waxman:1995vg,Waxman:2005id,Piran:2010yg} is found
to be a factor of 2 larger~\cite{Spruit:2013ud}. Only the brightest
AGNs and GRBs are known to satisfy the (\ref{LP-rigidity}) power
requirement while reaching $E_p \sim 10^{10}~{\rm
  GeV}$~\cite{Waxman:1995vg,Blandford:1999hi,Dermer:2010iz}.  We will
see in Sec.~\ref{sec:c_backyard}, however, that there are ways to
escape this constraint.

\subsection{Energy loss}
\subsubsection{Interaction rate of UHECRs on photon fields}

Thus far we have considered acceleration processes in astrophysical
environments without paying attention to the CR energy loss through
interactions with the source's photon backgrounds.  Moreover, as
UHECRs propagate {\it en route} to Earth they also lose energy
scattering off the pervasive radiation fields permeating the
universe. Before we estimate the UHECR mean free path of these
collisions we review some  basic phenomenology of collider physics.

For both an incident beam on a fixed target or two colliding beams,
the interaction rate $R$ is proportional to the number density of
particles $n_1$ and $n_2$ that approach each other with a certain
relative velocity.  The natural definition of the relativistic invariant flux is 
\begin{equation}
{\cal F} = (J_1 \cdot J_2) \ v_{\rm rel} \, ,
\end{equation}
where 
\begin{equation}
J_i = (n_i, n_i  \mathbf{v}_i) = n_i^0 (\gamma_i, \gamma_i \mathbf v_i) =
n_i^0  \ u_i \,,
\end{equation}
is a 4-vector current ($i = 1,2$), and where $n^0_i$ is the number density in the rest
frame and $u_i$ the 4-velocity $u_i = \gamma_i (1,\mathbf v_i)$, with $u_i^2 =
1$ and
$\gamma_i$  the Lorentz factor~\cite{Cannoni:2016hro}. One can check by inspection that this expression is a Lorentz scalar. For massless particles,
the velocity vector becomes the unitary vector in the direction of
propagation and if at least one massless particle is involved in the
collision then $v_{\rm rel} = 1$, yielding
\begin{equation}
{\cal F} = J_1 \cdot J_2 = n_1 n_2 (1 - \mathbf v_1 \cdot \mathbf v_2) \, .
\end{equation}
For two massless particles, the flux can be rewritten as ${\cal F} = n_1
n_2  \ (1 - \cos \theta)$, where $\theta$ is the angle between the
3-momenta $\hat{\mathbf{k}}_1$ and $\hat{\mathbf{k}}_2$ of the
incoming particles.   For collisions of a
massless with a massive particle, the  incident  flux is found to be
${\cal F} = n_1 n_2 (1 - v_2 \cos
\theta)$. The physical quantity that gives the intrinsic quantum
probability for an interaction is the cross section, defined by the
ratio $\sigma = R/{\cal F}$.

The interaction rate for a highly relativistic ($v \sim
1$) cosmic ray (with baryon number $A$ and energy $E = \gamma A~{\rm GeV}$)  propagating through an isotropic
photon background with energy $\varepsilon$ and spectrum
$n(\varepsilon)$, normalized so that the total number
of photons in a box is $\int n(\varepsilon) \
d\varepsilon$, can be derived assuming one of the densities in the
scattering process collapses into a delta function,
\begin{eqnarray} 
R & = & \frac{1}{4 \pi} \int_0^\infty  d\varepsilon  \ n(\varepsilon)   \int d\Omega  \
 \sigma(\varepsilon')    \ (1 - \cos \theta)  
 =  \frac{1}{2} \int_0^\infty d\varepsilon  \
n(\varepsilon)    \int_{-1}^{+1} d \cos \theta \ \,
 \sigma(\varepsilon')   \ (1 - \cos \theta) \nonumber \\
& = & \frac{1}{2 \gamma^2} \int_0^\infty
\frac{d\varepsilon}{\varepsilon^2} \  n(\varepsilon) \       \int_0^{2\gamma
  \varepsilon} d \varepsilon' \   {\varepsilon'}^2  \  \sigma
(\varepsilon') \,,
\label{Floyd}
\end{eqnarray}
where $\sigma (\varepsilon')$ is the cross section for UHECR
interaction with a photon of energy $\varepsilon'$ in the rest frame of the CR~\cite{Stecker:1969fw}.
The Mandelstam invariant, 
\begin{equation}
s = m^2 + 2 m \varepsilon' = m^2 + 2 (E \varepsilon - \mathbf{k} \cdot
\mathbf p) \,,
\label{Mandelstam}
\end{equation}
relates quantities in the rest frame of the UHECR  and the CMB frame,
respectively. This leads to
\begin{equation}
\varepsilon' = \frac{E \varepsilon - \mathbf{k} \cdot \mathbf{p}}{m} =
\frac{E}{m} \ \varepsilon ( 1 - \frac{p}{E} \cos \theta) = \gamma \varepsilon 
( 1 - v \cos \theta)
\end{equation}
and
\begin{equation}
d\varepsilon' = - \gamma \ \varepsilon \ v \ d \cos \theta \,,
\end{equation}
where $m = A~{\rm GeV}$, $k = \varepsilon$, $p/E = v$.

We begin discussing the energy loss by considering UHECR collisions
with the pervasive photon backgrounds permeating the universe. After
that, when we discuss the potential classes of sources which are able
to emit UHECRs in our cosmic backyard, we will address the limitation
of CR collisions with thermal photons inside the acceleration region.

\subsubsection{Opacity of the CMB to UHECR protons}
\label{CMBopacity} 

On the way to Earth, UHECR protons degrade their energy through
Bethe-Heitler (BH) pair production 
and photopion production, each
successively dominating as the proton energy increases. The fractional energy loss due to interactions with
the universal photon fields at a redshift $z=0$ is determined by the
integral of the proton energy loss per collision multiplied by the
probability per unit time for a proton collision in an isotropic gas
of photons~\cite{Stecker:1968uc}. Introducing the inelasticity in
(\ref{Floyd}),  this integral can be explicitly
written as 
\begin{eqnarray}
-\frac{1}{E} \frac{dE}{dt} & = & \frac{1}{2 \gamma^2}\,\sum_j\,   \int_0^{\infty} d\varepsilon \ 
n_j (\varepsilon) \  \frac{1}{\varepsilon^2} \,  \int_0^{\infty} d\varepsilon' \,\, 
\langle y (\varepsilon') \rangle_j  \, \,  
\sigma_j (\varepsilon')\, \, \varepsilon' \  \Theta (2 \gamma
\varepsilon - \varepsilon') \nonumber \\ &  = & \frac{1}{2 \gamma^2}\,\sum_j\, \int_0^{\infty} d\varepsilon' \,\, 
\langle y (\varepsilon') \rangle_j  \, \,  
\sigma_j (\varepsilon')\, \, \varepsilon' \, \int_{\frac{\varepsilon'}{2 \gamma}}^{\infty} d\varepsilon \ 
n_j (\varepsilon) \  \frac{1}{\varepsilon^2}  
\label{conventions}
\end{eqnarray} 
where $\varepsilon'$ is the photon energy in the rest frame of the
nucleon, $\langle y (\varepsilon')\rangle_j$ is the average fraction of the energy
lost by the nucleon for the
$j$th reaction channel, $n_j (\varepsilon) d\varepsilon$ stands for
the number density of photons with energy between $\varepsilon$ and
$\varepsilon+ d\varepsilon$, $\sigma_j(\varepsilon')$ is the total
cross section of the $j$th interaction channel, and $\gamma$ is the
Lorentz factor of the nucleon. The sum is carried out over all
relevant channels: $p \gamma \rightarrow p e^+ e^-$, $p \gamma
\rightarrow \pi^0 p$, $p \gamma \rightarrow \pi^+ n$, $p \gamma \to$
multi-$\pi p$, and $p \gamma \to$
multi-$\pi n$.

Pair production and photopion production processes are only of
importance for interactions with the CMB (collisions with optical and infrared
photons give a negligible contribution)~\cite{Berezinsky:1987ed,Berezinsky:1988wi}. For interactions
with the CMB, the photon density is that of a Planck spectrum 
\begin{equation}
n(\varepsilon) = (\varepsilon/\pi)^2 \left[e^{\varepsilon/T} - 1
\right]^{-1} \,,
\label{Bose-Einstein}
\end{equation}
and so the fractional energy loss is given by
\begin{equation}
-\frac{1}{E} \frac{dE}{dt} = - \frac{T}{2 \pi^2 \gamma^2}
 \int_{\varepsilon'_{{\rm th}}}^{\infty}  d\varepsilon' \,
\sigma_j  (\varepsilon') \, \langle y (\varepsilon') \rangle_j \, \varepsilon' \,
\ln \left[ 1 -  e^{-\varepsilon'/ (2 \gamma T)} \right]\,,
\label{phds!}
\end{equation}
where $\varepsilon'_{{\rm th}}$ is the threshold energy  
in the rest frame of the nucleon and $T = 2.7255(6)~{\rm K}$~\cite{Fixsen:2009ug} .

For $E \lesssim 10^{9}$~GeV, the BH pair production process proceeds
through the ``high-energy'' photons on the tail of the Planck
distribution. Hence, the inelasticity and the cross section can be
approximated by their values at threshold; i.e., $\langle y \rangle = 2
m_e/m_p$ and $\sigma (\varepsilon') = \tfrac{\pi}{12}\, \alpha\,\,
r_0^2 \,\, (\varepsilon'/m_e - 2)^3$, where $\alpha$ is the fine
structure constant, $r_0$ is the classical radius of the electron,
$m_e$ and $m_p$ the mass of the electron and the
proton~\cite{Berezinsky:1987ed,Berezinsky:1988wi}.  This leads to
\begin{equation}
-\frac{1}{E}\, \frac{dE}{dt}  = 
\frac{16 }{\pi}\, \frac{m_e}{m_p}\, 
\alpha\, r_0^2\, T^3 \, \left(\frac{\gamma T}{m_e}\right)^2\, 
\exp \left(-\frac{m_e}{\gamma T}\right) \, .
\label{uniden}
\end{equation}
At higher energies, say $E \gtrsim 10^{10}~{\rm GeV}$, the
characteristic time for the energy loss due to pair production is $-
E/(dE/dt) \approx 5\times 10^9~{\rm
  yr}$~\cite{Blumenthal:1970nn,Aharonian:1994nn}, and the photopion
production processes $p \gamma \rightarrow p \pi^0$ and $p \gamma
\rightarrow \pi^+ n$ give the main contribution to proton energy
loss. The cross sections of these processes are well known and the
kinematics is simple.

Photopion production turns on at a photon energy in the proton rest
frame of 145~MeV with a strongly increasing cross section at the
$\Delta (1232)$ resonance, which decays into the one pion channels
$\pi^+ n$ and $\pi^0 p$.  With increasing energy, heavier baryon
resonances occur and the proton (or instead a neutron) might reappear
only after successive decays of
resonances~\cite{Armstrong:1971ns}. The most important channel of this
kind is $p\gamma \rightarrow \Delta^{++} \pi^-$ with intermediate
$\Delta^{++}$ states leading finally to $\Delta^{++} \rightarrow
p\pi^+$. $\Delta^{++}$ examples in this category are the $\Delta
(1620)$ and $\Delta(1700)$ resonances. The cross section in this
region can be described by either a sum or a product of Breit-Wigner
distributions over the main resonances produced in $p \gamma$
collisions considering final states with pions, kaons, and a single
nucleon: $\pi N$, $\pi \pi N$ and $K\Lambda$ ($\Lambda \rightarrow N
\pi$)~\cite{Patrignani:2016xqp}. At high energies, $3.0\, {\rm GeV} <
\varepsilon' < 183 \, {\rm GeV}$, the CERN-HERA and COMPAS Groups have
made a fit to the $p\gamma$ cross section~\cite{Montanet:1994xu}.  The
parameterization is
\begin{equation}
\sigma (\varepsilon') = A + B \,\,\ln^2\left(\frac{\varepsilon'}{{\rm GeV}}\right) + C \,\,
\ln \left(\frac{\varepsilon'}{{\rm GeV}}\right) \,\,{\rm mb}\,,
\end{equation} 
where $A = 0.147 \pm 0.001$, $B = 0.0022\pm 0.0001$, 
and $C = -0.0170 \pm 0.0007$. 

We turn now to the kinematics of proton-photon interactions. The
inelasticity $\langle y \rangle$ depends not only on the outgoing particles but
also on the kinematics of the final state. Nevertheless, averaging
over final state kinematics leads to a good approximation of $\langle
y \rangle$.
The c.m. system quantities (denoted by $*$) are determined from the
relativistic invariance of the square of the total 4-momentum $p_\mu
p^\mu$ of the proton-photon system. From (\ref{Mandelstam}), this invariance leads to the
relation
\begin{equation}
s=(\varepsilon^* + E^*)^2 = m_p^2 + 2m_p \varepsilon'.
\end{equation}
The c.m. system energies of the particles are uniquely determined 
by conservation of energy and momentum.  For
$p  \gamma \rightarrow \rightarrow p  \pi$,   the mean energy of the
outgoing proton and pion are 
 given by
\begin{subequations}
\begin{equation}
\langle {E_p}^*  \rangle = \frac{(s + m_p^2 - m_\pi^2)}{2\, \sqrt{s}} 
\end{equation}
and
\begin{equation}
\langle {E_\pi}^*  \rangle = \frac{(s + m_\pi^2 - m_p^2)}{2\,
  \sqrt{s}}\, ,
\end{equation}
\end{subequations}
or in the lab frame by
\begin{subequations}
\begin{equation}
\langle E_p \rangle =  \gamma_{\rm c.m.}  \frac{(s + m_p^2 - m_\pi^2)}{2\,
  \sqrt{s}} =   \frac{E \, (s+m_p^2 -m_\pi^2)}{2\, s} 
\label{6}
\end{equation}
and
\begin{equation}
\langle E_\pi \rangle =  \gamma_{\rm c.m.}  \frac{(s + m_\pi^2 - m_p^2)}{2\,
  \sqrt{s}} =   \frac{E \, (s+m_\pi^2 -m_p^2)}{2\, s} \,,
\label{6pi}
\end{equation}
\end{subequations}
where $\gamma_{\rm c.m.} = (E + \varepsilon)/\sqrt{s} \simeq
E/\sqrt{s}$ is the Lorentz factor between the c.m. and  lab frames~\cite{Stecker:1968uc}.
The mean inelasticity is given by
\begin{equation}
\langle y (\varepsilon') \rangle = 1 -  \frac{\langle E_p\rangle}{E} =
\frac{1}{2} \left(1 + \frac{m_\pi^2 -m_p^2}{m_p^2 + 2 m_p
    \varepsilon'} \right)\, .
\label{averagey}
\end{equation}
It is well established
experimentally  that at very high energies ($\sqrt{s}
\gtrsim 3$~GeV) the incoming particles lose only one-half their
energy via pion photoproduction independently of the number of pions
produced, i.e., $\langle y \rangle \sim
1/2$~\cite{Golyak:1992cz}. This {\it leading particle effect} is
consistent with  (\ref{averagey}).

\begin{figure}
\postscript{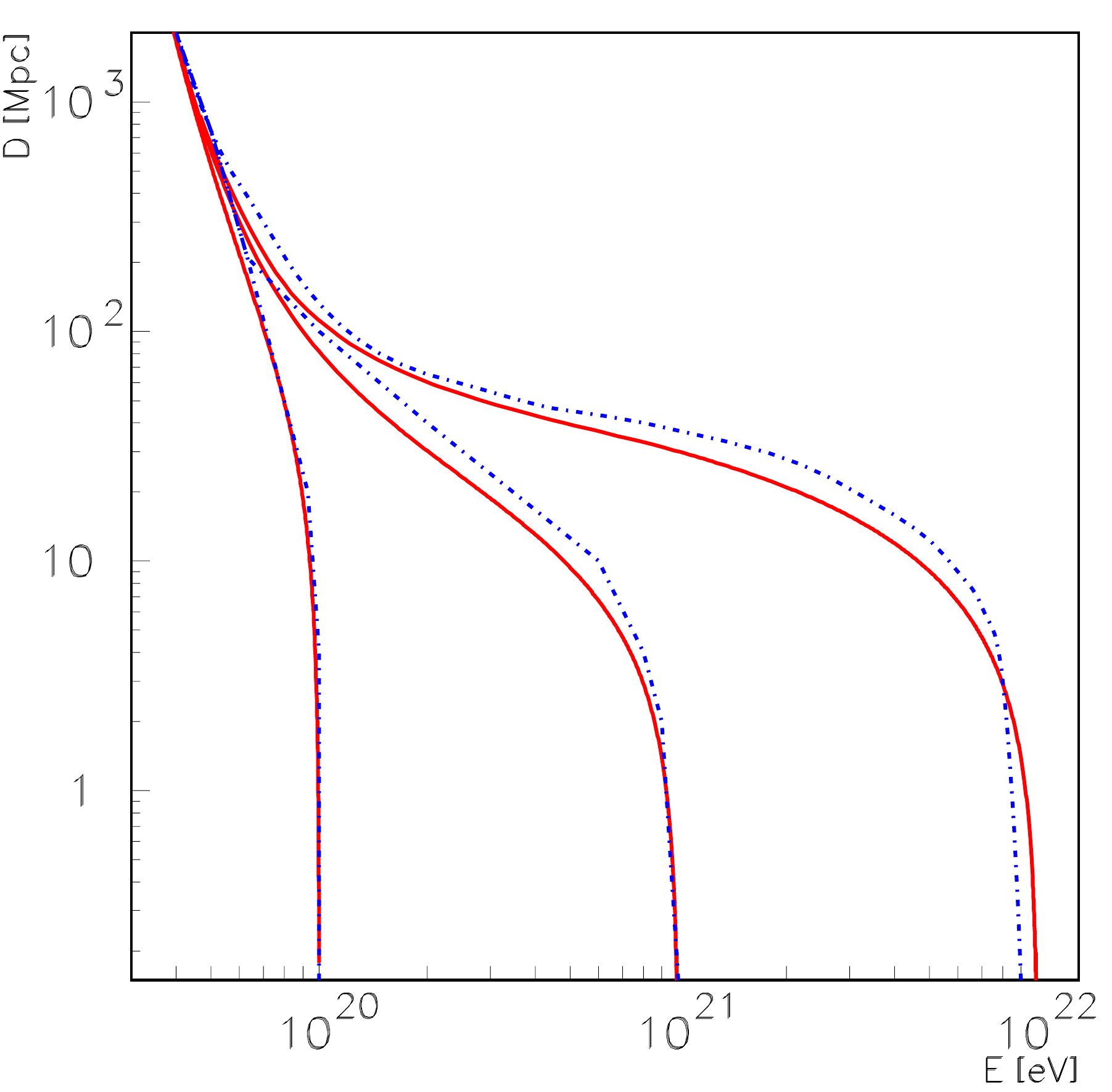}{0.5}
\caption{Energy attenuation length of protons in the intergalactic
medium. The solid-lines indicate the results from the continuous
energy loss approximation, whereas the dashed-lines correspond to the
results of a Monte Carlo simulation~\cite{Aharonian:1994nn}. 
Note that after a distance of $\sim 100$~Mpc, or propagation time 
$\sim 3 \times 10^{8}$~yr, the mean energy is essentially independent of the 
initial energy of the protons, with a critical energy around
$10^{20}~{\rm eV}$. 
From Ref.~\cite{Anchordoqui:1998ig}.} 
\label{fig:13}
\end{figure}

Numerical integration of~(\ref{phds!})  is carried out taking into account the aforementioned resonance decays and the production of multipion final states at high c.m.
energies. 
For $\sqrt{s} < 2$ GeV,  a $\chi^2$ fit of the numerical results,
using the exponential behavior
\begin{equation} 
- \frac{1}{E}\, \frac{dE}{dt} = A \, {\rm exp} [ - B / E ] 
\label{okop}
\end{equation}
derived in~\cite{Berezinsky:1988wi} approximating both the cross section
and the fractional energy loss by their threshold values,
gives
\begin{equation}
A = ( 3.66  \pm 0.08 )
\times 10^{-8} \, {\rm yr}^{-1} \ {\rm and} \  B = (2.87 \pm 0.03 )\times
10^{11}\, {\rm GeV} \, ,
\label{parame}
\end{equation} 
with $\chi^2/{\rm dof} = 3.9/10$~\cite{Anchordoqui:1997gs,Anchordoqui:1996ru}. The
fractional energy loss due to production of multipion final states at
higher c.m. energies ($\sqrt{s} \gtrsim 3$~GeV) is roughly a constant,
\begin{equation}
-\frac{1}{E}\, \frac{dE}{dt} = C = ( 2.42 \pm 0.03 ) 
\times 10^{-8}  \,\, {\rm yr}^{-1} \,.
\label{ce}
\end{equation}
From the values determined for the fractional energy loss, it is 
straightforward to compute the energy degradation of UHECRs in terms of their flight time. This is given by,
\begin{equation}
A \, t \, - \,  {\rm Ei}\,(B/E) 
+ \, {\rm Ei}\, (B/E_0) 
= 0\,,\,\,\,\,\,\,  {\rm for} \,\,10^{10}\,{\rm GeV} \lesssim E \lesssim 10^{12} \,{\rm GeV} \,,
\label{degradacion}
\end{equation}
and
\begin{equation}
E (t) = E_0 \exp[- \,C \ t \,]\,,\,\,\,\,\,\, {\rm for}\,\, E \gtrsim 10^{12} \, 
{\rm GeV}\,,
\end{equation}
where ${\rm Ei}(x)$ is the exponential integral~\cite{Abramowitz}. In
Fig.~\ref{fig:13} we show the proton energy degradation as a function of
the mean flight distance. It is evident from this figure that the CR energy degradation resulting from the
preceding semi-analytic calculation within the context of a 
continuous energy loss approximation  is
consistent  with that obtained through numerical simulations~\cite{Aharonian:1994nn}. Note that independent of the initial
energy, the mean energy values approach $10^{11}$~GeV
after the proton propagates a distance of about $100~{\rm Mpc}$.

\subsubsection{Photonuclear interactions}
\label{sec:323}

The relevant mechanisms for the energy loss that extremely high energy
nuclei suffer during their trip to Earth are: Compton interactions,
pair production in the field of the nucleus, photodisintegration, and
hadron photoproduction. The Compton interactions have no threshold
energy.  In the nucleus rest-frame, pair production has a threshold at
$\sim 1$~MeV, photodisintegration is particularly important at the
peak of the GDR (15 to 25~MeV), and photomeson production has a
threshold energy of $\sim 145$~MeV.

Compton interactions result in only a negligibly small energy loss for 
the nucleus given by
\begin{equation}
-\frac{dE}{dt} =  \frac{Z^4}{A^2} \epsilon_\gamma \left( \frac{E}{A m_p}
\right)^2 \, \,\, {\rm eV\, s^{-1}} \,
\end{equation}
where $\epsilon_\gamma$ is the energy density of the ambient photon field
in eV cm$^{-3}$~\cite{Puget:1976nz}.  The energy loss rate due to
photopair production is $Z^2/A$ times higher than for a proton of the
same Lorentz factor~\cite{Chodorowski}, whereas the energy loss rate
due to photomeson production remains roughly the same.  The latter is
true because the cross section for photomeson production by nuclei is
proportional to the baryon number $A$~\cite{Michalowski:1977eg}, while
the inelasticity is proportional to $1/A$. However, it is
photodisintegration rather than photopair and photomeson production
that determines the energetics of UHECR nuclei. During this process
some fragments of the nuclei are released, mostly single neutrons and
protons~\cite{Hayward:1963zz,Danos:1965yu}.  Experimental data of
photonuclear interactions are consistent with a two-step process:
photoabsorption by the nucleus to form a compound state, followed by a
statistical decay process involving the emission of one or more
nucleons, followed by immediate photo-emission from the excited
daughter nuclei. For brevity, we label the photonuclear process
$A_i
\gamma \to A_f^\ast N$, followed by $A_f^\ast \to A_f + \gamma$-ray
as ``$A^*$.''  In the energy region which
extends from threshold for single-nucleon emission $\sim 10~{\rm MeV}$
up to $\sim 30~{\rm MeV}$ the GDR dominates the $A^*$ process. The GDR
typically de-excites by the statistical emission of a single
nucleon. Above the GDR region, and up to the photo-pion production
threshold, the non-resonant $A^\ast$ processes provide a much smaller
cross section with a relatively flat dependence on energy.  The 
photodisintegration cross section for all the different nuclear
species has been obtained through a direct fit to
data~\cite{Puget:1976nz,Stecker:1998ib,Khan:2004nd}; the associated uncertainties
have been studied in~\cite{Boncioli:2016lkt}.

For $A > 4$ the cross section for losing one
nucleon can be described by a Breit-Wigner form
\begin{equation}
\sigma_A (\varepsilon') = \left\{\begin{array}{ll} \sigma_0 \,\, \dfrac{\varepsilon'^2 \,
\Gamma^2}{(\varepsilon^2_0 - \varepsilon'^2)^2 + \varepsilon'^2\, \Gamma^2} 
&{\rm for}~\varepsilon' \leq 30~{\rm MeV} \\
\dfrac{A}{8}~{\rm mb} & {\rm for} \ \varepsilon' > 30~{\rm MeV} \end{array} \right. \,,
\end{equation}
where $\Gamma = 8~{\rm MeV}$ is the width, $\varepsilon_0 =  42.65 \,
A^{-0.21}~{\rm MeV}$ is the central value of the GDR energy band, and
$\sigma_0 = 1.45 \ A~{\rm mb}$ is the
normalization~\cite{Karakula:1993he}. The GDR cross section can be safely
approximated by the single pole of the narrow-width approximation (NWA),
\begin{equation}
\sigma_A(\varepsilon') = \varpi \  
\delta(\varepsilon' - \varepsilon_0)\, ,
\label{sigma}
\end{equation}
where  $\varpi = \pi \, \sigma_0\, \Gamma/2$; the factor of 1/2 is introduced to match the integral (i.e. total cross section) of the Breit-Wigner and the delta function~\cite{Anchordoqui:2006pe}.

For $A\leq 4$, the photodisintegration cross section can be described
by the shifted log-normal distribution,
\begin{equation}
 \sigma_A (\varepsilon') = \sigma_0
\exp \left\{- \frac{ \ln^2 \left[(\varepsilon'-\varepsilon'_{\rm
      th})/(\varepsilon_0-\varepsilon'_{\rm th} )\right] }{2\Gamma^2} \right\}
\,,
\label{eq:shiftlognorm}
\end{equation}
where $\varepsilon_0$ is the central value of the GDR energy band (with
threshold $\varepsilon'_{\rm th}$), $\sigma_0$
is the cross section at $\varepsilon=\varepsilon_0$, and $\Gamma$
is a measurement of the dispersion around $\varepsilon_0$; the cross section
parameters are given in Table~\ref{tabla1}~\cite{Soriano:2018lly}.
For analytical order of magnitude estimates, the cross section can
take the form (\ref{sigma}) of the NWA, with 
$\varpi =  \sqrt{2\pi} \ \sigma_0 \
\Gamma \ (\varepsilon_0-\varepsilon'_{\rm th}) \ e^{\Gamma^2/2}$.

\begin{table}
\caption{Parameters of the photo-disintegration cross-section. \label{tabla1}}
\begin{tabular}{cccccc}
\hline
\hline
~~~~~~$A$~~~~~~&~~~~~~$\sigma_0$ (mb)~~~~~~&~~~~~~$\varepsilon_0$ (MeV)~~~~~~&~~~~~~$\varepsilon'_{\rm th}$
(MeV)~~~~~~&~~~~~~$\Gamma$~~~~~~ &~~~~~~$\varpi$ (mb MeV)~~~~~~ \\
\hline
4 & $3.22\pm 0.05$ & $26.6 \pm 0.4$ & $20.1 \pm 0.4$ & $0.94 \pm 0.08$
& $77\pm3 \phantom{0~}$
\\
3 & $1.82 \pm 0.05$ & $15.3 \pm 0.4$ & $5.1 \pm 0.2$ & $0.93 \pm 0.04$
& $67\pm 2\phantom{0~}$ \\
2 & $2.60 \pm 0.09$ & $3.87 \pm 0.09$ & $2.42 \pm 0.05$ & $1.48 \pm
0.04$ & $42.2 \pm 0.4$ \\
\hline
\hline
\end{tabular}
\end{table}

The general formula for the inverse photodisintegration mean-free-path (mfp) for a highly
relativistic nucleus with energy $E = \gamma A~{\rm GeV}$ propagating through an
isotropic photon background with energy $\varepsilon$ and spectrum $n(\varepsilon)$ is
given by  
\begin{equation}
\lambda^{-1} (A) = \frac{1}{2 \gamma^2} \int_0^\infty
\frac{n(\varepsilon)}{\varepsilon^2} \ d \varepsilon
\int_{\varepsilon'_{\rm th}}^{2 \gamma \varepsilon} \varepsilon' \sigma
(\varepsilon') \ d \varepsilon'  \,,
\label{lambdamenosuno}
\end{equation} 
where $\varepsilon'_{\rm th}$ is the GDR energy threshold. Note that
$\lambda (A) = \tau (A)$ is also the mean survival time, so
$\lambda^{-1} (A)$ is also the photodisintegration rate per nucleus
$R(A)$ given in (\ref{Floyd}).  The outer integral running over
$\varepsilon \in [0, \infty)$ can be splitted into two parts,
$\varepsilon \in [0, \varepsilon'_{\rm th}/(2 \gamma])$ and
$\varepsilon \in [\varepsilon'_{\rm th}/(2\gamma), \infty)$, and so
(\ref{lambdamenosuno}) can be rewritten as
\begin{equation}
\lambda^{-1} (A) = \frac{1}{2 \gamma^2} \int_0^{\frac{\varepsilon'_{\rm
    th}}{2 \gamma}}
\frac{n(\varepsilon)}{\varepsilon^2} \ d \varepsilon
\int_{\varepsilon'_{\rm th}}^{2 \gamma \varepsilon} \varepsilon' \sigma
(\varepsilon') \ d \varepsilon'  +  \frac{1}{2 \gamma^2}
\int_{\frac{\varepsilon'_{\rm th}}{2 \gamma}}  ^\infty
\frac{n(\varepsilon)}{\varepsilon^2} \ d \varepsilon
\int_{\varepsilon'_{\rm th}}^{2 \gamma \varepsilon} \varepsilon' \sigma
(\varepsilon') \ d \varepsilon'  \, .
\label{lambdaFloyd}
\end{equation}
Note that the first term has no contribution to $\lambda^{-1} (A)$,
because since $\varepsilon \in [0, \varepsilon'_{\rm th}/(2\gamma)]$
the upper limit of the  integral over $d\varepsilon'$ remains always below the cross
section energy threshold, i.e., $0 \leq 2 \gamma \varepsilon \leq
\varepsilon'_{\rm th}$.

\begin{figure}
\postscript{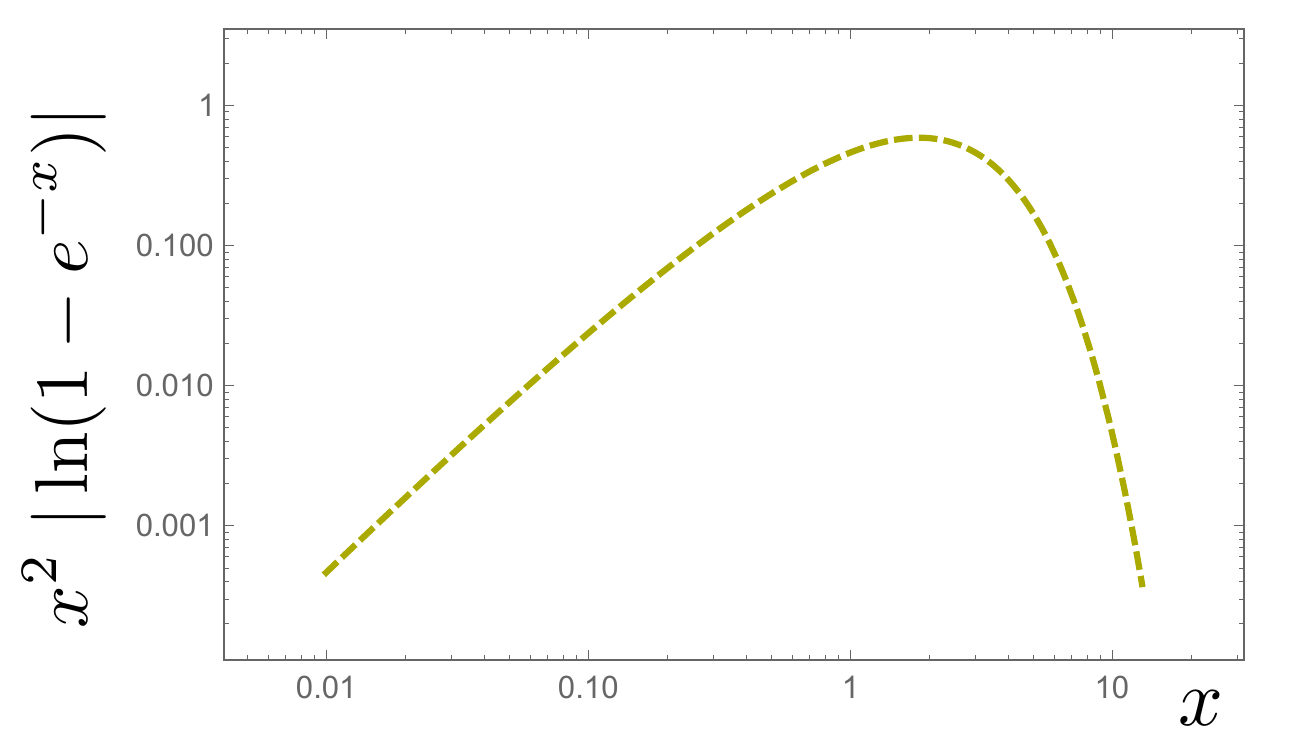}{0.5}
\caption{The scaling functions $x^2 |\ln (1 - e^{-x})|$, proportional
  to the photodisintegration rate in (\ref{RBE}). Adapted from~\cite{Anchordoqui:2006pd}.} 
\label{fig:14}
\end{figure}

Inserting  (\ref{sigma}) into (\ref{lambdaFloyd}) the integral over $d\varepsilon'$ becomes
\begin{eqnarray}
\int_{\varepsilon'_{\rm th}}^{2 \gamma \varepsilon} \varepsilon' \
\sigma (\varepsilon') \ d \varepsilon' & = & \int_{-\infty}^{+\infty}
\Theta (\varepsilon' -\varepsilon'_{\rm th}) \ \Theta (2 \gamma
\varepsilon - \varepsilon') \  \varepsilon'  \ \sigma(\varepsilon') \
d \varepsilon' \nonumber \\ & = & \varpi \ 
\int_{-\infty}^{+\infty} \Theta (\varepsilon' -\varepsilon'_{\rm th}) \ \Theta (2 \gamma
\varepsilon - \varepsilon') \  \varepsilon'  \ \delta(\varepsilon' -
\varepsilon_0) \ d \varepsilon' 
=  \varpi \, \Theta (2 \gamma \varepsilon - \varepsilon_0) \, \varepsilon_0
\, ,
\end{eqnarray}
because $\Theta (\varepsilon_0 - \varepsilon'_{\rm th}) = 1$ for
$\varepsilon_0 > \varepsilon'_{\rm th}$. This means that in the NWA the
reciprocal mfp is given by
\begin{equation}
\lambda^{-1} (A) = \frac{\varpi \ \varepsilon_0}{2 \gamma^2}
\int_{\frac{\varepsilon'_{\rm th}}{2\gamma}}^\infty
  \frac{n(\varepsilon)}{\varepsilon^2}  \ \Theta (2 \gamma \varepsilon
  - \varepsilon_0) \ d \varepsilon =  \frac{\varpi \ \varepsilon_0}{2 \gamma^2}
\int_{\frac{\varepsilon_0} {2\gamma}}^\infty
  \frac{n(\varepsilon)}{\varepsilon^2}  \ \ d \varepsilon\,
\label{world-cup}
\end{equation}
because $(2 \gamma \varepsilon - \varepsilon_0) > 0$ for $\varepsilon >
\varepsilon_0/ (2 \gamma)$. For a nucleus passing through a region
where the photon density is described by a Bose-Einstein distribution (\ref{Bose-Einstein}), the photodisintegration rate is given by
\begin{equation}
\lambda^{-1}_{\rm BE}  (A)  \approx  \frac{\varpi \,\varepsilon_0\,
\,T}{2 \gamma^2 \pi^2} \,\,
\left| \ln \left(1 - e^{-\epsilon_0/2 \gamma T}\right) \right| =
\frac{2 \ \varpi \ T^3 }{\pi^2 \varepsilon_0} \, x^2 \left|\ln \left(1
    - e^{-x}  \right) \right| \,,
\label{RBE}
\end{equation}
where we have defined a dimensionless scaling variable $x\equiv
\varepsilon_0/(2\,\gamma \, T)$.  From the pre-factor, we learn that
for $A>4$ the peak of $\lambda_{\rm BE}^{-1} (A)$ scales in $A$ as
$\sigma_0/\varepsilon_0\sim A^{1.21}$, and the value of $\gamma$ at
the peak scales as $\varepsilon_0\sim A^{-0.21}$.  The scaling
function $x^2\, |\ln (1 - e^{-x})|$ is shown in
Fig.~\ref{fig:14}. Approximations to the $|\ln|$~term yield $e^{-x}$
for $x>2$, and $|\ln x|$ for $x\ll 1$. Thus, the exponential
suppression of the process appears for $\varepsilon_0 > 4 \gamma T$, and the small $x$ region presents a mfp
that scales as $x^2\,|\ln x|$. The peak region provides the smallest
inverse mfp, and so this region dominates the $A^*$-process.  In the
peak region, $x$ is of order one, which implies that $2 \gamma \,
  T \sim \varepsilon_0$. When this latter relation between the nuclei boost
and the ambient photon temperature is met, then the
photo-disintegration rate is optimized.

\begin{figure}[tbp] 
\begin{minipage}[t]{0.49\textwidth}
    \postscript{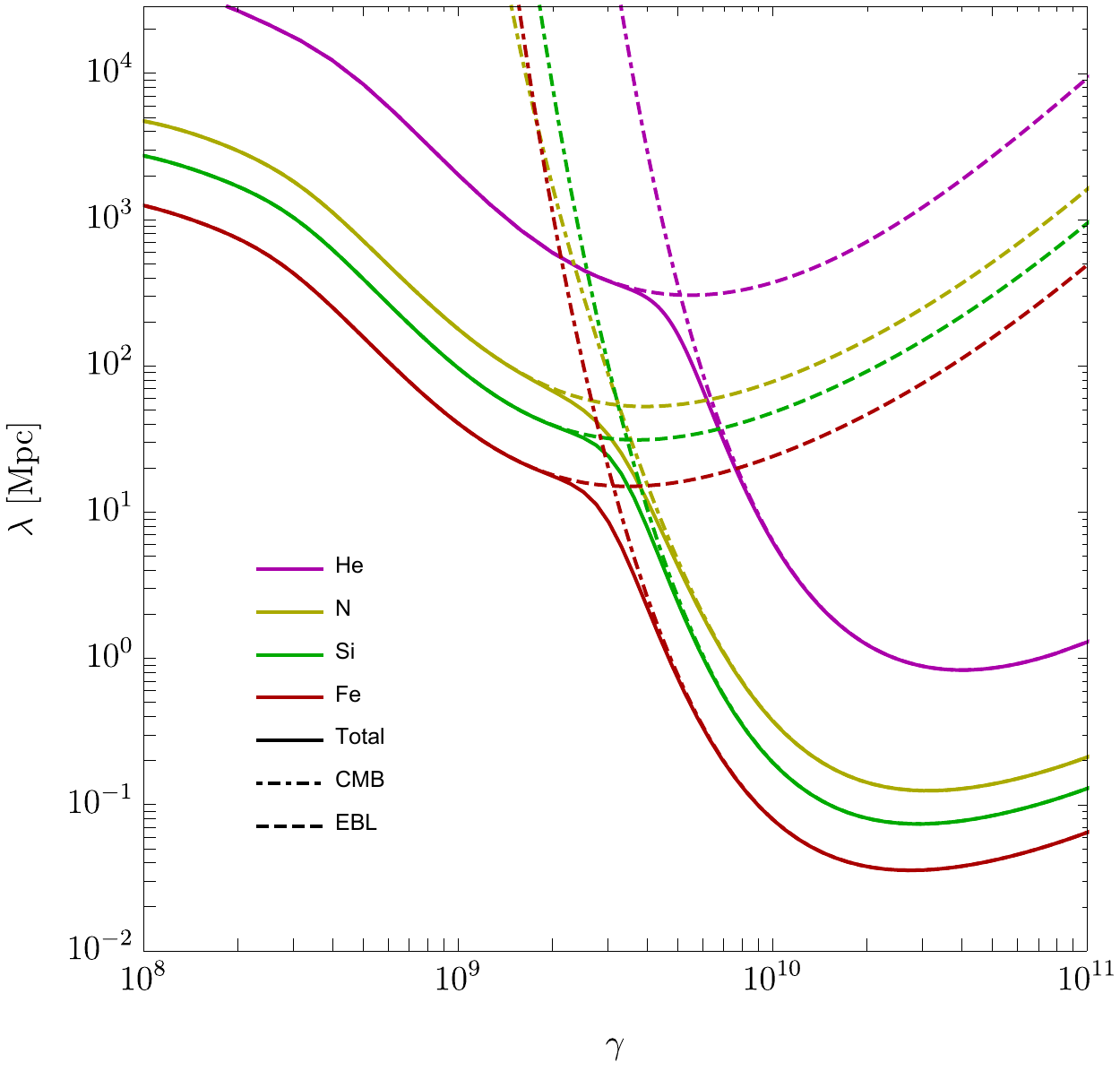}{0.92} 
\end{minipage} 
\hfill \begin{minipage}[t]{0.49\textwidth}
  \postscript{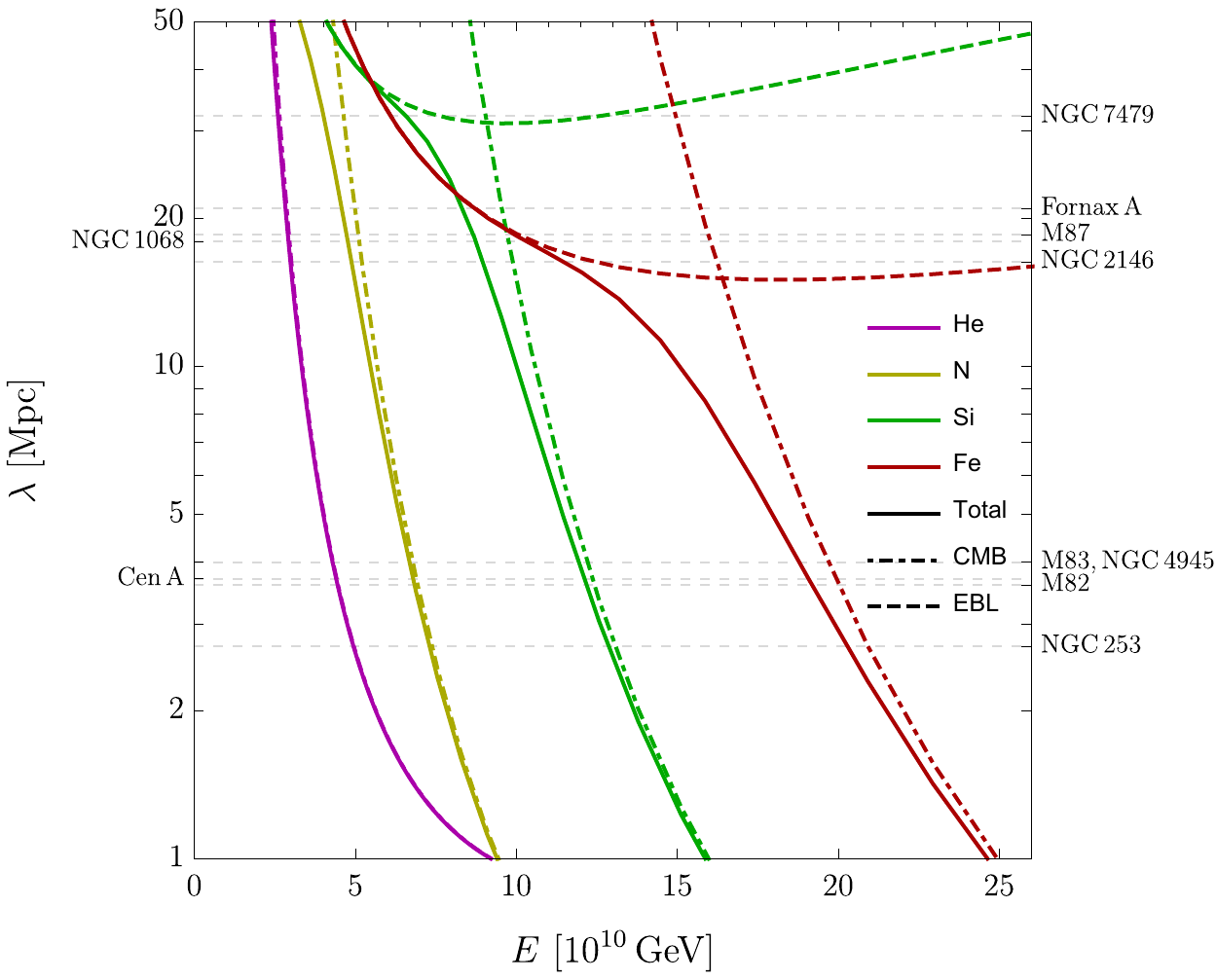}{1.1} 
\end{minipage}  
\caption{Photodisintegration mfp on the CMB and EBL as a function of
  the Lorentz factor $\gamma$ (left) and nucleus energy $E$
  (right). The horizontal dashed lines indicate the distance to nearby
  starbursts and radio-galaxies. This figure is courtesy of Jorge
  Fernandez Soriano.}
\label{fig:15} 
\end{figure}

In Fig.~\ref{fig:15} we show the nucleus mfp from a numerical integration of (\ref{lambdamenosuno}) 
using  precise cross section curves fitted to data for single-nucleon and
multi-nucleon emission. For the nitrogen, silicon, and iron
calculations we adopted a numerical integration of ``{\sc talys}-1.6
(restored)''  cross sections (as described in Appendix A of~\cite{Batista:2015mea}), using 
the optical and infrared backgrounds  
estimated in~\cite{Gilmore:2011ks}, as well as the CMB Planckian spectrum.  For helium, we used the fit to
photodisintegration cross section data of~\cite{Soriano:2018lly}. For
iron nuclei, $\varepsilon_0 \sim 18~{\rm MeV}$, and since the CMB
temperature is $T\sim 0.2348~{\rm meV}$ the maximum of the scaling
function $x \sim 1$ corresponds to a Lorentz factor $\gamma \sim
10^{10.6}$. The critical value of the scaling function, $x<2$, implies
that interactions with the CMB are exponentially suppressed for
$\gamma < 10^{10.3}$. These phenomenological estimates are visible in
Fig.~\ref{fig:15}. Actually, for $10^9 \lesssim \gamma \lesssim  10^{9.5}$ interactions with the
infrared  background dominate the $A^*$-rate for iron nuclei, and
the approximation given in (\ref{RBE}) breaks down. This corresponds to $E \lesssim 
 10^{11.2}~{\rm GeV}$. 

\begin{figure}
\postscript{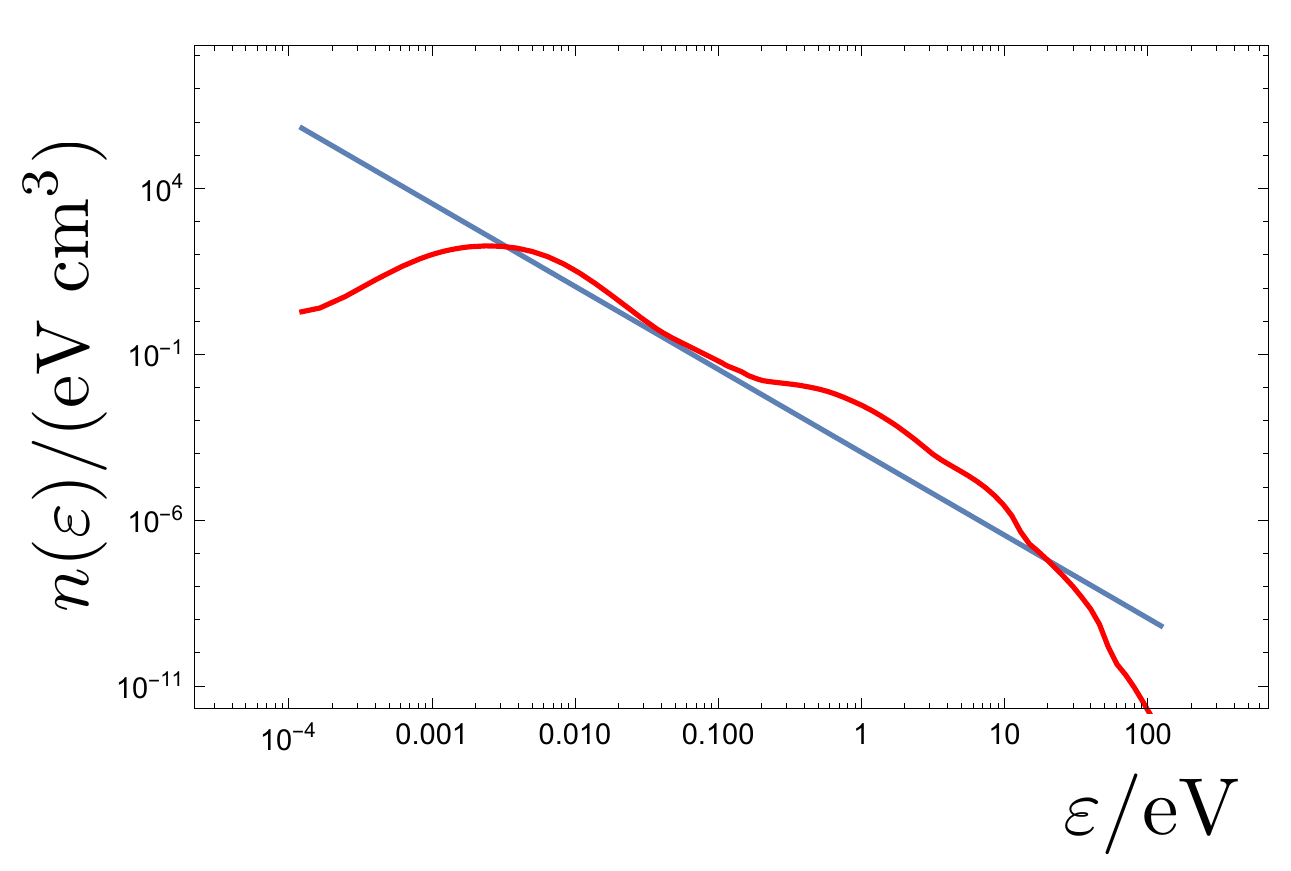}{0.5}
\caption{Comparison of the density of infrared photons as parametrized in
  (\ref{epele-roulet}) and estimated in~\cite{Gilmore:2011ks}.}
\label{fig:16}
\end{figure}

For  $2 \times 10^{-3} < \varepsilon/{\rm eV} < 0.8$, the spectral
density of the cosmic infrared background can be parametrized by~\cite{Epele:1998ia}
\begin{equation}
n(\varepsilon) \simeq 1.1 \times 10^{-4} \
\left(\frac{\varepsilon}{{\rm eV}} \right)^{-2.5}~{\rm cm^{-3} \
  eV^{-1}} \, .
\label{epele-roulet}
\end{equation}
A comparison of this parametrization and the estimate of
$n(\varepsilon)$ given in~\cite{Gilmore:2011ks} is shown in
Fig.~\ref{fig:16}.  Substitution of (\ref{epele-roulet}) into
(\ref{world-cup}) leads to the mfp of a nucleus scattering off the
infrared photons
\begin{equation}
\lambda^{-1} (A) \approx \frac{\varpi \ \varepsilon_0}{2 \gamma^2}
\int_{\frac{\varepsilon_0} {2\gamma}}^{\varepsilon_{\rm
      max}}
  \frac{n(\varepsilon)}{\varepsilon^2}  \ \ d \varepsilon \sim 8.4
  \times 10^{20} \bigg(\frac{\sigma_0}{{\rm
      cm}^2} \bigg) \, \left(\frac{\Gamma}{{\rm eV}} \right) \, \bigg(\frac{\varepsilon_0}{{\rm eV}}
  \bigg)^{-2.5} \, \gamma^{1.5}~~{\rm Mpc^{-1}} \,,
\end{equation}
where $\varepsilon_{\rm max} = 0.8~{\rm eV}$. Note that
$\varepsilon_{\rm min} = 2 \times 10^{-3}~{\rm eV}$ sets an upper
bound on the nucleus Lorentz factor, $\gamma < \varepsilon_0/(2
\varepsilon_{\min})$. For an iron nucleus, this translates to $\gamma
< 4.5 \times 10^{9}$ or equivalently $E < 10^{11.4}~{\rm GeV}$.

At this stage it is worthwhile to point out that since the Lorentz
factor is conserved in the photodisintegration process, the surviving
fragment $A_f$ sees the photons of the thermal background at the same
energy of the parent nucleus $A_i$. This implies that $A_f$ would
never reach the GDR threshold and the $A^*$ process must continue
until the primary nucleus photodisintegrates completely. In contrast,
the BH pair production and photopion production processes, involve the
creation of new particles that carry off energy, yielding baryons with
energies ever closer to the photopion or pair production
thresholds. This implies that if UHECRs are dominantly protons and the
observed suppression originates in the GZK interactions, then an
enhancement of the flux (visible as a bump in the spectrum) would
occur before the GZK cutoff as a consequence of the pile-up of energy
degraded protons recoiling down from higher energies and ending up
approximately below threshold to undergo further photoproduction
reactions~\cite{Hill:1983mk,Stecker:1989ti}. Although nuclei undergoing
photodisintegration never reach the threshold of the GDR, a bump in the
spectrum would be expected, too~\cite{Anchordoqui:1997rn}. The source
of the bump originates in a distinct property of the
photodisintegration process: after the same propagation distance, the
surviving fragments of two identical nuclei of baryon number $A$ emitted
with different energies can reach the Earth with the same energy, but
different baryon number. Therefore, the observation of a bump in the
spectrum (just before the cutoff) will be a clear indication that GZK
interactions are at play.

During the $A^*$ process, production of $e^+e^-$ pairs tend to reduce
the nucleus Lorentz factor and mitigate the rate of
photodisintegration. For $10^{10.5} \lesssim E/{\rm GeV} \lesssim
10^{11.5}$ and propagation distances $\lesssim 50~{\rm Mpc}$, the
effect of $e^+e^-$ pair and photopion production can be safely
neglected~\cite{Blumenthal:1970nn}. One sees in Fig.~\ref{fig:15} that
for \mbox{$E > E_{\rm supp}$}, the interaction mfp decreases rapidly with
increasing energy, and increases rapidly with increasing nuclear
composition:
\begin{itemize}[noitemsep,topsep=0pt]
\item at $E = 10^{10.7}$~GeV, the mfp for ionized helium ($^4$He) is
  about 3~Mpc, while at $10^{10.9}~{\rm GeV}$ it is nil; 
\item at  $E = 10^{10.9}~{\rm GeV}$, the mfp for ionized nitrogen ($^{14}$N) is about 4~Mpc, while at $10^{11}$~GeV it is nil; 
\item at $E =10^{11.1}$~GeV, the mfp for ionized silicon ($^{28}$Si)
  is about 2.5~Mpc, while at $10^{11.2}~{\rm GeV}$ it is nil;
\item etcetera, until finally we reach ionized iron ($^{56}$Fe) where
  the mfp at $E = 10^{11.3}$~GeV is about 3~Mpc, 
while at $10^{10.4}~{\rm GeV}$ it too is nil.
\end{itemize}
Altogether, the nearby Universe behaves as a {\it cosmic mass
  spectrometer}~\cite{Anchordoqui:2017abg}.  From sources at
increasing distance, fewer and heavier nuclei at highest energies are
expected to reach Earth.  The main features in the energy evolution of
the abundance of various nuclear species on Earth can be summarized as
follows:
\begin{itemize}[noitemsep,topsep=0pt]
\item the contribution of $^4$He should decrease with rising energy and then essentially disappear above
about $10^{10.8}~{\rm GeV}$; 
\item on average, only species heavier than $^{14}$N can contribute to the observed flux on 
Earth above $10^{11}~{\rm GeV}$, with nuclear species lighter than $^{28}$Si
highly suppressed at $10^{11.2}~{\rm GeV}$; 
\item the mean flux of iron nuclei becomes suppressed somewhat below $10^{11.4}~{\rm GeV}$. 
\end{itemize}
The three considerations enumerated above are similar to those
obtained assuming a continuous source distribution, with cutoff at
about 3~Mpc~\cite{Allard:2008gj,Allard:2011aa,Kotera:2011cp}. 

In the limit of small deflections (expected for nG field
strength~\cite{Pshirkov:2015tua}) the typical deflection of UHECRs in
the extragalactic magnetic field can be estimated to be
\begin{equation}
\delta \theta_{\rm extragalactic} \approx 1.5^\circ Z \  \sqrt{\frac{d}{3.8~{\rm Mpc}} \ \frac{\lambda}{0.1~{\rm Mpc}}} \ \left(\frac{B}{{\rm nG}}
\right) \ \left(\frac{E}{10^{10}~{\rm GeV}} \right)^{-1} \,,
\end{equation}
where $d$ is the source distance,  $Z$ is
the charge of the UHECR in units of the proton
charge, and $\lambda$ the magnetic field coherence length~\cite{Waxman:1996zn,Farrar:2012gm}. It is then reasonably to
assume that extragalactic deflections would generally be much smaller
than those arising from the Galactic magnetic field.

The Galactic magnetic field is not well constrained by current data, but if we
adopt our benchmark JF model~\cite{Jansson:2012pc,Jansson:2012rt,Unger:2017kfh}
typical values of the deflections of UHECRs crossing the Galaxy are
\begin{equation}
\delta \theta_{\rm Galactic} \sim 10^\circ \ Z  \ \left(\frac{E}{10^{10}~{\rm GeV}}
\right)^{-1} \,,
\label{deflection}
\end{equation} 
depending on the direction considered~\cite{Farrar:2017lhm}. When the
average magnetic field deflection is combined with the energy loss
during propagation shown in Fig.~\ref{fig:15} one can anticipate that
the energy dependence of the hot-spot contour produced by a
pure-proton source would be different from that of a source emitting a
mixed composition of nuclei~\cite{Anchordoqui:2017abg}. The difference
can be easily pictured when considering the main properties
characterizing the acceleration and propagation of a nucleus of charge
$Ze$ and baryon number $A$: {\it (i)}~the maximum energy of
acceleration capability of the sources grows linearly in $Z$, {\it
  (ii)}~the energy loss per distance traveled decreases with
increasing $A$, and {\it (iii)}~the bending of the cosmic ray
decreases as $Z/E$ with rising energy. This implies that the pointing
of UHECR nuclei to their nearest extragalactic sources would only be
expected for \mbox{$10^{10.6} \lesssim E/{\rm GeV} \lesssim 10^{11}$.}
Actually, the less deflected $^4$He nuclei could only contribute to
the anisotropy hot-spot if $E\lesssim 10^{10.8}~{\rm GeV}$. In
contrast, sources of UHECR protons exhibit anisotropy patterns which
become denser and compressed with rising energy. An intuitive
description with analogy to the peeling of an onion, which portrays
the hot-spot differences, is as follows: while peeling a hot-spot
produced by a pure-proton source one would be probing higher and
higher energies, whereas when peeling a hot-spot produced by source
that emits a mixed-composition of nuclei one would be probing lower
and lower energies. When effects of the regular component of the
Galactic magnetic field are included, the the hot-spots are elongated
depending on their location in the sky~\cite{Anjos:2018mgr}. Stock-still the variation in
(shape and size) of the expected ``squeezed onion layers'' provides a
powerful discriminator of the baryonic composition.

A point worth recalling at this juncture is that the TS
scan over the threshold energy in the most recent Auger anisotropy analysis gives a maximum
signal for SBGs when considering UHECRs with $E > 39~{\rm EeV}$ and
for $\gamma$AGNs when considering UHECRs with $E > 60~{\rm EeV}$~\cite{Aab:2018chp}. Then the
previous considerations on the energy dependence of the hot-spot
contour allows one to speculate that  $\gamma$AGNs are most
likely associated with pure-proton sources, while  SBGs are more likely sources
emitting a mixed composition of nuclei.

\subsection{Plausible sources in our cosmic backyard}
\label{sec:c_backyard}

\subsubsection{$\gamma$AGNs}

AGNs are composed of an accretion disk around a central super-massive
black hole and are sometimes associated with jets terminating in
lobes, which can be detected in radio. One can classify these objects
into two categories: radio-quiet AGN with no prominent radio emission
nor jets, and radio-loud objects presenting jets. All of the AGNs with
MeV-GeV-TeV $\gamma$-ray emission identified so far are 
radio-loud galaxies.  

Fanaroff-Riley II (FRII) galaxies~\cite{Fanaroff} are  a sub-class
of radio-loud sources. Localized regions of intense synchrotron
emission, known as hot-spots, are observed within their
lobes. These regions are presumably produced when the bulk kinetic
energy of the jets ejected by a central active nucleus is reconverted
into relativistic particles and turbulent fields at a {\it working
  surface} in the head of the jets~\cite{Blandford}. Specifically, the
speed $u_{\rm head}\approx \; u_{\rm jet}\,[ 1 + ( n_e /n_{\rm
  jet})^{1/2}]^{-1}$, with which the head of a jet advances into the
intergalactic medium of particle density $n_e$ can be obtained by
balancing the momentum flux in the jet against the momentum flux of
the surrounding medium; where $n_{\rm jet}$ and $u_{\rm jet}$ are the
particle density and the velocity of the jet flow, respectively (for
relativistic corrections, see~\cite{Rosen:1999jm}). For $n_e \geq
n_{\rm jet}$, $u_{\rm jet}> u_{\rm head}$ so that that the jet
decelerates. The result is the formation of a strong collisionless
shock, which is responsible for particle reacceleration and magnetic
field amplification~\cite{Begelman:1984mw}.  The acceleration of
particles up to ultrarelativistic energies in the hot-spots is the
result of repeated scattering back and forth across the shock front,
similar to that discussed in Sec.~\ref{fermi_acceleration}.  The
particle deflection in this mechanism is dominated by the turbulent
magnetic field with wavelength $k$ equal to the Larmor radius of the
particle concerned~\cite{Drury:1983zz}. A self-consistent (although
possibly not unique) specification of the turbulence is to assume that
the energy density per unit of wave number of MHD turbulence is of the
Kolmogorov type, $I(k) \propto k^{-5/3}$, just as for hydrodynamical
turbulence~\cite{Kolmogorov}. With this in mind, to order of magnitude
accuracy using effective quantities averaged over upstream (jet) and
downstream (hot-spot) conditions (considering that downstream counts a
fraction of 4/5) the acceleration timescale at a shock front is found
to be
\begin{equation}  
\left(\frac{1}{E}  \frac{dE}{dt} \right)^{-1} \sim \frac{20 \ D_K(E)}{u_{\rm jet}^2} \,, 
\label{acc}
\end{equation} 
where 
\begin{equation}
D_K (E) = \frac{2}{\pi\,  \varrho}\,\left(\frac{E}{ZeB}\right)^{1/3}\, R^{2/3}
\end{equation}
is the Kolmogorov diffusion coefficient, $\varrho$ is the ratio of
turbulent to ambient magnetic energy density in the region of the
shock (of radius $R$), and $B$ is the total magnetic field
strength~\cite{Rachen:1992pg}. Note that (\ref{acc}) follows from
(\ref{deltaEfermi2nd}), (\ref{Erate}), and (\ref{tcycle}) taking a
strong shock with compression ratio $\zeta = 4$ (corresponding to
$u_{\rm jet} = u_{\rm head}/4$) and $D_1 = D_2 = D_K$. On the basis of
the discussion given in Sec.~\ref{sec:323}, it is
reasonable to set $Z=1$ for $\gamma$AGNs.

The acceleration process will be efficient as long as the energy
losses by synchrotron radiation and/or pion-producing interactions do
not become dominant.  The synchrotron loss time for protons is given
by
\begin{equation}
 - \left(\frac{1}{E} \frac{dE}{dt}\right)^{-1}  \sim \frac{6\, \pi\, m_p^3\,c}{\sigma_{\rm T}\,m_e^2\,\gamma\,B^2}\,,
\label{tausyn}
\end{equation}
where  $\sigma_{\rm T}$ is the  Thomson
cross section  and  and $\gamma = E/m_p$ is 
Lorentz factor~\cite{Rybicki,Biermann:1987ep}. 

For typical hot-spot conditions, the number density of photons per
unit energy interval follows a power-law spectrum
\begin{equation}
n (\varepsilon) = \left\{\begin{array}{cl} (N_r/\varepsilon_r) \ \ (\varepsilon/\varepsilon_r)^{-2} ~~~~&~~~~ \varepsilon_r \leq \varepsilon \leq \varepsilon_g \\
0 ~~~&~~~~ {\rm otherwise} \end{array} \right.
\label{ngammaagn}
\end{equation} 
where $N_r$ is the normalization constant and $\varepsilon_r$ and
$\varepsilon_g$ correspond to radio and gamma rays energies,
respectively~\cite{Biermann:1987ep}. Straightforward substitution of
(\ref{ngammaagn}) into (\ref{conventions}) leads to the fractional energy loss due to
photopion production 
\begin{equation}
-\left( \frac{1}{E} \frac{dE}{dt} \right)^{-1} = \frac{4}{3} N_r
\frac{\varepsilon_r}{\varepsilon'_{\rm th}} \gamma \int_1^\infty \langle
  y(x\varepsilon'_{\rm th}) \rangle  \ \sigma (x \varepsilon'_{\rm
    th}) x^{-2} dx \sim \frac{2}{3} \ a \ \gamma \left[\frac{\langle \sigma_{\gamma
        p} \rangle}{\ln\left(\varepsilon_g/\varepsilon_r \right)} \right]
    \left(\frac{B^2}{m_p} \right) \,,
\end{equation}
where
\begin{equation}
a = \frac{N_r \ \varepsilon_r \ \ln(\varepsilon_g/\varepsilon_r)}{B^2/2}
\end{equation}
is the  ratio of photon to magnetic energy density, 
\begin{equation}
\int_1^\infty \langle
  y(x\varepsilon'_{\rm th}) \rangle  \ \sigma (x \varepsilon'_{\rm
    th}) x^{-2} dx \sim \frac{m_\pi}{m_p} \  \langle \sigma_{\gamma p}
  \rangle \sim
  900~\frac{m_\pi}{m_p}~\mu{\rm b} \, ,
\end{equation}
and $\varepsilon'_{\rm th} = m_\pi$~\cite{Biermann:1987ep}. The time
scale of the energy loss for protons, including synchrotron and photon
interaction losses, can be approximated by
\begin{equation}
 - \left(\frac{1}{E} \frac{dE}{dt}\right)^{-1}  \sim \frac{2}{3} \frac{
   m_p^4}{\sigma_{\rm T}\ m_e^2\ B^2\ (1+Aa)}\ E^{-1} \, ,
\end{equation} 
where 
\begin{equation}
A = \frac{\langle \sigma_{\gamma p} \rangle}{\sigma_{\rm T}} \,
\frac{(m_p/m_e)^2}{\ln (\varepsilon_g/\varepsilon_r)} \approx 1.6
\times 10^5 \frac{\langle \sigma_{\gamma p} \rangle}{\sigma_{\rm T} } \ \approx 200 \, .
\end{equation}
gives a measure of the relative strength of $\gamma p $ interactions
versus the synchrotron emission.  Note that $p \gamma \to \pi^+ n$
involves the creation of ultrarelativistic neutrons that can readily
escape the system and provide a directional signal for nearby
sources~\cite{Anchordoqui:2001nt,Anchordoqui:2011ks}. The maximum
attainable energy can be obtained by balancing the energy gains and
losses~\cite{Anchordoqui:2001bs}
\begin{equation}
E =1.4\times 10^{16}\,\left(\frac{B}{{\mu{\rm G}}}\right)^{-5/4}\,\,u_{\rm jet}^{3/2}
\,\,\varrho^{3/4}\  \left( \frac{R}{{\rm
      kpc}}\right)^{-1/2}\,\,(1+Aa)^{-3/4}~{\rm GeV} \,,
\label{ab}
\end{equation}
It is of interest to apply the acceleration conditions to the nearest
$\gamma$AGN. It is this that we now turn to study.

As the closest radio-loud galaxy to Earth, Cen A (identified at
optical frequencies with the galaxy NGC 5128) is the perfect {\it
  cosmic laboratory} to study the physical processes responsible for
UHECR acceleration.  Radio observations at different wavelengths have
revealed a rather complex morphology of this FRI
source~\cite{Israel:1998ws}. It comprises a compact core, a jet (with
subluminal proper motions $u_{\rm jet} \sim
0.5$~\cite{Hardcastle:2003ye}) also visible at $X$-ray frequencies, a
weak counter-jet, two inner lobes, a kpc-scale middle lobe, and two
giant outer lobes. The jet would be responsible for the formation of
the northern inner and middle lobes when interacting with the
interstellar and intergalactic media, respectively. There appears to
be a compact structure in the northern lobe, at the extrapolated end
of the jet. This structure resembles the hot-spots such as those
existing at the extremities of FRII galaxies. However, at Cen A it
lies at the side of the lobe rather than at the most distant northern
edge, and the brightness contrast (hot-spot to lobe) is not as
extreme~\cite{Burns}.

EGRET observations of the $\gamma$-ray flux for energies $>100~{\rm
  MeV}$ allow an estimate $L_{\gamma} \sim 10^{41}\ \es$ for Cen
A~\cite{Sreekumar:1999xw}. This value of $L_{\gamma}$ is consistent
with an earlier observation of photons in the TeV-range during a
period of elevated $X$-ray activity~\cite{Grindlay}, and is
considerably smaller than the estimated bolometric luminosity $L_{\rm
  bol}\sim 10^{43}\es$~\cite{Israel:1998ws}.  During the first
3-months of science operation, the {\it Fermi}-LAT confirmed the EGRET
detection of Cen A~\cite{Abdo:2009mg,Abdo:2009wu}. Besides, data from H.E.S.S. have
confirmed Cen A as a TeV $\gamma$-ray emitting
source~\cite{Aharonian:2009xn}. Extrapolating the spectrum measured
by {\it Fermi}-LAT in the GeV regime to very-high energies roughly matches the
H.E.S.S. spectrum~\cite{Falcone:2010fk}. Data from {\it Fermi}-LAT
established that a large fraction ($>1/2$) of the total $> 100~{\rm
  MeV}$ emission from Cen A emanates from the
lobes~\cite{Fermi-LAT:2010llz}. However, a combined analysis of
H.E.S.S. and {\it Fermi} data shows that the very-high energy $\gamma$-ray emission comes from the
core of Cen A with $12\sigma$ significance~\cite{Abdalla:2018agf}.

Estimates of the radio spectral index of
synchrotron emission in the hot-spot and the observed degree of linear
polarization in the same region suggests that the ratio of turbulent
to ambient magnetic energy density in the region of the shock is
$\varrho \sim 0.4$~\cite{Romero:1995tn}. The broadband radio-to-X-ray
jet emission yields an equipartition magnetic field $B
\sim 100~{\mu{\rm G}}$~\cite{Honda:2009xd}.\footnote{The usual way to estimate the
  magnetic field strength in a radio source is to minimize its total
  energy. The condition of minimum energy is obtained when the
  contributions of the magnetic field and the relativistic particles
  are approximately equal (equipartition condition). The corresponding
  $B$-field is commonly referred to as the equipartition magnetic
  field.}  The radio-visible size of the hot-spot can be directly
measured from the large scale map $R\simeq
2~{\rm kpc}$~\cite{Junkes}. The actual size can be larger because of
uncertainties in the angular projection of this region along the line
of sight.\footnote{For example, an explanation of the apparent absence
  of a counter-jet in Cen A via relativistic beaming suggests that the
  angle of the visible jet axis with respect to the line of sight is
  at most 36$^{\circ}$~\cite{Burns}, which could lead to a doubling of
  the hot-spot radius. It should be remarked that for a distance of
  3.4 Mpc, the extent of the entire source has a reasonable size even
  with this small angle.} Substituting these fiducial values in
(\ref{Hillas}) and (\ref{ab}) it is easily seen that if the ratio of photon to
magnetic energy density $a \lesssim 0.4$, it is plausible that Cen A
can accelerate protons up to the maximum observed energies:  $E_{p,{\rm max}} \sim  10^{11}$~GeV.

In order to ascertain the capability of Cen A to accelerate UHECR
protons up to $10^{11}~{\rm GeV}$ one must crosscheck the lower limit
on the rate at which the energy is carried by the out-flowing plasma
(\ref{LP-rigidity}), which must be provided by the source.  The minimum total power of
the jets inflating the giant lobes of Cen A is estimated to be
$\approx 8 \times 10^{43}~{\rm erg/s}$~\cite{Dermer:2008cy}.  This
argument provides a conservative upper limit for the magnetic field in
the jet with kpc-scale radius, $B \lesssim 50~\mu {\rm G}$, and through
(\ref{Hillas}) leads to $E \lesssim 10^{10.7}~{\rm GeV}$. The jet
power required to maintain these extreme $B$ values of ${\cal O}(\mu {\rm G})$
and $R$ of ${\cal O} ({\rm kpc})$ can be reached during flaring
intervals~\cite{Dermer:2008cy}. Acceleration of UHECR
protons up to somewhat beyond $10^{11}~{\rm GeV}$ is therefore in
principle possible during powerful episodes of jet activity.

Alternatively, shear acceleration~\cite{Rieger:2004jz} could help push
proton energies up to and beyond 50~EeV~\cite{Rieger:2009pm}. The
limb-brightening in the $X$-ray jet together with the longitudinal
magnetic field polarization in the large scale jet might be indicative
of internal jet stratification, {\em i.e.} a fast spine surrounded by
slower moving layers. Energetic particles scattered across such a
shear flow can sample the kinetic difference in the flow and will
naturally experience an additional increase in energy.  In particular,
protons that diffuse from the inner shock region into the outer shear
layers charge-exchange to produce neutrons with $E \sim 10^{11}~{\rm
  GeV}$, which would escape the source. Finally, it has been noted
that the jet powers in local radio galaxies could feasibly have been
different in the past~\cite{Matthews:2018laz}. Indeed, there seems to be evidence in Cen A for
enhanced activity within the last $\sim 100~{\rm Myr}$~\cite{Wykes:2013gba,Eilek:2014gya}.
Altogether, one can conclude that Cen A and other nearby radiogalaxies
(like M87 and Fornax A) can accelerate protons up to about $10^{11}~{\rm GeV}$.

We cannot go without noticing that AGN flares resulting from the tidal disruption of a star or from a
disk instability also meet the UHECR acceleration
requirements~\cite{Farrar:2008ex}. Interestingly, such tidal
disruption events can generate the luminosity required to account for
the full-sky UHECR intensity~\cite{Farrar:2014yla}, and may
accommodate the intermediate scale anisotropies which seem to be
emerging in TA and Auger data~\cite{Pfeffer:2015idq}. However, it is
not clear whether they can accommodate the observed mixed composition,
which appears to dominate the UHECR intensity above the
ankle~\cite{AlvesBatista:2017shr,Zhang:2017hom,Biehl:2017hnb}.

\subsubsection{SBGs}

Starbursts are galaxies (sometimes, the term also refers only to
particular regions of galaxies) undergoing a large-scale star
formation episode. The universal fast star formation in starburst
galaxies is directly correlated with the efficient ejection of gas,
which is the fuel for star formation. This happenstance generates a
galactic-scale superwind, which is powered by the momentum and energy
injected by massive stars in the form of supernovae, stellar winds,
and radiation~\cite{Heckman,Veilleux:2005ia}. Multi-wavelength
observations seem to indicate that these superwinds are genuinely
multi-phase: with hot, warm, cold, and relativistic (cosmic rays)
phases. These observations also suggest a pervasive development of the
hot ($T \sim 10^{7}~{\rm K}$) and warm diffuse ionized ($T \sim
10^4~{\rm K}$) phases. Namely, experiment shows that the hot and warm
large-scale supersonic outflows escalate along the rotation axis of
the disk to the outer halo area in the form of local chimneys.  Such a
supersonic outflow, however, does not extend indefinitely. As the
superwind expands adiabatically out beyond the confines of the
starburst region, its density decreases. At a certain radial distance
the pressure would become too small to further support a supersonic
flow. Whenever the flow is slowed down to subsonic speed a termination
shock stops the superwind. The shocked gas continues as a subsonic
flow. The termination shock would remain in steady state as long as
the starburst lasts.  This set up clearly provides a profitable arena
for acceleration of UHECRs~\cite{Anchordoqui:1999cu}.

Because of the high prevalence of supernovae, starbursts should
possess a large density of newly-born pulsars~\cite{Long:2014tsa}. Due to their important
rotational and magnetic energy reservoirs these young neutron stars,
with their metal-rich surfaces, have been explored as a potential
engine for UHECR acceleration~\cite{Blasi:2000xm,Arons:2002yj,Fang:2012rx}. As we
discussed already in Sec.~\ref{sec:unipolar}, the combination of the fast star rotation
and its strong magnetic field can induce large potential differences
in the out-flowing relativistic plasma for UHECR acceleration.

There are numerous indications that long GRBs are extreme
supernova events, which arise from the death of massive
stars~\cite{MacFadyen:1998vz}.  Starburst galaxies are characterized
by high star-formation rates per unit area, of the order of 15 to
20~$M_\odot~{\rm yr^{-1}} \, {\rm kpc}^{-2}$~\cite{Heckman}. This is up to several hundred times larger
than the characteristic value normally found in gas-rich 
galaxies like the Milky Way.  The observed supernova rate in
starbursts is also higher than average, and so it seems only natural to
expect a high rate of long GRBs too~\cite{Dreyer:2009pj,Biermann:2016xzl}. However, the
star formation rates per unit stellar mass of GRB host galaxies are
found to be higher than for typical nearby starburst
galaxies~\cite{Chary:2001yx}. Moreover, stronger and stronger
experimental evidence has been accumulating that implies GRB hosts are low mass
irregular galaxies and have low metallicity, see
e.g.~\cite{Stanek:2006gc,Modjaz:2007st,Jimenez:2013dka}. Altogether,
this makes the GBR $\leftrightharpoons$ (metal-rich) starburst
connection highly unlikely.\footnote{Similar considerations apply to
  hypernova host galaxies~\cite{Wang:2007ya}.}

The acceleration of particles in starburst galaxies may alternatively
proceed in a two-stage process~\cite{Anchordoqui:1999cu}. First, ions
are diffusively accelerated at single SNRs within the
nuclear region of the galaxy. Energies up to about $50~{\rm PeV}$ can be achieved in this
stage~\cite{Ptuskin:2010zn}. Collective plasma motions force the CR
gas to stream along from the starburst region. Some nuclei then escape
through the disk in opposite directions along the symmetry axis of the
system and experience supplementary acceleration at the terminal shock
of the galactic-scale superwind.  To picture the specifics of diffuse shock
acceleration in SBGs, consider a spherical cavity where core-collapse
supernovae and stellar winds inject kinetic energy. This kinetic
energy then thermalizes and drives a super-heated outflow that escapes
the sphere. To a first approximation we ignore gravity, radiative
cooling, and other effects~\cite{Chevalier:1985pc}. In this
approximation energy conservation leads to the asymptotic speed of the
outflow
\begin{equation}
u_\infty \approx \sqrt{\frac{2 \dot  E_{\rm sw}}{\dot M_{\rm sw}}} \sim 10^3
\sqrt{\frac{\epsilon}{\beta}}~~{\rm km} \ {\rm s^{-1}} \,, 
\label{vinfinity}
\end{equation}
where $\dot E_{\rm sw}$ and $\dot M_{\rm sw}$ are respectively the energy and mass
injection rates inside the spherical volume of the
starburst region, and where
$\beta$ is the mass loading factor, i.e. the ratio of the mass
injection rate to the star formation rate.  In the
second rendition we have scaled the energy injection rate expected
from core-collapse supernovae considering a thermalization efficiency
$\epsilon$.  For this order of magnitude calculation, we have assumed
that in total a $100 M_\odot$ star injects ${\cal O} (10^{51}~{\rm
  erg})$ into its surroundings during the wind phase. 

As the cavity expands adiabatically a strong shock front is formed on
the contact surface with the cold gas in the halo. At the region where
this occurs, the inward ram pressure is balanced by the pressure
inside the halo, $P_{\rm halo}$. A point worth noting at this juncture is that the
difference in pressure between the disk and the halo manifestly breaks the
symmetry, and so the outflowing fluid which escapes from the starburst
region features back-to-back chimneys with conic
profiles. Rather than considering a spherical shock we assume the outflow
cones fill a solid angle $\Omega$, and hence the ram pressure
at radius $r$ is found to be
\begin{equation}
P_{\rm ram} = \frac{\rho_{\rm sw} \  u_\infty^2}{2} = \frac{\dot
  p_{\rm sw}}{2 \Omega \ r^2} = \frac{\dot M_{\rm sw} \
  u_\infty}{2 \Omega \ r^2} = \frac{\sqrt{2 \ \dot E_{\rm sw} \
    \dot M_{\rm sw} }
}{2 \Omega \ r^2}  \,,
\label{rampi}
\end{equation}
where $\rho_{\rm sw} = \dot M_{\rm sw}/(\Omega u_\infty r^2)$ is the
density of the outflow and $\dot p_{\rm sw} = \dot M_{\rm sw} \,
u_\infty$ is the asymptotic momentum injection rate of the
superwind~\cite{Lacki:2013zsa}. The agitated superwind gas inside the shock
is in pressure equilibrium with the outside gas at a radius
\begin{equation}
R_{\rm sh} \sim \sqrt{\frac{\dot M_{\rm sw} \ u_\infty }{2 \Omega \ P_{\rm halo}}}
\, . 
\label{rampi-halo}
\end{equation}
In (\ref{rampi}) and (\ref{rampi-halo}) it was implicitly
assumed that the magnetic field is parallel to the shock
normal. Recall that for a flow-aligned field, the fluid motion
decouples from the field.
The termination shock is a steady-state feature, present even if the
starburst wind has always been active. 

To develop some sense of the orders of magnitude involved, we assume
that the prominent M82 typifies the nearby starburst population. For a
standard Kroupa initial mass function~\cite{Kroupa:2002ky}, our
archetypal starburst has a star formation rate $\sim 10 M_\odot~{\rm
  yr}^{-1}$ and a radius of about $400~{\rm pc}$.\footnote{The initial
  mass function is an empirical function that describes the
  distribution of initial masses for a population of stars.} Hard
X-ray observations provide direct observational evidence for a
hot-fluid phase.  The inferred gas temperature range is $10^{7.5}
\lesssim T/{\rm K} \lesssim 10^{7.9}$, the thermalization efficiency
$0.3 \lesssim \epsilon \lesssim 1$, and the mass loading factor $0.2
\lesssim \beta \lesssim 0.6$. Substituting for $\epsilon$ and $\beta$
into (\ref{vinfinity}) we obtain $1.4 \times 10^3 \lesssim
u_\infty/({\rm km \, s}^{-1}) \lesssim 2.2 \times
10^3$~\cite{Strickland:2009we}. The warm fluid has been observed
through nebular line and continuum emission in the vacuum ultraviolet,
as well as through mid- and far-infrared fine-structure line emission
excitations~\cite{Heckman:2000sj,Hoopes:2006mz,Beirao,Contursi:2012wa}. High-resolution
spectroscopic studies seem to indicate that the warm ($T \sim
10^4~{\rm K}$) gas has emission-line ratios consistent with a mixture
of photo-ionized gas by radiation leaking out of the starburst and
shock-heated by the outflowing superwind fluid generated within the
starburst~\cite{Heckman:1990fe}.  The kinematics of this gas, after
correcting for line-of-sight effects, yields an outflow speed of the
warm ionized fluid of roughly 600~${\rm km} \, {\rm s}^{-1}$. The
velocity field, however, shows rapid acceleration of the gas from the
starburst itself out to a radius of about 600~pc, beyond which the
flow speed is roughly constant.  The inferred speed from cold and warm
molecular and atomic gas observations~\cite{Veilleux:2009rb,Leroy} is
significantly smaller than those observed from the warm ionized phase.
This is also the case for the starburst galaxy NGC 253: ALMA
observations of CO emission imply a mass loading factor of at least 1
to 3~\cite{Bolatto:2013aqa}. However, it is important to stress that
the emission from the molecular and atomic gas most likely traces the
interaction of the superwind with detached relatively denser ambient
gas clouds~\cite{Heckman}, and as such it is not the best gauge to
characterize the overall properties of the superwind
plasma~\cite{Lacki:2013sda}. (See~\cite{Romero:2018mnb} for a
different perspective.) Herein, we adopt the properties of the hot gas
detected in hard X-rays to determine the shock terminal velocity. We
take an outflow rate of $\dot M_{\rm sw} \sim 3 M_\odot~{\rm
  yr}^{-1}$, which is roughly 30\% of the star-formation rate ($\beta
\sim 0.3$), yielding $\dot E_{\rm sw} \sim 3 \times 10^{42}~{\rm
  erg}\, {\rm s}^{-1}$~\cite{Heckman}. For $\Omega \sim \pi$, this leads to
$u_\infty \sim 1.8 \times 10^3~{\rm km \ s}^{-1}$ and $R_{\rm sh} \sim
8~{\rm kpc}$, where we have taken $P_{\rm halo} \sim 10^{-14}~{\rm
  erg} \, {\rm cm}^{-3}$~\cite{Shukurov:2002hh}.

Radio continuum and polarization observations of M82 provide an
estimate of the magnetic field strength in the core region of
$98~\mu{\rm G}$ and in the halo of $24~\mu{\rm G}$; averaging the
magnetic field strength over the whole galaxy results in a mean
equipartition field strength of $35~\mu{\rm
  G}$~\cite{Adebahr:2012ce}. Comparable field strengths have been estimated for
NGC 253~\cite{Beck,Heesen:2008cs,Heesen:2009sg,Heesen:2011kj} and
other starbursts~\cite{Krause:2014iza}. Actually, the field strengths
could be higher if the cosmic rays are not in equipartition with the
magnetic
field~\cite{Thompson:2006is,Paglione:2012ma,Lacki:2013ry}.  
If this were the case, e.g., the magnetic field strength in M82 and NGC 253
could be as high as $300~\mu$G~\cite{DomingoSantamaria:2005qk,delPozo:2009mh,Lacki:2013nda}.

The duration of the starburst phenomenon is subject to large
uncertainties. The most commonly cited timescale for a starburst is 5
to 10~Myr, comparable to the lifetime of massive
stars~\cite{Thornley:2000ib,Torres:2004ui,Torres:2012xk}. However, it has been suggested that the
starburst phenomenon can be a longer and more global event than
related by the lifetime of individual massive stars or pockets of
intense star formation~\cite{Meurer:2000yq,McQuinn:2009gc,McQuinn:2010kn}. In this
alternative viewpoint the short duration timescales are instead
interpreted as a measure of the {\it flickering} created by currently
active pockets of star formation that move around the
galaxy. Measuring the characteristics of just one of these flickers
reveals much about an individual star formation region but of course
does not measure the totality of the starburst phenomenon in the
galaxy. If starbursts are indeed a global phenomenon, then the events
are longer than the lifecycle of any currently observable massive star
or area of intense star formation and the bursts are not
instantaneous. An observation that measures currently observable star
formation activity will therefore only measure the {\it flickering}
associated with a starburst pocket and not the entire phenomenon. This
aspect, frequently denied or not yet sufficiently emphasized, may
bring still another rewarding dimension to the problem at hand.

A measurement of the starburst phenomenon in twenty nearby galaxies
from direct evaluation of their star formation histories reconstructed
using archival Hubble Space Telescope observations suggests the average
duration of a starburst  is between 450 and 
650~Myr~\cite{McQuinn:2010kn}. 

Since the large-scale terminal shock is far from the starburst region,
the photon field energy density in the acceleration region drops to
values of the order of the CMB. Now,
for $E \lesssim 10^{11}~{\rm GeV}$ and $Z \gtrsim 10$, the energy attenuation
length $\gtrsim 30~{\rm Mpc}$~\cite{Allard:2011aa}. Therefore, we will
restrict ourselves to $\tau \lesssim 100~{\rm Myr}$. This duration range
is in good agreement with the overall star formation history of
M82~\cite{deGrijs:2001ec,deGrijs:2002qk} and NGC
253~\cite{Davidge:2010qg,Davidge}, and it is also consistent with the
upper limit on the starburst age of these galaxies derived
in~\cite{Rieke:1980xt}.

In toto, substituting $u_\infty \sim 1.8 \times 10^3~{\rm km \, s}^{-1}$, $B \sim 300~\mu$G, and $\tau  \sim 40~{\rm Myr}$
into (\ref{ColoEngla}) we obtain~\cite{Anchordoqui:2018vji}
\begin{equation}
E \lesssim Z \,  10^{10}~{\rm GeV} \, .
\label{Emax2}
\end{equation}
Note  that
(\ref{Emax2}) is consistent with the Hillas
criterion~\cite{Hillas:1985is}, as the maximum energy of confined
baryons at a shock distance of $R_{\rm sh}$ is found to be
\begin{equation}
E \lesssim 10^9 \ Z \ \left(\frac{B}{\mu G} \right) \ \left(\frac{R_{\rm sh}}{{\rm
    kpc}}\right)~{\rm GeV} \ .
\end{equation}
To accommodate a hard emission spectrum,  as required by Auger data,
the maximum energy at the accelerator is driven by
UHECR leakage from the boundaries of the shock (a.k.a. direct
escape~\cite{Baerwald:2013pu}), which corresponds to $\varkappa(\zeta) = 1/12$~\cite{Anchordoqui:2018vji}. 

Note, however, that starburst large-scale superwinds struggle to meet the power
constraint (\ref{LP-rigidity}), because $\dot E_{\rm sw} <
10^{43}~{\rm erg} \, {\rm s}^{-1}$ and they have low shock velocities
($< 10^{3.5}~{\rm km} \, {\rm s}^{-1}$)~\cite{Matthews:2018laz}. Amplified magnetic fields
close to the shock, as observed in supernova
remnants~\cite{Vink:2002yx,Yamazaki:2003xq,Volk:2004vi}, may offer a
window to escape this constraint. Theoretical studies
seem to indicate that streaming CRs may excite MHD turbulence, at least in
principle, to amplify the magnetic field by orders of magnitude from
its initial seed value~\cite{Lucek:2000,Bell:2004,Bell:2005,Matthews:2017apu}. Even though many
complex, highly nonlinear microscopic processes remain to be explored,
there is the possibility that non-linear interactions between CRs and
the magnetic field could provide a peculiar scheme where the CRs
themselves provide the magnetic field necessary for their acceleration
to ultra-high energies. Whichever point of view one may find more
convincing, it seems most conservative at this point to depend on
experiment (if possible) to resolve the issue.

Superwinds of galactic-scale have also been observed in the other
nearby SBGs contributing to the UHECR correlation signal of Auger~\cite{Aab:2018chp} and
TA~\cite{Abbasi:2018tqo}. Indeed, it has long been suspected that the observed properties of NGC
4945 can best be understood in the framework of a starburst-driven
superwind scenario~\cite{Lipari}.  NGC 4945 also hosts of a very
peculiar X-ray-luminous AGN that is probably heavily obscured along
all lines of sight~\cite{Marconi:2000xn,Levenson:2002fq}. Despite the
fact this galaxy has an X-ray and H nuclear outflow cone similar to
NGC 253, it is lacking in diffuse X-ray or H emission when compared to
a starburst of the same {\it total} galactic bolometric luminosity,
$\log L_{X,{\rm tot}} / L_{\rm bol} = -
4.25$~\cite{Strickland:2003xk}. However, the AGN may dominate $L_{\rm
  bol}$. If we assume that the diffuse X-ray emission is due to a
starburst-driven wind alone, then the starburst must only account for
￼20\% of $L_{\rm bol}$. In general, the large-scale soft X-ray
emission in starburst composite galaxies is consistent with a
purely-starburst
origin~\cite{Levenson:2000qc,Levenson:2000qd,Levenson:2003ez}. Thus,
for a given total bolometric luminosity it appears that supernovae are
more effective at driving galactic-scale winds than
AGN~\cite{Strickland:2004qk}. This does not imply that AGN-driven
galactic winds do not exist. There clearly are galaxies with AGN but
lacking starbursts that have galactic-scale (i.e. of order 10 kpc)
outflows~\cite{Colbert:1995sz,Colbert:1997fa}, but their local space
density is lower than typical starburst superwind galaxies. Evidence
for a galactic-scale superwind has also been observed in NGC
1068~\cite{Krolik:1986}, M83~\cite{Soria:2002jk,Vogler:2005bg}, and
Circinus~\cite{Veilleux:1997nk,Elmouttie}. 

In the TA search for correlations of UHECRs and SBGs, IC 342 provides
the second relative source contribution weighted by the directional
exposure, whereas M82 provides the leading contribution to the
correlation signal~\cite{Abbasi:2018tqo}. IC 342 is a late-type spiral
galaxy and is located at a distance of about 4~Mpc, though derived
distances have varied between 2~Mpc and
4~Mpc~\cite{Luppino,Krismer,Buta,Tikhonov:2010in}. This discrepancy
arises because IC 342 is located close to the galactic disk, and so
its light is dimmed by the Milky Way's intervening clouds. The
discrepancy in derived distances to IC 342 must be kept in mind when
discussing distance-dependent quantities. High-resolution
interferometric observations of the CO and HCN molecules seem to
indicate that stellar winds and/or supernova shocks originating in the
central starburst region are pushing outward the in-falling molecular
gas~\cite{Schinnerer:2008pq}. X-ray observations with the ROSAT High
Resolution Imager support this picture, suggesting that IC 342 may be
a starburst galaxy early in its development~\cite{Bregman}.

\subsection{Fitting simultaneously the UHECR spectrum and its nuclear composition}

\label{UFA}

Thus far we have concentrated in semi-analytical calculations to
elucidate the underlying phenomenology. Given that Auger has been
taking high-quality data for over a decade, at this stage it becomes
necessary to resort to numerical simulations to acquire enough
theoretical precision to interpret the data. The subsequent discussion
is based on the ideas developed in~\cite{Unger:2015laa} to
simultaneously fit the UHECR spectrum and its nuclear composition, 
including effects during CR propagation on the source environment and on
the trip to Earth.

The flux and nuclear composition of UHECRs depend on the cosmic
distribution of their sources. As our knowledge of source
distributions and properties is limited, it is common practice to
assume spatially homogeneous and isotropic CR emissions, and compute a
mean spectrum based on this assumption. In reality, of course, this
assumption cannot be correct, especially at the highest energies where
the GZK effect severely limits the number of sources visible to us~\cite{Taylor:2011ta}. It
is always interesting to quantify the possible deviation from the mean
prediction based on the knowledge we do have on the source density and
the possible distance to the closest source populations. This next
statistical moment beyond the mean prediction is referred to as the
ensemble fluctuation~\cite{Ahlers:2012az}. It depends on, and thus
provides information on, the distribution of discrete local sources,
source composition, and energy losses during propagation. This
ensemble fluctuation in the energy spectrum is one manifestation of
the cosmic variance, which should also appear directly through
eventual identification of nearby source populations. In fact, once
statistics become sufficiently large, it will be interesting to try to
identify the ensemble fluctuations in the energy
spectrum~\cite{Ahlers:2013zxa}. For simplicity, herein consideration
will be given to a minimal model (i.e.  minimal number of free
parameters in the fit) adopting the canonical hypothesis of a uniform
source distribution.

The number of UHECR per unit volume and energy in the present universe
is equal to the number of particles accumulated during the entire
history of the universe and is comprised of both primary particles
emitted by the sources and secondaries produced in the
photodisintegration process. The co-moving space density of CRs
(\ref{nCR}) of mass $A$ from a population of uniformly distributed
sources with (possibly age-dependent) emission rate per volume ${\cal
  Q}(E^\prime,A^\prime,t)$ is given by
\begin{eqnarray}
n (E,A,A^\prime)  \equiv
\frac{dN}{dE\,dV}
 =   \int_E^\infty \! \!  \int_0^{t_H}
\!\! \frac{d\mathscr{P}_{AA^\prime}  (E^\prime, E, t)}{dE} \
{\cal Q}(E^\prime,A^\prime,t) \  \xi(t)  \ dE^\prime \ dt \,,
\label{nCR2}
\end{eqnarray}
where the variable $t$ characterizes a particular age of the universe
and $t_H$ indicates its present age, and where
$d\mathscr{P}_{AA^\prime}/dE$ is the expectation value for the number
of nuclei of mass $A$ in the energy interval ($E, E+dE)$ which derive
from a parent of mass $A^\prime$ and energy $E^\prime$ emitted at time
$t$. Here, $\xi(t)$ is the ratio of the product of co-moving source
density and ${\cal Q}(E^\prime,A^\prime,t)$, relative to the value of
that product today. A semi-analytical approximation of
$d\mathscr{P}_{AA^\prime}/dE$ indicates that not only the LO single nucleon
emission~\cite{Hooper:2008pm}, but also NLO corrections from
two-nucleon emission are relevant to interpret the high-quality
data~\cite{Ahlers:2010ty}. Note that $d\mathscr{P}_{AA^\prime}/dE$
includes propagation effects both at the source environment and on the
way to Earth.

Two additional assumptions will be exercised to fit the data. Firstly,
that the UHECR emission rate is the same for all sources and the
spectrum and composition is independent of the age of the universe, so
that evolution of the volumetric emission rate with cosmological time
can be described by an overall source evolution factor, $\xi(t)$.  The
cosmological evolution of the source density per co-moving volume is
parametrized as
\begin{equation}
 n_{\rm s} (z) = n_0\, \xi(z)
\end{equation}
with $\xi(z=0)=1$.
The evolution of sources
follows the star formation rate with
\begin{equation}
\xi(z) =  \frac{(1+z)^a}{1 + [(1+z)/b]^{c}}
\label{evolution}
\end{equation}
where $a = 3.26 \pm 0.21$, $b = 2.59 \pm 0.14$, and $c = 5.68 \pm
0.19$~\cite{Robertson:2015uda}.  Secondly,  that the
emission rate is fairly well described by a power-law spectrum. Under
these very general assumptions the source emission rate per volume takes
the form
\begin{equation}
{\cal Q} (E^\prime,A^\prime) = {\cal Q}_0 \left(\frac{E^\prime}{E_0}\right)^{-\gamma}
\exp\left(-\frac{E^\prime}{Z' E_p^{\rm max}}\right),
\label{injectionQ}
\end{equation}
where $E_p^{\rm max}$
is the maximal energy of emitted protons, i.e., maximum rigidity of the accelerator, $Z'$ is the nucleus' atomic
number, $E_0$ is some reference energy, and
\begin{equation}
 {\cal Q}_0 = \left\{\begin{array}{rl}  \dot{n}_0  \left.  \frac{dN_{A^\prime}}{dE^\prime}\right|_{E^\prime= E_0},  & ~~{\rm for \
       bursting \ sources} \\
n_0  \left. \frac{dN_{A^\prime}}{dE^\prime dt}\right|_{E^\prime=E_0},  &
~~{\rm for \ steady \ sources}
\end{array} \right. \,,
\end{equation}
and where $\dot{n}_0$ is the number of bursts per unit volume per unit
time and $dN_{A^\prime}/dE^\prime$ is the spectrum of particles
produced by each burst, or for a steady source $n_0$ is the number
density of sources at $z=0$, and \mbox{$Q_0 \equiv dN_{A^\prime}/dE^\prime dt$} is the UHECR production rate per unit energy per source.  The cosmic ray power
density above a certain energy $E^\prime_{\rm min}$ is given by
\begin{eqnarray}
   \dot{\epsilon}_{\rm CR} (> E_{\rm min}) & = & \int_{E^\prime_{\rm
        min}}^\infty E^\prime \, {\cal Q}(E^\prime,A^\prime) \, dE^\prime 
    = {\cal Q}_0 \int_{E^\prime_{\rm min}}^\infty
        E^\prime \left(\frac{E^\prime}{E_0}\right)^{-\gamma}
        \exp\left(-\frac{E^\prime}{Z^\prime E_p^{\rm max}}\right)\,
        dE^\prime  \nonumber \\
   & =& Z^\prime E_p^{\rm max} \left(\frac{Z^\prime E_p^{\rm max}}{E_0}\right)^{-\gamma+1}
       \int_{E^\prime_{\rm min} / (Z^\prime E_p^{\rm max})}^\infty
    t^{-\gamma+1} {\rm e}^{-t}\, dt  \nonumber \\
  &  = & {\cal Q}_0\, E_0^2 \left(\frac{Z^\prime E_p^{\rm max}}{E_0}\right)^{-\gamma+2}
        \Gamma\left(-\gamma+2,\, \frac{E^\prime_{\rm min}}{Z^\prime
            E_p^{\rm max}}\right), 
\label{tntpower}
\end{eqnarray}
where $\Gamma(x)$ denotes the upper incomplete gamma function.

For a given spectrum of injected nuclei of mass $A^\prime$, the space
density of cosmic rays at Earth with energy $E$ and mass $A$ is given
by (\ref{nCR2}). For an isotropic arrival direction distribution the
relation between the spectrum and the cosmic ray density follows from
(\ref{nCR}) and is given by
\begin{equation}
J (E, A, A^\prime) \equiv  \frac{dN}{dE \ dA \ dt \ d\Omega}
  =  \frac{1}{4\pi}  \ n  (E,A,A^\prime) \,,
\end{equation}
Note that  discretization of (\ref{nCR2}) allows a  numerical treatment of the
problem. For details of the discretization procedure, see~\cite{Unger:2015laa}.

Before discussing the results of the fit, we pause to discuss some
interesting phenomenological aspects of the source environment.  To visualize the
{\it high-pass filter} mechanism advertised in Sec.~\ref{high-low}, envision a
source in which the escape and
interaction times are both power laws in energy,
\begin{equation}
\tau_\mathrm{esc} = a\,(E/E_0)^\delta \quad {\rm and} \quad
\tau_\mathrm{int} = b\,(E/E_0)^\zeta.
\label{taus}
\end{equation}
Thus, a fraction
\begin{equation}
\eta_\mathrm{esc}(E) =  (1 + \tau_\mathrm{esc}/\tau_\mathrm{int})^{-1}
= \left[1 +
  R_0\,(E/E_0)^{\delta-\zeta}\right]^{-1} 
\end{equation}
of the particles escape without interaction and the rest interact
before escaping, so $\eta_\mathrm{int} = 1 - \eta_\mathrm{esc}$, with
$R_0=a/b$ the ratio of the escape and interaction time at reference
energy $E_0$.  Note that $\eta_\mathrm{esc}$ and $\eta_\mathrm{int}$
depend only on the ratio of the escape and interaction times, but not
on the absolute value of either of them.  When $\delta>\zeta$, the
source environment acts as a {\itshape low-pass filter} on the
particles injected from the accelerator, leading to a cutoff in the
escaping spectrum at high energies. This situation is typical of leaky
box models of diffuse acceleration at time-independent
shocks~\cite{Szabo:1994qx, Protheroe:1998pj} where $\delta>0$ because
the higher the energy of the particle, the longer it needs to stay in
the accelerator to reach its energy.  By contrast, if the escape time
decreases with energy, as in the case of diffusion in turbulent
magnetic fields outside the accelerator, then it is possible to have
$\delta<\zeta$ leading to a {\it high-pass filter} on the energy
spectrum of injected nuclei: the lower the energy, the more time the
nuclei have to interact before escaping, leading to a hardening of the
spectrum and lightening of the composition of nuclei escaping the
region surrounding the source.  The spallated nucleons have energies
of $E/A$; these nucleons are most abundant at low energies and
have a steeper spectrum than the parent nuclei.  Thus, the {\it
  high-pass filter} leads naturally to an ankle-like feature separating the
nucleonic fragments from the remaining nuclei.  The normalization and
slope of the spectrum of spallated nucleons relative to that of the
primary nuclei is determined by how thoroughly the primary nuclei are
disintegrated, which is governed by the ratio of escape and
interaction lengths of the most abundant primaries.

To obtain a more realistic treatment of the interaction time, one must
specify the shape of the spectrum of the target photons. The simple 
representative photon background of non-thermal emission adopted here is a broken power-law,
\begin{equation}
      n(\varepsilon) = n_*
        \begin{cases}
           (\varepsilon/\varepsilon_*)^\alpha & \varepsilon < \varepsilon_* \\
           (\varepsilon/\varepsilon_*)^\beta & \text{otherwise} 
        \end{cases} \,,
\label{app:eq:photonfield}
\end{equation}
where $\varepsilon$ is the photon energy and the maximum of the number
density is at an energy of $\varepsilon_*$~\cite{Szabo:1994qx}.  For
such peaky photon spectra, the interaction time does not have the
simple representation of (\ref{taus}) but it does have a rather
universal structure.  Substituting (\ref{app:eq:photonfield}) into
(\ref{world-cup}) yields:
\begin{equation}
\frac{1}{\tau_{\rm int} (E)} = \frac{1}{\tau_b}
\left\{\begin{array}{ll}  \,
(E_b / E)^{\beta +1} & ~ E \leq E_b  \\
(1-\beta)/(1-\alpha) \left[\left( E_b/E \right)^{\alpha +1} -
  \left(E_b/E\right)^2 \right] +
\left(E_b/E\right)^2 & ~ E > E_b
\end{array} \right. \, ,
\end{equation}
where
\begin{equation}
\tau_b = \frac{  E_b \ (1-\beta)} {c \, \varpi \
   \, A \, m_p \ 
   \ n_*} \quad {\rm and} \quad
E_b = \frac{\varepsilon_0 \ A \ m_p}{2 \varepsilon_*} .
\end{equation}
In the NWA  the photopion
production cross section can be described by (\ref{sigma}) with the following
parameters: $\sigma_0 \simeq 0.5\, A~{\rm mb}$, $\Gamma = 150~{\rm MeV}$, and
$\varepsilon_0 = (m_\Delta^2 - m_p^2)/(2 m_p) \simeq 340~{\rm
  MeV}$~\cite{Patrignani:2016xqp}.

\begin{figure}[t!]
   \postscript{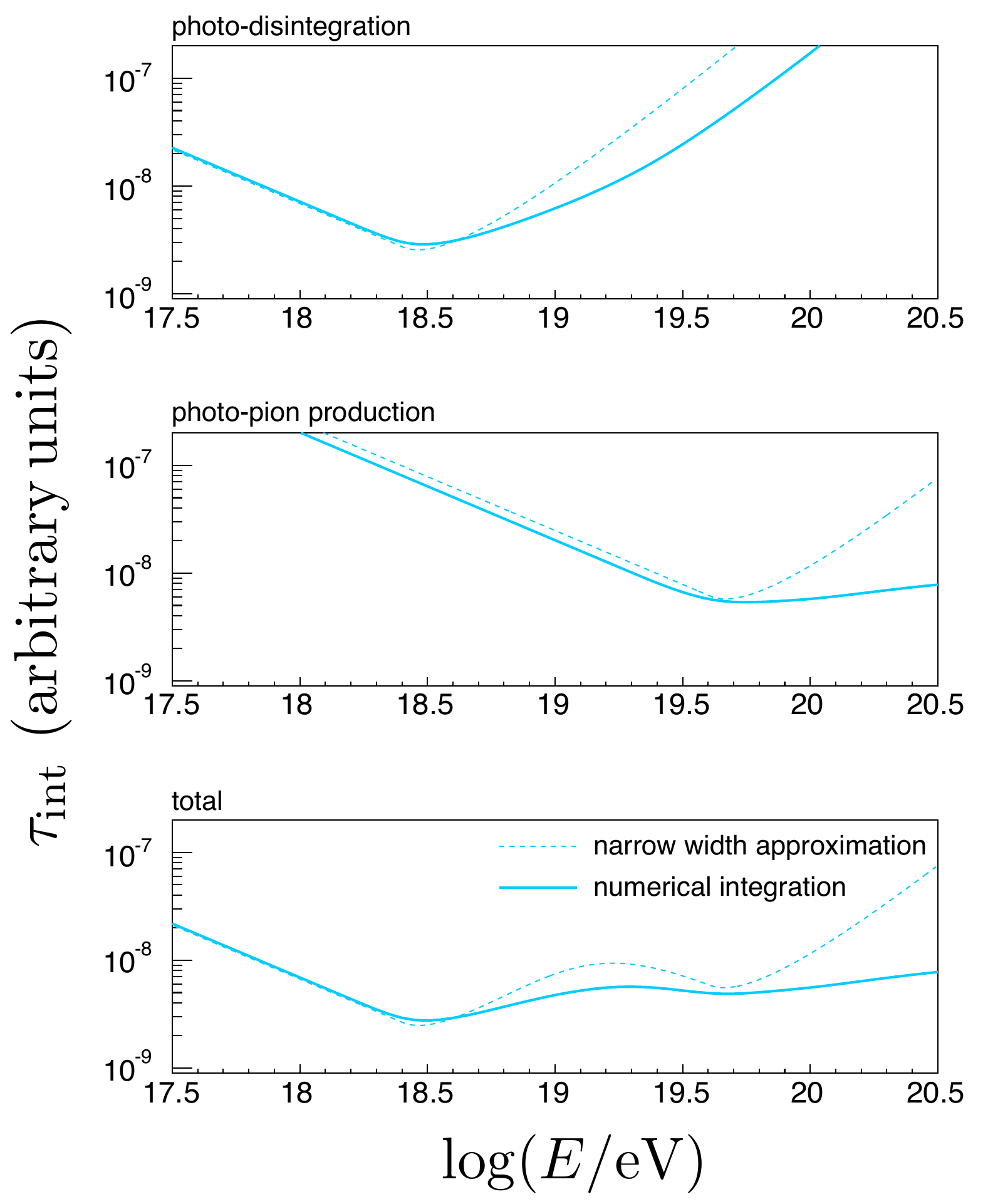}{0.6} 
\caption{Interaction
     times of $^{28}$Si in a broken power-law photon field with
     parameters $\alpha= 3/2$, $\beta=-1$ and $\varepsilon_* =
     0.11~{\rm eV}$. Top panel: photo-disintegration, middle panel:
     photo-pion production, bottom panel: sum of the two
     processes. The results of numerical integration using
     detailed cross sections are shown as thick solid lines, while those of the
     NWA  are displayed with thin dashed lines. From Ref.~\cite{Unger:2015laa}.}
\label{fig:17}
\end{figure}

As can be seen in Fig.\,\ref{fig:17} the folding of a single resonance
in the NWA with a broken power-law spectrum leads to a ``V'' shape
curve for $\tau_{\rm int}$ in a log-log plot for both
photo-disintegration (top panel) and photopion production (middle
panel).  Combining both processes in the NWA
yields an interaction time with a ``W'' shape, while  numerical integration of (\ref{lambdamenosuno}) 
using  precise cross section curves fitted to data (including the plateau for multi-pion production) softens
the ``W'' to what we shall refer to as an ``L'' shape for brevity, as
shown in the bottom panel of Fig.\,\ref{fig:17}.  As evident from
Fig.\,\ref{fig:17}, below the inflection point for photodisintegration
$E_{b}$, the NWA provides a good representation of the data, while from the
full numerical integration in the high-energy region $\tau_{\rm int}$
is roughly constant. Hence  we can write an
approximate representation of the interaction time 
\begin{equation}
\tau_{\rm int} (E) \approx \tau_b \left\{ \begin{array}{lr}
  (E/E_b)^{\beta+1} & E \leq E_b \\ 1 & E> E_b \\
\end{array} \right. \, .
\label{tautau}
\end{equation}
Returning to the discussion of $\tau_{\rm int} $, now  (\ref{taus}) with
(\ref{tautau}) together yield the fraction of nuclei which escape without
interaction in a peaky photon spectrum. It is straightforward to see
that if $\delta <0$ and the interaction time is described by an
L-shaped curve, then $\eta_{\rm esc}$ has the properties of a
high-pass filter.  These conclusions do not depend on the exact shape
of the photon spectrum. As one can guess from
Fig.~\ref{fig:14}, if the photon density is assumed to
follow a black body spectrum, then each interaction time in the NWA
would have a V shape and the total interaction time would flatten to
an L-curve as well, modulated by the cross section plateau of
 multi-pion production~\cite{Unger:2015laa}.

 Motivated by the energy dependence of the diffusion coefficient for
 propagation in a turbulent magnetic field, one can model
 $\tau_\mathrm{esc}$ as a power law in rigidity $E/Z$,
\begin{equation}
\tau_\mathrm{esc} = \tau_0 (E Z^{-1}/E_0)^\delta.
\end{equation}
Since only the ratio of escape and interaction times matters, and the
$\{E, A, Z\}$ dependence of this ratio is entirely determined once the
spectral index of the escape time $\delta$ is specified, the remaining
freedom in characterizing the source environment can be encoded by
specifying the ratio of escape to interaction time for a particular
choice of $\{E, A, Z\}$, say at $E = 10^{10}~{\rm GeV}$ for iron
nuclei, denoted $R_{10}^{\rm Fe}$.  In application to a particular
source candidate, $R_{10}^{\rm Fe}$ depends on the density of photons
and the properties of the turbulent magnetic field that delays the
escape of the UHECRs from the environment of their source.

\begin{figure}[t!]
   \postscript{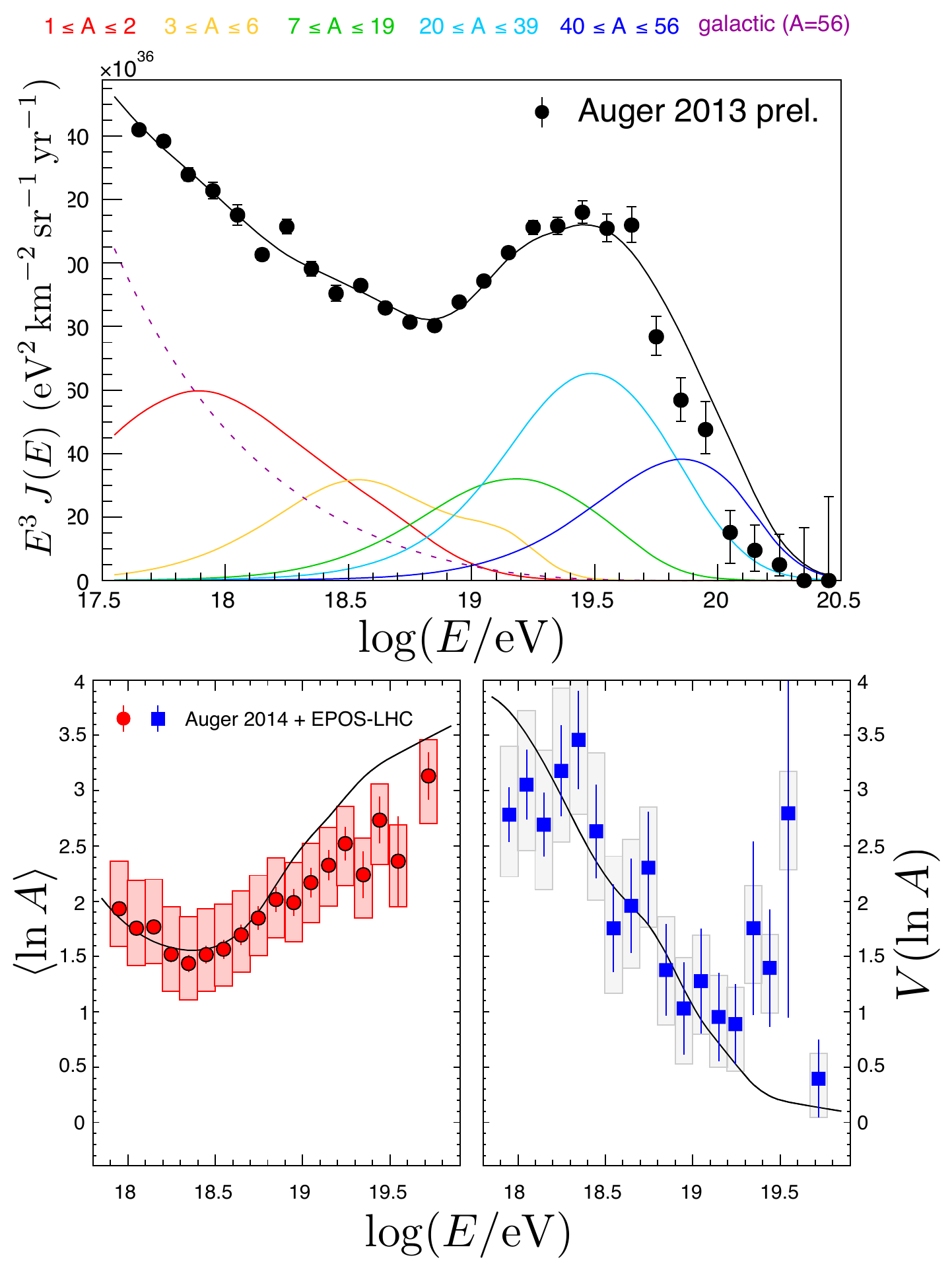}{0.6} 
   \caption{Spectrum and composition at Earth produced by a
     distribution of sources endowed with a {\it high-pass
       filter}. The data points are from the Pierre Auger
     Observatory~\cite{ThePierreAuger:2013eja, Aab:2014kda} {\it
       shifted by plus one sigma of systematic uncertainty for the
       energy scale and minus one sigma for the $X_{\rm max}$
       scale}. Error bars denote the statistical uncertainties and the
       shaded boxes illustrate the experimental systematic
       uncertainties of the composition. The composition estimates are
       based on an interpretation of air shower data with the {\sc
         epos}-LHC event generator~\cite{Pierog:2013ria}.
       From Ref.~\cite{Unger:2015laa}.}
\label{fig:18}
\end{figure}

{}Figure~\ref{fig:18} shows a comparison of the extragalactic
all-particle spectrum (obtained by combining contributions from
photodisintegration in the source environment and on the way to Earth)
with both the Auger measured flux above $10^{8.5}~{\rm
  GeV}$~\cite{ThePierreAuger:2013eja} and the mean and variance of the
distribution of the logarithm of the baryon number on top of the
atmosphere, $\langle \ln A \rangle$ and $V(\ln
A)$~\cite{Abreu:2013env,ThePierreAuger:2013eja, Aab:2014kda}.  The
source spectra are modelled using a mixed composition, which follows
the abundances of Galactic nuclei at a nucleus energy of $1~{\rm
  TeV}$. The corresponding nucleus fractions are: 0.365, 0.309, 0.044,
0.077, 0.019, 0.039, 0.039, 0.0096, 0.014, 0.084 for H, He, C, O,Ne,
Mg, Si, S, Ar+Ca, Fe, respectively. The best fit values of the free
parameters are: {\it (i)}~the power law index of the injected nuclei
$\gamma = 1.25 \pm 0.02$, {\it (ii)}~the cutoff rigidity
$\log(E_p^{\rm max}/{\rm GeV}) = 9.60\pm 0.01$, {\it (iii)}~the power
law index of the escape length $\delta=-1.01 \pm 0.03$, {\it (iv)}~the
ratio of interaction and escape time $\log(R_{10}^{\rm Fe}) = 2.57 \pm
0.02$, {\it (v)}~the flux fraction of Galactic cosmic rays $f = 0.686
\pm 0.01$, and {\it (vi)}~the power law index index of the Galactic
cosmic ray spectrum $\propto E^{-\gamma_{\rm gal}}$, assumed to be
  dominated by iron nuclei: $\gamma_{\rm gal} =
  3.71 \pm 0.02$~\cite{Unger:2015laa}. The energy of maximum of the photon field density
  was fixed  at $\varepsilon_* = 0.09~{\rm eV}$. Substituting the
  corresponding fit parameters in (\ref{tntpower}) one infers a comoving volumetric
  energy injection rate in CRs at $z=0$ of
\begin{equation}
  \dot{\epsilon}_{\rm CR} (> 10^{8.5}~{\rm GeV}) = 1.5 \times 10^{45}~{\rm erg}
  \, {\rm Mpc}^{-3} \, {\rm yr}^{-1} \, .
\end{equation}
There is a good overall agreement between the model and the data. The shape of the
spectrum is described well, including the ankle and the flux
suppression.  The model also qualitatively reproduces the increase of
the average logarithmic mass with energy and the decrease of its
variance. GRBs~\cite{Globus:2015xga},  SBGs~\cite{Anchordoqui:2017acf}, 
 and AGNs~\cite{Supanitsky:2018jje} have been proposed as astrophysical sites that can
accommodate a high-pass filter.

Alternative evolutions of the source luminosity density can be described by
the simple one-parameter functional form
\begin{equation}
\xi(z) =
\begin{cases}
    (1+z)^m & z < z_0 \\
    (1+z_0)^m \, \exp\left(-(z-z_0)\right) & \text{otherwise}
\end{cases} \,,
\end{equation}
with $z_0=2$ and $m$ ranging from $-4$ to $+4$.  $m=0$ yields a
uniform source luminosity distribution, $m=+4$ corresponds to a strong
evolution similar to the one of
AGNs~\cite{Hasinger:2005sb,Stanev:2008un}, and negative values result
in sources that are most abundant or most luminous within the
low-redshift universe~\cite{Taylor:2015rla}. One interesting source
class for which the number of objects increases at low redshifts are
low-luminosity $\gamma$-ray BL Lacertae (BL Lac)
objects.\footnote{From the viewpoint of AGN classification, BL Lacs are a
  blazar subtype. The blazar category encompasses all quasars oriented
  with the relativistic jet directed at the observer giving a unique
  radio emission spectrum. This includes both radio quiet AGN (BL
  Lacs) and optically violent variable quasars.} While
Fermi measurements have revealed that the number density of bright BL
Lacs peaks at a fairly high redshift of $z \simeq 1.2$, the more
numerous low-luminosity ($L_\gamma < 10^{44}~{\rm erg \, s^{-1}}$) and
high-synchrotron peaked members of this population exhibit negative
source evolution, and thus are overwhelmingly distributed at low
redshifts~\cite{Ajello:2013lka}. The resulting variations on the fit
parameters that simultaneously accommodate the shape of the UHECR spectrum and its
nuclear composition are as follows~\cite{Unger:2015laa}:
\begin{itemize}[noitemsep,topsep=0pt]
\item $\gamma = 2$ gives a poor
description of the data for $m\gtrsim 0$, but is a viable choice for
closeby sources;
\item  for positive values of $m$, a fixed value
of $\gamma=1$ gives a similar fit quality as the freely floating
$\gamma$, but the latter converges to values larger than $1$ for
source evolutions with $m>2$;
\item for the ``traditional'' source evolutions with $m \geq 0$ and
  fixed $\gamma=1$, most of the parameters exhibit only a minor
  variation with $m$, with the exception of the power-law index of the
  escape time $\delta$ and the power density $\dot{\epsilon}_{\rm CR}
  (> 10^{8.5}~{\rm GeV})$.
\end{itemize}
This implies that the fit to Auger data does not critically depend on
the choice of the source evolution, but that for a given choice of $m$
one can constrain the allowed values of $\gamma$, $\delta$ and
 $\dot{\epsilon}_{\rm CR} (> 10^{8.5}~{\rm GeV})$.

\subsection{Impact of $\nu$ and $\gamma$-ray observations on UHECR models}

In the late '90s, Waxman and Bahcall (WB) envisioned the CR engines as
machines where protons are accelerated and (possibly) permanently
confined by the magnetic fields of the acceleration
region~\cite{Waxman:1998yy}. The production of neutrons and pions and
subsequent decay produces neutrinos, $\gamma$-rays, and CRs. If the
neutrino-emitting source also produces high- or ultra-high-energy CRs,
then {\it pion production must be the principal agent for the high
  energy cutoff on the proton spectrum}.  Conversely, since the
protons must undergo sufficient acceleration, inelastic pion
production needs to be small below the cutoff energy; consequently,
the plasma must be optically thin. Since the interaction time for
protons is greatly increased over that of neutrons due to magnetic
confinement, the neutrons escape before interacting, and on decay give
rise to the observed CR flux. The foregoing can be summarized as three
conditions on the characteristic nucleon interaction time scale
$\tau_{\rm int}$; the neutron decay lifetime $\tau_n$; the
characteristic cycle time of confinement $\tau_{\rm cycle}$; and the
total proton confinement time $\tau_{\rm conf}$: $(i)\; \tau_{\rm
  int}\gg \tau_{\rm cycle}$; $(ii)\; \tau_n > \tau_{\rm cycle}$;
$(iii)\; \tau_{\rm int}\ll \tau_{\rm conf}$. The first condition
ensures that the protons attain sufficient energy.  Conditions $(i)$
and $(ii)$ allow the neutrons to escape the source before
decaying. Condition $(iii)$ permits sufficient interaction to produce
neutrons and neutrinos. These three conditions together define an
optically thin source~\cite{Ahlers:2005sn}. A desirable property to
reproduce the almost structureless energy spectrum is that a single
type of source will produce cosmic rays with a smooth spectrum across
a wide range of energy.

The UHECR intensity just below $E_{\rm supp}$ is often
summarized as ``one  particle per kilometer square per
year per steradian.'' This can be translated into an energy intensity~\cite{Gaisser:1997aw}
\begin{equation}
E \left\{ E \, J(E) \right\}  =  {10^{10.5}\,{\rm GeV} 
\over \rm (10^{10}\,cm^2)(10^{7.5}\,s) \, sr} 
  =   10^{-7}\rm\, GeV\ cm^{-2} \, s^{-1} \, sr^{-1} \,.
\end{equation}
From this we can derive the energy density $\epsilon_{\rm CR}$ in
UHECRs using intensity ${}={}$velocity${}\times{}$density, or
\begin{equation}
4\pi \int  dE \left\{ E \,  J (E)\right\} =  v \, \epsilon_{\rm CR}\,.
\end{equation}
This leads to
\begin{equation}
  \epsilon_{\rm CR} = {4\pi\over v} \int_{E_{\rm min}}^{E_{\rm max}} { 10^{-7}\over E} 
  dE \, {\rm {GeV\over cm^2 \, s}} \simeq   10^{-19} \, 
{\rm {TeV\over cm^3}} \,,
\end{equation}
taking the extreme energies of the accelerator(s) to be $E_{\rm min}
\simeq 10^{10}~{\rm GeV}$ and $E_{\rm max} = 10^{12}~{\rm GeV}$, and
$v \sim c$. The power required for a population of sources to generate
this energy density over the Hubble time ($t_H \approx
10^{10}$~yr) is: $\dot \epsilon_{\rm CR}^{[10^{10}, 10^{12}]} \sim 5
\times 10^{44}~{\rm TeV} \, {\rm Mpc}^{-3} \, {\rm yr}^{-1} \simeq 3
\times 10^{37}~{\rm erg} \, {\rm Mpc}^{-3} \, {\rm s}^{-1}$.  This
works out to roughly ({\it i}\,) $L \approx 3 \times 10^{39}$~erg
s${}^{-1}$ per galaxy, ({\it ii}\,) $L \approx 3 \times 10^{42}$~erg
s${}^{-1}$ per cluster of galaxies, ({\it iii}\,) $L \approx 2 \times
10^{44}$ erg s${}^{-1}$ per active galaxy, or ({\it iv}\,) $\int L \,
dt\approx 10^{52}$ erg per cosmological GRB~\cite{Gaisser:1997aw}.
The coincidence between these numbers and the observed output in
electromagnetic energy of these sources explains why they have emerged
as the leading candidates for accelerators of UHECR protons.

The energy production rate of protons derived professionally, assuming
a cosmological distribution of proton sources, with injection spectrum
$\propto E^{-2}$, is~\cite{Waxman:1995dg}
\begin{equation}
\dot \epsilon_{\rm CR}^{[10^{10}, 10^{12}]} \sim 5 \times 10^{44}~{\rm erg} \, {\rm Mpc}^{-3} \, {\rm yr}^{-1} \, .
\label{professionally}
\end{equation}
This is within a factor  ${\cal O}(1)$ of our back-of-the-envelope
estimate ($1~{\rm TeV} \simeq 1.6~{\rm erg}$). The energy-dependent generation rate of CRs is therefore given by 
\begin{equation}
E^2  \ \dot n (E) 
  =  \frac{\dot \epsilon_{\rm CR}^{[10^{10}, 10^{12}]}}{\ln(10^{12}/10^{10})} \nonumber \\
 \approx  10^{44}\,\rm{erg}\,\rm{Mpc}^{-3} \rm{yr}^{-1} \,\, .
\end{equation} 

The energy density of neutrinos produced through $p\gamma$
interactions of these protons can be directly tied to the injection
rate of CRs
\begin{equation}
E^2_{\nu}  \ n (E_\nu) 
\approx \frac{3}{8} \ \epsilon_\pi \  t_H \ E^2 \  \dot n (E) \,,
\end{equation}
where  $\epsilon_\pi$ is the
fraction of the energy which is injected in protons lost into photopion
interactions.  The factor of 3/8 comes from the fact that, close to
threshold, roughly half the pions produced are neutral, thus not
generating neutrinos, and one quarter of the energy of charged pion
decays ($\pi^+ \to e^+
\nu_e \nu_\mu \bar \nu_\mu$ and the conjugate process) goes to electrons rather than neutrinos. Namely, resonant $p \gamma$ interactions produce twice as many neutral pions as charged pions. Direct pion production via virtual  meson exchange contributes only about 20\% to the total cross section, but is almost exclusively into $\pi^+$. Hence, $p \gamma$ interactions produce roughly equal numbers of $\pi^+$ and 
$\pi^0$. The average  neutrino energy from the direct pion decay is
found to be $\langle E_{\nu}
  \rangle^\pi = (1-r)\,E_\pi/2 \simeq 0.22\,E_\pi$ and that of the
  muon is $\langle E_{\mu} \rangle^\pi = (1+r)\,E_\pi/2 \simeq
  0.78\,E_\pi$, where $r$ is the ratio of muon to the pion mass
  squared. In muon decay, all secondaries can be considered massless
  and so each of the neutrinos has about 1/3 of the muon energy. This
  gives an average neutrino energy of $\langle E_{\nu} \rangle^\mu =(1+r)E_\pi/6=0.26 \,
  E_\pi$, and so $\langle E_\nu \rangle \sim E_\pi/4$.

The WB-bound is defined by the condition $\epsilon_\pi
=1$
\begin{equation} 
E^2_{\nu} \ \Phi^{\rm WB}_{\nu_{\rm all}}  (E_\nu) \approx  \frac{3}{8}
  \,\xi_z\, \epsilon_\pi\, t_H \, \frac{c}{4\pi}\,E^2 \, 
    \dot n (E)   \approx  2.3
  \times 10^{-8}\,\epsilon_\pi\,\xi_z\, \rm{GeV}\,
  \rm{cm}^{-2}\,\rm{s}^{-1}\,\rm{sr}^{-1},
\label{wbproton}
\end{equation} 
where 
\begin{equation}
\xi_z  = \int_0^\infty dz  \frac{(1+z)^{-\gamma}}{\sqrt{\Omega_m (1+z)^3
    + \Omega_\Lambda}} \ \xi(z)
\end{equation}
accounts for the effects of source
evolution with redshift~\cite{Waxman:1998yy}. For $\gamma =2$ and no
source evolution in the local ($z<2$) universe,  $\xi (z) =1$ and
$\xi_z \simeq 0.5$~\cite{Ahlers:2018fkn}.  For sources  (with $\gamma =2$) following the star-formation rate,
$\xi (z)$ is given by (\ref{evolution}) and $\xi_z \sim 3$. For interactions
with the ambient gas (i.e.  $pp$ rather than $p \gamma$ collisions),
the average fraction of the total pion energy carried by charged pions
is about $2/3$, compared to $1/2$ in the photopion channel. In this
case, the upper bound given in (\ref{wbproton}) is enhanced by
33\%~\cite{Ahlers:2005sn}.  Electron antineutrinos can also be
produced through neutron $\beta$-decay~\cite{Anchordoqui:2003vc}. The
$\beta$-decay contribution to the diffuse neutrino flux, however,
turns out to be negligible.

The actual value of the neutrino flux depends on what fraction of the
proton energy is converted to charged pions (which then decay to
neutrinos), i.e.  $\epsilon_\pi$ is the ratio of charged pion energy to the {\em
  emerging} nucleon energy at the source.  For resonant
photoproduction, the inelasticity is kinematically determined by
requiring equal boosts for the decay products of the
$\Delta^+$, giving $\epsilon_\pi = E_{\pi^+}/E_n
\approx 0.28$, where $E_{\pi^+}$ and $E_n$ are the emerging charged pion
and neutron energies, respectively.  For $pp\rightarrow NN + {\rm
  pions},$ where $N$ indicates a final state nucleon, the inelasticity
is $\approx 0.6$~\cite{Frichter:1997wh}.  This then implies that
the energy carried away by charged pions is about equal to the
emerging nucleon energy, yielding (with our definition)
$\epsilon_\pi\approx 1.$

At production, if all muons decay, the neutrino flux consists of equal fractions of
$\nu_e$, $\nu_{\mu}$ and $\bar{\nu}_{\mu}$. Originally, the
WB-bound was presented for the sum of $\nu_{\mu}$ and
$\bar{\nu}_{\mu}$ (neglecting $\nu_e$), motivated by the fact that
only muon neutrinos are detectable as track events in neutrino
telescopes. Since oscillations in the neutrino sector mix the
different species, we chose instead to discuss the sum of all neutrino
flavors $\nu_{\rm all}$. When the effects of oscillations are accounted for, {\it
  nearly} equal numbers of the three neutrino flavors are expected at
Earth~\cite{Learned:1994wg}. 

If UHECR include nuclei heavier than hydrogen, then
the neutrino intensity expected from the cosmic ray sources may be
modified.  Nuclei undergoing acceleration can produce pions, just as
protons do, through interactions with the ambient gas, so the
WB argument would be unchanged in this case. However, if
interactions with radiation fields dominate over interactions with
matter, the neutrino flux would be suppressed if the cosmic rays are
heavy nuclei. This is because the photodisintegration of nuclei
dominates over pion production at all but the very highest energies.

The diffuse intensity of astrophysical neutrinos has an additional
component originating in the energy loss of UHECRs travelling to
Earth~\cite{Beresinsky:1969qj}. The accumulation of these neutrinos
over cosmological time is known as the cosmogenic neutrino flux.  The
GZK reaction chain generating cosmogenic neutrinos is well
known~\cite{Stecker:1978ah}. The intermediate state of the reaction $p
\gamma \to n \pi^+/p\pi^0$ is dominated by the $\Delta^+$
resonance, because the neutron decay length is smaller than the
nucleon mean free path on the CMB. Gamma-rays, produced via $\pi^0$
decay, subsequently cascade electromagnetically on intergalactic
radiation fields through $e^+ e^-$ pair production followed by inverse
Compton scattering. The net result is a pile up of $\gamma$-rays at
GeV-TeV energies, just below the threshold for further pair production
on the diffuse optical background. Meanwhile each $\pi^+$ decays to 3
neutrinos and a positron; the $e^+$ readily loses its energy through
inverse Compton scattering on the diffuse radio background or through
synchrotron radiation in intergalactic magnetic fields. As we have
seen, the neutrinos carry away about 3/4 of the $\pi^+$ energy,
therefore the energy in cosmogenic neutrinos is about 3/4 of that
produced in $\gamma$-rays.  The functional form of the cosmogenic
neutrino intensity depends on the source spectra, the source
evolution, and on the UHECR nuclear
composition~\cite{Hill:1983xs,Engel:2001hd,Fodor:2003ph,Hooper:2004jc,Ave:2004uj,Anchordoqui:2007fi,Kotera:2010yn,Ahlers:2012rz,AlvesBatista:2018zui}. For
proton primaries, the energy-squared-weighted intensity $E_\nu^2
\Phi_\nu (E_\nu)$ peaks between $10^{9.6}$ and $10^{10}~{\rm GeV}$,
where the magnitude is around 1 in WB units.\footnote{Recall that 1~WB $= 10^{-8}~{\rm
    GeV} \, ({\rm cm}^2 \, {\rm s} \, {\rm sr})^{-1}$.}  For heavy
nuclei, the peak is at much lower energy (around $10^{8.7}~{\rm GeV}$)
and the magnitude is about 0.1 to 0.01~WB, depending on
source evolution. The magnitude of the $\gamma$-ray pile up currently
provides the most stringent bound on the intensity of cosmogenic
neutrinos~\cite{Ahlers:2010fw}.

High- and ultra-high-energy neutrino detection has been one of the
experimental challenges in particle astrophysics. It is widely believed that one of the most appropriate
techniques for neutrino detection consists of measuring the Cherenkov
light from muons or showers produced by the neutrino interactions in
underground water or
ice~\cite{Gaisser:1994yf,Learned:2000sw,Halzen:2002pg,Anchordoqui:2009nf}. This
allows instrumentation of large enough volumes to compensate for both
the low neutrino cross section and the low fluxes expected. There are
several projects under way to build sufficiently large detectors to
measure the expected signals from a variety of neutrino sources. The
IceCube facility, deployed near the Amundsen-Scott station, is the
largest neutrino telescope in the world~\cite{Halzen:2007ip}.  It
comprises a cubic-kilometer of ultra-clear ice about a mile below the
South Pole surface, instrumented with long strings of sensitive photon
detectors which record light produced when neutrinos interact in the
ice. 

Neutrino (antineutrino) interactions in the Antarctic ice sheet can be
reduced to three categories: {\it (i)}~In charge current (CC) interactions the neutrino
becomes a charged lepton through the exchange of a $W^{\pm}$ with some
nucleon $N$, $\nu_\ell (\bar \nu_\ell) + N \to \ell^\pm + {\rm
  anything}$, where lepton flavor is labeled as $\ell
\in\{e,\mu,\tau\}$.  {\it (ii)}~In neutral current (NC) interactions the neutrino
interacts via a $Z$ transferring momentum to jets of hadrons, but
producing a neutrino rather than a $\ell^\pm$ in the final state:
$\nu_\ell (\bar \nu_\ell)+ N \to \nu_\ell (\bar \nu_\ell) + {\rm
  anything}$.  The scattered $\nu_\ell$ exits the detector, carrying
away energy, and so the observed energy presents a lower bound for the
incident $\nu_\ell$ energy.  All three neutrino flavors exhibit a NC.
These two possibilities are then projected onto two kinds of IceCube
topologies to yield the three final possibilities: {\it
  (i)}~``Shower'' ($\oplus$) events result from all three flavors of
NC events, and from the CC events of the electron and tau neutrinos
below $\sim 2$~PeV.  Shower events (also called ``cascade'' events)
refer to the fact that energy is deposited no charged tracks (produced
by muons or taus) are observed. {\it (ii)}~Below a few PeV, ``track''
($\odot$) events are produced only by the muon neutrino CC.  The
$\numu$~CC creates a muon and a hadronic shower within the IceCube
detector, the muon track contributes to the deposited energy, but then
the muon is seen to exit the detector as a single track of unknown
energy.  The deposited energy is only a lower bound to the incident
muon neutrino energy. {\it (iii)}~At $\nu_\tau$ energies above 3~PeV, $\nu_{\tau}$
CC interactions begin to produce separable {\it double bang}
events~\cite{Learned:1994wg}, with one smaller-energy shower produced
by the initial $\nu_\tau$ collision in the ice, and the second
larger-energy shower resulting from the subsequent $\tau$ decay.

UHECR experiments, like Auger, provide a complementary technique for
ultra-high-energy cosmic neutrino (UHEC$\nu$) detection by searching
for deeply--developing, large zenith angle ($>75^\circ$)
showers~\cite{Capelle:1998zz}. At these large angles, hadron-induced
showers traverse the equivalent of several atmospheres before reaching
detectors at the ground.  Beyond about 2 atmospheres, most of the
electromagnetic component of a shower is extinguished and only very
high energy muons survive.  Consequently, a hadron-induced shower
front is relatively flat and the shower particles arrive within a
narrow time window. In contrast, a neutrino shower exhibits
characteristics similar to those of a vertical shower, which has a
more curved front and a wider distribution in particle arrival times
due to the large number of lower energy electrons and photons.
Furthermore, the ``early'' part of the shower will tend to be
dominated by the electromagnetic component, while ``late'' portion
will be enriched with tightly bunched muons. Using these
characteristic features, it is possible to distinguish neutrino
induced events from background hadronic showers. Moreover, because of
full flavor mixing, tau neutrinos are expected to be as abundant as
other species in the cosmic flux. Tau neutrinos can interact in the
Earth's crust, producing $\tau$ leptons which may decay above the
ground-based
detectors~\cite{Domokos:1997ve,Domokos:1998hz,Bertou:2001vm,Feng:2001ue,Fargion:2000iz}.
Details on how such neutrino events can be selected at the Pierre Auger Observatory
are discussed in~\cite{Abraham:2007rj,Abraham:2009uy,Abreu:2011zze,Abreu:2012zz,Abreu:2013zbq,Aab:2015kma}.

Because the shape of the astrophysical neutrino intensity is unknown,
it is convenient to define a procedure to set  model-independent
limits on the total neutrino flux~\cite{Anchordoqui:2002vb}.  To this end,  we first write a generic expression
for the neutrino event rate
\begin{equation}
N = \sum_{i,X} \int dE_i\, N_A \, \Phi_i(E_i) \, \sigma_{i
N \to X} (E_i) \, {\cal E} (E_i)\ , \label{numevents}
\end{equation}
where the sum is over all neutrino species $i = \nu_e, \bar{\nu}_e,
\nu_{\mu}, \bar{\nu}_{\mu}, \nu_{\tau}, \bar{\nu}_{\tau}$, and all
final states $X$. $N_A = 6.022 \times 10^{23}$ is Avogadro's number,
and $\Phi_i $ is the source flux of neutrino species $i$, $\sigma$ as usual denotes 
the cross section, and ${\cal E}$ is the exposure measured in cm$^3$
w.e. sr time. We assume the simplest scenario in which there are no events  that unambiguously pass all the experimental cuts, with zero events expected from 
background. This implies an upper bound of 2.4 events at 90\%CL from
neutrino fluxes~\cite{Patrignani:2016xqp}. Poisson intervals for more complex combinations of
detection and background events are summarized in~\cite{Feldman:1997qc}.
Note that if the number of events integrated over
energy is bounded by 2.4, then it is certainly true bin by bin in
energy. Thus, using (\ref{numevents}) one obtains
\begin{equation}
\sum_{i,X} \int_{\Delta} dE_i\, N_A \, \Phi^i (E_i) \,
\sigma_{i N \to X} (E_i) \, {\cal E} (E_i)\  < 2.4\ ,
\label{bound}
\end{equation}
at 90\% CL for some interval $\Delta$. Here, the sum over $X$ takes
into account charge and neutral current processes.  In a logarithmic
interval $\Delta$ where a single power law approximation
\begin{equation}
\Phi_i(E_i)\, \sigma_{i N \to X} (E_i) \, {\cal E} (E_i)
\sim E_i^{\alpha}
\end{equation}
is valid, a straightforward calculation shows that
\begin{equation}
\int_{\langle E\rangle e^{-\Delta/2}}^{\langle E\rangle e^{\Delta/2}}
\frac{dE_i}{E_i} \,
E_i\, \Phi_i \, \sigma_{i N \to X}  \, {\cal
E}  =   \langle \sigma_{i N\rightarrow X}\,
{\cal E} \, E_i\, \Phi_i \rangle \, \frac{\sinh \delta}{\delta}\, \Delta \,,
\label{sinsh}
\end{equation}
where $\delta=(\alpha+1)\Delta/2$ and $\langle A \rangle$ denotes the
quantity $A$ evaluated at the center of the logarithmic interval.  The
parameter $\alpha = 0.363 + \beta - \gamma$, where 0.363 is the
power law index of the SM neutrino cross
section~\cite{Gandhi:1998ri} and $\beta$ and $-\gamma$ are the power
law indices (in the interval $\Delta$) of the exposure and flux
$\Phi_i$, respectively.  Since $\sinh \delta/\delta >1$, a
conservative bound may be obtained from (\ref{bound}) and
(\ref{sinsh}):
\begin{equation}
N_A\, \sum_{i,X} \langle \sigma_{i N\rightarrow X} (E_i)
\rangle \, \langle {\cal E} (E_i)\rangle\, \langle E_i
\Phi^i \rangle < 2.4/\Delta\ . \label{avg}
\end{equation}
By taking $\Delta =1$ as a likely interval in which the single power
law behavior is valid (this corresponds to one $e$-folding of energy),
it is straightforward to obtain upper limits on the neutrino flux.
The model-independent upper limits on the total neutrino flux, derived
using the nine-year sample of IceCube data are collected in
Table~\ref{tabla3}~\cite{Aartsen:2018vtx}. The sensitivity of existing
neutrino-detection facilities is about to reach $1~{\rm WB}$,
challenging cosmic-ray models for which the highest energies are
proton-dominated~\cite{Ahlers:2009rf,Aloisio:2015ega,Heinze:2015hhp,Supanitsky:2016gke,Aartsen:2016ngq}.

\begin{table}
\begin{center}
\caption{All-flavor differential 90\% CL upper limit based on
the nine-year sample of IceCube data~\cite{Aartsen:2018vtx}. \label{tabla3}}
\begin{tabular}{cc}
\hline
\hline
~~~~~~~~~~~~~ $\log_{10}(E_\nu/{\rm GeV})$ ~~~~~~~~~~~~ &~~~~~~~~~~~~
$\log_{10}[E_\nu^2 \, \Phi_{\nu_{\rm all}} (E_\nu)/({\rm GeV} {\rm cm}^{-2} \,
{\rm s}^{-1}
\, {\rm sr}^{-1})]$ ~~~~~~~~~~~~~ \\
\hline
7.0                      &        $-7.86$ \\ 
7.5                     &        $-7.71$ \\
8.0                        &      $-7.69$ \\
8.5                       &      $-7.75$ \\ 
9.0                          &    $-7.66$ \\
9.5                         &    $-7.45$ \\
10.0                          &  $ -7.20$ \\
\hline
\hline
\end{tabular}
\end{center}
\end{table}

The multi-messenger program will also help discriminate among UHECR
acceleration models. As an illustration, we consider the two main
mechanisms proposed to accelerate UHECRs in starburst galaxies:
unipolar induction in newly-born pulsars and Fermi shock acceleration
in the galactic scale superwind. On the one hand, UHECRs crossing the
supernova ejecta surrounding neutron stars would experience an
effective optical depth to hadronic interactions which is larger than
unity, and so one expects guaranteed fluxes of neutrinos in the energy
range $10^8 \lesssim E_\nu/{\rm GeV} \lesssim
10^9$~\cite{Fang:2013vla}. Actually, the differential upper limits on
the diffuse neutrino flux from IceCube already constrain models of
UHECR acceleration in the core of starburst
galaxies~\cite{Fang:2015xhg}. Recall that for SBGs, the anisotropy
signal has been observed for only a fraction $f_{\rm sig} = (10 \pm 4)
\%$ of the UHECR sample, so a
small window of the parameter space still remains opened. On the other
hand, if UHECRs are accelerated at the terminal shock of the starburst
superwind, we expect the maximum energy to be constrained by direct
escape of the nuclei, and so the flux of photons and neutrinos
accompanying the starburst UHECR emission would be strongly
suppressed. Indeed, if this were the case, the neutrino emission from
starbursts would cutoff somewhat above $10^7~{\rm GeV}$, as
entertained in~\cite{Loeb:2006tw}.

After this discussion it appears evident the importance of
multi-messenger observations to narrow the search for the UHECR
origin(s).  Next generation IceCube detector will play a key role in this endeavor~\cite{Aartsen:2014njl}.

\subsection{Grand unified spectrum of diffuse extragalactic background radiation}
\label{sec:DEBRA}

The diffuse extragalactic background radiation (DEBRA) is an indicator
of the integrated luminosity of the universe~\cite{Ressell:1989rz}.  The analysis of the
different components of DEBRA leads to the grand unified spectrum,
covering roughly 34 decades of energy. This spectrum is continuously
updated thanks to the numerous space- and ground-based multi-frequency
experiments observing different cosmic messengers. In this section we review
a series of simultaneous and coordinated multi-frequency observations
which can help elucidate the UHECR origin(s).

\begin{figure}[tpb] 
  \postscript{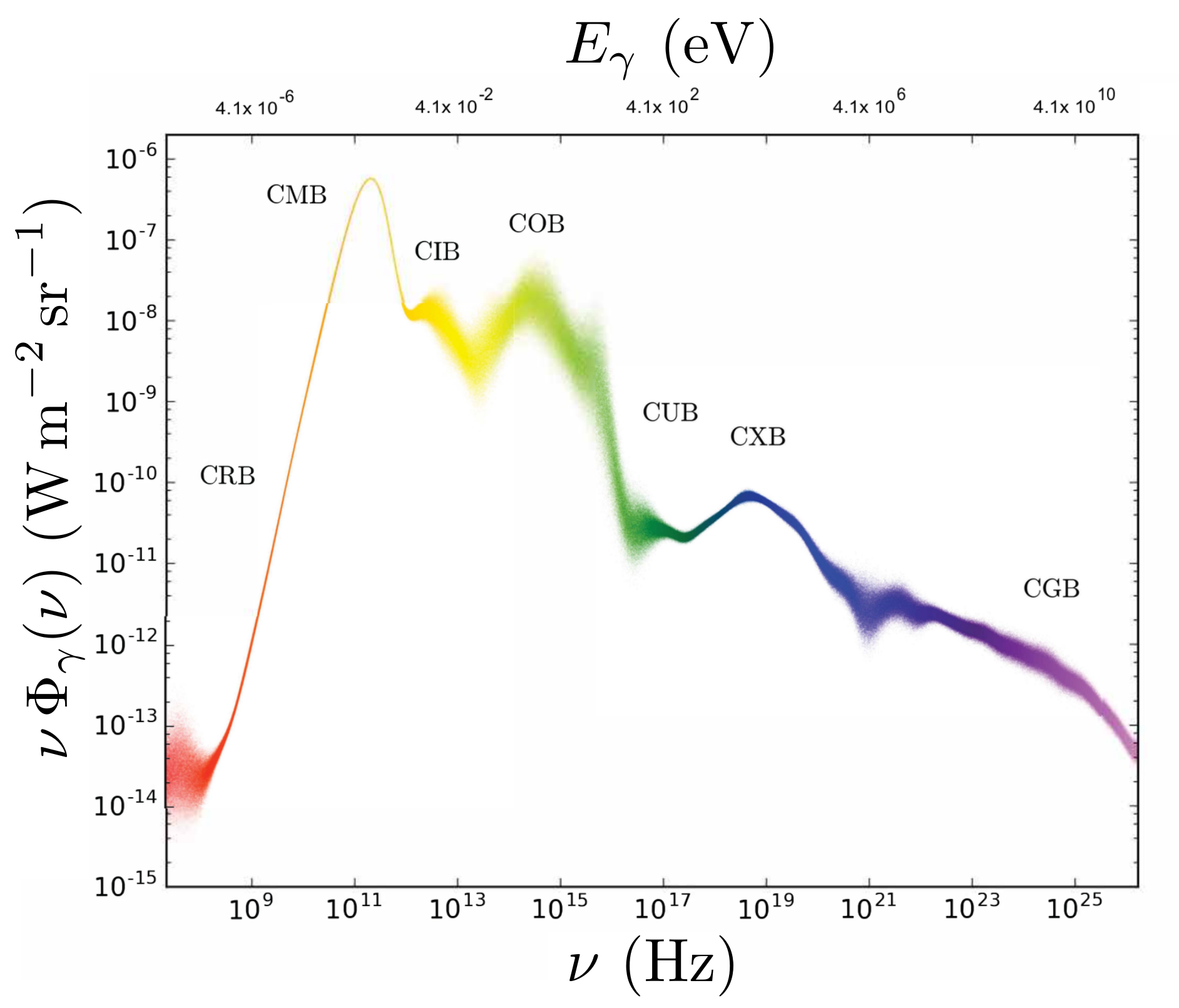}{0.6} \caption{Frequency-weighted intensity of the EBL as
    measured in radio (CRB), microwave,
    infrared (CIB), optical (COB), UV (CUB), X-rays (CXB), and
    $\gamma$-rays
    (CGB). From Ref.~\cite{DeAngelis:2018lra}. \label{fig:19}}
\end{figure}

The intensity of the extragalactic background light (EBL) spans close
to 20 decades in photon frequency; see Fig~\ref{fig:19}.  Across
this whole range, the EBL spectrum captures cosmological backgrounds
associated with either primordial phenomena, such as the CMB, or
photons emitted by stars, galaxies, and AGNs due to nucleosynthesis or
other radiative processes, including dust scattering, absorption and
reradiation. It has been known since the early '60s that high-energy $\gamma$-rays
from sources at cosmological distances will be absorbed along the way
by the diffuse background of softer photons via electron-positron pair
production~\cite{Nikishov,Gould:1966pza,Gould:1967zza}. Roughly
speaking, photons originating at a redshift $z$ will be absorbed above about
an energy   $\sim 100 (1 + z)^{-2}~{\rm
  TeV}$~\cite{Stecker:1969gamma,Fazio:1970pr}. This implies that the
intensity of extragalactic $\gamma$-rays has to be suppressed above
about 1~TeV, see Fig.~\ref{fig:19}.

In 2012,  the IceCube Collaboration famously announced the
observation of two $\sim 1$~PeV  neutrinos discovered in a search for
the nearly guaranteed cosmogenic neutrinos~\cite{Aartsen:2013bka}. The search
technique was refined to extend the neutrino sensitivity to lower
energies~\cite{Schonert:2008is,Gaisser:2014bja}, resulting in the discovery of
additional ``high-energy starting events'' (HESEs), i.e.\ events
initiated within the IceCube detector volume by entering
neutrinos~\cite{Aartsen:2013jdh,Aartsen:2014gkd,Aartsen:2014muf}. At
the time of writing, 82 HESEs (including showers and tracks) have been reported from six years of
IceCube data taking (2078 days between $2010$ to early
2016)~\cite{Aartsen:2017mau}. The HESE signal has been confirmed via a
complementary measurement using CC interactions of $\nu_\mu + \bar
\nu_\mu$, for which the interaction vertex can be outside IceCube instrumented
volume~\cite{Aartsen:2015knd,Aartsen:2015rwa,Aartsen:2016xlq}.

\begin{figure}[ht]
\postscript{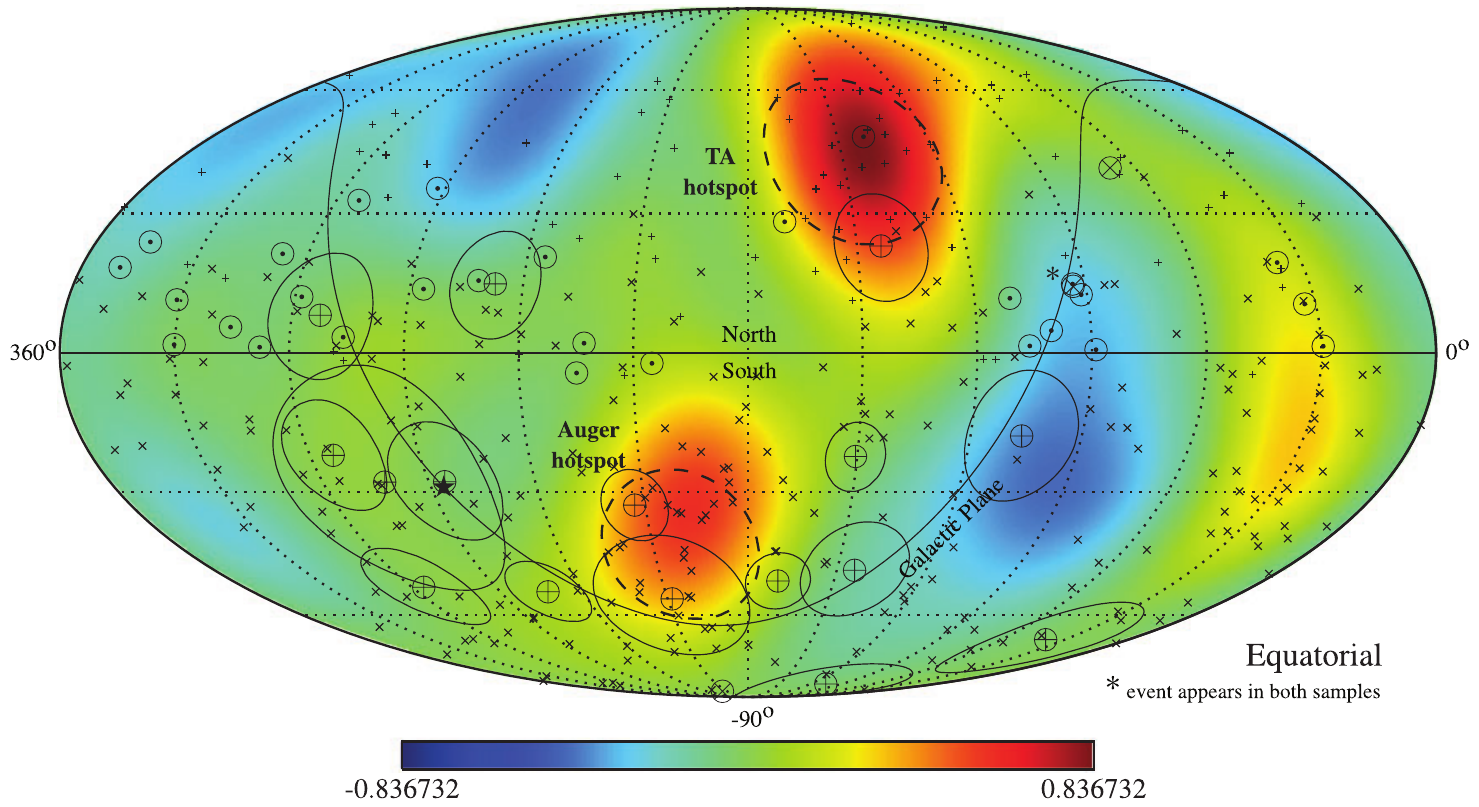}{0.9} \caption{Mollweide
    projection of the arrival direction of IceCube neutrinos and
    UHECRs. The neutrino sample is from  the six-year upgoing track
    analysis and  the four-year HESE
    analysis  (tracks $\odot$ and cascades $\oplus$). Cascade events
    are indicated together with their median angular
    uncertainty (thin circles). One event ($*$) appears in both event
    samples.  The Auger sample $(\times)$ consists of events recorded 
between 1 January 2004 and 31 March 2014 with $E > 52~{\rm EeV}$ and
    $\theta < 80^\circ$~\cite{PierreAuger:2014yba}. The TA sample
    ($+$) consists of events recorded between 11 May 2008 and 4 May 2013
    with $E > 57~{\rm EeV}$ and $\theta < 55^\circ$~\cite{Abbasi:2014lda}. The background
    shows the anisotropy of the combined UHECR map derived by
    smoothing the events at  best-fit position $\mathbf{\hat n}_i$ over the sphere
    with unit vector $\mathbf{\hat n}$ following a Fisher-Von Mises distribution $\mathfrak{F} (\mathbf{\hat n},
\mathbf {\hat n}_i; \varkappa)$. The parameter $\varkappa$ is fixed to 11.5 such that 50\% of
the distribution is contained within an opening angle of $20^\circ$. From the smoothed event distribution
$f(\mathbf{n}) = \sum_i \mathfrak{F} (\mathbf{\hat n},
\mathbf {\hat n}_i; \varkappa)$, the anisotropy is define as $\delta
J(\mathbf{\hat n}) = f(\mathbf{\hat n})/\langle f(\delta (\mathbf{\hat
  n})) \rangle - 1$, where $\langle f(\delta (\mathbf{\hat
  n})) \rangle$ is the average of the distribution $f$ in each declination bin $\delta$.
This simple procedure ensures that spurious anisotropies coming from
the detector exposures that depend mostly on declination are corrected
from the map. The dashed circles indicate excess regions found in Auger (sampling
radius of $15^\circ$; post-trial $p$-value of $1.4 \times 10^{-2}$)
and TA (sampling radius of $20^\circ$; post-trial $p$-value of $3.7
\times 10^{-4}$) data samples. From Ref.~\cite{Ahlers:2017wkk}.}
\label{fig:20}
\end{figure}

A myriad of models have been proposed to interpret the
data~\cite{Anchordoqui:2013dnh,Ahlers:2015lln,Meszaros:2017fcs}, but
the origin of IceCube events remains unknown. The distribution of
arrival directions is compatible with isotropy, without any hints of
concentrations towards either the Galactic center or plane~\cite{Aartsen:2017ujz}.\footnote{See,
however,~\cite{Anchordoqui:2013qsi,Neronov:2013lza,Anchordoqui:2014rca,Neronov:2015osa,Neronov:2016bnp,Neronov:2018ibl}.} Neither is
there any significant correlation with the arrival direction of UHECRs
observed by Auger and/or TA~\cite{Aartsen:2015dml}; see
Fig.~\ref{fig:20}. Even though there is no significant correlation
with any type of extragalactic sources, the isotropicity of the
neutrino flux strongly suggests the working assumption that it is of
extragalactic origin.\footnote{IceCube has identified the blazar  TKS 0506+056 as a
  neutrino source~\cite{IceCube:2018cha}. However,  IceCube searches constrain the maximum
  contribution of blazars in the {\it Fermi}-LAT 2LAC
catalogue~\cite{Fermi-LAT:2011xmf} to the observed astrophysical neutrino flux to be 27\% or less
between around 10 TeV and 2 PeV, assuming equipartition of flavors at
Earth and a single power-law spectrum $\propto E_\nu^{-2.5}$~\cite{Aartsen:2016lir}.  The search
also excludes that the 2LAC blazars (and sub-populations) emit more
than 50\% of the observed neutrinos for a harder spectrum
$\propto E_\nu ^{-2.2}$ in the same energy range.} The Earthly flavor ratio
$\nu_e:\nu_\mu:\nu_\tau$ is compatible with a
$1:1:1$
distribution~\cite{Mena:2014sja,Chen:2014gxa,Aartsen:2015ivb,Palomares-Ruiz:2015mka,Vincent:2016nut}. This
seems to indicate that IceCube neutrinos originate via pion decay in
optically thin sources and experience vacuum oscillations across
cosmological distances.

The most immediate impact of the discovery of astrophysical neutrinos
is that the intensity level observed is exceptionally high by
astronomical standards. The magnitude of the observed flux per
steradian is at a level of the WB bound, which applies to neutrino
production in proton sources that are also responsible for
UHECRs. However, it is important to {\it stress} that the IceCube
events have energies $E_\nu \lesssim 5~{\rm PeV}$, and therefore 
the energy per nucleon of the parent CRs must be $E_p \lesssim 500~{\rm
  PeV}$. Recall that efficient production of neutrinos at this energy
would prevent acceleration of the parent protons to ultra-high
energies, because of the 4th constraint (interaction losses) on
particle acceleration; see Sec.~\ref{sec:pheno}. Under the optimal
combination of parameters after fine tuning, the 3 conditions 
that define an optically thin source $(\tau_{\rm int}\gg \tau_{\rm
  cycle}$, $\tau_n > \tau_{\rm cycle}$, $\tau_{\rm
  int}\ll \tau_{\rm conf})$ would of course hold. Only for such a
particular case, the sources producing the neutrinos observed by
IceCube would also emit UHECR protons.  Such cosmic accelerators
produce equal numbers of neutral, positive and negatively charged
pions in the proton-proton beam dump. The neutral pions accompanying
the charged parents of the neutrinos decay promptly into photons that
could only be observed indirectly after propagation in the
extragalactic background light. Losing energy, these photons cascade
down to energies below 1~TeV where they can be observed with the {\it
  Fermi}-LAT satellite.  The uncomplicated assumptions that IceCube's
neutrinos are produced via pion decay in optically thin sources and
the neutrino spectrum follows an unbroken power law creates a tension
between the relative magnitudes of the diffuse $\gamma$-ray flux
detected by the {\it Fermi}-LAT satellite and the high energy neutrino
flux per steradian detected at the South
Pole~\cite{Anchordoqui:2016ewn}. The tension can be somewhat relaxed
if there is a break in the spectrum, but still most of the
$\gamma$-ray energy in the non-thermal universe must be produced in
the hadronic accelerators responsible for IceCube's neutrinos. This
further constrains cosmogenic neutrino models~\cite{Ahlers:2010fw} and
the proton fraction of UHECRs, as {\it Fermi}-LAT photons cannot be
counted twice.  Mixed composition models, however, are safely below
the bounds~\cite{Muzio,Kachelriess:2017tvs}.

Adding to the story, during its first observing run on 2015, the
Advanced LIGO detectors recorded gravitational waves from the
coalescence of two stellar-mass black holes (BBHs), GW150914 and
GW151226, with a third candidate LVT151012 also likely to be a BBH
system~\cite{Abbott:2016blz,TheLIGOScientific:2016wfe,Abbott:2016nmj,TheLIGOScientific:2016pea}.
GW150914 and GW151226 have comparable luminosity distance estimates:
$D_L = 420^{+150}_{-180}~{\rm Mpc}$ ($z = 0.09^{+0.03}_{-0.04}$) and
$D_L = 440^{+180}_{-190}~{\rm Mpc}$ ($z = 0.09^{+0.03}_{-0.04}$),
respectively. LVT151012 is the quietest signal and is inferred to be
at a greater distance $D_L = 1000^{+500}_{-500}~{\rm Mpc}$ ($z =
0.20^{+0.09}_{-0.09}$).  The broadband frequency spans the spectrum
from below $35~{\rm Hz}$ to above $450~{\rm Hz}$.  BBH mergers have extremely high gravitational-wave
luminosities: for GW150914, LVT151012, and GW151226 the peak values
are respectively $3.6^{+0.5}_{-0.4} \times 10^{56}~{\rm erg/s}$,
$3.1^{+0.8}_{-1.8} \times 10^{56}~{\rm erg/s}$ , and
$3.3^{+0.8}_{-1.6} \times 10^{56}~{\rm erg/s}$. A direct comparison of
these luminosities with the power requirements of (\ref{LP-rigidity})
seems to indicate that binary BBH mergers could provide a profitable
arena for UHECR acceleration, provided there are magnetic fields and
disk debris remaining from the formation of the black holes. Moreover,
it is also reasonable to suspect that if this were the case, then the
accelerated CRs would interact with the surrounding matter or
radiation to produce UHEC$\nu$~\cite{Kotera:2016dmp,Murase:2016etc,Anchordoqui:2016dcp}. Unfortunately,
a targeted search for these neutrinos yielded no candidates in Auger
data~\cite{Aab:2016ras}.

During the second observing run, the LIGO-Virgo detector network
observed a gravitational-wave signal (GW170817) from the inspiral of
two low-mass compact objects consistent with a binary neutron star
merger~\cite{TheLIGOScientific:2017qsa}. Nearly simultaneously, the
{\it Fermi}-LAT and INTEGRAL telescopes detected a gamma-ray transient,
GRB170817A, both spatially and temporally coincident with
GW170817~\cite{Monitor:2017mdv}. Within about 12 hours of the
gravitational wave trigger the host galaxy and post-merger
electromagnetic transient was
identified~\cite{Coulter:2017wya,Soares-Santos:2017lru,Valenti:2017ngx,Arcavi:2017xiz,Lipunov:2017dwd},
and intensive programs of X-ray, optical/infrared, and radio
observations soon
followed~\cite{Tanvir:2017pws,Pian:2017gtc,Troja:2017nqp,Haggard:2017qne,Hallinan:2017woc,Kasliwal:2017ngb}. Light
curves measured over the next several weeks identified the hallmarks
of a kilonova, i.e.  emission driven by neutron-rich ejected material
undergoing r-process
nucleosynthesis~\cite{Li:1998bw,Metzger:2010sy}. However, no neutrino
events signaling the acceleration of CRs have been
observed~\cite{ANTARES:2017bia}.\footnote{Neutrino emission from binary neutron
star mergers has been studied in~\cite{Biehl:2017qen}. It was also noted that nucleus photodisintegration within
binary neutron star merger remnants could produce the population of CRs  below the ``ankle''~\cite{Rodrigues:2018bjg}.}

\begin{figure}[tpb] \postscript{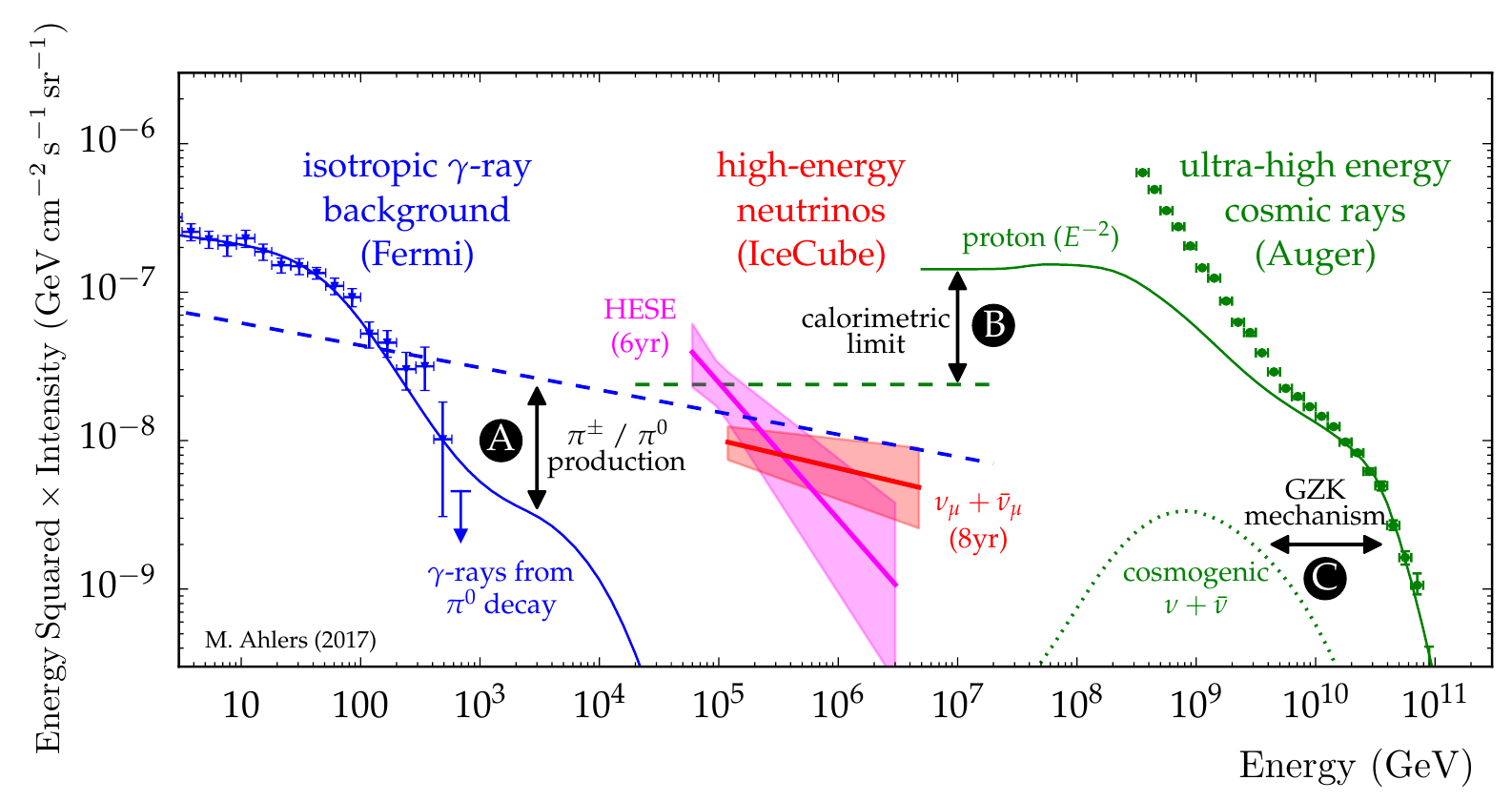}{0.77} \caption{
Multi-messenger interfaces.
The neutrino intensity $\Phi_\nu(E_\nu)$ inferred from the eight-year
upgoing track analysis (red fit) and the six- year HESE analysis
(magenta fit) compared to the intensity of unresolved extragalactic
$\gamma$-ray sources $\Phi_\gamma(E_\gamma)$ (blue data) and UHECRs $J(E)$  (green
data). The neutrino spectra are indicated by the best-fit power-law
(solid line) and $1\sigma$ uncertainty range (shaded range). Various
multi-messenger interfaces are specified. {\bf A}: The joined
production of charged pions ($\pi^\pm$) and neutral pions ($\pi^0$) in
CR interactions leads to the emission of neutrinos (dashed blue) and
$\gamma$-rays (solid blue), respectively. {\bf B}: CR emission models
(solid green) of the most energetic cosmic rays imply a maximal
WB-intensity (calorimetric limit) of neutrinos from the same sources
(green dashed). {\bf C}: UHECR models also predict the guaranteed
emission of cosmogenic neutrinos from the collision with cosmic
background photons (GZK effect). From Ref.~\cite{Ahlers:2018fkn}.}
\label{fig:21}
\end{figure}

The latest chapter in the story is courtesy of Auger. As alluded to
already in Sec.~\ref{DoAD}, in 2017 the Auger Collaboration reported the observation of a
large scale dipole anisotropy for events with $E> 8~{\rm EeV}$~\cite{Aab:2017tyv}. 
The direction of the reconstructed dipole  lies
about $125^\circ$ from the Galactic center. This suggests an
extragalactic origin of UHECRs. The high and ultra-high energy
components of DEBRA, together with the interconnections among the cosmic
messengers, are summarized in Fig~\ref{fig:21}.  Note that the lower-energy CRs are Galactic
and so would not be measured in intergalactic space.

\section{Phenomenology of UHECR air showers}
\label{sec:4}

\subsection{Nature's calorimeter}

\begin{figure}[tbp]
\postscript{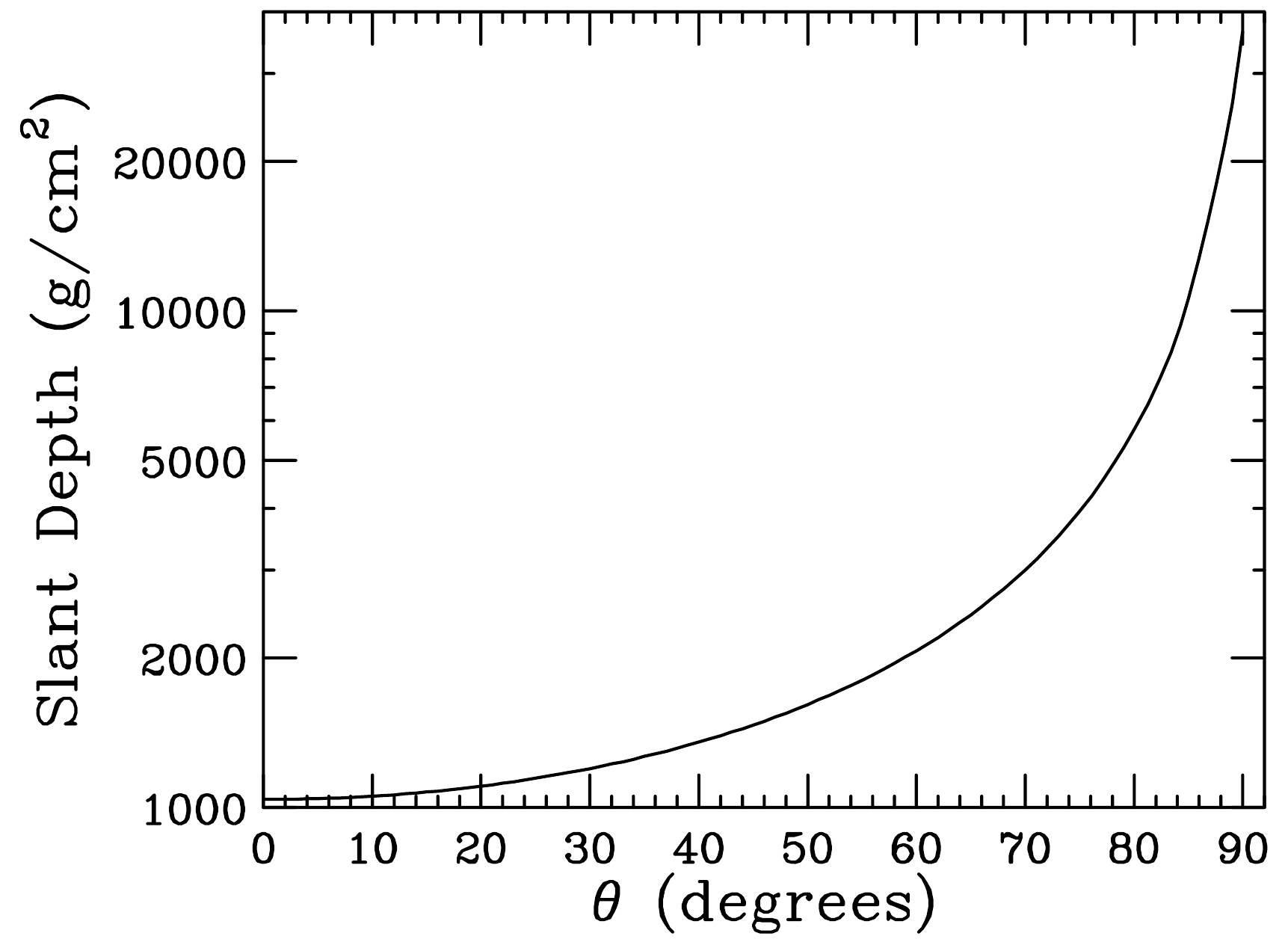}{0.56}
\caption{Slant depths corresponding to various zenith angles
$\theta$ considering the curvature of the Earth. From Ref.~\cite{Anchordoqui:2004xb}.}
\label{fig:22}
\end{figure}

In this section we focus on UHECR phenomenology from the top of the atmosphere to the Earth's surface.
Unlike man-made calorimeters, the atmosphere is a calorimeter whose 
properties vary in a predictable way with altitude,  and in a relatively unpredictable
way with time.  Beginning with the easier of the two variations, we note
that the density and pressure depend strongly on the height, while the temperature
does not change by more than about 30\% over the range 0--100~km above sea level.
Therefore we can get a reasonable impression of the density variation by assuming
an isothermal atmosphere, in which case the density $\rho_{\rm atm} (h) \approx \rho_0 e^{-h/h_0}$,
where $\rho_0 \approx 1.225$~kg/m$^3$ and $h_0 = R\,T /(\mu\,g) \approx 8.4$~km 
is known as the scale-height of the 
atmosphere,  $R$ being the ideal gas constant, $\mu$ the average molecular weight of 
air, $g$ the acceleration due to gravity and $T \approx 288$~K.
Of course, reading out such a natural calorimeter is complicated by the 
effects of varying aerosol and molecular attenuation and scattering.

The quantity that most intuitively describes the varying density of the 
atmospheric medium is the vertical atmospheric depth,
$X_v (h) = \int_h^\infty \rho_{\rm atm}(z) \,\,dz$, where $z$ is the height. 
However, the quantity 
most relevant in air shower simulations is the 
slant depth, $X$, which defines the actual amount of air traversed by
the shower.  The variation of the 
slant depth with zenith angle is shown in Fig.~\ref{fig:22}.
If the Earth curvature is not taken into account, then $X = X_v(h)/ \cos 
\theta$, where $\theta$ is the zenith angle of the shower axis. For 
$\theta \lesssim 80^\circ,$ the error associated with this  
approximation is less than 4\%.

The atmospheric 
medium is endowed with a magnetic field. In general, the geomagnetic field is 
described by 3 parameters, 
its strength $B,$ its inclination $\iota,$ and its declination 
$\delta$. The 
inclination is defined as the angle between the local horizontal plane and the 
$\mathbf{B}$-field. The declination is defined as the angle between the horizontal 
component of the field $B_\perp$ (i.e., perpendicular to the arrival 
direction of the air shower) and the geographical North 
(direction of the local meridian). The angle $\iota$ is 
positive when $\mathbf{B}$ points downward and $\delta$ is positive 
when $B_\perp$ is inclined towards the East.

\subsection{Systematic uncertainties in air shower measurements from  hadronic interaction models}
\label{hadronic}

Uncertainties in hadronic interactions at ultra-high energies constitute one of the most problematic sources of systematic error  in the analysis of air showers.  This section will explain the  two principal schools of thought for extrapolating collider data to  ultrahigh energies.

Soft multiparticle production with small transverse momenta with respect to 
the collision axis is a dominant feature of most hadronic events at c.m. energies 
$10~{\rm GeV} < \sqrt{s} < 62~{\rm GeV}$ (see e.g.,~\cite{Capella:yb,Predazzi:1998rp}). 
Despite the fact that strict calculations based on ordinary QCD perturbation 
theory are not feasible, there are some phenomenological models that 
successfully take into account the main properties of the soft diffractive 
processes. These models, inspired by $1/N$ QCD expansion are also 
supplemented with generally accepted theoretical principles like duality, 
unitarity, Regge behavior, and parton structure. The 
interactions are no longer described by single particle exchange, but by 
highly complicated modes known as Reggeons. Up to about 62~GeV, the slow growth of the cross 
section with $\sqrt{s}$ is driven by a dominant contribution of a special 
Reggeon, the Pomeron. 

At higher energies, semihard (SH) interactions arising from 
the hard scattering of partons that carry only a very small fraction of the 
momenta of their parent hadrons can compete successfully with soft 
processes~\cite{Cline:1973kv,Ellis:1973nb,Halzen:1974vh,Pancheri:sr,Gaisser:1984pg,DiasdeDeus:1984ip,
Pancheri:ix,Pancheri:qg}. These semihard interactions lead 
to the ``minijet'' phenomenon, i.e.  jets with transverse energy 
($E_T = |p_{_T}|$) 
much smaller than the total c.m. energy.  Such low-$p_{_T}$ processes
cannot be identified by jet finding algorithms, but  (unlike soft 
processes) still they can be calculated using perturbative QCD.
The cross section for SH interactions is described  by
\begin{equation}
\sigma_{\rm QCD}(s,p_{{_T}}^{\rm min}) = \sum_{i,j} \int 
\frac{dx_1}{x_1}\, \int \frac{dx_2}{x_2}\,
\int_{Q_{\rm min}^2}^{\hat{s}/2} \, d|\hat t|\,\, 
\frac{d\hat{\sigma}_{ij}}{d|\hat t|}\,\,
x_1 f_i(x_1, |\hat t|)\,\,\, x_2 f_j(x_2, |\hat t|) \,\,\,,
\label{sigmaminijet}
\end{equation}
where $x_1$ and $x_2$ are the fractions of the momenta of the parent hadrons 
carried by the partons which collide,
$d\hat{\sigma}_{ij}/d|\hat t|$ is the cross section for scattering of 
partons of types $i$ and $j$ according to elementary QCD diagrams, 
$f_i$ and $f_j$ are parton distribution functions (PDFs), 
$\hat{s} = x_1\,x_2 s$ 
and $-\hat{t} = \hat{s}\, (1 - \cos \vartheta^*)/2 =  Q^2$ 
are the Mandelstam variables for this parton-parton process,
and the sum is over all parton species. Here,
\begin{equation}
p_{_T} = E_{\rm jet}^{\rm lab} \,\,\sin \vartheta_{\rm jet} = \frac{\sqrt{\hat s}}{2}\,\, 
\sin \vartheta^*\,,
\end{equation}
and
\begin{equation}
p_{_\parallel} = E_{\rm jet}^{\rm lab} \,\,\cos \vartheta_{\rm jet}\,,
\end{equation}
where $E_{\rm jet}^{\rm lab}$ is the energy 
of the jet in the lab frame,   
$\vartheta_{\rm jet}$ the angle of the jet with respect to the beam 
direction in the lab frame, and $\vartheta^*$ is the angle of the jet with respect to the beam direction 
in the c.m. frame of the elastic parton-parton collision. This implies that for 
small $\vartheta^*,$ $p_{_T}^2 \approx Q^2$. The integration limits satisfy 
\begin{equation}
Q_{\rm min}^2 < |\hat t| < \hat{s}/2 \,,
\end{equation}
 where $Q_{\rm min} = 1 - 2~{\rm GeV}$ is the 
minimal momentum transfer. The measured minijet cross sections
indicate that the onset of SH interactions has just occurred by CERN
SPS  energies, $\sqrt{s} > 200~{\rm GeV}$~\cite{Albajar:1988tt}.

A first source of uncertainty in modeling cosmic ray interactions at
ultra-high energy is encoded in the extrapolation of the measured
parton densities several orders of magnitude down to low $x$.  Primary
protons that impact on the upper atmosphere with energy $\sim
10^{11}$~GeV yield partons with $x \equiv 2 p^*_{_\parallel}/\sqrt{s}
\sim m_\pi/\sqrt{s} \sim 10^{-7},$ whereas current data on quark and
gluon densities are only available for $x \gtrsim 10^{-5}$ to within
an experimental accuracy of few percent for $Q^2 \approx
100$~GeV$^2$~\cite{Patrignani:2016xqp}.  In addition, extrapolation of
HERA and LHC data to UHECR interactions assumes universality of the
PDFs. This assumption, based on the QCD factorization conjecture (the
cross section (\ref{sigmaminijet}) can always be written in a form
which factorizes the parton densities and the hard interaction
processes irrespective of the order in perturbation theory and the
particular hard process) holds in the limit $Q^2 \gg \Lambda_{\rm
  QCD}$, where $\Lambda_{\rm QCD} \sim 200$~MeV is the QCD
renormalization scale.

For large $Q^2$ and not too small $x$, the 
Dokshitzer-Gribov-Lipatov-Altarelli-Parisi (DGLAP) 
equations~\cite{Gribov:rt,Gribov:ri,Dokshitzer:sg,Altarelli:1977zs} 
\begin{equation}
\frac{\partial}{\partial \ln Q^2} {q(x, Q^2) \choose g(x,Q^2)}  =
\frac{\alpha_s(Q^2)}{2 \pi} {P_{qq} \,\,\,\, P_{qg} \choose P_{gq} 
\,\,\,\, P_{gg}}  \otimes  {q(x, Q^2) \choose g(x,Q^2)} 
\end{equation} 
successfully 
predict the $Q^2$ dependence of the quark and gluon densities ($q$ and $g,$ 
respectively). Here, $\alpha_s = g^2/(4\pi),$ with  $g$ the strong 
coupling constant. The splitting functions $P_{ij}$ indicate the probability
of finding a daughter parton $i$ in the parent parton $j$ with a given 
fraction of parton $j$ momentum. This probability will depend on the number 
of splittings allowed in the approximation. In the double-leading-logarithmic approximation, that is $\lim_{x \to
  0} \ln(1/x)$ and $\lim_{Q^2 \to \infty}\ln(Q^2/\Lambda_{\rm QCD})$,
the 
DGLAP equations predict a steeply rising gluon density, $xg \sim x^{-0.4},$
which dominates the quark density at low $x$. HERA data are found to be consistent with a 
power law, $xg(x,Q^2) \sim x^{-\Delta_{\rm H}},$ with an exponent $\Delta_{\rm H}$ 
between 0.3 and 0.4~\cite{Dittmar:2009ii,Engel:ac}.

The high energy minijet cross section is then 
determined by  the dominant gluon distribution 
\begin{equation}
\sigma_{\rm QCD} (s,p_{{_T}}^{\rm min})   
\approx \int \frac{dx_1}{x_1}\, \int \frac{dx_2}{x_2} \,
\int_{Q_{\rm min}^2}^{\hat{s}/2} \, d|\hat t|\,\,\, 
\frac{d\hat \sigma}{d|\hat t|}\,\,\,
x_1 g(x_1, |\hat t|)\,\,\, x_2 g(x_2, |\hat t|) \,\,,
\label{gluon}
\end{equation}
where the integration limits satisfy
\begin{equation}
x_1 x_2 s > 2 |\hat t| > 2 Q^2_{\rm min} \, .
\end{equation}
Furthermore, because $d\hat \sigma/d|\hat t|$ is peaked at the low end of the $|\hat t|$ integration (see  e.g~\cite{Anchordoqui:2009eg}), the high energy behavior of $\sigma_{\rm QCD}$ is controlled (via the lower limits of the $x_1, \, x_2$ integrations) by the small-$x$ behavior of the gluons~\cite{Kwiecinski:1990tb}
\begin{eqnarray}
\sigma_{\rm QCD} (s)   
\propto  \int_{2\,Q_{\rm min}^2/s}^{1} \frac{dx_1}{x_1}\,\,
x_1^{-\Delta_{\rm H}}\, \int_{2 \,Q_{\rm min}^2/x_1s}^{1} \frac{dx_2}{x_2} 
\,\,x_2^{-\Delta_{\rm H}} \sim 
s^{\Delta_{\rm H}}\, \ln s \, \sim_{_{\!\!\!\!\! \! \!\!\!\! s \to \infty}}  s^{\Delta_{\rm H}} \, .
\label{KH}
\end{eqnarray}
This estimate is, of course, too simplistic. At sufficiently small $x$, gluon shadowing corrections suppress the singular $x^{-\Delta_H}$  behavior of $xg$ and hence suppress the power growth of $\sigma_{\rm QCD}$ with increasing $s$.

Although we have shown that the onset of semihard processes is an unambiguous prediction of QCD, in practice it is difficult to isolate these contributions from the soft interactions. 
Experimental evidence indicates that  SH interactions can essentially
be neglected up to and throughout the CERN ISR energy regime,
$\sqrt{s} < 62~{\rm GeV}$. Therefore,  measurements made in this
energy region can be used to model the soft interactions. A reasonable
approach introduced in~\cite{DiasdeDeus:1987yw}  is to base the
extrapolation of the soft interactions on the assumption of
geometrical scaling~\cite{DiasDeDeus:1987bf}, which is observed to be
true throughout the ISR energy
range~\cite{Amaldi:1979kd,Castaldi:1985ft}. We adopt the standard partial-wave amplitude in impact-parameter space $f(s,b)$, which is the Fourier transform of the elastic $pp$ (or $p\bar p$) scattering amplitude. (We neglect any difference between $pp$ and $p \bar p$ for $\sqrt{s} > 200~{\rm GeV}$.) Geometrical scaling (GS) corresponds to the assumption that $f$, which {\em a priori} is a function of two  dimensional variables $b$ and $s$, depends only upon one dimensional  variable $\beta = b/R(s)$, where $R$ is the energy dependent radius of the proton, {\it i.e.}
\begin{equation}
f(s,b) = f_{\rm GS} \left(\beta = b/R(s) \right) \, .
\end{equation}
Physically, this means that the opaqueness of the proton remains constant with rising energy  and that the increase of the total cross section, $\sigma_{\rm tot}$, in the ISR energy range reflects a steady growth of the radius $R(s)$. An immediate obvious consequence of GS is that the partial wave at $b=0$ should be independent of energy
\begin{equation}
f(s,b=0) = f_{\rm GS} (\beta =0) \, .
\end{equation}
Another consequence is that the ratio of elastic scattering to total cross section, $\sigma_{\rm el} (s)/\sigma_{\rm tot} (s)$, should be energy independent. This follows from
\begin{subequations}
\begin{equation}
\sigma_{\rm tot}  =  8 \pi \int {\rm Im} f(s,b) \, b \, db 
 =_{\!\!\!\!\!\!\!{_{\rm GS}}}  8 \pi R^2(s) \, \int {\rm Im}  f_{\rm GS}(\beta) \, \beta \, d\beta, 
\end{equation}
and
\begin{equation}
\sigma_{\rm el}  =  8 \pi \int \left|f(s,b) \right|^2 \, b \, db 
 =_{\!\!\!\!\!\!\!{_{\rm GS}}}  8 \pi R^2(s) \, \int \left|f_{\rm GS}(\beta) \right|^2 \, \beta \, d\beta \, . 
\end{equation}
\end{subequations}
To determine the gross features at high energies  we can assume that the elastic amplitude has a simple form 
\begin{equation}
F(s,t) = i \ \sigma_{\rm tot} (s) \ \ e^{Bt/2} \, ,
\label{giusti}
\end{equation}
with $B$  the  slope parameter that measures the size of the proton~\cite{Block:1984ru}. This is a reasonable assumption: the amplitude is predominantly imaginary, and the exponential behavior observed for $|t| \lesssim 0.5~{\rm GeV}^2$ gives the bulk of the elastic cross section. Now, the Fourier transform $f(s,b)$ of the elastic amplitude $F(s,t)$ given by (\ref{giusti}) has a Gaussian form in impact parameter space
\begin{equation}
f(s,b) = \frac{i \sigma_{\rm tot}(s)}{8 \pi B} e^{-b^2/2B} \, ,
\label{percudani}
\end{equation}
and it follows that
\begin{equation}
{\rm Im}  f(s, b=0) = \frac{\sigma_{\rm tot}}{8 \pi B} = \frac{2 \sigma_{\rm el}}{\sigma_{\rm tot}} \, .
\label{bochini}
\end{equation}
Equation (\ref{bochini}) offers a very clear way to see the breakdown of GS and to identify semihard interactions from the growth of the central partial wave.

In general unitarity requires ${\rm Im}  f(s,b) \leq \frac{1}{2}$,
which in turm implies $\sigma_{\rm el}/\sigma_{\rm tot} \leq
\frac{1}{2}$~\cite{Block:1984ru}. This seems to indicate that the
Gaussian form (\ref{percudani}) may not longer be applicable at
ultra-high energies, but rather it is expected that the proton will
approximate a ``black disk'' of radius $b_0$, {\em i.e.}  $f(s,b) =
\frac{i}{2}$ for $0<b\lesssim b_0$ and zero for $b \gtrsim b_0$. Then
$\sigma_{\rm el} \simeq \frac{1}{2} \sigma_{\rm tot} \simeq \pi
b_0^2$. In order to satisfy the unitarity constraints, it is convenient to introduce the eikonal $\chi$ defined by
\begin{equation}
f(s,b) = \frac{i}{2} \left\{ 1 - {\rm exp} \left[i \chi (s,b) \right] \right\} \,,
\end{equation}
where ${\rm Im}  \, \chi \geq 0$. If we neglect for the moment the shadowing corrections to the PDFs and take $xg \propto x^{-\Delta_{\rm H}}$ in the small-$x$ limit we obtain, as explained above,  power growth of the cross section for SH interactions, $\sigma_{\rm QCD} \sim s^{\Delta_{\rm H}}$ and 
${\rm Im}  \chi(s, b=0) \gg 1$ as $s \to \infty$. Indeed we expect ${\rm Im}  \, \chi \gg 1$ (and unitarity to be saturated) for a range
of $b$ about $b=0$. Then we have
\begin{equation}
\sigma_{\rm tot}  =  4 \pi \int_0^\infty b \ db \,  \Theta (b_0 -  b) 
  \simeq  4 \pi \int_0^{b_0(s)} b \,  db = 2 \pi b_0^2 \,,
\end{equation}
with $\chi \simeq \chi_{_{\rm SH}}$ and where $b_0(s)$ is such that
\begin{equation}
{\rm Im} \, \chi_{_{\rm SH}} (s, b_0(s)) \simeq 1 \, .
\label{handset}
\end{equation}

Hereafter, we ignore the small real part of the scattering amplitude, which is good approximation at high energies. The unitarized elastic, inelastic, and total cross sections (considering now a real eikonal function) are given by~\cite{Glauber:1970jm,L'Heureux:jk,Durand:prl,Durand:cr}
\begin{subequations}
\begin{equation}
\sigma_{\rm el}=2\pi \int d b\, b\
\left\{1-\exp\left[ -\chi_{_{\rm soft}}(s, b)
-\chi_{_{\rm SH}}(s, b)\right]\right\}^2\ \,,
\label{elastic}
\end{equation}
\begin{equation}
\sigma_{\rm inel}= 2 \pi \int d b\, b \,
\left\{1-\exp\left[ -2\chi_{_{\rm soft}}(s, b)
-2\chi_{_{\rm SH}}(s, b)\right]\right\}\ ,
\label{inelastic}
\end{equation}
\begin{equation}
\sigma_{\rm tot}= 4 \pi \int d b\, b\,
\left\{1-\exp\left[-\chi_{_{\rm soft}}(s, b)
-\chi_{_{\rm SH}}(s, b)\right]\right\}\ ,
\label{total}
\end{equation}
\end{subequations}
where the scattering is compounded as a sum of QCD ladders
via SH and soft processes through the
eikonals $\chi_{_{\rm SH}}$ and $\chi_{_{\rm soft}}$. 

Now, if the eikonal function, 
$\chi (s, b) \equiv \chi_{_{\rm soft}}(s, b) + 
\chi_{_{\rm SH}}(s, b) =\lambda/2,$
indicates the mean number of partonic interaction pairs at impact parameter 
$b,$ the probability $p_n$ for having $n$ independent partonic 
interactions using Poisson statistics reads, 
$p_n = (\lambda^n/n!) \, e^{-\lambda}$.
Therefore, the factor $1-e^{-2\chi} = \sum_{n=1}^\infty p_n$ in (\ref{inelastic}) 
can be interpreted semi-classically as the probability 
that at least 1 of the 2 protons is broken up in a collision at impact 
parameter $b$.
With this in mind, the inelastic cross section is simply the integral
over all collision impact parameters of the probability of having at 
least 1 interaction, yielding a mean minijet multiplicity of 
\mbox{$\langle n_{\rm minijet} \rangle \approx \sigma_{\rm QCD}/
\sigma_{\rm inel}$~\cite{Gaisser:1988ra}.} The leading
contenders to approximate the (unknown) cross sections at
cosmic ray energies, {\sc sibyll}~\cite{Fletcher:1994bd} and 
{\sc qgsjet}~\cite{Kalmykov:te}, share the eikonal
approximation but differ in their {\em ans\"atse} for the
eikonals. In both cases, the core of dominant scattering at
very high energies is the SH cross section given in (\ref{sigmaminijet}),
\begin{equation}
\chi_{_{\rm SH}} = \frac{1}{2} \, 
\sigma_{\rm QCD}(s,p_{{_T}}^{\rm min})\,\, A(s,\vec b) \,,
\label{hard}
\end{equation}
where the normalized profile function, 
$2 \pi \int_0^\infty d b \, b \,A(s, b) = 1,$
 indicates the distribution of partons in the plane transverse to the 
collision axis. 

In the {\sc qgsjet}-like 
models, the  core of the SH eikonal 
is dressed with a soft-pomeron pre-evolution factor. This amounts to
taking a parton distribution which is Gaussian in the transverse
coordinate distance $b,$ 
\begin{equation}
A(s, b) = \frac{e^{- b^2/R^2(s)}}{\pi R^2(s)} \, ,
\label{a}
\end{equation}
with  $R$ being a parameter. For a QCD cross section dependence, $\sigma_{\rm QCD} \sim 
 s^{\Delta_{\rm H}},$ one gets for a Gaussian profile
\begin{equation}
b_0^2(s) \sim R^2 \Delta_{\rm H} \, \ln s
\end{equation}
 and at high energy 
\begin{eqnarray}
\sigma_{\rm inel}  =   2 \pi \int_0^{b_0(s)} d b \, b  \sim  \pi R^2 \Delta_{\rm H} \, \ln s \, .
\label{qsig}
\end{eqnarray}
If the effective radius $R$ (which controls parton shadowing) is energy-independent, the cross section increases only logarithmically with rising energy. However, the parameter $R$ itself depends on the collision energy through a convolution with the parton momentum fractions, $R^2(s) \sim  R_0^2 + 4 \, \alpha^\prime_{\rm eff} \,\ln^2 s$, with 
$\alpha^\prime_{\rm eff} \approx 0.11$~GeV$^{-2}$~\cite{Alvarez-Muniz:2002ne}. Thus, the {\sc qgsjet}  cross section exhibits a faster than $\ln \, s$ rise,
\begin{equation}
 \sigma_{\rm inel } \sim 4\pi \, \alpha^\prime_{\rm eff} \,\,\Delta_{\rm H}\,\,
\ln^2 s \, .
\label{beber}
\end{equation}

In  {\sc sibyll}-like models, the transverse density
distribution is taken as the Fourier transform of the proton electric
form factor, resulting in an energy-independent exponential 
(rather than Gaussian) fall-off of the parton density profile 
 for large $b$,
\begin{eqnarray}
A(b)  =  \frac{\mu^2}{96 \pi} (\mu b)^3 \,   K_3 (\mu b) 
  \sim  e^{-\mu b} \,,
\end{eqnarray} 
where $K_3 (x)$ denotes the modified Bessel function of the third kind and $\mu^2 \approx 0.71  {\rm GeV}^{2}$~\cite{Fletcher:1994bd}. Thus, (\ref{hard}) becomes
\begin{equation}
\chi_{_{\rm SH}} \sim e^{-\mu b}  \,  s^{\Delta_{\rm H}} ,
\end{equation}
 and  (\ref{handset}) is satisfied when
\begin{equation}
b_0(s) = \frac{\Delta_{\rm H}}{\mu} \, \ln \, s \, .
\end{equation}
Therefore, for {\sc sibyll}-like models, the  growth of the inelastic
cross section also saturates the $\ln^2s$ Froissart 
bound~\cite{Froissart:ux},
but with a multiplicative constant which is larger 
than the one in {\sc qgsjet}-like models
\begin{equation}
\sigma_{\rm inel} \sim \pi c \,\, \frac{\Delta_{\rm H}^2}{\mu^2} \,  \ln^2 s \,\,,
\label{ssig}
\end{equation}
where the coefficient $c \approx 2.5$ is found numerically~\cite{Alvarez-Muniz:2002ne}.

The main characteristics of the $pp$ cascade spectrum
resulting from these choices are readily predictable: the harder 
form of the {\sc sibyll}
form factor allows a greater retention of energy by the leading
particle, and hence less available for the ensuing 
shower. Consequently, on average {\sc sibyll}-like models predict a smaller  
multiplicity than {\sc qgsjet}-like models~\cite{Anchordoqui:1998nq}. 

In {\sc qgsjet}-like models, both the soft and hard 
processes are formulated in terms of Pomeron exchanges. To describe the 
minijets, the soft Pomeron  mutates into a ``semihard Pomeron'', 
an ordinary soft Pomeron with the middle piece replaced by a QCD parton 
ladder, as sketched in the previous paragraph.  This is generally referred to 
as the ``quasi-eikonal'' model.  In
contrast, {\sc sibyll}  follows a ``two channel'' eikonal model, where the soft and  
the semi-hard regimes are demarcated by a sharp cut in the transverse momentum: 
{\sc sibyll}  uses a cutoff parametrization inspired in the double leading logarithmic approximation 
of the DGLAP equations,
\begin{equation}
p_{{_T}}^{\rm min} (\sqrt{s}) =  p_{{_T}}^0 + 0.065~{\rm GeV}\, \exp[0.9\,\sqrt{\ln s}]\,,
\end{equation}
where  $p_{{_T}}^0 = 1$~GeV~\cite{Ahn:2009wx}.

The transition process from asymptotically free partons to colour-neutral 
hadrons is described in all codes by string fragmentation 
models~\cite{Sjostrand:1987xj}.  Different choices of fragmentation functions 
can lead to some differences in the hadron multiplicities. 
However, the main difference in the predictions of {\sc qgsjet}-like and 
{\sc sibyll}-like models arises from different assumptions in extrapolation of the parton 
distribution function to low energy. 

The proton-air collisions of interest for air shower development cause
additional headaches for event generators.  Both {\sc sibyll} and {\sc
  qgsjet} adopt the Glauber formalism~\cite{Glauber:1970jm}, which is
equivalent to the eikonal approximation in nucleon-nucleon scattering,
except that the nucleon density functions of the target nucleus are
folded with that of the nucleon.  The inelastic and production cross
sections read:
\begin{equation}
\sigma_{\rm inel}^{p{\rm -air}} \approx 2 \pi \int d b\, b \,
\left\{1-\exp\left[\sigma_{\rm tot} 
\,\, A T_N( b) \right]\right\}\ \,,
\label{inelasticn}
\end{equation}
\begin{equation}
\sigma_{\rm prod}^{p{\rm -air}} \approx 2 \pi \int d b\, b\,
\left\{1-\exp\left[\sigma_{\rm inel} 
\,\, A T_N(b) \right]\right\}\ \,,
\label{prod}
\end{equation}
where $T_N(b)$ is the transverse distribution function of a nucleon
inside a nucleus. Here, $\sigma_{\rm inel}$ and $\sigma_{\rm tot}$ are
given by (\ref{inelastic}) and (\ref{total}), respectively. The
$p$-air inelastic cross section is the sum of the ``quasi-elastic''
cross section, which corresponds to cases where the target nucleus
breaks up without production of any new particles, and the production
cross section, in which at least one new particle is generated.
Clearly the development of EASs is mainly sensitive to the production
cross section.  Overall, the geometrically large size of nitrogen and
oxygen nuclei dominates the inclusive proton-target cross section, and
as a result the disagreement from model-dependent extrapolation is not
more than about 15\%.  More complex nucleus-nucleus interactions are
discussed in~\cite{Engel:vf}.

Adding a greater challenge to the determination of the proton air
cross section at ultra-high energies is the lack of direct measurements
in a controlled laboratory environment. The measured shower
attenuation length, $\Lambda_m$, is not only sensitive to the
interaction length of the protons in the atmosphere, $\lambda_{p{\rm
    -air}}$, with
\begin{equation}
\Lambda_m = k \lambda_{p{\rm -air}} = k { 14.4~m_p \over \sigma_{\rm prod}^{p{\rm -air}}} \,,  \label{eq:Lambda_m}
\end{equation}
but also depends on the rate at which the energy of the primary proton
is dissipated into the electromagnetic (EM) shower energy observed in the experiment. Here,
$\Lambda_m$ and $\lambda_{p{\rm -air}}$ are in g\,cm$^{-2}$, the
proton mass $m_p$ is in g, and the inelastic production cross section
$\sigma_{\rm prod}^{p{\rm -air}}$ is in mb.  The value of $k$ depends
critically on the inclusive particle production cross section and its
energy dependence in nucleon and meson interactions on the light
nuclear target of the atmosphere. The measured depth $X_{\rm max}$ at
which a shower reaches maximum development in the atmosphere has been
the basis of cross section measurements from experiments prior to
HiRes and Auger. However, $X_{\rm max}$ is a combined measure of the
depth of the first interaction (which is determined by the inelastic
cross section) and of the subsequent shower development (which has to
be corrected for).  The model dependent rate of shower development and
its fluctuations are the origin of the deviation of $k$ from unity in
(\ref{eq:Lambda_m}). As can be seen in Table~\ref{ktable}, there
is a large range of $k$ values (from 1.6 for a very old model where
the inclusive cross section exhibited Feynman scaling, to 1.15 for
modern models with large scaling violations) that make the published
values of $\sigma_{p{-\rm air}}$ unreliable.

\begin{table}
\caption{Different $k$-values used in cosmic ray experiments.\label{ktable}} 
\begin{center}
 \begin{tabular}[b]{cc}
\hline
\hline
~~~~~~~~~~~~~~~~~~~~~~~~Experiment~~~~~~~~~~~~~~~~~~~~~~~~&~~~~~~~~~~~~~~~~~~~~~~~~~~~~~~~$k$~~~~~~~~~~~~~~~~~~~~~~~~~~~~~~~\\ \hline
~~~~~~~~~~Fly's Eye~~~~~~~~~~&~~~~~~~~~~~~~~~~~~~~~~~~~~~~~~~ 1.6 ~~~~~~~~~~~~~~~~~~~~~~~~~~~~~~~\\ 
Akeno&~1.5~\\ 
Yakutsk-99 &~1.4~\\ 
EAS-TOP & ~~1.15\\ \hline
\hline
\end{tabular}
\end{center}
\end{table}

The HiRes Collaboration developed a quasi-model-free method of
measuring $\sigma_{\rm prod}^{p{\rm -air}}$
directly~\cite{Belov:2006mb}.  This is accomplished by folding a
randomly generated exponential distribution of first interaction
points into the shower development program, and therefore fitting the
entire distribution and not just the trailing edge.  Interestingly,
the measured $k = 1.21^ {+0.14}_{-0.09}$ by the HiRes group is in
agreement with the one obtained by tuning the data to the
theory~\cite{Block:1999ub,Block:2007rq}.

\begin{figure}[tbp]
\begin{minipage}[t]{0.48\textwidth}
\postscript{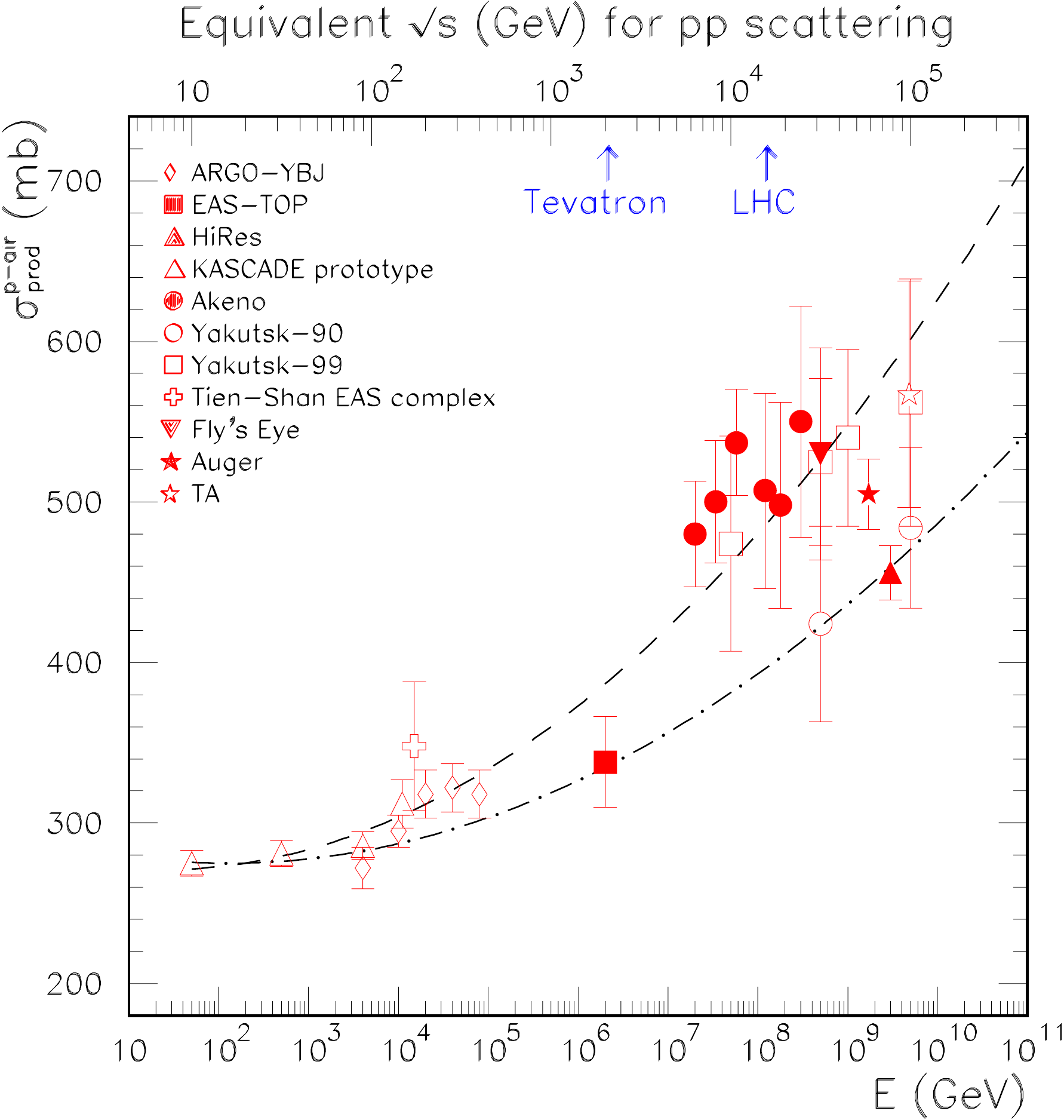}{0.99}
\end{minipage}
\hfill
\begin{minipage}[t]{0.48\textwidth}
\postscript{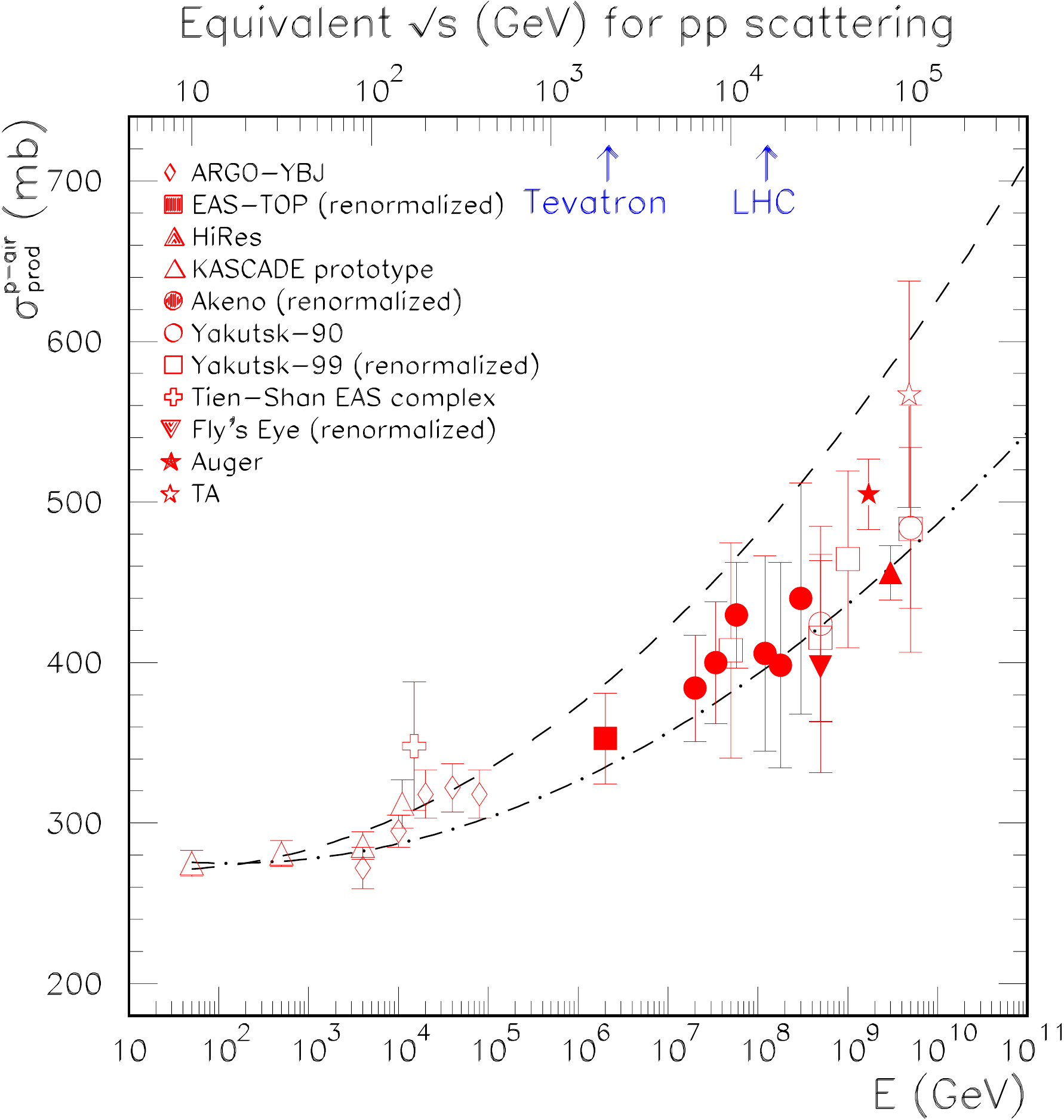}{0.99}
\end{minipage}
\caption{Compilation of proton-air production cross section from
  cosmic ray measurements (ARGO-YBJ~\cite{Aielli:2009ca},
  EAS-TOP~\cite{Aglietta:2009zza}, HiRes~\cite{Belov:2006mb}, KASCADE
  prototype~\cite{Mielke:1994un}, Akeno~\cite{Honda:1992kv},
  Yakutsk-90~\cite{Dyakonov:1990cd},
  Yakutsk-99~\cite{Knurenko:1999cr}, Tien-Shan EAS
  complex~\cite{Nam:1975xk}, Fly's Eye~\cite{Baltrusaitis:1984ka},
  Auger~\cite{Collaboration:2012wt}, and TA~\cite{Abbasi:2015fdr}). The data are compared to the parametrizations discussed in the text; case-$i$ corresponds to the dashed line and case-$ii$ to the dot-dashed line. \label{fig:23}}
\end{figure}

A compilation of published proton-air cross section measurements is shown in Fig.~\ref{fig:23}. In the left panel we show the data without any modification. In the right panel, the published values of $\sigma_{\rm prod}^{p{\rm -air}}$ for Fly's Eye~\cite{Baltrusaitis:1984ka}, Akeno~\cite{Honda:1992kv}, Yakutsk-99~\cite{Knurenko:1999cr}, and EAS-TOP~\cite{Aglietta:2009zza} collaborations have been renormalized using the {\em common} value of  $k = 1.264 \pm 0.033 (\rm stat) \pm 0.013 (\rm syst)$~\cite{Block:2007rq}.  We have parametrized the rise of the cross section using a functional form that saturates the   Froissart bound,
\begin{equation}
\sigma_{\rm prod}^{p{\rm -air}} = {\cal A} - {\cal B} \ln  (E/{\rm GeV}) + {\cal C} \ln^2 (E/{\rm GeV})~{\rm mb} \, .
\end{equation}
The curve with a fast rise, hereafter referred to as case-$i$, 
corresponds to ${\cal A} = 280$, ${\cal B} = 5.7$, and ${\cal C} =
0.9$. The slow rise of case-$ii$ has the following parameters: ${\cal
  A} = 290$, ${\cal B} = 6.2,$ and ${\cal C} = 0.64.$ 

In summary, high energy hadronic interaction models are still being
refined and therefore the disparity between them can vary even from
version to version~\cite{dEnterria:2011twh}. At the end of the day,
however, the relevant parameters boil down to two: the mean free path,
$\lambda_{{\rm CR-air}} = (\sigma_{\rm prod}^{{\rm CR-air}} \,n_{\rm
  atm})^{-1},$ and the inelasticity, $y_{_{{\rm CR-air}}} = 1 - E_{\rm
  lead}/E_{\rm proj}$, where $n_{\rm atm}$ is the number density of
atmospheric target nucleons, $E_{\rm lead}$ is the energy of the most
energetic hadron with a long lifetime, and $E_{\rm proj}$ is the
energy of the projectile particle. The first parameter characterizes
the frequency of interactions, whereas the second one quantifies the
energy lost per collision. Overall, {\sc sibyll} has a shorter mean
free path and a smaller inelasticity than {\sc qgsjet}.  Since a
shorter mean free path tends to compensate for a smaller inelasticity,
the two codes generate similar predictions for an air shower which has
lived through several generations.  Both models predict the same
multiplicity below about $10^{7}$~GeV, but the predictions diverge
above that energy. Such a divergence readily increases with rising
energy. While {\sc qgsjet} predicts a power law-like increase of the
number of secondaries up to the highest energy, {\sc sibyll}
multiplicity exhibits a logarithmic growth. As it is extremely
difficult to observe the first interactions experimentally, it is not
straightforward to determine which model is closer to reality.

\subsection{Electromagnetic processes}  
\label{EMP}

The evolution of an extensive air shower is dominated by
EM processes. The interaction of a baryonic cosmic ray
with an air nucleus high in the atmosphere leads to a cascade of
secondary mesons and baryons. The first few generations of charged
pions interact again, producing a hadronic core, which continues to
feed the EM and muonic components of the shower. Up to
about $50$~km above sea level, the density of atmospheric target
nucleons is $n_{\rm atm} \sim 10^{20}$~cm$^{-3},$ and so even for
relatively low energies, say $E_{\pi^{\pm}}\approx 1$~TeV, the
probability of decay before interaction falls below 10\%.  Ultimately,
the EM cascade dissipates around 90\% of the primary
particle's energy, and hence the total number of EM 
particles is very nearly proportional to the shower
energy~\cite{Barbosa:2003dc}.

Roughly speaking, at $10^{11}$~GeV, baryons and charged pions have
interaction lengths of the order of $40~{\rm g}/{\rm cm}^2,$
increasing to about $60~{\rm g}/{\rm cm}^2$ at $10^7$~GeV.
Additionally, below $10^{10}$~GeV, photons, electrons, and positrons
have mean interaction lengths of $37.6~{\rm g}/{\rm cm}^2$.  Altogether,
the atmosphere acts as a natural colorimeter with variable density,
providing a vertical thickness of 26 radiation lengths and about 15
interaction lengths. Amusingly, this is not too different from the
number of radiation and interaction lengths at the LHC detectors. For
example, the CMS electromagnetic calorimeter is $\gtrsim 25$ radiation
lengths deep, and the hadron calorimeter constitutes 11 interaction
lengths.

By the time a vertically incident $10^{11}$~GeV proton shower
reaches the ground, there are about $10^{11}$ secondaries with energy above
90~keV in the the annular region extending 8~m to 8~km from the shower core.
Of these, 99\%  are photons, electrons, and positrons, with a typical ratio 
of $\gamma$ to $e^+  e^-$ of 9 to 1. Their mean energy  is 
around 10~MeV and they transport 85\% of the total energy at ground level. Of course, 
photon-induced showers are even more dominated by the electromagnetic channel, 
as the only significant muon generation mechanism in this case is the decay of
charged pions and kaons produced in $\gamma$-air  interactions~\cite{Mccomb:tp}. 

It is worth mentioning that these figures dramatically change for the case of 
very inclined showers. For a primary zenith angle, $\theta > 70^{\circ},$ the 
electromagnetic component becomes attenuated exponentially with atmospheric 
depth, being almost completely absorbed at ground level. Note 
that the vertical atmosphere is approximately $1,000~{\rm g/cm}^{2}$, and is about 36 
times deeper for completely horizontal showers~\cite{Anchordoqui:2004xb}. 
As a result, most of the energy at ground level from an inclined shower is
carried by muons.

In contrast to  hadronic collisions, the electromagnetic 
interactions of shower particles can be calculated very accurately from 
quantum electrodynamics. Electromagnetic interactions are thus not a 
major source of systematic errors in shower simulations. The first 
comprehensive treatment of electromagnetic showers was elaborated by 
Rossi and Greisen~\cite{Rossi:1941zza}.  This treatment was recently cast in a more 
pedagogical form by Gaisser~\cite{Gaisser:1990vg}, which we summarize in the 
subsequent paragraphs.

The generation of the electromagnetic component is 
driven by electron bremsstrahlung and pair production~\cite{Bethe:1934za}.
Eventually the average energy per particle drops below a critical energy,
$\epsilon_0$, at which point ionization takes over from bremsstrahlung and pair
production as the dominant energy loss mechanism. The $e^\pm$ energy loss rate due to
bremsstrahlung radiation is nearly proportional to their energy, whereas the 
ionization loss rate varies only logarithmically with the $e^\pm$ energy.
Though several different definitions of the critical energy appear in the 
literature~\cite{Patrignani:2016xqp}, throughout  we take the critical energy 
to be that at which the ionization loss per radiation length
is equal to the electron energy; this leads to $\epsilon_{0} = 710~{\rm MeV}/(Z_{\rm eff} +0.92) \sim 
$~86~MeV~\cite{Rossi:book}.\footnote{For altitudes up to 90~km above sea level, the air is a mixture 
of 78.09\% of N$_2$, 20.95\% of O$_2$, and 0.96\% of other gases~\cite{Weast}. 
Such a mixture is generally modeled as an homogeneous substance with atomic charge
and mass numbers $Z_{\rm eff} = 7.3$ and $A_{\rm eff} = 14.6,$ respectively.}  The changeover 
from radiation losses to ionization losses depopulates the shower.
One can thus categorize the shower development in three phases: the growth phase, in which all the particles 
have energy $> \epsilon_0$; the shower maximum, $X_{\rm max}$; and the shower 
tail, where the particles only lose energy, get absorbed or decay. 

The relevant quantities participating in the development
of the electromagnetic cascade are the probability for an electron of
energy $E$ to radiate a photon of energy $k=y_{_{\rm brem}}E$ and
the probability for a photon to produce a pair $e^+e^-$
in which one of the particles (hereafter $e^-$) has energy $E=y_{_{\rm pair}}k$.
These probabilities are determined by the properties of the air and 
the cross sections of the two processes. 

In the energy range of interest, the impact parameter of the electron
or photon is larger than an atomic radius, so the nuclear field is
screened by its electron cloud.  In the case of complete screening,
where the momentum transfer is small, the cross section for
bremsstrahlung can be approximated by
\begin{equation}
\frac{d\sigma_{e \rightarrow \gamma}}{dk} \approx \frac{A_{\rm eff}}{X_{_{\rm EM}} N_A k}
\left(\frac{4}{3}-\frac{4}{3}y_{_{\rm brem}}+y_{_{\rm brem}}^2\right)\,\,,\label{brem}
\end{equation}
where $A_{\rm eff}$ is the effective mass number of the air, $X_{_{\rm EM}}$ is a constant,
and $N_A$ is Avogadro's number~\cite{Tsai:1973py}. In the infrared limit ({\it i.e.} $y_{_{\rm brem}} \ll 1$) this approximation 
is inaccurate at the level of about 2.5\%, which is small compared to typical experimental 
errors associated with cosmic air shower detectors.  Of course, the approximation fails 
as $y_{_{\rm brem}}  \rightarrow 1$, when nuclear screening becomes incomplete, and as $y_{_{\rm brem}}  \rightarrow 0$, at which 
point the LPM and dielectric suppression effects become important, as we discuss below.

Using similar approximations, the cross section
for pair production can be written as~\cite{Tsai:1973py}
\begin{equation}
\frac{d\sigma_{\gamma \rightarrow e^+e^-}}{dE} \approx \frac{A_{\rm eff}}{X_{_{\rm EM}} N_A}
\left(1-\frac{4}{3}y_{_{\rm pair}}+\frac{4}{3}y_{_{\rm pair}}^2\right) \,. \label{pair}
\end{equation}
The similarities between this expression and (\ref{brem}) are to be expected, 
as the Feynman diagrams for pair production and bremsstrahlung are variants of one another.

The probability for an electron to radiate a photon 
with energy in the range $(k,k+dk)$ in traversing $dt=dX/X_{_{\rm EM}}$ of atmosphere is 
\begin{equation} \label{p_b}
\frac{d\sigma_{e \rightarrow \gamma}}{dk}\,\frac{X_{_{\rm EM}} N_A}{A_{\rm eff}}\,dk\,dt
\approx \left(y_{_{\rm brem}} +\frac{4}{3}\,\,\frac{1-y_{_{\rm brem}} }{y_{_{\rm brem}} }\right)\,dy_{_{\rm brem}} \,dt \,\,,
\end{equation}
whereas the corresponding probability density for a photon producing 
a pair, with electron energy in the 
range $(E,E+dE)$, is
\begin{equation} \label{p_pp}
\frac{d\sigma_{\gamma \rightarrow e^+e^-}}{dE}\,\frac{X_{_{\rm EM}} N_A}{A_{\rm eff}}\,dE\,dt
\approx \left(1-\frac{4}{3}y_{_{\rm pair}}+\frac{4}{3}y_{_{\rm pair}}^2\right)\,dy_{_{\rm pair}}\,dt \,\,.
\end{equation}
The total probability for pair production per unit of $X_{_{\rm EM}}$ follows from 
integration of (\ref{p_pp}), 
\begin{equation}
\int \frac{d\sigma_{\gamma \rightarrow e^+e^-}}{dE}\,\,\,
\frac{X_{_{\rm EM}} N_A}{A_{\rm eff}}\, dE\, \approx \int_0^1 
\left(1-\frac{4}{3}y_{_{\rm pair}} + \frac{4}{3}y_{_{\rm pair}}^2\right)
\,d y_{_{\rm pair}} = \frac{7}{9}\,.
\end{equation}

As can be seen from (\ref{p_b}), the total probability for 
bremsstrahlung radiation is logarithmically divergent.  However, this 
infrared divergence is eliminated by the interference of 
bremsstrahlung amplitudes from multiple scattering centers.  
This collective effect of the electric potential of several
atoms is known as the LPM
effect~\cite{Landau:um,Migdal:1956tc}. Of course,
the LPM suppression of the cross section results in an 
effective increase of the mean free path of electrons and photons. This 
effectively retards the development of the electromagnetic component of the shower. 
It is natural to introduce 
an energy scale, $E_{\rm LPM}$, at which the inelasticity is low enough that the LPM effect becomes 
significant~\cite{Stanev:au}.
Below $E_{\rm LPM}$, the energy loss rate due to bremsstrahlung is roughly
\begin{equation}
\frac{dE}{dX} \approx -\frac{1}{X_{_{\rm EM}}} \, \int_0^1 y_{_{\rm brem}} \ E\,  
\left(y_{_{\rm brem}} +\frac{4}{3}\,\,\frac{1-y_{_{\rm brem}} }{y_{_{\rm brem}} }\right) \,dy_{_{\rm brem}}  \, = -\frac{E}{X_{_{\rm EM}}} \,.
\end{equation}
With this in mind, we now identify the constant $X_{_{\rm EM}} \approx 36.7$~g cm$^{-2}$ 
with the radiation length in air  defined as the mean distance over which a high-energy 
electron loses $1/e$ of its energy, 
or equivalently $7/9$ of the mean free path for pair production by a high-energy 
photon~\cite{Patrignani:2016xqp}.

The most evident signatures of the LPM effect on shower development
are a shift in the position of the shower maximum $X_{\rm max}$ and
larger fluctuations in the shower development. When considering the
LPM effect in the development of air showers produced by UHECRs, one
has to keep in mind that the suppression in the cross sections is
strongly dependent on the atmospheric depth.\footnote{The same occurs
  for dielectric suppression, although the influence is not as
  important as for the LPM effect~\cite{Cillis:1998hf}.} Since the
upper atmosphere is very thin, the LPM effect becomes noticeable only
for photons and electrons with energies $E_{\rm LPM} \gtrsim
10^{10}$~GeV.  For baryonic primaries, the LPM effect does not become
important until the primary energy exceeds $10^{12}$GeV.  This is
because the electromagnetic shower does not commence until after a
significant fraction of the primary energy has been dissipated through
hadronic interactions. At energies at which the LPM effect is important (viz. $E > E_{\rm LPM}$), 
$\gamma$-ray showers will have already commenced in the geomagnetic field at almost 
all latitudes: primary photons with $E > 10^{10}$~GeV convert into $e^{+}e^{-}$ 
pairs, which in turn emit synchrotron photons. This reduces 
the energies of the primaries that reach the atmosphere, and thereby  
compensates for  the tendency of the LPM effect to retard the shower development~\cite{Vankov:2002cb}.

The muonic component of an EAS differs from the electromagnetic component for two main reasons~\cite{Cillis:2000xc}.
First, muons are generated through the decay of cooled 
($E_{\pi^\pm} \lesssim 1$~TeV) charged pions, and thus the muon content is sensitive to the initial
baryonic content of the primary particle.  Furthermore, since there is no ``muonic cascade'', the number
of muons reaching the ground is much smaller than the number of electrons. Specifically, there are
about $5\times 10^{8}$ muons above 10~MeV at ground level for a vertical $10^{11}$~GeV proton 
induced shower.
Second,  the muon has a much smaller cross section for radiation and pair production than the electron, 
and so the muonic component of an EAS develops differently than does the electromagnetic component.
The smaller multiple scattering suffered by muons leads to 
earlier arrival times at the ground for muons than for the electromagnetic component.  

The ratio of electrons to muons depends strongly on the distance from
the core; for example, the $e^+ e^-$ to $\mu^+ \mu^-$ ratio for a
$10^{11}$~GeV vertical proton shower varies from 17 to 1 at 200~m from
the core to 1 to 1 at 2000~m.  The ratio between the electromagnetic
and muonic shower components behaves somewhat differently in the case
of inclined showers.  For zenith angles greater than $60^{\circ}$, the
$e^+ e^-$/$\mu^+ \mu^-$ ratio remains roughly constant at a given
distance from the core.  As the zenith angle grows beyond $60^\circ$,
this ratio decreases, until at $\theta = 75^{\circ}$, it is 400 times
smaller than for a vertical shower.  Another difference between
inclined and vertical showers is that the average muon energy at
ground level changes dramatically.  For horizontal showers, the lower
energy muons are filtered out by a combination of energy loss
mechanisms and the finite muon lifetime: for vertical showers, the
average muon energy is 1~GeV, while for horizontal showers it is about
2 orders of magnitude greater.  The muon densities obtained in shower
simulations using {\sc sibyll}-like models fall more rapidly with lateral distance
to the shower core than those obtained using {\sc qgsjet}-like models.  This can
be understood as a manifestation of the enhanced leading particle
effect in {\sc sibyll}, which can be traced to the relative hardness
of the electromagnetic form factor profile function.  The curvature of
the distribution $(d^2\rho_{\mu}/dr^2)$ is measurably different in the
two cases, and, with sufficient statistics, could possibly serve as a
discriminator between hadronic interaction models, provided the
primary species can be determined from some independent
observable(s)~\cite{Anchordoqui:2003gm}.

\subsection{Paper-and-pencil air shower modeling}
\label{ppm}

Most of the general features of an electromagnetic cascade can be
understood in terms of the toy model due to Heitler~\cite{Heitler}. In
this model, the shower is imagined to develop exclusively via
bremsstrahlung and pair production, each of which results in the
conversion of one particle into two.  As was previously discussed,
these physical processes are characterized by an interaction length
$X_{_{\rm EM}} \approx 37.6~{\rm g/cm}^2$. One can thus imagine the
shower as a particle tree with branches that bifurcate every $X_{_{\rm
    EM}}$, until they fall below a critical energy, $\epsilon_0
\approx 86~{\rm MeV}$, at which point energy loss processes dominate.
Up to $\epsilon_0$, the number of particles grows geometrically, so
that after $n = X/X_{_{\rm EM}}$ branchings, the total number of
particles in the shower is $N \approx 2^n$.  At the depth of shower
maximum $X_{\rm max}$, all particles are at the critical energy,
$\epsilon_0$, and the energy of the primary particle, $E_0$, is split
among all the $N_{\rm max} = E_0 / \epsilon_0$ particles.  Putting
this together, we get:
\begin{equation} \label{heitler}
X_{\rm max} \approx X_{_{\rm EM}} \,\, \frac{\ln(E_0/\epsilon_0)}{\ln 2} \,\,.
\end{equation}
Changes in the mean mass composition of the CR flux as a function of energy will 
manifest as changes in the mean values of $X_{\rm max}$. This change of $X_{\rm max}$ 
with energy is commonly known as the elongation rate~\cite{Linsley:P3}:
\begin{equation} 
D_e = \frac{\delta X_{\rm max}}{\delta \ln E} \,\,.
\end{equation}
For purely electromagnetic showers, $X_{\rm max}(E) \approx X_{_{\rm
    EM}}\, \ln(E/\epsilon_0)$, and hence $D_e \approx X_{\rm EM}$.
For convenience, the elongation rate is often written in terms of
energy decades, $D_{10} = \partial \langle X_{\rm max}
\rangle/\partial \log E$, where \mbox{$D_{10} = 2.3 D_e.$} The elongation
rate obtain from the Heitler model, $D_{10}^\gamma \approx 84~{\rm
  g/cm}^2$, is in very good agreement with the results from Monte
Carlo simulations. However, the prediction for the particle number at
maximum is overestimated by a factor of about 2 to 3. Moreover,
Heitler's model predicts a ratio of electron to photons of 2, whereas
simulations and direct cascade measurements in the air show a ratio of
the order of 1/6. This difference is due to the fact that multiple
photons are emitted during bremsstrahlung and that electrons lose
energy much faster than photons do.

As we have seen, baryon-induced showers are also dominated by
electromagnetic processes, thus Heitler's toy model is still
enlightening for such cases. For proton primaries, the multiplicity
rises with energy, and the resulting elongation rate becomes smaller.
This can be understood by noting that, on average, the first
interaction is determined by the proton mean free path in the
atmosphere, $\lambda_{p{\rm -air}} = X_0$.  In this first interaction
the incoming proton splits into $\langle n(E) \rangle$ secondary
particles, each carrying an average energy $E/\langle n(E) \rangle$.
Assuming that $X_{\rm max}(E)$ depends dominantly on the first
generation of $\gamma$ subshowers, the depth of maximum is obtained as
in (\ref{heitler}),
\begin{equation}
X_{\rm max}(E) \approx X_0 + X_{_{\rm EM}}\, \ln[E/\langle n(E) \rangle]\,\, .
\label{Xmax-Matthews1}
\end{equation}
For a proper evaluation of $X_{\rm max}$, it would be necessary to sum
each generation of subshowers carefully from their respective points
of origin, accounting for their attenuation near and after the
maxima. If we now further assume a multiplicity dependence $\langle
n(E) \rangle \approx n_0 E^{\Delta}$, then the elongation rate
becomes,
\begin{equation} 
\frac{\delta X_{\rm max}}{\delta \ln E}= X_{_{\rm EM}}\,\left[1-\frac{\delta \ln \langle n(E) 
\rangle}{\delta \ln E} \right] + \frac{\delta X_0}{\delta \ln E}
\end{equation}
which corresponds to the form given by Linsley and Watson ~\cite{Linsley:1981gh},
\begin{equation} 
D_e = X_{_{\rm EM}} \,\left[ 1-\frac{\delta \ln \langle n(E) \rangle}{\delta \ln E} + 
\frac{X_0}{X_{_{\rm EM}}} \frac{\delta \ln(X_0)}{\delta \ln E} \right] =  X_{_{\rm EM}}\,(1-B) \,\,,
\label{ERLW}
\end{equation}
where
\begin{equation}
B \equiv \Delta - \frac{X_0}{X_{_{\rm EM}}} \,\,\frac{\delta \ln X_0}{\delta \ln E} \, .
\label{theB}
\end{equation} 

A precise calculation of a proton shower evolution has been carried
out by Matthews~\cite{Matthews:2005sd}, using the simplifying
assumption that hadronic interactions produce exclusively pions,
$2N_\pi$ charged and $N_\pi$ neutral. $\pi^0$'s decay immediately and
feed the electromagnetic component of the shower, whereas $\pi^\pm$'s
soldier on. The hadronic shower continues to grow, feeding the
electromagnetic component at each interaction, until charged pions
reach a characteristic energy at which decay is more likely than a new
interaction.  The interaction length and the pion multiplicity
($3N_\pi$) are energy independent in the Heitler-Matthews model. The
energy is equally shared by the secondary pions. For pion energy
between 1~GeV and 10~TeV, a charged multiplicity of 10 ($N_\pi$ = 5)
is an appropriate number.

 The first interaction diverts 1/3 of the available energy ($E_0/3$)
 into the EM component while the remaining 2/3 continue as
 hadrons. The number of hadrons increases through subsequent
 generation of particles and in each generation about 30\% of the
 energy is transferred to the EM cascade. Therefore the longer it
 takes for pions to reach the characteristic energy $\xi^{\pi^\pm}_{\rm c}
 \sim 20~{\rm GeV}$ (below which they will decay into muons), the
 larger will be the EM component. Consequently, in long developing
 showers the energy of the muons from decaying pions will be smaller.
 In addition, because of the density profile of the atmosphere,
 $\xi^{\pi^\pm}_{\rm c}$ is larger high above ground than at sea level and
 deep showers will produce fewer muons.

 This positive correlation  introduces a link between the primary
 cosmic ray interaction cross section on air and the muon
 content at ground level. According to those principles,
 primaries with higher cross
 sections will have a larger muon to electron ratio at ground level.  

 To obtain the number of muons in the shower, one  simply assumes
 that all charged pions decay into muons when they reach the critical energy:
 $N_\mu = (2N_\pi)^{n_c}$, where $n_c = \ln(E_0/\xi_{\rm c}^{\pi^\pm})/\ln(3N_\pi)$
 is the number of steps needed for the pions to reach
 $\xi^\pi_{\rm c}$. Introducing $\beta=\ln(2N_\pi)/\ln(3N_\pi)$
 we have
 \begin{equation}\label{hs} 
N_\mu = (E_0/\xi^{\pi^\pm}_{\rm c})^\beta \, . 
\end{equation} 
 For $N_\pi=5$, $\beta = 0.85$. Unlike the electron number, the muon multiplicity does not grow
 linearly with the primary energy, but at a slower rate.
 The precise value of $\beta$ depends on the average pion
 multiplicity used.  It also depends on the inelasticity of the hadronic interactions.
 Assuming that only half of the available energy goes into the pions
 at each step (rather than all of it, as done above) would lead
 to $\beta=0.93$. Detailed simulations give values of $\beta$ in
 the range 0.9 to 0.95~\cite{Alvarez-Muniz:2002ne}.

 The first interaction yields $N_\gamma = 2 N_{\pi^0} =
 N_{\pi^\pm}$. Each photon initiates an EM shower of energy $E_0/(3
 N_{\pi^\pm}) = E_0/(6 N_\pi).$ Using $pp$ data~\cite{Patrignani:2016xqp}, we parametrized the charged particle
 production in the first interaction as $N_{\pi^{\pm}} =
 41.2(E_0/1~{\rm PeV})^{1/5}$.  Now, from the approximation in
 (\ref{Xmax-Matthews1}), based on the sole evolution of the EM cascade
 initiated by the first interaction, we obtain
\begin{equation}
X_{\rm max}^p   =  X_0 + X_{_{\rm EM}}  \, \ln[E_0/( 6 N_\pi \epsilon_0)]  
                           =  (470 + 58 \, {\rm log_{10}} [E_0/1~{\rm PeV}])~{\rm g/cm}^2 \, .
\label{Xmax-Matthews2}
\end{equation}
This falls short of the full simulation value by about 100~${\rm g/cm}^2$~\cite{Matthews:2005sd}.

A good approximation of the elongation rate can be obtained
 when introducing the cross section and multiplicity energy dependence.
 Using a $p$-air cross section of 550~mb at 10$^{9}$~GeV and a rate
 of change of about 50~mb per decade of energy leads to~\cite{Ulrich:2009zq}
\begin{equation}\label{lambda-i}
 X_0 \simeq 90 - 9 \log{(E_0/{\rm EeV})}~{\rm g/cm}^2 \, .
\end{equation}
 Now, assuming (as in~\cite{Matthews:2005sd}) that the first interaction
 initiates $2N_\pi$ EM cascades of energy $E_0/6N_\pi$, with
 $N_\pi\propto (E_0/{\rm PeV})^{1/5}$ for the evolution of the first interaction
 multiplicity with energy, we can calculate the elongation rate 
\begin{equation}
 D_{10}^ p=\frac{dX_{\rm max}}{d\log{E_0}}= \frac{d(X_0 \, \ln{2}+X_{_{\rm EM}} \, \ln[E_0/(6N_\pi \epsilon_0)]}{d\log{E_0}} 
=\frac{4}{5}D_{10}^\gamma - 9\ln{2} \simeq 62~{\rm g/cm}^2 \, \ .
\end{equation} 
 This result is quite robust as it only depends on the cross section
 and multiplicity evolution with energy. It is in good agreement with
 Monte Carlo simulation~\cite{Alvarez-Muniz:2002ne}.

To extend this discussion to heavy nuclei, we can apply the superposition principle as 
a reasonable first approximation. In this approximation, we pretend that the nucleus
comprises unbound nucleons, such that the point of first interaction of one nucleon is
independent of all the others.  Specifically, a shower produced by a nucleus with energy 
$E_0$ and baryon number $A$ is modeled by a collection of $A$ proton showers, each with $A^{-1}$ of the
nucleus energy. Modifying Eq.~(\ref{heitler}) accordingly one easily obtains
$X_{\rm max} \propto \ln (E_0/A)$. Assuming that $B$ is not changing with energy, one obtains for mixed primary 
composition~\cite{Linsley:1981gh}
\begin{equation}  
D_e =\, X_0\,(1-B)\,
\left[1 - \frac{\partial \langle \ln A \rangle }{\partial \ln E} \right]\, .
\label{er}
\end{equation}
Thus, the elongation rate provides a measurement of the change of the 
mean logarithmic mass with energy.  One caveat of the procedure discussed above is 
that (\ref{ERLW}) accounts for the energy dependence of the cross section and violation of Feynman 
scaling only for the first interaction. Note that subsequent interactions are assumed to be 
characterized by Feynman scaling and constant interaction cross sections; see Eq.~(\ref{theB}).  
Above $10^{7}$~GeV, these secondary interactions play a more important role, and thus the predictions of 
Eq.~(\ref{er}), depending on the hadronic interaction model assumed, may
vary by up to 20\%~\cite{Alvarez-Muniz:2002ne}.

The muon content of an EAS at ground level $N_\mu,$ as well as the ratio $N_{\mu}/N_{e}$, 
are sensitive to primary composition (here, $N_e$ is the electron content at ground level).
To estimate the ratio of the  muon content of nucleus-induced to proton-induced showers, we can resort again to the principle of superposition. 
Using $\beta = 0.93$ we find that the total 
number of muons produced by the superposition of $A$ individual proton showers 
is, $N_{\mu}^A \propto A (E_{_A}/A)^{0.93}$. Consequently, in a vertical shower, 
one expects a cosmic ray nucleus to produce about $A^{0.07}$  more 
muons than a proton.  This implies that a shower initiated by 
an iron nucleus produces about 30\% more muons than a proton shower.  

Over the past few decades, it has been suspected that the number of
registered muons at the surface of the Earth is by some tens of
percentage points higher than expected with extrapolations of existing
hadronic interaction models~\cite{AbuZayyad:1999xa,Aab:2014pza}.  The
latest study from the Auger Collaboration has strengthened this
suspicion, using a novel technique to mitigate some of the measurement
uncertainties of earlier methods~\cite{Aab:2016hkv}. The new analysis
of Auger data suggests that the hadronic component of showers (with
primary energy $10^{9.8} < E/{\rm GeV} < 10^{10.2}$) contains about
$30\%$ to $60\%$ more muons than expected. The significance of the
discrepancy between data and model prediction is somewhat above $2.1
\sigma$. The TA Collaboration also reported a muon signal which is
larger in the data than in the air shower Monte Carlo prediction~\cite{Abbasi:2018fkz}. Many models
have been proposed to explain this
anomaly~\cite{Farrar:2013sfa,Anchordoqui:2016oxy,Tomar:2017mgc,Soriano:2017bvs}.

While the toys models discussed above are very useful for imparting a
first intuition regarding global shower properties, the details of
shower evolution are far too complex to be fully described by a simple
analytical model. Full Monte Carlo simulation of interaction and
transport of each individual particle is required for precise modeling
of the shower development. For details on the various Monte Carlo
packages and their predictions, see
e.g.,~\cite{Knapp:2002vs,Ulrich:2010rg}.

\section{UHECR as probes of particle physics beyond the electroweak scale}
\label{sec:5}

\subsection{Testing models of the early universe via 
  top-down production of cosmic rays and neutrinos}

In an epic paper, well ahead of its time, Lema\^{\i}tre -- a
forerunner of the Big Bang hypothesis -- introduced the idea that the
entire material filling the universe, as well as the universe's
expansion, originated in the super-radioactive disintegration of a
``Primeval Atom'', which progressively decayed into atoms of smaller
and smaller atomic weight~\cite{Lemaitre:1931zzb}. The CRs were introduced in
this picture as the energetic particles emitted in intermediate stages
of the decay-chain. Echoing Lema\^{\i}tre, in the so-called ``top-down
models'', extreme energy ($\gg 10^{11}~{\rm GeV}$) cosmic rays and
neutrinos arise in the decay of topological
defects~\cite{Hill:1982iq,Chudnovsky:1986hc,Hill:1986mn,Bhattacharjee:1989vu,Bhattacharjee:1990js,Bhattacharjee:1991zm,Bhattacharjee:1994pk,Berezinsky:1997kd,Berezinsky:1997td,Bhattacharjee:1998qc} and
super-heavy elementary
$X$-particles~\cite{Kuzmin:1997cm,Kuzmin:1998uv,Dubovsky:1998pu,Berezinsky:1998ed,Birkel:1998nx,Sarkar:2001se,Kuzmin:1999zk}.

To maintain an appreciable decay rate today, it is necessary to tune
the $X$ lifetime to be longer (but not too much longer) than the age
of the universe, or else ``store'' short-lived $X$ particles in
topological vestiges of early universe phase transitions (such as
magnetic monopoles, cosmic strings, cosmic necklaces, etc.). Discrete
gauged
symmetries~\cite{Hamaguchi:1998wm,Hamaguchi:1998nj,Hamaguchi:1999cv}
or hidden sectors~\cite{Ellis:1990iu,Benakli:1998ut} are generally
introduced to stabilize the $X$ particles. Higher dimensional
operators, wormholes, and instantons are then invoked to break the new
symmetry super-softly to maintain the long
lifetime~\cite{Berezinsky:1997hy,Kuzmin:1997cm} (collisional
annihilation has been considered too~\cite{Blasi:2001hr}).  Quanta
associated with these fields are typically of the order of the
symmetry-breaking scale, which in Grand Unified Theories (GUTs) can be
$\sim 10^{16} - 10^{19}$~GeV.  Arguably, these metastable super-heavy
relics may constitute (a fraction of) the dark matter in galactic
haloes.

The cascade decay to cosmic ray particles is driven by the ratio of
the number density of the $X$-particle $n_X = \rho_c \Omega_X /M_X$
to its decay time $\tau_X$, where $M_X$ is the mass of the
$X$-particle and $\rho_c \simeq 1.054 \times 10^{-4} h^2~{\rm GeV} \,
{\rm cm^{-3}}$ is the critical density in terms of the present Hubble
parameter $h \equiv H_0/100~{\rm km} \, {\rm s^{-1}} \, {\rm
  Mpc^{-1}}$. This cascade is very model dependent, as neither the cosmic
average mass density contributed by the relics $\Omega_X$, nor
$\tau_X$ is known with any degree of confidence. In addition, the
precise decay modes of the $X$'s and the detailed dynamics of the
first generation of secondaries depend on the exact nature of the $X$
particles under consideration. However, one expects the bulk flow
of outgoing particles to be almost independent of such details,
enabling one to infer from the ``known'' evolution of quarks and
leptons the gross features of the $X$ particle decay: the strongly
interacting quarks would fragment into jets of hadrons containing
mainly pions together with a 3\% admixture of
nucleons~\cite{Coriano:2001rt,Barbot:2002gt,Barbot:2003cj}. This
implies that the injection spectrum is a rather hard
fragmentation-type shape (with an upper limit usually fixed by the GUT
scale) and dominated by $\gamma$-rays and neutrinos produced via pion
decay.  Therefore, the $\nu/p$ and $\gamma/p$ ratios can be used as a
diagnostic tool in determining the possible contribution of $X$
particle decay to the UHECR spectrum without violating any
observational flux measurements or limits at higher or lower
energies~\cite{Aharonian:1992qf, Sigl:1998vz}. In particular, neutrino
and $\gamma$-ray fluxes depend on the energy released integrated over
redshift, and thus on the specific top-down model. Recall that the
electromagnetic energy injected into the Universe above the pair
production threshold on the CMB is recycled into a generic cascade
spectrum below this threshold on a short time scale compared with the
Hubble time. Therefore, it can have several potential observable
effects, such as modified light element abundances due to $^4$He
photodisintegration, or induce spectral distortions of universal
$\gamma$-ray and neutrino
backgrounds~\cite{Sigl:1995kk,Sigl:1996gm}. Additionally, measurements
of the diffuse GeV $\gamma$-ray flux, to which the generic cascade
spectrum would contribute directly, limit significantly the parameter
space in which $X$'s can contribute to the UHECR intensity, especially
if there is already a significant contribution to this background from
conventional sources such as unresolved $\gamma$-ray
blazars~\cite{Protheroe:1996pd,Protheroe:1996zg}. EGRET data in the
energy interval $10~{\rm MeV} < E_\gamma < 100~{\rm GeV}$, with a
spectrum $\propto E_\gamma^{-2.10 \pm 0.03}$~\cite{Sreekumar:1997un},
constrain the energy density of the cascade photons $\omega_{\rm cas}
< 5 \times 10^{-6}~{\rm eV/cm^3}$~\cite{Berezinsky:1998ft}.  The first
10 months of {\it Fermi}-LAT observations have provided an stronger
limit on the isotropic diffuse gamma-ray
background~\cite{Abdo:2010nz}. For the $200~{\rm MeV} < E_\gamma <
120~{\rm GeV}$, the more steep power-law spectrum $\propto
E_\gamma^{-(2.41 \pm 0.05)}$ leads to a lower cascade energy density
$\omega_{\rm cas} = 5.8 \times 10^{-7}~{\rm
  eV/cm^3}$~\cite{Berezinsky:2010xa}. A more stringent limit can be
extracted from the 50 months observation of {\it
  Fermi}-LAT~\cite{Ackermann:2014usa}. The limit becomes stronger due
to the highest energy bin $580 < E_\gamma/{\rm GeV} < 820$, where the {\it
  Fermi}-LAT intensity is particularly low. The current upper bound on the
energy density of cascade radiation, $\omega_{\rm cas} < 8.3 \times
10^{-8}~{\rm eV/cm^3}$~\cite{Berezinsky:2016feh}, limits significantly
the parameter space in which cosmologically distant $X$'s can generate
on decay UHECRs and UHEC$\nu$s~\cite{Berezinsky:2011cp}.\footnote{The cascade
  limit on $\Phi_\gamma$ also constrains the photon flux produced by
  extreme energy neutrinos interacting in the local universe via the
  $Z$-burst
  mechanism~\cite{Weiler:1982qy,Weiler:1997sh,Fargion:1997ft}.}

Super-heavy $X$-particles could behave as cold dark matter and cluster
efficiently in all gravitational potential wells. If this were the
case, their abundance in our galactic halo would be enhanced above
their cosmological abundance by a factor $f_{\rm cos} \equiv n_X^{\rm
  halo}/n_X^{\rm cos}$. If for simplicity we assume an spherical halo
of radius $R_{\rm halo} \sim 100~{\rm kpc}$ and density $\rho_{\rm
  halo} \sim 0.3~{\rm GeV} \, {\rm cm}^{-3}$, then $f_{\rm cos} \sim 3
\times 10^4 h^{-2}$~\cite{Birkel:1998nx}. The actual density of dark
matter in the halo must of course fall off as $r^{-2}$ to account for
the flat rotation curve of the disk but we do not consider it
necessary herein to investigate realistic mass models. All in all, the
universal density of $X$-particles is smaller than the halo density by
about the same numerical factor by which the distance to the horizon
($\sim 3000\, h^{-1}~{\rm Mpc}$) exceeds the halo radius. Therefore,
in this scenario ultra-high-energy photons and nucleons from the halo
of our Galaxy would provide the dominant contribution to the intensity
observed on Earth. This is because photons from the decay of
$X$-particles which are clustered on the Galactic halo are not
degraded in energy. The extragalactic component of UHECRs is
suppressed by the smaller extragalactic density of $X$-particles and
so the cascade photon limit is
relaxed~\cite{Berezinsky:1997hy}. Currently, the most restrictive
constraint on the lifetime $\tau_X$ of super-heavy relics clustered on
the Galactic halo comes from Auger upper limits on the intensity of
ultra-high-energy photons; see Table~\ref{tabla1}. Auger data place a
lower bound $\tau_X \gtrsim 10^{22}~{\rm
  yr}$~\cite{Aloisio:2015lva}. Next generation of UHECR observatories
will be able to effectively study supermassive $X$'s, with possible
lifetime detections or constraints reaching values as high as $\tau_X
\sim 10^{24}~{\rm yr}$~\cite{Aloisio:2015lva}.

Extreme-energy CR and $\nu$ physics provides a framework to search for
cosmic strings complementary to those based on the gravitational
effects of strings, including structure formation, CMB data,
gravitational radiation, and gravitational lensing. For strings with a
symmetry breaking energy scale $\Lambda$, the strongest bound due
to lensing effects is $G\Lambda^2 \lesssim
10^{-7}$~\cite{Christiansen:2010zi}, and the bound from millisecond
pulsar observations is $G\Lambda^2 \lesssim 4 \times 10^{-9}$, where $G$ is
Newton's constant. Next generation gravitational wave detectors are
expected to probe $G \Lambda^2 \sim
10^{-12}$~\cite{Damour:2004kw,Olmez:2010bi}.  Remarkably, next
generation UHECR detectors will be able to detect extreme energy
neutrinos from strings with $G \Lambda^2$ values as small as $\sim
10^{-20}$~\cite{Berezinsky:2011cp}.

In summary,  at the present level of knowledge we can argue that astrophysical sources cannot
accelerate CRs to energies $\gtrsim 10^{12}~{\rm GeV}$, with the
maximum neutrino energy an order of magnitude lower. Therefore, detection
of neutrinos with $E_\nu > 10^{11}~{\rm GeV}$ would be momentous
discovery, and a clean signature that top-down models are at play.

Super-heavy right-handed neutrinos ($\nu_R$'s) are also interesting
dark matter candidates, particularly in minimal extensions of the SM,
i.e. constructs with the usual gauge group $SU(3)_C\otimes SU(2)_L
\otimes U(1)_Y$ and the usual matter fields but including a
right-handed neutrino in each
generation~\cite{Boyle:2018tzc,Boyle:2018rgh}.  If one of the
$\nu_R$'s contributes to the dark matter sector, then its
non-gravitational couplings do not necessarily have to vanish, but
have to be small enough so that the $\nu_{R}$ has a lifetime
$\tau_{\nu_{R}} \gg H_0^{-1} = 9.778 \, h^{-1}~{\rm Gyr}$.\footnote{IceCube data set a lower limit on the lifetime of
  the right-handed neutrino of ${\cal O} (10^{29}~{\rm
    s})$~\cite{Aartsen:2018mxl}.} This opens up the possibility to
indirectly observe $\nu_{R}$ through its decay products. For two-body
decays, conservation of angular momentum forces the $\nu_{R}$ to decay
into a Higgs boson and a light Majorana neutrino, i.e. $\nu_R \to H
\nu_L$.  This decay mode is particularly interesting for recent
observations of the ANITA experiment.\footnote{The Antarctic Impulsive
  Transient Antenna (ANITA) is an experiment which has completed four
  long-duration balloon flights above Antarctica. ANITA searches for
  impulsive radio-Cherenkov emission arising from the Askaryan charge
  excess which develops in ultra-high energy neutrino-induced particle
  cascades in the Antarctic ice.  The large radio transparency of ice
  allows for the radio-Cherenkov pulse from these cascades to be
  recorded by a cluster of balloon-borne antennas, flying at an
  altitude of 35 to 40~km. The details of the ANITA instrument are
  given in e.g.~\cite{Gorham:2008dv}. ANITA Collaboration reported the
  most stringent upper limits to date on the intensity of extreme
  energy neutrinos, e.g., for $E_\nu \sim 10^{12}~{\rm GeV}$, the
  energy-squared-weighted intensity is at the level of 20
  WB~\cite{Allison:2018cxu}.}

The three balloon flights of the ANITA experiment have resulted in the
observation of two unusual upgoing showers with energies of ${\cal O}
(100)~{\rm PeV}$~\cite{Gorham:2016zah,Gorham:2018ydl}.  In principle, these events could originate in
the atmospheric decay of an upgoing $\tau$-lepton produced through a
charged current interaction of $\nu_\tau$ inside the Earth. However,
the relatively steep arrival angles of these events (about $30^\circ$
above the horizon) create a tension with the SM
neutrino-nucleon interaction cross section.   It is compelling that the two ANITA
events are similar in energy and were observed at roughly the same
angle above the horizon. This fueled speculation that the two events have
similar energies because they result from the two-body decay of a new
quasi-stable relic, itself gravitationally trapped inside the Earth~\cite{Anchordoqui:2018ucj}.

A dense population of $\nu_{R}$ is expected at the center of the
Earth because as the Earth moves through the halo, the $\nu_{R}$
scatter with Earth matter, lose energy and become gravitationally
trapped. An accumulated $\nu_{R}$ then decays into a Higgs and an
active neutrino that propagates through the Earth and produces a
$\tau$ lepton near the Earth's surface.  The particular angle of the
ANITA events is a combination of the dark matter distribution in the
Earth, the neutrino interaction cross section, and the $\tau$ survival
probability.  The non-gravitational couplings have to be chosen to
produce a long lifetime and the needed abundance of right-handed
neutrinos in the Earth to yield the two ANITA events. To achieve a
sizable dark matter density in the Earth self-interactions may be
invoked.\footnote{The number of right-handed neutrinos intercepted by the Earth during
its lifetime is $ \sim t_\oplus \rho_{\rm DM} v_\oplus \pi R_\oplus^2/
  M_{\nu_R} \sim 10^{33.6}$, where $t_\oplus$ = 4.55~Gyr  is the age of the
Earth~\cite{Patterson}, $\rho_{\rm DM} \simeq 0.5~({\rm GeV}/c^2)~{\rm
  cm}^{-3}$ is the dark matter mass density in the Galactic
plane~\cite{Bienayme:2014kva,Piffl:2014mfa,McKee:2015hwa,Sivertsson:2017rkp}, $v_\oplus \simeq 200~{\rm km/s}$ is the average velocity of the Earth relative to the
dark matter particles, and $M_{\nu_R} \sim 5~{\rm PeV}$~\cite{Boyle:2018tzc,Boyle:2018rgh}. The
average number density within the Earth is $\sim 3 f_{\rm cap} t_\oplus \rho_{\rm DM} v_\oplus/ (4
  R_\oplus M_{\nu_R}) \sim 10^{21.7} f_{\rm cap}~{\rm km^{-3}}$,
  where $f_{\rm cap}$ is the fraction of $\nu_R$'s captured by the Earth~\cite{Neufeld:2018slx}. Thus, for $\tau_{\nu_R}
  \sim 10^{29}~{\rm s}$~\cite{Aartsen:2018mxl}, a decay rate on the
  order of 1 per
km$^3$ per yr would require $f_{\rm cap}$ to be ${\cal O}(1)$.}

The event rate integrated over the entire Earth at a particular time  is 
\begin{equation}
{\rm Rate} \equiv \frac{dN}{dt} = 4\pi\int_0^{R_\oplus} r^2\,dr\ \frac{n(r,t)}{\tau_{\nu_{R}}}\,,\nonumber
\label{wholeEarth}
\end{equation}
where $n(r,t)$ is the number density of $\nu_{R}$ at time $t$ and
$R_\oplus$ is the Earth's radius~\cite{Anchordoqui:2018ucj}.  The observable rate
today ($t=t_0$), as a function of nadir angle $\theta_n$ 
is given by
\begin{equation}
A_{\rm eff}\frac{d\,{\rm Rate}}{d\, |\cos\theta_n|}   =  2\pi A_0\times 2\pi \int^{R_\oplus}_{ {R_\oplus} {\sin\theta_n} }\,
r^2 dr \, \frac{n(r,t_0)}{\tau_{\nu_{R}}}  \  \left( e^{-(l_+/\lambda)} + e^{-(l_-/\lambda)} \right) \,
{\cal E}(\theta_n)\,,
\label{ratenu}
\end{equation}
where the effective area $A_{\rm eff}=A_0\Eps(\theta_n)$ defines the
experimental efficiency $\Eps$ that includes the target area
dependence on $\theta_n$ but not the $e^{-l/\lambda}$ suppression
factor, which is given explicitly in the integrand, and where
\mbox{$\lambda = 1.7 \times 10^7/(\sigma/{\rm pb}) ~{\rm km \, w.e.}$}
is the mean-free-path, with $\sigma$ the neutrino-nucleon
CC cross section.  Here, $l_\pm$ are the roots of
$R_\oplus^2+l^2-2 R_\oplus l \cos\theta_n = r^2$, i.e., $l_\pm =
R_\oplus [\cos\theta_n \pm \sqrt{(r/R_\oplus )^2 -\sin^2 \theta_n }
]$. The fact that for fixed $r$, we have two special values for $l$,
i.e. $l_\pm$, can be easily seen as follows: Draw a circle at constant
$r$ less than $R_\oplus$ about the earth's center.  Then draw a line
through the circle, and intersecting the ANITA detector.  This line
represents the trajectory of the traveling particle to ANITA, and
necessarily crosses the fixed circle at two special points, at
trajectory distances which here we have named $l_\pm$.  The quadratic
equation for $l$ derives from the cosine theorem. Of course, if $r$
were too small, then the trajectory at fixed $\theta_n$ would not
intersect the circle at fixed radius $r$ at all; this is the origin of
the lower limit in the integration of $dr$.  Note that (\ref{ratenu})
has a factor of $2\pi$ and not the original $4\pi$ of
(\ref{wholeEarth}) because we do not integrate over $d \cos \theta_n$;
this angle is fixed by the experimental observation. The second factor
of $2\pi$ comes from the fact that the decay of the right-handed
neutrino is isotropic.  For ANITA, $\Eps(\theta_n)$ vanishes for
$\theta_n<35^\circ$, peaks at about $75^\circ$, and vanishes above
$85^\circ$.  The two unusual ANITA events occur at similar angles
above the horizon, so we may set the peak of the distribution at $\sim
30^\circ$ above the horizon, corresponding to a nadir angle of
$\theta_n \sim 60^\circ$. So, taking the view that the event
distribution is maximized at $\theta_n=60^\circ$ by a combination of
ANITA's efficiency and the dark matter distribution in the Earth, it
follows that
\begin{equation}
\left. \frac{d^2\,{\rm Rate}}{d\, |\cos\theta_n|^2} \right|_{\cos \theta_n
  = \half}  = 0  \,.
\label{condition}
\end{equation} 
This result becomes a constraint on the model parameters
in~(\ref{ratenu}), suggesting an atypical dark matter density
distribution in the Earth.  Integrating over the duration of the
experiment yields the event number as opposed to the event rate.

Data from the fourth ANITA flight is currently being analyzed and may
lead to further enlightenment.  The second generation of the Extreme
Universe Space Observatory (EUSO) instrument, to be flown aboard a
super-pressure balloon (SPB) in 2022 will monitor the night sky of the
Southern hemisphere for upgoing showers emerging at large angles below
the horizon, providing an important test of the unusual ANITA
events~\cite{Adams:2017fjh}.

\subsection{Search for Lorentz invariant breaking effects }

At present, there is no reason to anticipate the existence of a
universal scale below which our present notion of flat spacetime
geometry is not valid. However, Lorentz invariance should not be
accepted on faith but rather as a plausible hypothesis subject to
experimental test.  It is possible to introduce the notion of Lorentz
invariance violation either with or without accompanying anomalous
kinematics.  If no anomalous kinematics is involved~\cite{Redei:1966,Redei:1967zz,Anchordoqui:1995im}, any search for
Lorentz invariant breaking effects will require testing  length scales
below $10^{-20}$~cm or less. However,
introducing anomalous kinematical constraints allows tiny departures
from Lorentz invariance, which would be undetectable at the
electroweak scale, to be magnified rapidly with rising energy.  For
example, if Lorentz invariance is broken in the form of non-standard
dispersion relations for free particles, 
\begin{equation}
E^2 = \mathbf{p}^2 (1 + 2 \delta) + m^2 
\label{disprel}
\end{equation}
then absorption and energy loss processes for UHECR interactions would
be modified~\cite{Coleman:1998ti}.  Recall that the GZK interactions
(photopion production and nucleus photodisintegration) are
characterized by well defined energy thresholds (near the excitation
of the $\Delta^+ (1232)$ and the GDR, respectively), which can be
predicted on the basis of Lorentz invariance. Therefore, the
experimental confirmation that UHECR processes occur at the expected
energy thresholds can be considered as an indirect piece of evidence
supporting Lorentz symmetry under colossal boost transformations~\cite{Coleman:1998en}.

The canonical formalism to explore observable consequences of Lorentz
invariant breaking effects was developed by Coleman and Glashow (CG),
assuming renormalizable and gauge invariant perturbations to the SM
Lagrangian that are rotationally invariant in a preferred frame, but
not Lorentz invariant~\cite{Coleman:1998ti}. By shifting both the
renormalized mass by the small amount $m \to m/(1 + 2\delta)$ and the
velocity by the amount $c_{\rm mav} = \sqrt{1 + 2 \delta} \simeq 1 +
\delta$ in (\ref{disprel}) one recovers the standard form of the
dispersion relations
\begin{equation}
E^2 = \mathbf{p}^2 c^2_{\rm mav} + m^2 c^4_{\rm mav} \,,
\end{equation}
where $c_{\rm mav}$ is identified as the maximum attainable velocity
of the free particle in the CMB frame. This implies that in the CG
framework different particles may have different maximum attainable
velocities, which in principle can all be different from 1 and also
different from one another. In such a case, the possible departure
from Lorentz invariance can be phrased in terms of the difference
between the particle maximum attainable velocities
\begin{equation}
\delta_{ij} = c_i - c_j \,,
\label{deltalight}
\end{equation}
where $c_i$ denotes the maximum attainable velocity of a particle
species $i$. From  (\ref{disprel}) and (\ref{deltalight}), a dispersion relation can be
constructed for a particle species $i$,
\begin{equation}
E^2 = \mathbf{p}^2 (1 + 2 \delta_i) + m_i^2 \,,
\label{disrel2}
\end{equation}
where $\delta_i$ is the difference between the maximum attainable
velocity for particle species $i$ and the speed of light in the low
momentum limit where $c = 1$.  As long as all limiting velocities are
less than or equal to the limiting velocity of the photon, causality
is preserved: new ``lightcones'' appear inside the lightcone. To be
generic, it is feasible to add small energy-dependent Lorentz-violating terms in the
free particle Lagrangian that are suppressed by powers of some quantum
gravity energy scale ${\cal O}(M_{\rm Pl})$~\cite{Aloisio:2000cm,Jankiewicz:2003sm,Galaverni:2007tq,Galaverni:2008yj,Mattingly:2009jf}.  This leads to dispersion
relations having a series of smaller and smaller terms proportional to
$\mathbf{p}^{n+2}/M_{\rm Pl}^n \simeq E^{n+2}/M_{\rm Pl}^n$,
\begin{equation}
\delta_i = \sum_{n=0}^\infty \eta_i^{(n)}   \left(\frac{E}{M_{\rm Pl}}
  \right)^n \, .
\label{LIV-ext}
\end{equation} 
In our discussion we will adopt the GC formalism that truncates the
series (\ref{LIV-ext}) considering only the first $n=0$ term.  For a
comprehensive review on astrophysical constraints on Lorentz symmetry,
see e.g.,~\cite{Bietenholz:2008ni,Stecker:2017gdy}.

As we have seen in Sec.~\ref{CMBopacity}  if one assumes Lorentz invariant
kinematics, the energy threshold of photopion production via
interactions of UHECR protons (with initial laboratory energy $E$) and
low energy photons of the CMB (with laboratory energy $\varepsilon$)
is determined by the relativistic invariance of the square of the
total four-momentum of the proton-photon system, and is given by
(\ref{gzkEth}). Now, (\ref{gzkEth}) together with (\ref{6pi})
evaluated at threshold, $s = (m_p + m_\pi)^2$, lead to the threshold condition
for head on collisions in the laboratory frame,
\begin{equation}
4 \varepsilon E_\pi = \frac{m_\pi^2 ( 2m_p + m_\pi)}{m_p + m_\pi} \, .
\label{unbk}
\end{equation}
If Lorentz invariance is broken because $c_\pi > c_p$, then the threshold energy for photopion
production is modified. Namely, using (\ref{disprel}), (\ref{deltalight}) and
(\ref{unbk}) it is easily seen that the square of the four-momentum is
shifted from its Lorentz invariant form and the threshold
condition (\ref{unbk}) becomes~\cite{Coleman:1998ti}
\begin{equation}
4 \varepsilon E_\pi = \frac{m_\pi^2 ( 2m_p + m)}{m_p + m_\pi} + 2 \delta_{\pi p} E_\pi^2 \, .
\label{con}
\end{equation}
If Lorentz symmetry is unbroken, $\delta_{\pi p} =0$ and (\ref{con}) leads to the conventional threshold for a head-on collision (\ref{unbk}). Otherwise, (\ref{con}) 
is a quadratic form in $E_\pi$, with real roots if 
\begin{equation}
\delta_{\pi p} \leq  \frac{2 \varepsilon^2 (m_p +
  m_\pi)}{m^2 (2 m_p + m_\pi)} \simeq \frac{\varepsilon^2}{m_\pi^2}
\simeq 3.23 \times 10^{-24} \left(\frac{\varepsilon}{2.35 \times 10^{4}~{\rm
    eV}} \right)^2  \, .
\label{LLI-condition}
\end{equation}
If one hypothesizes a Lorentz invariance violation with $\delta_{\pi
  p} > 0$, pion photoproduction would only proceed through
interactions with CMB photons that can satisfy
(\ref{LLI-condition}). As noted above, the dominance of photopion
production --via excitation of $\Delta^+ (1232)$-- is near the
photopion production threshold. Hence, combining (\ref{con}) and
(\ref{LLI-condition}) it is straightforward to see that for
$\delta_{\pi p} > 0$ photopion interactions leading to the GZK
suppression would occur for ``low energy'' protons interacting with
CMB photons on the Wien tail of the Planck distribution. Note,
however, that for ``high energy'' protons, which would normally
interact with ``low energy'' photons, the photopion production process
will be forbidden. Thus, the observed UHECR spectrum may exhibit the
characteristics of GZK suppression near the normal GZK energy threshold, but
the UHECR spectrum can {\it recover} at higher energies, because
photopion interactions at higher proton energies may be forbidden. The
observed spectrum, which is shown in Fig.~\ref{fig:2}, has no signal
of the GZK recovery. Indeed, the best fit to the data considering
Lorentz invariant violation effects and uniform distribution of proton
sources yields $\delta_{\pi p} = 3.0^{+1.5}_{-3.0} \times 10^{-23}$,
consistent with an upper limit of $4.5 \times
10^{-23}$~\cite{Scully:2008jp,Stecker:2009hj}.

Studies of the CR nuclear composition, which are also shown in
Fig.~\ref{fig:2}, indicate that there is a significant fraction of
nuclei at the high-end of the energy spectrum.  The dispersion
relation for nuclei can be written assuming a superposition model for
nuclei, i.e. considering them as the combination of $A$ nucleons of
energy $E/A$~\cite{Saveliev:2011vw,Anchordoqui:2017pmf}. Actually,
since we expect nuclear physics to have negligible Lorentz effects it
is reasonable to assume $\delta_{A,Z} =\delta/A^2$, where $\delta$ regulates
deviations from Lorentz symmetry in the nucleon.

Duplicating the procedure established to simultaneously fit the
spectrum and nuclear composition~\cite{Aab:2016zth} the Auger Collaboration performed
a search for Lorentz invariant breaking effects~\cite{Aab:2017njo}. In the Auger analysis
the UHECR sources are assumed to be identical and homogeneously
distributed in a co-moving volume, and the nuclear composition at the
sources is assumed to be a mix of $^1$H, $^4$He, $^{14}$N, and $^{28}$Si. The
source emission rate per volume is described by 
\begin{equation}
{\cal Q}_0 (E',A') = f(A') \  {\cal Q}_0 \left(\frac{E'}{{\rm EeV}}\right)^{-\gamma}
f_{\rm cut}(E',Z',E_p^{\rm max}),
\end{equation}
where $f(A')$ is the fraction of isotopes of type $A'$ emitted with 
 $E = 10^{9}~{\rm GeV}$.  The cutoff of the source spectra is
 modulated by 
\begin{equation}
f_{\rm cut} (E',Z',E_p^{\rm max}) = \left\{ \begin{array}{ll}
1 &~~~~~E' < Z' E_p^{\rm max} \\
\exp [ 1 - E'/(Z'E_p^{\rm max})] &~~~~~E' > Z' E_p^{\rm max} 
\end{array} \right. \, ;
\end{equation}
note a minor difference with the cutoff function in (\ref{injectionQ}).
The free parameters of the fit are the spectral index $\gamma$, the
cutoff rigidity $E_p^{\rm max}$, the normalization ${\cal Q}_0$, and
three of the fractions $f(A')$, the fourth being fixed by $\sum_{A'}
f(A') = 1$.  Since the effect of enhancing $\delta$ is to increase the
interaction length of the particles, a way to investigate an extreme
case is to switch off all the interactions with background
photons~\cite{Boncioli:2015cqa}. The maximal Lorentz invariance
violation, $\delta_{\rm max}$, is simulated with a simplified version
of the propagation code, where only the adiabatic energy losses due
to the expansion of the Universe are taken into account. Both the
spectrum and composition are fitted at energies $\log_{10}(E/{\rm
  GeV}) > 9.7$, i.e. above the ankle.

Because the intensity $J$ and the $X_{\rm max}$ distribution
are independent measurements, the  likelihood function can be written as $\mathscr{L} = \mathscr{L}_J \ 
\mathscr{L}_{X_{\rm max}}$. The goodness-of-fit is assessed with a generalized
$\chi^2$ (the deviance, $D$), defined as the negative log-likelihood ratio of
a given model and the saturated model that perfectly describes the
data~\cite{Aab:2016zth}
\begin{equation}
D = D(J) + D(X_{\rm max}) = - 2 \ln (\mathscr{L}/\mathscr{L^{\rm
  sat}}) = - 2 \ln (\mathscr{L}_J/\mathscr{L}_J^{\rm
  sat}) -  2 \ln (\mathscr{L}_{X_{\rm max}}/\mathscr{L}_{X_{\rm max}}^{\rm
  sat}) \, .
\end{equation}
The parameter $\delta$ characterizing the level of Lorentz invariance
violation is taken to be the same for the photopion and
photodisintegration process. The simulations are performed for various
values of $\delta$ and the corresponding best-fit parameters are given
in Table~\ref{tablaLIV}. The best-fit parameters are found to be
almost independent of $\delta$: the spectral index is hard and the
rigidity cutoff is low, so as to reproduce the low level of $A$
mixture at each energy.  For $\delta_{\rm max}$, a visible difference
appears in the proton fraction with respect to other analyzed values
of $\delta$. This is because protons must be present already at the
source so as to compensate for the absence of interactions.  By
comparing the values of the deviance at the minimum, $\delta_{\rm
  max}$ is disfavored at more than $3\sigma$ over $\delta
=0$~\cite{Aab:2017njo}.

\begin{table}
\caption{Best fit parameters. Fractions are defined at fixed energy $=
  10^9~{\rm GeV}$~\cite{Aab:2017njo}. \label{tablaLIV}}
\begin{center}
\begin{tabular}{cccccccccc}
\hline \hline
$\delta$  & $\gamma$ & $\log_{10}(E_p^{\rm max}/{\rm GeV})$   & H (\%) & He(\%) & N(\%)
 & Si(\%) & $D(J)$ & $D(X_{\rm max})$ & $D$ \\
\hline
0 & 0.96 & 9.68 & 0. & 67.3 & 28.1 & 4.6 & 13.3 & 161.1 & 174.4 \\
$5 \times 10^{-24} $& 0.91 & 9.65 & 0. & 71.8 & 23.9 & 4.3 & 15.1 &
163.5 & 178.5 \\
$1 \times 10^{-23}$ & 0.91 & 9.65 & 0. & 71.4 & 24.3 & 4.3 & 14.9 &
163.6 & 178.5 \\
$1 \times 10^{-22}$ & 0.94 & 9.65 & 0. & 72.8 & 22.7 & 4.6 & 18.2 &
163.6 & 181.8 \\
$\delta_{\rm max}$ & 0.95 & 9.40 & 62.3 & 32.2 & 5.4 & 0.08& 27.3 &
162.0 & 189.3 \\
\hline \hline
\end{tabular}
\end{center}
\end{table}

\subsection{Delve into the electroweak sector in search for new
  physics at subfermi distances}
\label{beyondSM}

If new physics interactions occur at LHC energies, then CR collisions
with c.m. energies ranging up to $250$~TeV would obviously involve new
physics as well. The question is, can new physics be detected by CR
experiments?  At ultra-high energies, the cosmic ray luminosity $ \sim
7 \times 10^{-10}~(E/{\rm PeV})^{-2}$~cm$^{-2}$ s$^{-1}$ (taking a
single nucleon in the atmosphere as a target and integrating over $2
\pi$~sr) is about 50 orders of magnitude smaller than the LHC
luminosity.  This renders the hunt for physics beyond the electroweak
scale futile in hadronic cosmic ray interactions occurring in the
atmosphere.\footnote{See, however, \cite{Pavlidou:2018yux}.} However,
there is still a possibility of uncovering new physics at sub-fermi
distances in cosmic neutrino interactions.

Neutrinos are unique probes of new physics, as their interactions are
uncluttered by the strong and electromagnetic forces and, upon arrival
at the Earth, they may experience collisions with c.m. 
energies up to $\sqrt{s} \lesssim 250$~TeV.  However, rates for new
physics processes are difficult to test since the flux of UHEC$\nu$
 is virtually unknown.  Interestingly, it is possible in
principle to disentangle the unknown flux and new physics processes by
using multiple
observables~\cite{Kusenko:2001gj,Anchordoqui:2001cg}.

For example, possible deviations of the neutrino--nucleon cross
section due to new non-perturbative interactions can be uncovered at
UHECR facilities by combining information from Earth-skimming and
quasi-horizontal showers.\footnote{Herein we use this term to describe
  neutrino interactions in which the final state energy is dominated
  by the hadronic component. We are {\em not} considering here new
  ``perturbative'' physics {\em e.g.} (softly broken) supersymmetry at
  the TeV scale which would have quite different signatures in cosmic
  neutrino showers.}  In particular, if an anomalously large rate is found
for deeply developing quasi-horizontal showers, it may be ascribed
either to an enhancement of the incoming neutrino flux, or an
enhancement in the neutrino-nucleon cross section (assuming
non-neutrino final states dominate).  However, these possibilities can
be distinguished by comparing the rates of Earth-skimming and
quasi-horizontal events.  For instance, an enhanced flux will increase
both quasi-horizontal and Earth-skimming event rates, whereas an
enhanced interaction cross section will also increase the former but
{\em suppress} the latter, because the hadronic decay products cannot
escape the Earth's crust.  Essentially this approach constitutes a
straightforward counting experiment, as the detailed shower properties
are not employed to search for the hypothesized new physics. Hence,
this constitutes an entirely general approach to searching for
non-perturbative interactions without any dependence on what
hypothetical mechanism might actually cause the ``hadrophilia.''

Consider first a flux of Earth-skimming tau neutrinos with energy
$E_0$.  Given the high energies required for detection, the most
relevant energies are $10^9 \lesssim E_0/{\rm GeV} \lesssim
10^{10}~{\rm GeV}$, and we may therefore limit the discussion to this
rather narrow band of energy. The neutrinos can convert to $\tau$
leptons in the Earth via the CC interaction
$\nu_{\tau^{\pm}} N \rightarrow \tau^{\pm} X$.  In the (perturbative)
SM, the interaction path length for the neutrino is
\begin{equation}
L_{\rm CC}^{\nu} = \left[ N_A \rho_s \sigma_{\rm CC}
\right] ^{-1} \ ,
\end{equation}
where $\sigma_{\rm CC}$ is the CC cross section for a neutrino energy $E_{\nu} = E_0$. The density 
of the material through which the neutrinos pass, $\rho_s$,
is about $2.65~{\rm g}/{\rm cm}^3$ for the Earth's crust.  Here we have neglected NC interactions, which at
these energies only reduce the neutrino energy by
approximately 20\%, which is within the systematic uncertainty.  For
$E_0 \sim 10^{10}~{\rm GeV}$, $L_{\rm CC}^{\nu} \sim {\cal
  O}(100)$~km.  Let us assume some hypothetical non-perturbative physics
process enhances the $\nu N$ cross section.  Then the interaction path length 
becomes
\begin{equation}
L_{\rm tot}^{\nu} = \left[ N_A \rho_s (\sigma_{\rm CC} +
\sigma_{\rm NP}) \right] ^{-1} \ ,
\end{equation}
where $\sigma_{\rm NP}$ is the non-perturbative contribution to the
cross section for $E_{\nu} = E_0$.

Once a $\tau$ is produced by a CC interaction, it can 
be absorbed in the Earth or escape and possibly decay,
generating a detectable air shower.  For $E_\nu \gtrsim 10^{8.5}~{\rm GeV}$,
the $\tau$ propagation length in the Earth is 
dominated by energy loss rather than the finite $\tau$ 
lifetime.  The energy loss can be expressed as
\begin{equation}
\label{energyloss}
\frac{dE_{\tau}}{dz} = -(\alpha_{\tau} + \beta_{\tau} E_{\tau})
\rho_s \ ,
\end{equation}
where $\alpha_\tau$ characterizes energy loss due to ionization and
$\beta_\tau$ characterizes losses through bremsstrahlung, pair 
production and hadronic interactions.  At these energies, 
energy losses due to ionization turn out to be negligible, 
while   
$\beta_\tau \simeq 0.8 \times 10^{-6}$~${\rm cm}^2 / {\rm g}$~\cite{Anchordoqui:2005is}.
{}From~\eqref{energyloss}, we observe that the maximum path 
length for a detectable $\tau$ can be written
\begin{equation}
L^{\tau} = \frac{1}{\beta_{\tau} \rho_s} \ln \left( E_{\rm max} /
E_{\rm min} \right) \ ,
\label{ltau}
\end{equation}
where $E_{\rm max} \approx E_0$ is the energy at which the $\tau$ is
created, and $E_{\rm min}$ is the minimal energy at which a $\tau$ can
produce a shower big enough to be detected.
For $E_{\rm max} / E_{\rm min} = 10$, $L^{\tau} = 11$~km.

The probability for a neutrino with incident nadir angle $\theta$
to emerge as a detectable $\tau$ is
\begin{equation}
P(\theta) = \int_0^l \frac{dz}{L_{\rm CC}^{\nu}}
e^{-z/L_{\rm tot}^{\nu}} \
\Theta \left[ z - (l - L^{\tau} ) \right],
\label{P}
\end{equation}
where $l = 2 R_{\oplus} \cos\theta$ is the chord length of the
intersection of the neutrino's trajectory with the Earth. Note that we
have neglected the possibility that non-perturbative processes could
lead to a detectable signal, since the hadrons which dominate the
final state will be absorbed in the Earth. The step function in
\eqref{P} reflects the fact that a $\tau$ will only escape the Earth
if $z + L^\tau > l$~\cite{Anchordoqui:2001cg}.

Assuming an isotropic tau neutrino flux, the number of
taus that emerge from the Earth with sufficient energy to be detected
is proportional to an ``effective solid angle''
\begin{equation}
\Omega_{\rm eff} \equiv \int P(\theta)\, \cos\theta\, d\cos\theta\, d\phi.
\end{equation}
Evaluation of the integrals~\cite{Kusenko:2001gj} yields the unfortunate expression
\begin{eqnarray}
\Omega_{\rm eff}  =  2 \pi
\frac{L_{\rm tot}^{\nu}}{L_{\rm CC}^{\nu}}
\left[ e^{L^{\tau} / L_{\rm tot}^{\nu}} - 1 \right] 
\left[ \left( \frac{L_{\rm tot}^{\nu}}{2 R_{\oplus}} \right)^2 
-  \left( \frac{L_{\rm tot}^{\nu}}{2 R_{\oplus}} +
\left( \frac{L_{\rm tot}^{\nu}}{2 R_{\oplus}} \right)^2 \right)
e^{-2R_{\oplus} / L_{\rm tot}^{\nu}} \right] \ .
\label{Omegaeff}
\end{eqnarray}
At the relevant energies, however, the neutrino interaction length satisfies
$L_{\rm tot}^{\nu} \ll R_{\oplus}$.  In addition, if the hypothesized
non-perturbative cross section enhancement is less than typical
hadronic cross sections, we have $L_{\rm tot}^{\nu} \gg L^{\tau}$.
With these approximations, \eqref{Omegaeff} simplifies to~\cite{Anchordoqui:2001cg}
\begin{equation}
\Omega_{\rm eff} \approx
2\pi \frac{L_{\rm tot}^{\nu\, 2} L^{\tau}}{4 R_{\oplus}^2
L_{\rm CC}^{\nu}} \ .
\label{omegaeffapprox}
\end{equation}

Equation~\eqref{omegaeffapprox} describes the functional dependence of the
Earth-skimming event rate on the non-perturbative cross section.  This rate is, of
course, also proportional to the neutrino flux $\Phi_{\nu_{\rm all}}$ at
$E_0$. Thus, the number of Earth-skimming neutrinos is given by
\begin{equation}
N_{\rm ES} \approx
C_{\rm ES}\, \frac{\Phi_{\nu_{\rm all}}}{\Phi_{\nu_{\rm all}}^{\rm WB}}
\frac{\sigma^{\nu\, 2}_{\rm CC}}
{\left( \sigma^{\nu}_{\rm CC} + \sigma^{\nu}_{\rm NP} \right)^2} \ ,
\label{ES}
\end{equation}
where $C_{\rm ES}$ is the number of Earth-skimming events
expected for some benchmark flux $\Phi_{\nu_{\rm all}}^{\rm WB}$ in the absence
of new physics. In contrast to~\eqref{ES}, the rate for quasi-horizontal showers
has the form
\begin{equation}
N_{\rm QH} = C_{\rm QH} \frac{\Phi_{\nu_{\rm all}}}{\Phi_{\nu_{\rm
      all}}^{\rm WB}}
\frac{\sigma^{\nu}_{\rm CC} + \sigma^{\nu}_{\rm NP}}
{\sigma^{\nu}_{\rm CC}} \ ,
\label{QH}
\end{equation}
where  $C_{\rm QH}$ is the number of quasi-horizontal events
expected for flux $\Phi_{\nu_{\rm all}}^{\rm WB}$.

\begin{figure}[tb]
\centering
\includegraphics[width=0.99\textwidth]{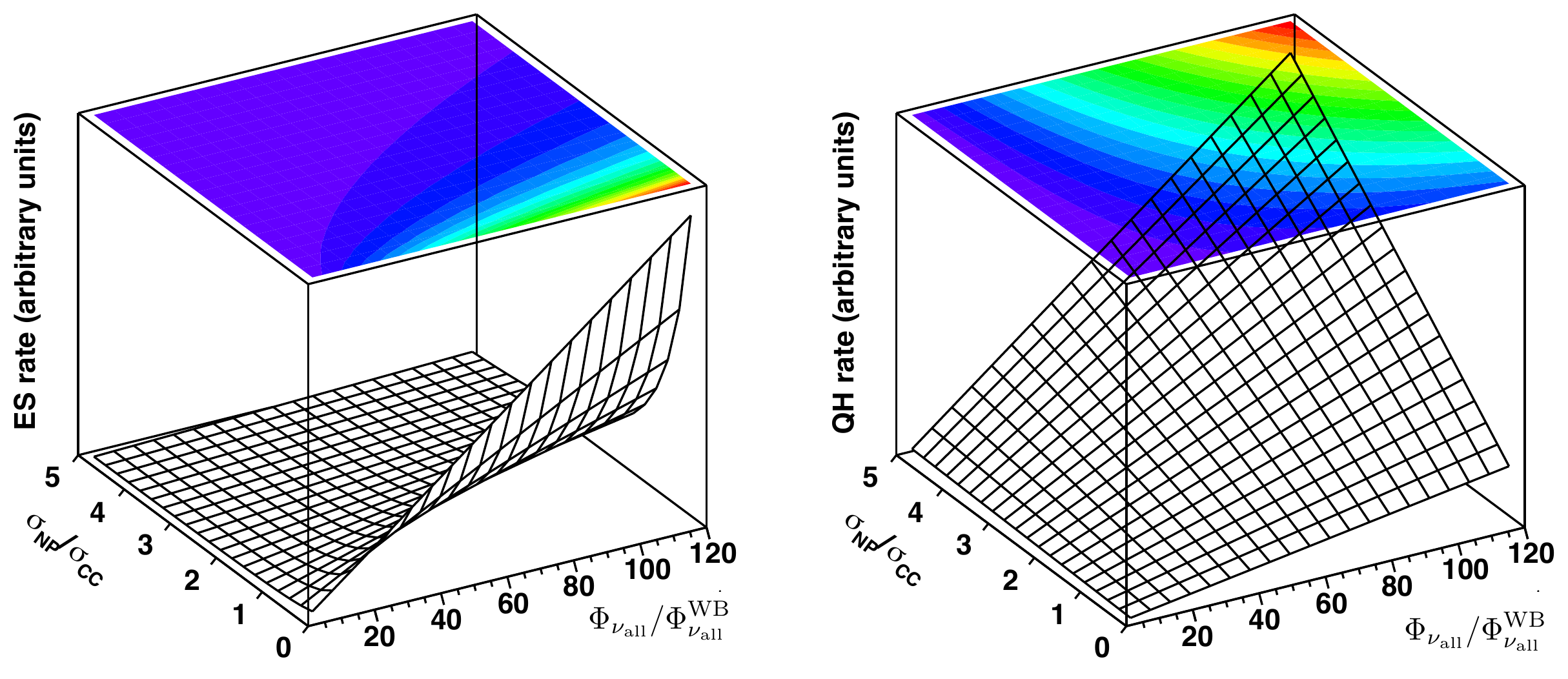}
\caption{Event rates for Earth-skimming (left) and 
quasi-horizontal (right) events in the 
$\Phi_{\nu_{\rm all}}/\Phi_{\nu_{\rm all}}^{\rm WB}-\sigma_{\rm NP}/\sigma_{\rm CC}$ 
plane.  Note that the contours are roughly orthogonal, and so the two
types of event provide complementary information about flux and cross
section. From Ref.~\cite{Anchordoqui:2011gy}.}
\label{fig:24}
\end{figure}

Given a flux $\Phi_{\nu_{\rm all}}$ and new non-perturbative physics
cross section $\sigma_{\rm NP}$, both $N_{\rm ES}$ and $N_{\rm
  QH}$ are determined.  On the other hand, given just a
quasi-horizontal event rate $N_{\rm QH}$, it is impossible to
differentiate between an enhancement of the cross section due to
non-perturbative physics and an increase of the flux.  However, in the
region where significant event rates are expected, the contours of
$N_{\rm QH}$ and $N_{\rm ES}$, given by \eqref{ES} and \eqref{QH}, are
more or less orthogonal and provide complementary information. This is
illustrated in Fig.~\ref{fig:24}. With measurements of $N_{\rm
  QH}^{\rm obs}$ and $N_{\rm ES}^{\rm obs}$, both $\sigma_{\rm
  NP}$ and $\Phi_{\nu_{\rm all}}$ may be determined independently, and
neutrino interactions beyond the (perturbative) SM may be
unambiguously identified~\cite{Anchordoqui:2001cg}. Event rate estimates assuming a neutrino
flux at the level of the WB bound suggest that existing ground-based
experiments~\cite{Anchordoqui:2005ey,Anchordoqui:2010hq} and future sapce-based
missions~\cite{PalomaresRuiz:2005xw} would be within reach of testing
possible enhancement of the neutrino-nucleon cross section.

\section{Looking ahead}
\label{sec:6}

Thanks to a prodigious experimental effort the origin of the highest
energy particles in the Universe are beginning to be revealed.
Nonetheless, 60 years after their discovery much remains a mystery.
Even the reason for the sharp suppression on the region of the
expected GZK effect remains uncertain.  Resolving the UHECR origin(s)
and investigating particle physics above accelerator energies, will
require both enhanced experimental techniques implemented at the
existing observatories, as well as a significant increase in exposure
to catch the exceedingly rare highest energy events.  Even before we
know the results from the upcoming generation of UHECR observatories
it seems clear that still larger aperture observatories with much
better energy and $X_{\rm max}$ resolution will be called for, in
order to measure the spectra of individual sources. The way forward is
clear and practical. The next-generation UHECR observatories will have
three primary goals:
\begin{itemize}[noitemsep,topsep=0pt]
\item {\it Increased statistics in both Northern and Southern
    hemispheres.}  A large increase in statistics is obviously
  important to increase the significance and resolution of all
  results.  In particular, it will improve the chances for anisotropy searches, allow a more
  sensitive measurement of the spectral suppression, and potentially
  establish variations in the spectrum in different regions of the
  sky. Furthermore, increased statistics will aid in reducing
  systematic uncertainties (of all sorts) for all measurements.
\item {\it  Composition-tagging for each individual event.}  Probabilistic
composition-tagging for all events will address the question of how the
baryonic composition evolves with energy, thereby clarifying the nature of the
spectral cutoff and the acceleration mechanism(s).  It will also aid in source
identification by allowing events to be backtracked through the
Galactic magnetic field, with reduced
ambiguity from their charge assignment, and allow correlation studies to be restricted
to proton-like events with smaller deflections. 
\item {\it Detailed observations of UHECR showers.}  It is essential
  to have reliable shower-development measurements to be able to
  understand hadronic interactions in the ultra-high-energy regime and
  to infer the nuclear composition from the shower properties.  UHECRs are also
  Nature's highest energy particle beam and thus present an
  opportunity to explore particle physics beyond collider energies.
\end{itemize}

Moving beyond existing technologies, it is inspiring to note that some
5 million UHECRs with energies above about $5.5 \times 10^{10}~{\rm GeV}$ strike the
Earth's atmosphere each year, from which we currently collect only
about 50 or so with present observatories. In this sense, there exists
some 5 orders of magnitude room for improvement! It may well be that
the best hope to make inroads in this area is to take the search for
UHECR sources into space. 

At present the most advanced project in pursuit of this objective is
the Probe Of Extreme Multi-Messenger Astrophysics (POEMMA) satellites,
selected by NASA for an in-depth probe mission concept study in
preparation for the next decadal survey~\cite{Olinto:2017xbi}. POEMMA
will boldly go where no UHECR observatory has gone before, combining
the well-developed Orbiting Wide-field Light-collectors (OWL)
concept~\cite{Stecker:2004wt} with the recently proposed CHerenkov
from Astrophysical Neutrinos Telescope (CHANT)
concept~\cite{Neronov:2016zou} to form a multi-messenger probe of the
most extreme environments in the universe. In addition to its
unprecedented physics potential, POEMMA will serve as a pathfinder for
future space-based missions, establishing feasibility and
cost-effectiveness, uncovering challenges and opportunities, and
stimulating development of second-generation technology for more
ambitious projects. An optimist might even imagine an eventual
constellation of satellites eating their way into the 5 orders of
magnitude of thus-far untapped UHECR luminosity.

In summary, pursuing improved ground-based detection techniques and
pioneering space-based observation will offer complementary tools to
piece together challenging astrophysical puzzles to unmask the UHECR
origin(s). Moreover, with the combined power of the space- and
ground-based approaches, a few decades from now we may even have
harnessed the study of UHECR showers to explore particle physics at
energies inaccessible to terrestrial accelerators. As exemplified time
and again, the development of novel ways of looking into space
invariably results in the discovery of unanticipated phenomena.

\section*{Acknowledgments} 
\addcontentsline{toc}{section}{Acknowledgments } 

I would like to thank Rasha Abbasi, Jim Adams, Felix Aharonian, Markus
Ahlers, Roberto Aloisio,  Rita dos Anjos, Vernon Barger, John Beacom,
Marty Block, Mark Christl, Jorge Combi, Rafael Colon,  Alessandro De
Angelis, Hans Dembinski, Peter Denton, Tere Dova, Loyal Durand, Ralph
Engel, Luis Epele, Glennys Farrar, Anatoli Fedynitch, Jonathan Feng,
Jorge Fernandez Soriano, Tom Gaisser, Carlos Garc\'{i}a Canal, Haim
Goldberg, Daniel G\'omez Dumm, Concha Gonzalez Garcia, Juan Guerra,
Phuock Ha, Francis Halzen, Tao Han, Dan Hooper, Albrecht Karle, John
Krizmanic, Alex
Kusenko, John Learned, Manuela Mallamaci, Danny Marfatia, Tom
McCauley, Rob Moncada, Teresa Montaruli, Kohta Murase, Marco Muzio, Andrii Neronov,
Matt O'Dowd, Angela Olinto, Tim Paglione, Sandip Pakvasa, Sergio
Palomares-Ruiz, Tom Paul, Santiago Perez Bergliaffa, Hallsie Reno, Felix Riehn,
Andreas Ringwald, Gustavo Romero, Subir Sarkar, Sergio Sciutto, Dmitri
Semikoz, Al Shapere, Todor Stanev, Andy Taylor, Diego Torres, Peter
Tinyakov, Michael Unger, Tom Weiler, and Lawrence Wiencke
for valuable discussions and permission to reproduce some of the
figures. I would also like to thank my colleagues of the Pierre Auger
and POEMMA collaborations for valuable discussions, and the TA
Collaboration for permission to reproduce Fig.~\ref{fig:4}. The
research of L.A.A.\ is supported by the U.S. National Science
Foundation (NSF Grant PHY-1620661) and the National Aeronautics and
Space Administration (NASA Grant 80NSSC18K0464).  Any opinions,
findings, and conclusions or recommendations expressed in this
material are those of the author and do not necessarily reflect the
views of the NSF or NASA.

\appendix

\section{}
\label{app1}

In the absence of a well justified model for the CR intensity, one
may prefer performing an interpolation to the data. This is not a
choice free approach, since one has to decide the functions used to
interpolate. In this Appendix a method is described using cubic splines to
achieve this task~\cite{Dembinski:2017zsh}.

In the GSF model, the CR intensity is divided into four
baryonic groups, which cover roughly equal ranges in logarithmic
baryon number $\ln A$, because air-shower measurements are sensitive
to changes in $\ln A$ rather than $A$. Each group has a leading
element $L$ that contributes most of the intensity per energy
interval. For each particle type, the interpolating intensity is
written as a function of the rigidity $J_L({\mathscr R})$, with
$J(\mathscr{R}) = \sum_{L=1}^4 J_L(\mathscr{R})$.\footnote{Rigidity,
  energy divided by charge, is a relevant parameter for CR propagation
  in a magnetic field. Rigidity is properly measured in units of GV but
  when only magnetic deflections and not energy losses are of concern,
  GV and GeV may be used interchangeably since knowing the deflection
  of a proton of energy $E_p$ in GeV specifies the deflection of any CR
  with the same value of $E_p = E/Z$.}  Note that if two
elements have the same abundance in intensity per rigidity interval
$J(\mathscr{R}) \propto dN/(d\mathscr{R} \, dA \, dt \, d\Omega)$, the
element with the higher charge contributes more to the intensity per
energy interval $dN/(dE\, dA\, dt\, d\Omega)$. As a consequence, the
leading elements are the heaviest abundant elements in each group;
namely proton, helium, oxygen, and iron. The oxygen and iron groups
contain many sub-leading elements. In the oxygen group, carbon
contributes nearly as much as oxygen. In the GSF model, the flux
$J_i(\mathscr{R})$ of a sub-leading element $i$ is kept in a constant
ratio $f_{iL}$ to the leading element $L \in \{p,\ {\rm He,\ O,\
  Fe}\}$ of its group, $J_i(\mathscr{R}) = f_{iL} \times
J_L(\mathscr{R})$. To remove the major power law contribution to the
intensity, it is convenient to consider a function that exhibits a softer
dependance on $\mathscr{R}$ than $J_L$; namely,
\begin{equation} f_L(x)\equiv
  \left(\frac{\mathscr{R}}{\mathrm{GV}}\right)^3
  J_L(\mathscr{R}),\end{equation}where $x\equiv\ln( \mathscr{R}/\mathrm{GV})$.

Let $\mathcal D=\{x_i,y_i\}$ be a set of $N+1$ experimental points,
and $\Omega_k=[x_k,x_{k+1}]$ the intervals between contiguous pairs of
points. A degree $n$ spline interpolator for $\mathcal D$ is a
piecewise function \begin{equation} f(x)=f_k(x)\mbox{ if }
  x\in\Omega_k,\end{equation} exhibiting some smoothness properties at
the internal ($k\neq0,N$) points, where $f_k(x)$ are polynomials of
degree at most $n$. For a fixed set of points, fixed degree $n$, and
fixed smoothness properties at the points, the set of all possible
splines forms a vector space. A B-spline is an element of a basis
$\{b_k\}$ of the vector space. Then, the spline function $f$ can be
written as \begin{equation} f(x)=\sum_{k=0}^{N-1} \alpha_k
  b_k(x).\label{eq:3}\end{equation} For cubic splines, the smoothness
properties require $f$ to be $C^3$ at the points. The cubic basis
functions are obtained by recurrence from the lower order basis
functions. One can write $b_k(x)=B_{k,3}(x)$, where
\begin{subequations}\begin{equation}
B_{k,0}(x) := \left\{
\begin{array}{lc}
1 & \mathrm{if} \quad x\in\Omega_k, \\
0 & \mathrm{otherwise}, 
\end{array}
\right.
\end{equation}
\begin{equation}B_{k,i}(x) := \frac{x - x_k}{x_{k+i} - x_k} B_{k,i-1}(x) + \frac{x_{k+i+1} - x}{x_{k+i+1} - x_{k+1}} B_{k+1,i-1}(x).\end{equation}
\end{subequations}

This would give a set of $N$ polynomials completely determined by the
data points and the smoothness conditions at them. There is still a
freedom on the spline interpolator, written as a linear combinations
of these B-splines. In order to find the coefficients $\alpha_k$ in
(\ref{eq:3}), one performs a least squares fit,
minimizing \begin{equation} U(\alpha_1,...,\alpha_{N-1})=\sum_{j=0}^N
  w_j\left(y_j-\sum_{k=0}^{N-1}\alpha_k
    b_k(x_j)\right)^2,\end{equation} where $w_j$ is the weight
assigned to each point. If the data is accompanied with a series of
uncertainties $\sigma_i$ for each value $y_i$, one can choose the
weight to be $w_i=1/\sigma_i^2$, which would make
$U(\alpha_1,...,\alpha_{N-1})$ be the familiar quantity 
\begin{equation}
  \chi^2=\sum_{j=0}^N\left(\frac{y_j-f(x_j)}{\sigma_j}\right)^2.
\end{equation}

Note that $J(\mathscr{R})$ parametrizes the differential flux of
nuclei per rigidity interval. Air-shower measurements of the CR flux
are reported as the differential flux of nuclei per energy interval
$J(E)$. The latter is computed from the former as $J(E) =
J(\mathscr{R}) \, d\mathscr{R}/dE$. The relation between total energy
$E$ and rigidity $\mathscr{R}$ depends on the number of nucleons $A$
and protons $Z$ and must be computed individually for each
element. Note that a primary CR with rigidity $\mathscr{R}/{\rm GV}$
could be a proton with energy $E_p/{\rm GeV}$ or a nucleus with energy
$Z E_p/{\rm GeV}$, with $\mathscr{R}/{\rm GV} = E_p/{\rm
  GeV}$. Thereupon, we will specify, all through, the CR rigidity of a
nucleus of charge $Ze$ using the proton energy $E_p$. Air-shower
measurements describe the flux of mass groups. In the GSF model a sum
of the flux of all elements in each group is carried out when
comparing the model to such measurements.

For the GSF model shown in Fig.~\ref{fig:2}, the $\chi^2 = 385.2$ for
724 degrees of freedom, which indicates a good
fit~\cite{Dembinski:2017zsh}.  The good agreement of the fit (when
systematic uncertainties are taken into account) in turn implies that
the data sets are overall consistent. The fit shown in
Fig.~\ref{fig:2} has been carried out discarding the proton-helium
data of ARGO-YBJ beyond $10^6~{\rm GeV}$ and assigning 10\% to 20\%
systematic uncertainty to results where none were
reported.\footnote{For ARGO-YBJ, the reported proton+helium intensity
  drops sharply for $E> 10^6~{\rm GeV}$, in contradiction to three
  other data sets.} One can refer to the original
paper~\cite{Dembinski:2017zsh} for a careful assessment of the
adjustment of the energy scales within systematic uncertainties.

\section{}
\label{app2}

This Appendix contains a guide, as complete as possible, to the use of
statistical likelihood-based methods in data analysis. For further
details, see e.g.~\cite{Cowan:2010js,Cousins:2018tiz}.

The probability density function for a random variable, $x$,
conditioned on a set of parameters,
$\boldsymbol{\theta}=\{\theta_1,...,\theta_m\}$, is denoted
$f(x ; \boldsymbol{\theta})$.  This function identifies the
data-generating process that underlies an observed sample of data and,
at the same time, provides a mathematical description of the data that
the process will produce. The joint density of $N$ independent and
identically distributed observations from this process, $\mathbf
x=\{x_1,...,x_N\}$, is the product of the individual densities,
\begin{equation}
f(x_1 \dots x_N ; \boldsymbol{\theta}) = \prod_{i=1}^N f (x_i ;
\boldsymbol{\theta} )= \mathscr{L}(\boldsymbol
  \theta ; \mathbf x) \, .
\end{equation}
This joint density is the likelihood function, defined as a function
of the unknown parameter vector, $\boldsymbol{\theta}$. Note that
the joint density is written as a function of the data conditioned on the
parameters, whereas when one forms the likelihood function,  
the function is written in reverse, i.e. as a function of the parameters, conditioned
on the data. Though the two functions are the same, it is to be
emphasized that the likelihood function is written in this fashion to
highlight the interest in the parameters, and the information about
them that is contained in the observed data. However, it is understood
that the likelihood function is not meant to represent a probability
density for the parameters. In this classical estimation framework,
the parameters are assumed to be fixed constants that one awaits to learn
about from the data.

Extension to a multivariate density, $\mathbf f=\{f_1,...,f_N\}$,  is straightforward.
Note that in general the experimental data do not need to be of the same
  kind, but rather each point in the data-sample may follow a different statistical
  model $f_i$. Strictly speaking, for a given set of $m$ parameters $\boldsymbol{\theta}$, a function $f_i$
assigns the probability density of a random variable $x$ to take the a
value $x_i$, and so $f_i(x_i;\boldsymbol{\theta})$
gives \emph{the probability of the data point $x_i$ given the
  individual model $\{f_i,\boldsymbol\theta\}$}, i.e. $
f_i(x_i;\boldsymbol{\theta}) \equiv \mathcal
P(x_i|\mathrm{model}_i)$. The likelihood function, 
\begin{equation}\mathscr{L}(\boldsymbol
  \theta;\mathbf x)=\prod_{i=1}^N f_i(x_i;\boldsymbol\theta) \,, \end{equation}
is \emph{the
  probability of all the data, $\mathbf x$, given the complete model
  $\{\mathbf f,\boldsymbol\theta\}$}, i.e. $\mathcal P(\mathbf
x|\mathrm{model})$. 

The estimated values $\boldsymbol\theta=\hat{\boldsymbol\theta}$ of the parameters are
  obtained by finding the global maximum of the likelihood
  function,
\begin{equation}\left.\frac{\partial\mathscr{ L}( \boldsymbol\theta;
      \mathbf
      x)}{\partial\theta_i}\right|_{\boldsymbol\theta=\hat{\boldsymbol\theta}}=0,\quad
  {\rm with} \quad 1\leq i\leq m.\label{eq:app3}\end{equation}
In practice, it is often more convenient to work with the
  logarithm of the likelihood function, called the log-likelihood and
  to search for the minimum of the negative log-likelihood function:
\begin{equation}
- \ln \mathscr{ L} (\boldsymbol{\theta}, \mathbf{x}) = - \sum_{i=1}^N
\ln f (x_i, \boldsymbol{\theta}) \, .
\end{equation}
Unless the minimum occurs at the boundary of the allowed range of values for $\boldsymbol{\theta}$,
a necessary condition for the minimum is that the negative
log-likelihood satisfies the following $m$ equations:
\begin{equation}
\left. -\frac{\partial \ln \mathscr{L} (\boldsymbol{\theta}; \mathbf
    x)}{\partial\theta_i}\right|_{\boldsymbol\theta=\hat{\boldsymbol\theta}}=0
\, .
\end{equation}
The likelihood function must be constructed using normalized
probability density functions: $\int f(x;\boldsymbol{\theta}) \ dx
= 1$, so that $\int \mathscr{ L} (\boldsymbol \theta; \mathbf x) \ dx_1 \cdots dx_N
=1$. In other words it is essential that the integral of the
likelihood function does not depend on the parameters
$\boldsymbol{\theta}$.

Likelihood maximization methods can be used to find the parameters
$\hat{\boldsymbol\theta}$ maximizing the likelihood, as well as to
find the confidence region(s) of certain parameter(s) in
$\boldsymbol\theta$ around $\hat{\boldsymbol\theta}$. Consider a set
of parameters $\boldsymbol\mu\subseteq\boldsymbol\theta$ whose study
is of interest, and a set of nuisance parameters collectively denoted
by $\boldsymbol\nu=\boldsymbol\theta\setminus\boldsymbol\mu$. To make
the splitting of $\boldsymbol\theta$ explicit, the likelihood is
written hereafter as $\mathscr{ L} ( \boldsymbol\mu,\boldsymbol\nu;
\mathbf x)$. For the evaluation of the confidence regions, it is
practical to use the profile likelihood ratio,
\begin{equation}
  \lambda(\boldsymbol\mu)\equiv\frac{\mathscr{
      L}(\boldsymbol\mu,\hat{\hat{\boldsymbol\nu}}(\boldsymbol\mu);\mathbf
    x )}{\mathscr{L}   (\hat{\boldsymbol\mu},\hat{\boldsymbol\nu}; \mathbf
    x ) } \,,\label{eq:app4}\end{equation}
where in the numerator  there is a profile likelihood
function in which $\hat{\hat{\boldsymbol\nu}}$ is the value of
$\boldsymbol \nu$ maximizing $\mathscr{L}$ for the assumed
$\boldsymbol \mu$, namely
\begin{equation}\left.\frac{\partial \mathscr{L}
      (\boldsymbol\mu,\boldsymbol\nu;\mathbf
      x)}{\partial
      \nu_i}\right|_{\boldsymbol\nu=\hat{\hat{\boldsymbol\nu}}(\boldsymbol\mu)}=0,\quad
  {\rm with} \quad
  1\leq i\leq \dim\boldsymbol\nu \,;\label{eq:app5}
\end{equation} 
i.e. the likelihood is maximised only in the parameters $\hat
{\boldsymbol \nu}$ for each $\boldsymbol \mu$. Note that
$\hat{\hat{\boldsymbol\nu}}$ is the conditional maximum likelihood
estimator of $\boldsymbol{\nu}$ and consequently is a function of
$\boldsymbol \mu$ itself.  The denominator, instead, is maximized in
an unconstrained way, thus $\hat {\boldsymbol \mu}$ and $\hat{
  \boldsymbol \nu}$ are the true maximum likelihood estimators.  By
definition then the profile likelihood ratio is comprised between $ 0
\le \lambda (\boldsymbol \mu) \leq 1$.  The upper limit is picked up
when the hypothesized $\boldsymbol \nu$ coincides with $\hat
{\boldsymbol \nu}$, showing therefore great compatibility between the data
and the hypothesis. The lower limits is instead picked up when the
assumed $\boldsymbol{\nu}$ is at odd with $\hat {\boldsymbol \nu}$,
denoting in this way a high degree of incompatibility between the data
and the hypothesis.

In the large-sample limit, where the likelihood approaches a Gaussian,
$-2\ln \lambda(\boldsymbol\mu)$ follows a $\chi^2$ distribution with
$d=\dim\boldsymbol\mu$ degrees of freedom. This condition is usually referred
to as the Wilks’ theorem~\cite{Wilks:1938dza}.  One can then use the
quantiles $\chi^2_c(d,\alpha)$ of the $\chi^2$ distribution to
evaluate $\alpha$-confidence regions,
\begin{equation}
\alpha=\int_0^{\chi_c^2(d,\alpha)}f_{\chi^2}(z;d) \  dz \,,
\end{equation} 
where
 \begin{equation}f_{\chi^2}(z;d)=\frac{z^{d/2-1}e^{-z/2}}{2^{d/2}\Gamma\left(\frac
      d 2\right)} \end{equation} is the probability density for a random variable $z$ following a $\chi^2$ distribution of $d$ degrees of freedom. 
These quantiles define the rise in $-2 \ln λ(\boldsymbol \mu)$ corresponding to the points
of $\boldsymbol \mu$ on the border of the confidence region,
\begin{equation}
  -2\ln \lambda(\boldsymbol\mu)\leq\chi_c^2(d,\alpha) \, . 
\label{eq:app6}
\end{equation}
With this construction, the region $[0,\chi^2_c(d,\alpha)]$ contains
the values of $-2\ln \lambda(\boldsymbol\mu)$ that allows one to
phrase the statement: \emph{$\boldsymbol\mu$ is in the region defined
  by (\ref{eq:app6})} with a confidence level $\alpha$.

If a statistical hypothesis to be tested can be expressed in terms of
the parameters $\boldsymbol\theta$ in the likelihood, one can use the
function defined in (\ref{eq:app4}) to build a TS and asses the
confidence an experiment gives to say that the hypothesis is true or
false. In general, the hypothesis to be tested, referred as \emph{null
  hypothesis}, may be expressed in terms only of a subset
($\boldsymbol\mu$) of all the involved parameters. Let
$H_0:\boldsymbol\mu=\boldsymbol\mu_0$ define the null hypothesis, to
be confronted with its negation, the \emph{alternative hypothesis}
$H_1:\boldsymbol\mu\neq\boldsymbol\mu_0$. According to the method
developed above, the condition \begin{equation} -2\ln \lambda
  (\boldsymbol \mu_0) \equiv -2\ln
  \lambda_0=\chi^2_0(d,\alpha),\end{equation} 
gives the $\alpha$-confidence in the exclusion of the null hypothesis, with
\begin{equation}\alpha=\int_0^{-2\ln \lambda_0}f_{\chi^2}(z;d) \ 
  dz.\label{eq:app9}\end{equation}

\begin{figure}[tpb]
\postscript{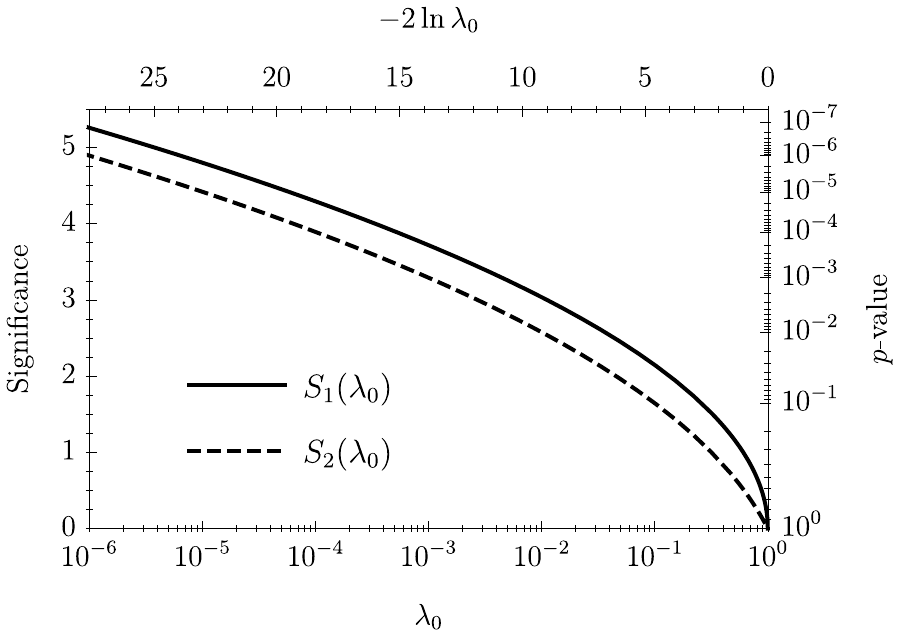}{0.7}
\caption{Relation between the significance and $\lambda_0$ for $d=1$
  and $d=2$.}
\label{fig:25}
\end{figure}

One can measure the significance associated with the previous
$\alpha$-confidence using the standard method to relate significance and
$p$-values, with a unit normal distribution. For a confidence level $\alpha$, the significance $S$ is
given by
 \begin{equation}\alpha=\int_{\mu-S\sigma}^{\mu+S \sigma}f_\mathcal
  N (z;\mu,\sigma) \ 
  dz=\mathrm{erf}\left(\frac{S}{\sqrt2}\right).
\label{eq:app9b}
\end{equation}
Comparing  (\ref{eq:app9}) with (\ref{eq:app9b})  the significance is
found to be  
\begin{equation}
  S_d=\sqrt2\,\mathrm{erf}^{-1}\left(\int_0^{-2\ln\lambda_0}f_{\chi^2}(z;d) \
    dz\right).
\label{eq:app9c}
\end{equation}
For one and two degrees of freedom, which are  particularly
relevant cases in this review, (\ref{eq:app9c}) becomes
\begin{subequations}\begin{equation}
    S_1=\sqrt2\,\mathrm{erf}^{-1}\left[\mathrm{erf}\left(\sqrt{\frac{-2\ln\lambda_0}{2}}\right)\right]=\sqrt{-2\ln\lambda_0},\end{equation}\begin{equation}S_2=\sqrt
    2\, \mathrm{erf}^{-1}\left[1-e^{-\frac{-2\ln \lambda_0}{2}}\right]=\sqrt 2\,\mathrm{erf}^{-1}(1-\lambda_0).\end{equation}\end{subequations}
The significance for one and two degrees of freedom as a function of
$\lambda_0$ is display in Fig.~\ref{fig:25}.

A useful application of the statistical methods described above is the
rejection of \emph{background only hypothesis} against a
\emph{background and source hypothesis}. Consider an experiment
(possibly combined with simulations) that provides a number of
observed events $N_{\mathrm{on}}$ in a certain region (or sample)
where the hypotheses will be analyzed, and a number of events
$N_{\mathrm{off}}$ in another region (or in a Monte Carlo sample)
where the null hypothesis is understood to be true. A comparison of
the sizes of those two regions (or samples) is necessary to infer the
background in the region of interest from the background in the
auxiliary region, assuming that its nature in both regions is the
same. To this end it is useful to introduce the parameter $\eta =
N/N_{\rm sim}$, where $N$ is the total number of observed events in
the data-sample and $N_{\rm sim}$ is the number of events in the Monte
Carlo sample. Then,  in the notation introduced above: \begin{itemize}[noitemsep,topsep=0pt]
\item $\mathbf x=\{N_{\mathrm{on}},N_{\mathrm{off}}\}$,
\item $\boldsymbol\theta=\{\left<N_S\right>,\left<N_B\right>\}$,\begin{itemize}[noitemsep,topsep=0pt]
	\item[$\cdot$] $\boldsymbol\mu=\{\left<N_S\right>\}$,
	\item[$\cdot$] $\boldsymbol\nu=\{\left<N_B\right>\}$,
	\end{itemize}
\item $\mathbf
  f=\{p(\cdot,\left<N_S\right>+\left<N_B\right>),p(\cdot,\left<N_S\right>/\eta)\}$ \,,
\end{itemize}
where $\left<N_S\right>$ and $\left<N_B\right>$ are the expected
number of source and background events in the \emph{on} region,
respectively. The functions $p(\cdot,N)$ are Poisson distribution
functions with mean $N$, given by 
\begin{equation} p(x,N)=\frac{e^{-N} N^x}{x!}.
\end{equation} 
The likelihood is given by 
\begin{equation}\mathscr{ L}(\avg{N_S},\avg{N_B};\mathbf
  x)=p(N_{\mathrm{on}},\avg{N_S}+\avg{N_B})\,p(N_{\mathrm{off}},\avg{N_B}/\eta)
  \, .\end{equation}
The maximum likelihood conditions (\ref{eq:app3})
yield \begin{subequations}\begin{equation}\hat{\avg{N_S}}=N_{\mathrm{on}}-\eta
    N_{\mathrm{off}} \end{equation}
and
\begin{equation}\hat{\avg{N_B}}=\eta
  N_{\mathrm{off}}.\end{equation}\end{subequations} 
The condition (\ref{eq:app5}) gives \begin{equation} \hat{\hat{\avg{N_B}}}(\avg{N_S})=
  \frac{N_{\mathrm{on}}+N_{\mathrm{off}}-\left<N_S\right>   \varkappa
    \pm\sqrt{\left[N_{\mathrm{on}}+N_{\mathrm{off}}-\left<N_S\right>\varkappa\right]^2+4\left<N_S\right>
      \varkappa N_{\mathrm{off}}}}{2 \varkappa} \,,\end{equation} 
where $\varkappa = 1 + 1 /\eta$. For the null hypothesis, $\avg{N_S}=0$, and so
\begin{equation}\hat{\hat{\avg{N_B}}}_0\equiv\hat{\hat{\avg{N_B}}}(0)=
\frac{N_{\mathrm{on}}+N_{\mathrm{off}}}{1+1/\eta}.\end{equation} Then, 
\begin{eqnarray} \lambda_0=\frac{\mathscr L(
    0,\hat{\hat{\avg{N_B}}}_0;\mathbf
    x )}{\mathscr L (\hat{\avg{N_S}},\hat{\avg{N_B}} ; \mathbf
    x  )  } & = &
  \frac{p(N_\mathrm{on},(N_\mathrm{on}+N_\mathrm{off})/(1+1/\eta))}{p(N_\mathrm{on},N_\mathrm{on})}\frac{p(N_\mathrm{off},(N_\mathrm{on}+N_\mathrm{off})/(\eta+1))}{p(N_\mathrm{off},N_\mathrm{off})}
    \nonumber \\
  & = & \left[\frac{\eta
      (N_\mathrm{on}+N_\mathrm{off})}{(\eta+1)N_\mathrm{on}}\right]^{N_\mathrm{on}}\left[\frac{N_\mathrm{on}+N_\mathrm{off}}{(\eta+1)N_\mathrm{off}}\right]^{N_\mathrm{off}},\end{eqnarray}
and for $d=1$,  the statistical Li-Ma significance is given by (\ref{Li-Ma})~\cite{Li:1983fv}.

\section{}
\label{app3}

This Appendix provides an overview of some generalities of the Rayleigh
distribution. The statistical properties of the estimators $\{\hat
a_m, \hat b_m \}$ in (\ref{Rayleigh-estimators}) derive from the
Poissonian nature of the sampling of $N$ points over the circle,
matching the underlying angular distribution. Namely, the first and
second moments of $\delta \hat J (\alpha)$ in (\ref{jofalpha})
averaged over a large realization of events are
\begin{subequations}
\begin{equation}
\langle \delta \hat  J(\alpha) \rangle_{\rm P} = \delta J(\alpha) \,,
\end{equation}
and
\begin{equation}
\langle \delta \hat J(\alpha) \ \delta \hat  J (\alpha')
  \rangle_{\rm P} = \delta J(\alpha) \delta J(\alpha') + \delta
  J(\alpha) \delta (\alpha,\alpha') \, ,
\end{equation} 
\end{subequations}
respectively. The mean and root-mean-square of $\hat a_m$ and
$\hat b_m$ can be calculated propagating these
properties into (\ref{orthogonality}) while taking $\hat a_0$ constant (the latter
is a very precise approximation  in most practical cases). All in all,
it is easily seen that the estimators are unbiased, $\langle
\hat a_m \rangle_{\rm P} = a_m$  and $\langle
\hat b_m \rangle_{\rm P} = b_m$, and obey the covariance
matrix coefficients given by~\cite{Deligny:2018blo}
\begin{subequations}
\begin{eqnarray}
{\rm cov}(\hat a_i, \hat a_j) & = & \frac{J_0 \omega_0}{2\pi^3 a_0^2} \int
\frac{d\alpha}{1 + \delta \omega (\alpha)}  \ \delta J (\alpha) \ \cos
(i\alpha) \ \cos (j\alpha) \,,   \\
{\rm cov}(\hat b_i, \hat b_j) & = & \frac{J_0 \omega_0}{2\pi^3 a_0^2} \int
\frac{d\alpha}{1 + \delta \omega (\alpha)}  \ \delta J (\alpha) \ \sin
(i\alpha) \ \sin (j\alpha) \, . 
\end{eqnarray}
\end{subequations}
For small anisotropies, $|a_m| \ll 1$ and $|b_n| \ll 1$, and so the
uncertainties of the estimators are given by
\begin{subequations}
\begin{eqnarray}
\sigma (\hat a_m) & = & \left[ \frac{2}{\pi {\cal N}_\alpha} \int
  \frac{d \alpha}{1 + \delta \omega (\alpha) } \cos^2 (m \alpha)
\right]^{1/2} \,,  \\
\sigma (\hat b_m) & = & \left[ \frac{2}{\pi {\cal N}_\alpha} \int
  \frac{d \alpha}{1 + \delta \omega (\alpha) } \sin^2 (m \alpha)
\right]^{1/2} \, .
\end{eqnarray}
\label{B3}
\end{subequations}
Since $\delta \omega (\alpha)$ is in practice always small and smooth,
the integrals in (\ref{B3}) approximate very well near $\pi$, yielding
$\sigma (\hat a_m) = \sigma (\hat b_m) = \sqrt{2/ {\cal
    N_\alpha}}$. The coefficients $\hat a_m$ and $\hat b_m$ are
endowed with Gaussian probability density functions, $p_{A_m}$ and
$p_{B_m}$, which derive from the central limit theorem and are fully
determined by the following parameter pairs: $\{\langle \hat a_m
\rangle_{\rm P}, \sigma^2\}$ and $\{\langle \hat b_m \rangle_{\rm P}, \sigma^2\}$, with $\sigma^2 = 2/{\cal N}_\alpha$.

 For any given data-sample containing $N$
events, $\hat a_m$ and $\hat b_m$ are random
variables. Therefore, in the limit of large statistics the joint
probability distribution function  $p_{A_m,B_m}$ can be factorized as a
product of $p_{A_m}$ and $p_{B_m}$. For any harmonic $m$, the joint
probability distribution function of the estimated $\hat r_m$
and $\hat \varphi_m$ can be
derived using the pertinent Jacobian transformation~\cite{Linsley:1975kp} 
\begin{equation}
p_{R_m,\varPhi_m} (\hat r_m \hat \varphi_m; r_m \varphi_m) = \frac{\hat
  r_m}{2 \pi \sigma^2} \ \exp\{-[\hat r_m^2 +
  r_m^2 - 2 \hat r_m r_m \cos (\hat \varphi_m - \varphi_m)]/2
  \sigma^2 \}\, .
\label{jaco}
\end{equation}
The probability distribution function of the amplitude, $p_{R_m}$, is
obtained by marginalizing (\ref{jaco}) over the phase
\begin{equation}
p_{R_m} (\hat r_{m}; r) = \frac{\hat r_m}{\sigma^2} \ \exp \left( -
  \frac{\hat r_m^2 + r_m^2}{2 \sigma^2} \right) \ I_0 \left(\frac{
    \hat r_m r_m}{\sigma^2} \right) \,,
\end{equation}
with $I_0 (x)$ the modified Bessel function of first kind with order
zero. Likewise, the probability distribution function of the phase, $p_{\varPhi_m}$, is
obtained by marginalizing (\ref{jaco}) over the amplitude
\begin{eqnarray}
p_{\varPhi_m} (\hat \varphi_m; r_m,\varphi_m) & = & \frac{1}{2 \pi} \ \exp
  \left(- \frac{r_m^2}{2 \sigma^2} \right) \left\{ 1 +
    \sqrt{\frac{\pi}{2}} \ \frac{r_m}{\sigma} \ \cos (\psi_m)
    \  \exp\left(\frac{r^2_m \ \cos^2 \psi_m}{2 \sigma^2} \right) \right.
    \nonumber \\
& \times & \left. \left[1 +
        \xi_m \  {\rm erf} \left( \frac{\xi_m \ r_m \ \cos
            (\psi_m)}{\sqrt{2} \ \sigma}
        \right)\right] \right\} \,, 
\end{eqnarray}
where $\psi_m = \hat \varphi_m - \varphi_m$, and $\xi = 1$ if $|\psi_m|
\leq \pi/2$ and $-1$ otherwise.

Note that if the underlying distribution is isotropic the
$p_{\varPhi_m}$ is uniform, whereas $p_{R_m}$ reduces to the Rayleigh
distribution. This lets out a  genuine estimation of the
probability that an observed amplitude $\hat r_m$
arises from pure statistical fluctuations as
\begin{equation}
p (\geq \hat r_m) = \int_0^\infty d \hat r_m' \ p_{R_m} (\hat r'_m; r_m =
0) = \exp \left(- \frac{{\cal N}_\alpha \ \hat r_m^2}{4} \right)
\end{equation}
For a non-zero amplitude $r_m$, depending on the signal-to-noise ratio
parameter $r_m/\sigma$, both $p_{R_m}$ and $p_{\varPhi_m}$ smoothly
evolve from the Rayleigh and uniform distributions to bell curves
well-defined about the values of $r_m$ and $\varphi_m$. For
$r_m/\sigma \to \infty$, the bell curves are identical to 
Gaussian ones.


\section*{Bibliography}
    \addcontentsline{toc}{section}{Bibliography}
\begin{thebibliography}{99}
 


\bibitem{GBM:2017lvd} 
  B.~P.~Abbott {\it et al.} [LIGO Scientific and Virgo and Fermi GBM and INTEGRAL and IceCube and IPN and Insight-Hxmt and ANTARES and Swift and Dark Energy Camera GW-EM and Dark Energy Survey and DLT40 and GRAWITA and Fermi-LAT and ATCA and ASKAP and OzGrav and DWF (Deeper Wider Faster Program) and AST3 and CAASTRO and VINROUGE and MASTER and J-GEM and GROWTH and JAGWAR and CaltechNRAO and TTU-NRAO and NuSTAR and Pan-STARRS and KU and Nordic Optical Telescope and ePESSTO and GROND and Texas Tech University and TOROS and BOOTES and MWA and CALET and IKI-GW Follow-up and H.E.S.S. and LOFAR and LWA and HAWC and Pierre Auger and ALMA and Pi of Sky and DFN and ATLAS Telescopes and High Time Resolution Universe Survey and RIMAS and RATIR and SKA South Africa/MeerKAT Collaborations and AstroSat Cadmium Zinc Telluride Imager Team and AGILE Team and 1M2H Team and Las Cumbres Observatory Group and MAXI Team and TZAC Consortium and SALT Group and Euro VLBI Team and Chandra Team at McGill University],
    {\color{rossoCP3}  Multi-messenger observations of a binary neutron star merger},
  Astrophys.\ J.\  {\bf 848}, no. 2, L12 (2017)
  doi:10.3847/2041-8213/aa91c9
  [arXiv:1710.05833 [astro-ph.HE]].


\bibitem{IceCube:2018dnn} 
  M.~G.~Aartsen {\it et al.} [IceCube and Fermi-LAT and MAGIC and AGILE and ASAS-SN and HAWC and H.E.S.S. and INTEGRAL and Kanata and Kiso and Kapteyn and Liverpool Telescope and Subaru and Swift NuSTAR and VERITAS and VLA/17B-403 Collaborations],
     {\color{rossoCP3}  Multimessenger observations of a flaring blazar coincident with high-energy neutrino IceCube-170922A},
  Science {\bf 361}, no. 6398, eaat1378 (2018)
  doi:10.1126/science.aat1378
  [arXiv:1807.08816 [astro-ph.HE]].




\bibitem{Aab:2018chp} A.~Aab {\it et al.} [Pierre Auger
  Collaboration], {\color{rossoCP3} Indication of anisotropy in
    arrival directions of ultra-high-energy cosmic rays through
    comparison to the flux pattern of extragalactic gamma-ray
    sources}, Astrophys.\ J.\ Lett.\ doi:10.3847/2041-8213/aaa66d
  [arXiv:1801.06160 [astro-ph.HE]].

\bibitem{Anchordoqui:1999cu} 
  L.~A.~Anchordoqui, G.~E.~Romero and J.~A.~Combi,
    {\color{rossoCP3} Heavy nuclei at the end of the cosmic ray spectrum?},
  Phys.\ Rev.\ D {\bf 60}, 103001 (1999)
  doi:10.1103/PhysRevD.60.103001
  [astro-ph/9903145].


\bibitem{Abbasi:2007sv} 
  R.~U.~Abbasi {\it et al.} [HiRes Collaboration],
  {\color{rossoCP3} First observation of the Greisen-Zatsepin-Kuzmin suppression},
  Phys.\ Rev.\ Lett.\  {\bf 100}, 101101 (2008)
  doi:10.1103/PhysRevLett.100.101101
  [astro-ph/0703099].


\bibitem{Abraham:2008ru} 
  J.~Abraham {\it et al.} [Pierre Auger Collaboration],
  {\color{rossoCP3} Observation of the suppression of the flux of cosmic rays above $4\times 10^{19}$~eV},
  Phys.\ Rev.\ Lett.\  {\bf 101}, 061101 (2008)
  doi:10.1103/PhysRevLett.101.061101
  [arXiv:0806.4302 [astro-ph]].

\bibitem{Abraham:2010mj} 
  J.~Abraham {\it et al.} [Pierre Auger Collaboration],
  {\color{rossoCP3} Measurement of the energy spectrum of cosmic rays above $10^{18}$~eV using the Pierre Auger Observatory},
  Phys.\ Lett.\ B {\bf 685}, 239 (2010)
  doi:10.1016/j.physletb.2010.02.013
  [arXiv:1002.1975 [astro-ph.HE]].





\bibitem{Greisen:1966jv}
  K.~Greisen,
 {\color{rossoCP3} End to the cosmic ray spectrum?},
  Phys.\ Rev.\ Lett.\  {\bf 16}, 748 (1966)
  doi:10.1103/PhysRevLett.16.748.

\bibitem{Zatsepin:1966jv}
  G.~T.~Zatsepin and V.~A.~Kuzmin,
  {\color{rossoCP3} Upper limit of the spectrum of cosmic rays},
  JETP Lett.\  {\bf 4}, 78 (1966)
  [Pisma Zh.\ Eksp.\ Teor.\ Fiz.\  {\bf 4}, 114 (1966)].

\bibitem{Ade:2015xua} 
  P.~A.~R.~Ade {\it et al.} [Planck Collaboration],
  {\color{rossoCP3} Planck 2015 results. XIII. Cosmological parameters},
  Astron.\ Astrophys.\  {\bf 594}, A13 (2016)
  doi:10.1051/0004-6361/201525830
  [arXiv:1502.01589 [astro-ph.CO]].



\bibitem{Patrignani:2016xqp} 
  M.~Tanabashi {\it et al.} [Particle Data Group],
  {\color{rossoCP3} Review of Particle Physics},
  Phys.\ Rev.\ D  {\bf 98},  030001 (2018).


\bibitem{Anchordoqui:2011gy} 
  L.~A.~Anchordoqui,
    {\color{rossoCP3} Ultra-high-energy cosmic rays: facts, myths, and legends},
  doi:10.5170/CERN-2013-003.303
  arXiv:1104.0509 [hep-ph].

\bibitem{Anchordoqui:2013eqa} 
  L.~A.~Anchordoqui {\it et al.},
   {\color{rossoCP3}  Roadmap for ultra-high energy cosmic ray physics and astronomy (whitepaper for Snowmass 2013)},
  arXiv:1307.5312 [astro-ph.HE].


\bibitem{Anchordoqui:2002hs} 
  L.~Anchordoqui, T.~C.~Paul, S.~Reucroft and J.~Swain,
  {\color{rossoCP3}  Ultrahigh-energy cosmic rays: The state of the art before the Auger Observatory},
  Int.\ J.\ Mod.\ Phys.\ A {\bf 18}, 2229 (2003)
  doi:10.1142/S0217751X03013879
  [hep-ph/0206072].


\bibitem{Torres:2004hk} 
  D.~F.~Torres and L.~A.~Anchordoqui,
  {\color{rossoCP3}  Astrophysical origins of ultrahigh energy cosmic rays},
  Rept.\ Prog.\ Phys.\  {\bf 67}, 1663 (2004)
  doi:10.1088/0034-4885/67/9/R03
  [astro-ph/0402371].


\bibitem{Anchordoqui:2004xb} 
  L.~Anchordoqui, M.~T.~Dova, A.~G.~Mariazzi, T.~McCauley, T.~C.~Paul, S.~Reucroft and J.~Swain,
  {\color{rossoCP3}  High energy physics in the atmosphere: Phenomenology of cosmic ray air showers},
  Annals Phys.\  {\bf 314}, 145 (2004)
  doi:10.1016/j.aop.2004.07.003
  [hep-ph/0407020].











\bibitem{Hess:1912srp} V.~F.~Hess, 
{\color{rossoCP3} \"Uber
    Beobachtungen der durchdringenden Strahlung bei sieben
    Freiballonfahrten [Observation of penetrating radiation in seven
    free balloon flights]}, 
 Phys.\ Z.\ {\bf 13}, 1084 (1912).



\bibitem{Auger:1938ef} P.~Auger, R.~Maze and T.~Grivet-Mayer,
  {\color{rossoCP3} Grandes gerbes cosmiques atmosph\'eriques
    contenant des corpuscules ultrap\'en\'etrants [Extensive cosmic
    showers in the atmosphere containing ultra-penetrating
    particles]}, 
  Compt.\ Rend.\ Hebd.\ Seances Acad.\ Sci.\ {\bf 206}, 1721 (1938).


\bibitem{Auger:1939sh} 
  P.~Auger, P.~Ehrenfest, R.~Maze, J.~Daudin and R.~A.~Fr\'eon,
  {\color{rossoCP3} Extensive cosmic ray showers},
  Rev.\ Mod.\ Phys.\  {\bf 11}, 288 (1939).
  doi:10.1103/RevModPhys.11.288

\bibitem{Clark:1961mb} 
  G.~W.~Clark, J.~Earl, W.~L.~Kraushaar, J.~Linsley, B.~B.~Rossi, F.~Scherb and D.~W.~Scott,
   {\color{rossoCP3} Cosmic-ray air showers at sea level},
  Phys.\ Rev.\  {\bf 122}, 637 (1961).
  doi:10.1103/PhysRev.122.637


\bibitem{Linsley:1963km} 
  J.~Linsley,
  {\color{rossoCP3}  Evidence for a primary cosmic-ray particle with
    energy $10^{20}~{\rm eV}$}
  Phys.\ Rev.\ Lett.\  {\bf 10}, 146 (1963).
  doi:10.1103/PhysRevLett.10.146




\bibitem{Penzias:1965wn} 
  A.~A.~Penzias and R.~W.~Wilson,
  {\color{rossoCP3}  A measurement of excess antenna temperature at 4080-Mc/s},
  Astrophys.\ J.\  {\bf 142}, 419 (1965).
  doi:10.1086/148307



\bibitem{Stecker:1969fw} 
  F.~W.~Stecker,
  {\color{rossoCP3}  Photodisintegration of ultrahigh-energy cosmic rays by the universal radiation field},
  Phys.\ Rev.\  {\bf 180}, 1264 (1969).
  doi:10.1103/PhysRev.180.1264


\bibitem{Puget:1976nz} 
  J.~L.~Puget, F.~W.~Stecker and J.~H.~Bredekamp,
  {\color{rossoCP3}  Photonuclear interactions of ultrahigh-energy cosmic rays and their astrophysical consequences},
  Astrophys.\ J.\  {\bf 205}, 638 (1976).
  doi:10.1086/154321




\bibitem{Nagano:ve}
M.~Nagano and A.~A.~Watson,
{\color{rossoCP3} Observations and implications of the ultrahigh-energy cosmic rays},
Rev.\ Mod.\ Phys.\  {\bf 72}, 689 (2000).

\bibitem{Kampert:2012vi} 
  K.~H.~Kampert and A.~A.~Watson,
    {\color{rossoCP3} Extensive air showers and ultra-high-energy cosmic rays: a historical review},
  Eur.\ Phys.\ J.\ H {\bf 37}, 359 (2012)
  doi:10.1140/epjh/e2012-30013-x
  [arXiv:1207.4827 [physics.hist-ph]].

\bibitem{Baltrusaitis:1985mx}
  R.~M.~Baltrusaitis {\it et al.},
  {\color{rossoCP3} The Utah Fly's Eye detector},
  Nucl.\ Instrum.\ Meth.\  A {\bf 240}, 410 (1985).



\bibitem{AbuZayyad:2000uu}
  T.~Abu-Zayyad {\it et al.},
  {\color{rossoCP3} The prototype high-resolution Fly's Eye cosmic ray detector},
  Nucl.\ Instrum.\ Meth.\  A {\bf 450}, 253 (2000).




\bibitem{AbuZayyad:2012kk} 
  T.~Abu-Zayyad {\it et al.} [Telescope Array Collaboration],
   {\color{rossoCP3} The surface detector array of the Telescope Array experiment},
  Nucl.\ Instrum.\ Meth.\ A {\bf 689}, 87 (2013)
  doi:10.1016/j.nima.2012.05.079
  [arXiv:1201.4964 [astro-ph.IM]].


\bibitem{Tokuno:2012mi} 
  H.~Tokuno {\it et al.},
   {\color{rossoCP3} New air fluorescence detectors employed in the Telescope Array experiment},
  Nucl.\ Instrum.\ Meth.\ A {\bf 676}, 54 (2012)
  doi:10.1016/j.nima.2012.02.044
  [arXiv:1201.0002 [astro-ph.IM]].




\bibitem{Abraham:2004dt}
  J.~Abraham {\it et al.}  [Pierre Auger Collaboration],
  {\color{rossoCP3} Properties and performance of the prototype instrument for the Pierre Auger
   Observatory},
  Nucl.\ Instrum.\ Meth.\  A {\bf 523}, 50 (2004).

\bibitem{Aab:2017njo} 
  A.~Aab {\it et al.} [Pierre Auger Collaboration],
   {\color{rossoCP3} The Pierre Auger Observatory: Contributions to the 35th International Cosmic Ray Conference (ICRC 2017)},
  arXiv:1708.06592 [astro-ph.HE].




\bibitem{Abraham:2010zz}
  J.~Abraham {\it et al.}  [Pierre Auger  Collaboration],
  {\color{rossoCP3} Trigger and aperture of the surface detector array of the Pierre Auger
  Observatory},
  Nucl.\ Instrum.\ Meth.\  A {\bf 613}, 29 (2010).

\bibitem{Abraham:2009pm}
  J.~A.~Abraham {\it et al.}  [Pierre Auger Collaboration],
  {\color{rossoCP3} The fluorescence detector of the Pierre Auger Observatory},
  Nucl.\ Instrum.\ Meth.\  A {\bf 620}, 227 (2010)
  [arXiv:0907.4282].

\bibitem{:2010zzl}
  P.~Abreu {\it et al.}  [Pierre Auger  Collaboration],
  {\color{rossoCP3} The exposure of the hybrid detector of the Pierre Auger Observatory},
  Astropart.\ Phys.\  {\bf 34}, 368 (2011)
  [arXiv:1010.6162 [astro-ph.HE]].








\bibitem{Rybicki}
G. B. Rybicki and A. P. Lightman, 
 {\color{rossoCP3} Radiative Processes in Astrophysics}, 
(John Wiley \& Sons, Massachusetts, 1979) ISBN 978-0-471-82759-7.



\bibitem{Kachelriess:2008ze} 
  M.~Kachelriess,
   {\color{rossoCP3} Lecture notes on high energy cosmic rays},
  arXiv:0801.4376 [astro-ph].



\bibitem{Maurin:2013lwa} 
  D.~Maurin, F.~Melot and R.~Taillet,
    {\color{rossoCP3} A database of charged cosmic rays},
  Astron.\ Astrophys.\  {\bf 569}, A32 (2014)
  doi:10.1051/0004-6361/201321344
  [arXiv:1302.5525 [astro-ph.HE]].


\bibitem{Anchordoqui:1998nq} 
  L.~A.~Anchordoqui, M.~T.~Dova, L.~N.~Epele and S.~J.~Sciutto,
 {\color{rossoCP3} Hadronic interactions models beyond collider energies},
  Phys.\ Rev.\ D {\bf 59}, 094003 (1999)
  doi:10.1103/PhysRevD.59.094003
  [hep-ph/9810384].


\bibitem{GarciaCanal:2009xq} 
  C.~A.~Garcia Canal, S.~J.~Sciutto and T.~Tarutina,
   {\color{rossoCP3} Testing hadronic interaction packages at cosmic ray energies},
  Phys.\ Rev.\ D {\bf 79}, 054006 (2009)
  doi:10.1103/PhysRevD.79.054006
  [arXiv:0903.2409 [astro-ph.HE]].


\bibitem{Gondolo:1995fq} 
  P.~Gondolo, G.~Ingelman and M.~Thunman,
   {\color{rossoCP3} Charm production and high-energy atmospheric muon and neutrino fluxes},
  Astropart.\ Phys.\  {\bf 5}, 309 (1996)
  doi:10.1016/0927-6505(96)00033-3
  [hep-ph/9505417].



\bibitem{Aab:2016hkv} 
  A.~Aab {\it et al.} [Pierre Auger Collaboration],
  {\color{rossoCP3} Testing hadronic interactions at ultrahigh energies with air showers measured by the Pierre Auger Observatory},
  Phys.\ Rev.\ Lett.\  {\bf 117}, 192001 (2016)
  doi:10.1103/PhysRevLett.117.192001
  [arXiv:1610.08509 [hep-ex]].



\bibitem{Linsley:1981gh} 
  J.~Linsley and A.~A.~Watson,
   {\color{rossoCP3} Validity of scaling to $10^{20}~{\rm eV}$ and high-energy cosmic ray composition},
  Phys.\ Rev.\ Lett.\  {\bf 46}, 459 (1981).
  doi:10.1103/PhysRevLett.46.459





\bibitem{Landau:um}
L.~D.~Landau and I.~Pomeranchuk,
  {\color{rossoCP3} Limits of applicability of the theory of Bremsstrahlung electrons and pair
production at high-energies},
Dokl.\ Akad.\ Nauk Ser.\ Fiz.\  {\bf 92}, 535 (1953).

\bibitem{Migdal:1956tc} 
  A.~B.~Migdal,
 {\color{rossoCP3}  Bremsstrahlung and pair production in condensed media at high-energies},
  Phys.\ Rev.\  {\bf 103}, 1811 (1956).
  doi:10.1103/PhysRev.103.1811



\bibitem{Aab:2016agp} 
A.~Aab {\it et al.} [Pierre Auger  Collaboration], 
{\color{rossoCP3} Search for photons with energies
    above 10$^{18}$~eV using the hybrid detector of the Pierre Auger
    Observatory}, JCAP {\bf 1704} (2017) no.04, 009
  doi:10.1088/1475-7516/2017/04/009 [arXiv:1612.01517 [astro-ph.HE]].



\bibitem{Abraham:2006ar} 
  J.~Abraham {\it et al.} [Pierre Auger Collaboration],
    {\color{rossoCP3} An upper limit to the photon fraction in cosmic rays above $10^{19}$-eV from the Pierre Auger Observatory},
  Astropart.\ Phys.\  {\bf 27}, 155 (2007)
  doi:10.1016/j.astropartphys.2006.10.004
  [astro-ph/0606619].


\bibitem{Aglietta:2007yx} 
  J.~Abraham {\it et al.} [Pierre Auger Collaboration],
    {\color{rossoCP3} Upper limit on the cosmic-ray photon flux above $10^{19}$~eV using the surface detector of the Pierre Auger Observatory},
  Astropart.\ Phys.\  {\bf 29}, 243 (2008)
  doi:10.1016/j.astropartphys.2008.01.003
  [arXiv:0712.1147 [astro-ph]].

\bibitem{Abraham:2009qb} 
  J.~Abraham {\it et al.} [Pierre Auger Collaboration],
    {\color{rossoCP3} Upper limit on the cosmic-ray photon fraction at EeV energies from the Pierre Auger Observatory},
  Astropart.\ Phys.\  {\bf 31}, 399 (2009)
  doi:10.1016/j.astropartphys.2009.04.003
  [arXiv:0903.1127 [astro-ph.HE]].





\bibitem{Abraham:2010yv} 
  J.~Abraham {\it et al.} [Pierre Auger Collaboration],
    {\color{rossoCP3} Measurement of the depth of maximum of extensive air showers above $10^{18}$~eV},
  Phys.\ Rev.\ Lett.\  {\bf 104}, 091101 (2010)
  doi:10.1103/PhysRevLett.104.091101
  [arXiv:1002.0699 [astro-ph.HE]].


\bibitem{Aab:2014kda} 
  A.~Aab {\it et al.} [Pierre Auger Collaboration],
    {\color{rossoCP3} Depth of maximum of air-shower profiles at the Pierre Auger Observatory I: Measurements at energies above $10^{17.8}$~eV},
  Phys.\ Rev.\ D {\bf 90}, no. 12, 122005 (2014)
  doi:10.1103/PhysRevD.90.122005
  [arXiv:1409.4809 [astro-ph.HE]].


\bibitem{ObservatoryMichaelUngerforthePierreAuger:2017fhr} 
  M.~Unger [Pierre Auger Collaboration],
    {\color{rossoCP3} Highlights from the Pierre Auger Observatory},
  PoS ICRC {\bf 2017}, 1102 (2017)
  [arXiv:1710.09478 [astro-ph.HE]].




\bibitem{Abbasi:2014sfa} 
  R.~U.~Abbasi {\it et al.},
  {\color{rossoCP3} Study of ultrahigh energy cosmic ray composition using Telescope Array’s Middle Drum detector and surface array in hybrid mode},
  Astropart.\ Phys.\  {\bf 64}, 49 (2014)
  doi:10.1016/j.astropartphys.2014.11.004
  [arXiv:1408.1726 [astro-ph.HE]].





\bibitem{Abbasi:2015xga} 
  R.~Abbasi {\it et al.} [Pierre Auger and Telescope Array Collaborations],
  {\color{rossoCP3} Report of the working group on the composition of ultrahigh energy cosmic rays},
  JPS Conf.\ Proc.\  {\bf 9}, 010016 (2016)
  doi:10.7566/JPSCP.9.010016
  [arXiv:1503.07540 [astro-ph.HE]].


\bibitem{Abbasi:2018nun} 
  R.~U.~Abbasi {\it et al.} [Telescope Array Collaboration],
  {\color{rossoCP3} Depth of ultra-high energy cosmic ray induced air shower maxima measured by the Telescope Array Black Rock and Long Ridge FADC fluorescence detectors and surface array in hybrid mode},
  arXiv:1801.09784 [astro-ph.HE].

\bibitem{Hanlon:2018dhz} 
  W.~Hanlon {\it et al.},
    {\color{rossoCP3} Report of the working group on the mass composition of ultrahigh energy cosmic rays},
  JPS Conf.\ Proc.\  {\bf 19}, 011013 (2018).
  doi:10.7566/JPSCP.19.011013



\bibitem{Aab:2016htd} 
  A.~Aab {\it et al.} [Pierre Auger Collaboration],
    {\color{rossoCP3} Evidence for a mixed mass composition at the ``ankle'' in the cosmic-ray spectrum},
  Phys.\ Lett.\ B {\bf 762}, 288 (2016)
  doi:10.1016/j.physletb.2016.09.039
  [arXiv:1609.08567 [astro-ph.HE]].

\bibitem{Aab:2017cgk} 
  A.~Aab {\it et al.} [Pierre Auger Collaboration],
    {\color{rossoCP3} Inferences on mass composition and tests of hadronic interactions from 0.3 to 100~EeV using the water-Cherenkov detectors of the Pierre Auger Observatory},
  Phys.\ Rev.\ D {\bf 96}, no. 12, 122003 (2017)
  doi:10.1103/PhysRevD.96.122003
  [arXiv:1710.07249 [astro-ph.HE]].




\bibitem{Ivanov:2015pqx} 
  D.~Ivanov,
    {\color{rossoCP3} TA spectrum summary},
  PoS ICRC {\bf 2015}, 349 (2016).



\bibitem{Engelmann:1990zz}
  J.~J.~Engelmann, P.~Ferrando, A.~Soutoul, P.~Goret and E.~Juliusson,
    {\color{rossoCP3} Charge composition and energy spectra of cosmic-ray for elements from Be to NI: Results from HEAO-3-C2},
  Astron.\ Astrophys.\  {\bf 233} (1990) 96.


\bibitem{Juliusson} E. Juliusson,
{\color{rossoCP3}  Charge composition and energy spectra of cosmic-ray nuclei at energies above 20 GeV per nucleon}
Astrophys.\ J.\  {\bf 191}, 331 (1974)
 doi:10.1086/152972.


\bibitem{Adriani:2011cu} 
  O.~Adriani {\it et al.} [PAMELA Collaboration],
    {\color{rossoCP3} PAMELA measurements of cosmic-ray proton and helium spectra},
  Science {\bf 332}, 69 (2011)
  doi:10.1126/science.1199172
  [arXiv:1103.4055 [astro-ph.HE]].


\bibitem{Adriani:2013as} 
  O.~Adriani {\it et al.},
    {\color{rossoCP3} Time dependence of the proton flux measured by PAMELA during the July 2006 - December 2009 solar minimum},
  Astrophys.\ J.\  {\bf 765}, 91 (2013)
  doi:10.1088/0004-637X/765/2/91
  [arXiv:1301.4108 [astro-ph.HE]].



\bibitem{Aguilar:2015ooa} 
  M.~Aguilar {\it et al.} [AMS Collaboration],
    {\color{rossoCP3} Precision measurement of the proton flux in primary cosmic rays from rigidity 1 GV to 1.8 TV with the Alpha Magnetic Spectrometer on the International Space Station},
  Phys.\ Rev.\ Lett.\  {\bf 114}, 171103 (2015).
  doi:10.1103/PhysRevLett.114.171103



\bibitem{Aguilar:2015ctt} 
  M.~Aguilar {\it et al.} [AMS Collaboration],
    {\color{rossoCP3} Precision measurement of the helium flux in primary cosmic rays of rigidities 1.9 GV to 3 TV with the Alpha Magnetic Spectrometer on the International Space Station},
  Phys.\ Rev.\ Lett.\  {\bf 115}, no. 21, 211101 (2015).
  doi:10.1103/PhysRevLett.115.211101



\bibitem{Aguilar:2017hno} 
  M.~Aguilar {\it et al.} [AMS Collaboration],
  Phys.\ Rev.\ Lett.\  {\bf 119}, no. 25, 251101 (2017).
  doi:10.1103/PhysRevLett.119.251101


\bibitem{Ahn:2009tb} 
  H.~S.~Ahn {\it et al.},
    {\color{rossoCP3} Energy spectra of cosmic-ray nuclei at high energies},
  Astrophys.\ J.\  {\bf 707}, 593 (2009)
  doi:10.1088/0004-637X/707/1/593
  [arXiv:0911.1889 [astro-ph.HE]].



\bibitem{Yoon:2011aa} 
  Y.~S.~Yoon {\it et al.},
  {\color{rossoCP3}  Cosmic-ray proton and helium spectra from the first CREAM flight},
  Astrophys.\ J.\  {\bf 728}, 122 (2011)
  doi:10.1088/0004-637X/728/2/122
  [arXiv:1102.2575 [astro-ph.HE]].


\bibitem{Maestro:2009zz} 
  P.~Maestro {\it et al.},
    {\color{rossoCP3} Measurements of cosmic-ray energy spectra with the 2nd CREAM flight},
  Nucl.\ Phys.\ Proc.\ Suppl.\  {\bf 196}, 239 (2009)
  doi:10.1016/j.nuclphysbps.2009.09.045
  [arXiv:1003.5757 [astro-ph.HE]].


\bibitem{Yoon:2017qjx} 
  Y.~S.~Yoon {\it et al.},
    {\color{rossoCP3} Proton and helium spectra from the CREAM-III flight},
  Astrophys.\ J.\  {\bf 839}, no. 1, 5 (2017)
  doi:10.3847/1538-4357/aa68e4
  [arXiv:1704.02512 [astro-ph.HE]].



\bibitem{deNolfo:2006qj} 
  G.~A.~de Nolfo {\it et al.},
    {\color{rossoCP3} Observations of the Li, Be, and B isotopes and constraints on cosmic-ray propagation},
  Adv.\ Space Res.\  {\bf 38}, 1558 (2006)
  doi:10.1016/j.asr.2006.09.008
  [astro-ph/0611301].

\bibitem{Lave}
K. A. Lave {\it et  al.},
   {\color{rossoCP3} Galactic cosmic-ray energy spectra and composition during the 2009-2010 solar minimum period},
 Astrophys.\ J.\  {\bf 770}, 117 (2013)
  doi:10.1088//0004-637X/770/2/117.

\bibitem{Swordy} S. P. Swordy, D. Mueller, P. Meyer, J. L'Heureux,
  J. M.  Grunsfeld, 
 {\color{rossoCP3} Relative abundances of secondary and primary cosmic
   rays at high energies}
Astrophys.\ J.\  {\bf 349}, 625 (1990)
 doi:10.1086/168349.

\bibitem{Mueller} D. Mueller, S. P.  Swordy, P. Meyer, J. L'Heureux,
  J. M. Grunsfeld, 
 {\color{rossoCP3} Energy spectra and composition of primary cosmic rays}
Astrophys.\ J.\  {\bf 374}, 356 (1991)
 doi:10.1086/170125.


\bibitem{Aharonian:2007zja} 
  F.~Aharonian {\it et al.} [H.E.S.S. Collaboration],
    {\color{rossoCP3} First ground based measurement of atmospheric Cherenkov light from cosmic rays},
  Phys.\ Rev.\ D {\bf 75}, 042004 (2007)
  doi:10.1103/PhysRevD.75.042004
  [astro-ph/0701766 [ASTRO-PH]].



\bibitem{Montini:2016fvq} 
  P.~Montini {\it et al.} [ARGO-YBJ Collaboration],
    {\color{rossoCP3} The bending of the proton plus helium flux in primary cosmic rays measured by the ARGO-YBJ experiment in the energy range from 20 TeV to 5 PeV},
  arXiv:1608.01389 [hep-ex].


\bibitem{Prosin:2014dxa} 
  V.~V.~Prosin {\it et al.},
    {\color{rossoCP3} Tunka-133: Results of 3 year operation},
  Nucl.\ Instrum.\ Meth.\ A {\bf 756}, 94 (2014).
  doi:10.1016/j.nima.2013.09.018



\bibitem{Korosteleva:2007ek} 
  E.~E.~Korosteleva, V.~V.~Prosin, L.~A.~Kuzmichev and G.~Navarra,
    {\color{rossoCP3} Measurement of cosmic ray primary energy with the atmospheric Cherenkov light technique in extensive air showers},
  Nucl.\ Phys.\ Proc.\ Suppl.\  {\bf 165}, 74 (2007).
  doi:10.1016/j.nuclphysbps.2006.11.012



\bibitem{Aartsen:2015awa} 
  M.~G.~Aartsen {\it et al.} [IceCube Collaboration],
    {\color{rossoCP3} The IceCube neutrino observatory contributions to ICRC 2015 Part III: cosmic rays},''
  arXiv:1510.05225 [astro-ph.HE].


\bibitem{Apel:2012rm} 
  W.~D.~Apel {\it et al.} [KASCADE-Grande Collaboration],
    {\color{rossoCP3} The spectrum of high-energy cosmic rays measured with KASCADE-Grande},
  arXiv:1206.3834 [astro-ph.HE].



\bibitem{Schoo:2015oxd} 
  S.~Schoo {\it et al.} [KASCADE-Grande Collaboration],
    {\color{rossoCP3} The energy spectrum of cosmic rays in the range from $10^{14}$ to $10^{18}\mathrm{eV}$},
  PoS ICRC {\bf 2015}, 263 (2016).





\bibitem{Dembinski:2017zsh} 
  H.~P.~Dembinski, R.~Engel, A.~Fedynitch, T.~Gaisser, F.~Riehn and T.~Stanev,
    {\color{rossoCP3} Data-driven model of the cosmic-ray flux and mass composition from 10~GeV to $10^{11}$~GeV},
  PoS ICRC {\bf 2017}, 533 (2017)
  [arXiv:1711.11432 [astro-ph.HE]].

\bibitem{Gleeson:1968zza} 
  L.~J.~Gleeson and W.~I.~Axford,
    {\color{rossoCP3} Solar modulation of Galactic cosmic rays},
  Astrophys.\ J.\  {\bf 154}, 1011 (1968).
  doi:10.1086/149822

\bibitem{Usoskin} 
  I.~G.~Usoskin, G. A.~Bazilevskaya, and G.~A.~Kovaltsov, 
   {\color{rossoCP3} Solar modulation parameter for cosmic rays since1936
reconstructed from ground-based neutron monitors and ionization chambers},
  J.\ Geophys.\ Res.\ Space Phys.\  {\bf 116}, A02104 (2011)
  doi:10.1029/2010JA016105.
  

\bibitem{Abbasi:2018ygn} 
  R.~U.~Abbasi {\it et al.},
    {\color{rossoCP3} Evidence for declination dependence of ultra-high-energy cosmic ray spectrum in the Northern hemisphere},
  arXiv:1801.07820 [astro-ph.HE].


\bibitem{Abbasi:2017vru} 
  R.~U.~Abbasi {\it et al.} [Telescope Array Collaboration],
    {\color{rossoCP3} Search for anisotropy in the ultra-high-energy cosmic ray spectrum using the Telescope Array surface detector},
  [arXiv:1707.04967 [astro-ph.HE]].


\bibitem{Aab:2014aea} 
  A.~Aab {\it et al.} [Pierre Auger Collaboration],
    {\color{rossoCP3} Depth of maximum of air-shower profiles at the Pierre Auger Observatory II: Composition implications},''
  Phys.\ Rev.\ D {\bf 90}, no. 12, 122006 (2014)
  doi:10.1103/PhysRevD.90.122006
  [arXiv:1409.5083 [astro-ph.HE]].




\bibitem{Aab:2015bza} 
  A.~Aab {\it et al.} [Pierre Auger Collaboration],
    {\color{rossoCP3} The Pierre Auger Observatory contributions to the 34th International Cosmic Ray Conference (ICRC 2015)},
  arXiv:1509.03732 [astro-ph.HE].







\bibitem{Peters}
B. Peters,
{\color{rossoCP3}   Primary cosmic radiation and extensive air showers}, 
Nuovo Cim.\ {\bf 22}, 800 (1961)
doi:10.1007/BF02783106

\bibitem{Abbasi:2018xsn} 
  R.~U.~Abbasi {\it et al.},
    {\color{rossoCP3} The cosmic-ray energy spectrum between 2~PeV and 2~EeV observed with the TALE detector in monocular mode},
  arXiv:1803.01288 [astro-ph.HE].





\bibitem{Hillas:1967} M. Hillas,
  {\color{rossoCP3} The energy spectrum of cosmic rays in an evolving universe},
  Phys. Lett. A {\bf 24}, 677 (1967)
doi:10.1016/0375-9601(67)91023-7.



\bibitem{Berezinsky:2002nc} 
  V.~Berezinsky, A.~Z.~Gazizov and S.~I.~Grigorieva,
  {\color{rossoCP3} On astrophysical solution to ultrahigh-energy cosmic rays},
  Phys.\ Rev.\ D {\bf 74}, 043005 (2006)
  doi:10.1103/PhysRevD.74.043005
  [hep-ph/0204357].




\bibitem{Aloisio:2013hya}
  R.~Aloisio, V.~Berezinsky and P.~Blasi,
  {\color{rossoCP3} Ultrahigh energy cosmic rays: implications of Auger data for source spectra and chemical composition},
  JCAP {\bf 1410}, no. 10, 020 (2014)
  doi:10.1088/1475-7516/2014/10/020
  [arXiv:1312.7459 [astro-ph.HE]].




\bibitem{Unger:2015laa} 
  M.~Unger, G.~R.~Farrar and L.~A.~Anchordoqui,
 {\color{rossoCP3} Origin of the ankle in the ultrahigh energy cosmic ray spectrum, and of the extragalactic protons below it},
  Phys.\ Rev.\ D {\bf 92}, no. 12, 123001 (2015)
  doi:10.1103/PhysRevD.92.123001
  [arXiv:1505.02153 [astro-ph.HE]].


\bibitem{Anchordoqui:2014pca} 
  L.~A.~Anchordoqui,
  {\color{rossoCP3} Neutron $\beta$-decay as the origin of IceCube’s PeV (anti)neutrinos},
  Phys.\ Rev.\ D {\bf 91}, 027301 (2015)
  doi:10.1103/PhysRevD.91.027301
  [arXiv:1411.6457 [astro-ph.HE]].

\bibitem{Farrar:2015ikt} 
  G.~R.~Farrar, M.~Unger and L.~Anchordoqui,
    {\color{rossoCP3} The origin of the ankle in the UHECR spectrum, and of the extragalactic protons below it},
  PoS ICRC {\bf 2015}, 513 (2016)
  doi:10.22323/1.236.0513
  [arXiv:1512.00484 [astro-ph.HE]].


\bibitem{Aab:2016zth} 
  A.~Aab {\it et al.} [Pierre Auger Collaboration],
    {\color{rossoCP3} Combined fit of spectrum and composition data as measured by the Pierre Auger Observatory},
  JCAP {\bf 1704}, 038 (2017)
  Erratum: [JCAP {\bf 1803},  E02 (2018)]
  doi:10.1088/1475-7516/2018/03/E02, 10.1088/1475-7516/2017/04/038
  [arXiv:1612.07155 [astro-ph.HE]].




\bibitem{Sommers:2000us} 
  P.~Sommers,
 {\color{rossoCP3}   Cosmic ray anisotropy analysis with a full-sky observatory},
  Astropart.\ Phys.\  {\bf 14}, 271 (2001)
  doi:10.1016/S0927-6505(00)00130-4
  [astro-ph/0004016].


\bibitem{diMatteo:2018vmr} 
  A.~di Matteo {\it et al.},
   {\color{rossoCP3}  Arrival directions of cosmic rays at ultra-high energies},
  JPS Conf.\ Proc.\  {\bf 19}, 011020 (2018).
  doi:10.7566/JPSCP.19.011020


\bibitem{Abbasi:2014lda} 
  R.~U.~Abbasi {\it et al.} [Telescope Array Collaboration],
 {\color{rossoCP3}  Indications of intermediate-scale anisotropy of cosmic rays with energy greater than 57 EeV in the Northern sky measured with the surface detector of the Telescope Array Experiment},
  Astrophys.\ J.\  {\bf 790}, L21 (2014)
  doi:10.1088/2041-8205/790/2/L21
  [arXiv:1404.5890 [astro-ph.HE]].


\bibitem{Abraham:2007bb} 
  J.~Abraham {\it et al.} [Pierre Auger Collaboration],
  {\color{rossoCP3}  Correlation of the highest energy cosmic rays with nearby extragalactic objects},
  Science {\bf 318}, 938 (2007)
  doi:10.1126/science.1151124
  [arXiv:0711.2256 [astro-ph]].





\bibitem{PierreAuger:2014yba} 
  A.~Aab {\it et al.} [Pierre Auger Collaboration],
   {\color{rossoCP3}  Searches for anisotropies in the arrival directions of the highest energy cosmic rays detected by the Pierre Auger Observatory},
  Astrophys.\ J.\  {\bf 804}, no. 1, 15 (2015)
  doi:10.1088/0004-637X/804/1/15
  [arXiv:1411.6111 [astro-ph.HE]].

\bibitem{Li:1983fv} 
  T.-P.~Li and Y.-Q.~Ma,
   {\color{rossoCP3}  Analysis methods for results in gamma-ray astronomy},
  Astrophys.\ J.\  {\bf 272}, 317 (1983).
  doi:10.1086/161295

\bibitem{Matthews}
J. N. Matthews,
  {\color{rossoCP3}  Highlights from the Telescope Array},
PoS ICRC {\bf 2017}, 1096 (2017).

\bibitem{Abbasi:2018qlh} R.~U.~Abbasi {\it et al.}, 
{\color{rossoCP3}  Evidence of intermediate-scale energy spectrum anisotropy of
    cosmic rays $E \geq 10^{19.2}~{\rm eV}$ with the Telescope Array
    surface detector}, arXiv:1802.05003 [astro-ph.HE].


\bibitem{Anchordoqui:2003bx} 
  L.~A.~Anchordoqui, C.~Hojvat, T.~P.~McCauley, T.~C.~Paul, S.~Reucroft, J.~D.~Swain and A.~Widom,
    {\color{rossoCP3} Full-sky search for ultra-high-energy cosmic ray anisotropies},
  Phys.\ Rev.\ D {\bf 68}, 083004 (2003)
  doi:10.1103/PhysRevD.68.083004
  [astro-ph/0305158].



\bibitem{Denton:2014hfa} 
  P.~B.~Denton and T.~J.~Weiler,
    {\color{rossoCP3} The fortuitous latitude of the Pierre Auger Observatory and Telescope Array for reconstructing the quadrupole moment},
  Astrophys.\ J.\  {\bf 802}, no. 1, 25 (2015)
  doi:10.1088/0004-637X/802/1/25
  [arXiv:1409.0883 [astro-ph.HE]].


\bibitem{Aab:2014ila} 
  A.~Aab {\it et al.} [Telescope Array and Pierre Auger Collaborations],
    {\color{rossoCP3} Searches for large-scale anisotropy in the arrival directions of cosmic rays detected above energy of $10^{19}$~eV at the Pierre Auger Observatory and the Telescope Array},
  Astrophys.\ J.\  {\bf 794}, no. 2, 172 (2014)
  doi:10.1088/0004-637X/794/2/172
  [arXiv:1409.3128 [astro-ph.HE]].


\bibitem{Denton:2015bga} 
  P.~B.~Denton and T.~J.~Weiler,
    {\color{rossoCP3} Sensitivity of full-sky experiments to large scale cosmic ray anisotropies},
  JHEAp {\bf 8}, 1 (2015)
  doi:10.1016/j.jheap.2015.06.002
  [arXiv:1505.03922 [astro-ph.HE]].

\bibitem{Linsley:1975kp} 
  J.~Linsley,
   {\color{rossoCP3} Fluctuation effects on directional data},
  Phys.\ Rev.\ Lett.\  {\bf 34}, 1530 (1975).
  doi:10.1103/PhysRevLett.34.1530



\bibitem{Wittkowski:2017nfd} 
  D.~Wittkowski and K.~H.~Kampert,
   {\color{rossoCP3} On the anisotropy in the arrival directions of ultra-high-energy cosmic rays},
  Astrophys.\ J.\  {\bf 854}, no. 1, L3 (2018)
  doi:10.3847/2041-8213/aaa2f9
  [arXiv:1710.05617 [astro-ph.HE]].


\bibitem{Aublin:2005nv} J.~Aublin and E.~Parizot, {\color{rossoCP3}
    Generalized 3D-reconstruction method of a dipole anisotropy in
    cosmic-ray distributions}, Astron.\ Astrophys.\ {\bf 441}, 407
  (2005) doi:10.1051/0004-6361:20052833 [astro-ph/0504575].


\bibitem{Abreu:2011ve} 
  P.~Abreu {\it et al.} [Pierre Auger Collaboration],
   {\color{rossoCP3} Search for first harmonic modulation in the right ascension distribution of cosmic rays detected at the Pierre Auger Observatory},
  Astropart.\ Phys.\  {\bf 34}, 627 (2011)
  doi:10.1016/j.astropartphys.2010.12.007
  [arXiv:1103.2721 [astro-ph.HE]].

\bibitem{ThePierreAuger:2014nja} 
  A.~Aab {\it et al.} [Pierre Auger Collaboration],
    {\color{rossoCP3}  Large scale distribution of ultra high energy cosmic rays detected at the Pierre Auger Observatory with zenith angles up to $80^\circ$},
  Astrophys.\ J.\  {\bf 802}, no. 2, 111 (2015)
  doi:10.1088/0004-637X/802/2/111
  [arXiv:1411.6953 [astro-ph.HE]].


\bibitem{Aab:2017tyv} 
  A.~Aab {\it et al.} [Pierre Auger Collaboration],
    {\color{rossoCP3} Observation of a large-scale anisotropy in the arrival directions of cosmic rays above $8 \times 10^{18}$~eV},
  Science {\bf 357}, no. 6537, 1266 (2017)
  doi:10.1126/science.aan4338
  [arXiv:1709.07321 [astro-ph.HE]].


\bibitem{Jansson:2012pc} 
  R.~Jansson and G.~R.~Farrar,
    {\color{rossoCP3} A new model of the Galactic magnetic field},
  Astrophys.\ J.\  {\bf 757}, 14 (2012)
  doi:10.1088/0004-637X/757/1/14
  [arXiv:1204.3662 [astro-ph.GA]].


\bibitem{Jansson:2012rt} 
  R.~Jansson and G.~R.~Farrar,
   {\color{rossoCP3} The Galactic magnetic field}
  Astrophys.\ J.\  {\bf 761}, L11 (2012)
  doi:10.1088/2041-8205/761/1/L11
  [arXiv:1210.7820 [astro-ph.GA]].


\bibitem{Unger:2017kfh} 
  M.~Unger and G.~R.~Farrar,
 {\color{rossoCP3}  Uncertainties in the magnetic field of the Milky Way},
 PoS ICRC {\bf 2017}, 558 (2017)  
[arXiv:1707.02339 [astro-ph.GA]].

\bibitem{Aab:2018mmi}
  A.~Aab {\it et al.} [Pierre Auger Collaboration],
 {\color{rossoCP3}  Large-scale cosmic-ray anisotropies above 4 EeV measured by the Pierre Auger Observatory},
  Astrophys.\ J.\  {\bf 868} (2018) no.1,  4
  doi:10.3847/1538-4357/aae689
  [arXiv:1808.03579 [astro-ph.HE]].


\bibitem{Lemaitre}
G. Lemaitre and M. S. Vallarta,
  {\color{rossoCP3} On Compton's latitude effect of cosmic radiation},
Phys.\ Rev.\  {\bf 43}, 87 (1933).
doi:10.1103/PhysRev.43.87

\bibitem{Swann:1933zz} 
  W.~F.~G.~Swann,
    {\color{rossoCP3} Application of Liouville's theorem to electron orbits in the Earth's magnetic field},
  Phys.\ Rev.\  {\bf 44}, 224 (1933).
  doi:10.1103/PhysRev.44.224


\bibitem{Compton:1935wde} 
  A.~H.~Compton and I.~A.~Getting,
    {\color{rossoCP3} An apparent effect of Galactic rotation on the intensity of cosmic rays},
  Phys.\ Rev.\  {\bf 47},  817 (1935).
  doi:10.1103/PhysRev.47.817


\bibitem{Kogut:1993ag} 
  A.~Kogut {\it et al.},
    {\color{rossoCP3} Dipole anisotropy in the COBE DMR first year sky maps},
  Astrophys.\ J.\  {\bf 419}, 1 (1993)
  doi:10.1086/173453
  [astro-ph/9312056].


\bibitem{Hinshaw:2008kr} 
  G.~Hinshaw {\it et al.} [WMAP Collaboration],
    {\color{rossoCP3} Five-year Wilkinson Microwave Anisotropy Probe (WMAP) observations: Data processing, sky maps, and basic results},
  Astrophys.\ J.\ Suppl.\  {\bf 180}, 225 (2009)
  doi:10.1088/0067-0049/180/2/225
  [arXiv:0803.0732 [astro-ph]].




\bibitem{Adam:2015rua} 
  R.~Adam {\it et al.} [Planck Collaboration],
    {\color{rossoCP3} Planck 2015 results I: Overview of products and scientific results},
  Astron.\ Astrophys.\  {\bf 594}, A1 (2016)
  doi:10.1051/0004-6361/201527101
  [arXiv:1502.01582 [astro-ph.CO]].





\bibitem{Kachelriess:2006aq} 
  M.~Kachelriess and P.~D.~Serpico,
    {\color{rossoCP3} The Compton-Getting effect on ultra-high energy cosmic rays of cosmological origin},
  Phys.\ Lett.\ B {\bf 640}, 225 (2006)
  doi:10.1016/j.physletb.2006.08.006
  [astro-ph/0605462].


\bibitem{Globus:2017fym} 
  N.~Globus and T.~Piran,
    {\color{rossoCP3} The extragalactic ultra-high energy cosmic-ray dipole},
  Astrophys.\ J.\  {\bf 850}, no. 2, L25 (2017)
  doi:10.3847/2041-8213/aa991b
  [arXiv:1709.10110 [astro-ph.HE]].

\bibitem{Globus:2007bi} 
  N.~Globus, D.~Allard and E.~Parizot,
    {\color{rossoCP3} Propagation of high-energy cosmic rays in extragalactic turbulent magnetic fields: resulting energy spectrum and composition},
  Astron.\ Astrophys.\  {\bf 479}, 97 (2008)
  doi:10.1051/0004-6361:20078653
  [arXiv:0709.1541 [astro-ph]].

\bibitem{Kronberg:1993vk} 
  P.~P.~Kronberg,
    {\color{rossoCP3} Extragalactic magnetic fields},
  Rept.\ Prog.\ Phys.\  {\bf 57}, 325 (1994).
  doi:10.1088/0034-4885/57/4/001


\bibitem{Blasi:1999hu} 
  P.~Blasi, S.~Burles and A.~V.~Olinto,
   {\color{rossoCP3} Cosmological magnetic fields limits in an inhomogeneous universe},
  Astrophys.\ J.\  {\bf 514}, L79 (1999)
  doi:10.1086/311958
  [astro-ph/9812487].




\bibitem{Pshirkov:2015tua} 
  M.~S.~Pshirkov, P.~G.~Tinyakov and F.~R.~Urban,
   {\color{rossoCP3} New limits on extragalactic magnetic fields from rotation measures},
  Phys.\ Rev.\ Lett.\  {\bf 116}, no. 19, 191302 (2016)
  doi:10.1103/PhysRevLett.116.191302
  [arXiv:1504.06546 [astro-ph.CO]].




\bibitem{Parizot:2004wh} 
  E.~Parizot,
    {\color{rossoCP3} GZK horizon and magnetic fields},
  Nucl.\ Phys.\ Proc.\ Suppl.\  {\bf 136}, 169 (2004)
  doi:10.1016/j.nuclphysbps.2004.10.034
  [astro-ph/0409191].

\bibitem{Huchra:2011ii} 
  J.~P.~Huchra {\it et al.},
    {\color{rossoCP3} The 2MASS redshift survey: Description and data release},
  Astrophys.\ J.\ Suppl.\  {\bf 199}, 26 (2012)
  doi:10.1088/0067-0049/199/2/26
  [arXiv:1108.0669 [astro-ph.CO]].

\bibitem{Erdogdu:2005wi} 
  P.~Erdogdu {\it et al.},
    {\color{rossoCP3} The dipole anisotropy of the 2 Micron All-Sky Redshift Survey},
  Mon.\ Not.\ Roy.\ Astron.\ Soc.\  {\bf 368}, 1515 (2006)
  doi:10.1111/j.1365-2966.2006.10243.x
  [astro-ph/0507166].



\bibitem{Harari:2015hba} 
  D.~Harari, S.~Mollerach and E.~Roulet,
    {\color{rossoCP3} Anisotropies of ultrahigh energy cosmic ray nuclei diffusing from extragalactic sources},
  Phys.\ Rev.\ D {\bf 92}, no. 6, 063014 (2015)
  doi:10.1103/PhysRevD.92.063014
  [arXiv:1507.06585 [astro-ph.HE]].

\bibitem{Clay:2003pv} 
  R.~W.~Clay [Pierre Auger Collaboration],
   {\color{rossoCP3}  The anisotropy search program for the Pierre Auger Observatory},
  astro-ph/0308494.



\bibitem{Ackermann:2015uya} 
  M.~Ackermann {\it et al.} [Fermi-LAT Collaboration],
    {\color{rossoCP3} 2FHL: The second catalog of hard \textit{Fermi}-LAT sources},
  Astrophys.\ J.\ Suppl.\  {\bf 222}, no. 1, 5 (2016)
  doi:10.3847/0067-0049/222/1/5
  [arXiv:1508.04449 [astro-ph.HE]].



\bibitem{Ackermann:2012vca} 
  M.~Ackermann {\it et al.} [Fermi-LAT Collaboration],
    {\color{rossoCP3} GeV observations of star-forming galaxies with \textit{Fermi}-LAT},
  Astrophys.\ J.\  {\bf 755}, 164 (2012)
  doi:10.1088/0004-637X/755/2/164
  [arXiv:1206.1346 [astro-ph.HE]].




\bibitem{Tang:2014dia} 
  Q.~W.~Tang, X.~Y.~Wang and P.~H.~Thomas Tam,
    {\color{rossoCP3} Discovery of GeV emission from the direction of the luminous infrared galaxy NGC 2146},
  Astrophys.\ J.\  {\bf 794}, no. 1, 26 (2014)
  doi:10.1088/0004-637X/794/1/26
  [arXiv:1407.3391 [astro-ph.HE]].


\bibitem{Peng:2016nsx} 
  F.~K.~Peng, X.~Y.~Wang, R.~Y.~Liu, Q.~W.~Tang and J.~F.~Wang,
    {\color{rossoCP3} First detection of GeV emission from an
      ultraluminous infrared galaxy: Arp 220 as seen with the {\it Fermi} Large Area Telescope},
  Astrophys.\ J.\  {\bf 821}, no. 2, L20 (2016)
  doi:10.3847/2041-8205/821/2/L20
  [arXiv:1603.06355 [astro-ph.HE]].


\bibitem{Hayashida:2013wha} 
  M.~Hayashida {\it et al.},
    {\color{rossoCP3} Discovery of GeV emission from the Circinus
      galaxy with the {\it Fermi} Large Area Telescope},
  Astrophys.\ J.\  {\bf 779}, 131 (2013)
  doi:10.1088/0004-637X/779/2/131
  [arXiv:1310.1913 [astro-ph.HE]].


\bibitem{Acciari:2009wq} 
  V.~A.~Acciari {\it et al.},
    {\color{rossoCP3} A connection between star formation activity and cosmic rays in the starburst galaxy M82},
  Nature {\bf 462}, 770 (2009)
  doi:10.1038/nature08557
  [arXiv:0911.0873 [astro-ph.CO]].

\bibitem{Abdalla:2018nlz} 
  H.~Abdalla {\it et al.} [H.E.S.S. Collaboration],
   {\color{rossoCP3} The starburst galaxy NGC 253 revisited by H.E.S.S. and Fermi-LAT},
  [arXiv:1806.03866 [astro-ph.HE]].



\bibitem{Fisher}
R. A. Fisher, 
{\color{rossoCP3}{Dispersion on a sphere}}
Proc. Roy. Soc. London Ser. A. {\bf 217} 295 (1953).

\bibitem{Abreu:2010ab} 
  P.~Abreu {\it et al.} [Pierre Auger Collaboration],
    {\color{rossoCP3} Update on the correlation of the highest energy cosmic rays with nearby extragalactic matter},
  Astropart.\ Phys.\  {\bf 34}, 314 (2010)
  doi:10.1016/j.astropartphys.2010.08.010
  [arXiv:1009.1855 [astro-ph.HE]].



\bibitem{Harari:2008zp} 
  D.~Harari, S.~Mollerach and E.~Roulet,
    {\color{rossoCP3} Kolmogorov-Smirnov test as a tool to study the distribution of ultra-high energy cosmic ray sources},
  Mon.\ Not.\ Roy.\ Astron.\ Soc.\  {\bf 394}, 916 (2009)
  doi:10.1111/j.1365-2966.2008.14327.x
  [arXiv:0811.0008 [astro-ph]].



\bibitem{Becker:2009hw} 
  J.~K.~Becker, P.~L.~Biermann, J.~Dreyer and T.~M.~Kneiske,
    {\color{rossoCP3} Cosmic Rays VI: Starburst galaxies at multiwavelengths},
  arXiv:0901.1775 [astro-ph.HE].


\bibitem{Acero:2015hja} 
  F.~Acero {\it et al.} [Fermi-LAT Collaboration],
    {\color{rossoCP3} {\it Fermi} Large Area Telescope third source catalog},
  Astrophys.\ J.\ Suppl.\  {\bf 218}, no. 2, 23 (2015)
  doi:10.1088/0067-0049/218/2/23
  [arXiv:1501.02003 [astro-ph.HE]].



\bibitem{Anchordoqui:2014yva} 
  L.~A.~Anchordoqui, T.~C.~Paul, L.~H.~M.~da Silva, D.~F.~Torres and B.~J.~Vlcek,
     {\color{rossoCP3}   What IceCube data tell us about neutrino emission from star-forming galaxies (so far)},
  Phys.\ Rev.\ D {\bf 89}, no. 12, 127304 (2014)
  doi:10.1103/PhysRevD.89.127304
  [arXiv:1405.7648 [astro-ph.HE]].


\bibitem{Fang:2014uja} 
  K.~Fang, T.~Fujii, T.~Linden and A.~V.~Olinto,
     {\color{rossoCP3}   Is the ultra-high energy cosmic-ray excess observed by the Telescope Array correlated with IceCube neutrinos?},
  Astrophys.\ J.\  {\bf 794}, no. 2, 126 (2014)
  doi:10.1088/0004-637X/794/2/126
  [arXiv:1404.6237 [astro-ph.HE]].


\bibitem{He:2014mqa} 
  H.~N.~He, A.~Kusenko, S.~Nagataki, B.~B.~Zhang, R.~Z.~Yang and Y.~Z.~Fan,
    {\color{rossoCP3} Monte Carlo Bayesian search for the plausible source of the
  Telescope Array hot spot},
  Phys.\ Rev.\ D {\bf 93}, 043011 (2016)
  doi:10.1103/PhysRevD.93.043011
  [arXiv:1411.5273 [astro-ph.HE]].

\bibitem{Pfeffer:2015idq} 
  D.~N.~Pfeffer, E.~D.~Kovetz and M.~Kamionkowski,
     {\color{rossoCP3} Ultrahigh-energy cosmic ray hotspots from tidal disruption events},
  Mon.\ Not.\ Roy.\ Astron.\ Soc.\  {\bf 466}, no. 3, 2922 (2017)
  doi:10.1093/mnras/stw3337
  [arXiv:1512.04959 [astro-ph.HE]].

\bibitem{Attallah:2018euc} 
  R.~Attallah and D.~Bouchachi,
    {\color{rossoCP3} Ultra-high-energy cosmic rays from nearby starburst galaxies},
  Mon.\ Not.\ Roy.\ Astron.\ Soc.\  {\bf 478}, no. 1, 800 (2018)
 doi:10.1093/mnras/sty986
 [arXiv:1804.06603 [astro-ph.HE]].


\bibitem{Anchordoqui:2017abg} 
  L.~A.~Anchordoqui, V.~Barger and T.~J.~Weiler,
    {\color{rossoCP3} Cosmic mass spectrometer},
  JHEAp {\bf 17}, 38 (2018)
  doi:10.1016/j.jheap.2017.12.001
  [arXiv:1707.05408 [astro-ph.HE]].


\bibitem{Globus:2016gvy} 
  N.~Globus, D.~Allard, E.~Parizot, C.~Lachaud and T.~Piran,
   {\color{rossoCP3} Can we reconcile the TA excess and hotspot with Auger observations?},
  Astrophys.\ J.\  {\bf 836}, no. 2, 163 (2017)
  doi:10.3847/1538-4357/836/2/163
  [arXiv:1610.05319 [astro-ph.HE]].


\bibitem{Anchordoqui:2002dj} 
 L.~A.~Anchordoqui, H.~Goldberg and D.~F.~Torres,
  {\color{rossoCP3}  Anisotropy at the end of the cosmic ray spectrum?},
  Phys.\ Rev.\ D {\bf 67}, 123006 (2003)
  doi:10.1103/PhysRevD.67.123006
  [astro-ph/0209546].




\bibitem{LetessierSelvon:2011dy} 
  A.~Letessier-Selvon and T.~Stanev,
    {\color{rossoCP3} Ultrahigh energy cosmic rays},
  Rev.\ Mod.\ Phys.\  {\bf 83}, 907 (2011)
  doi:10.1103/RevModPhys.83.907
  [arXiv:1103.0031 [astro-ph.HE]].


\bibitem{Gorbunov:2008ef} 
  D.~S.~Gorbunov, P.~G.~Tinyakov, I.~I.~Tkachev and S.~V.~Troitsky,
    {\color{rossoCP3} On the interpretation of the cosmic-ray anisotropy at ultra-high energies},
  arXiv:0804.1088 [astro-ph].



\bibitem{Matthews:2018laz} 
  J.~H.~Matthews, A.~R.~Bell, K.~M.~Blundell and A.~T.~Araudo,
   {\color{rossoCP3}  Fornax A, Centaurus A and other radio galaxies as sources of ultra-high energy cosmic rays},
  doi:10.1093/mnrasl/sly099
  arXiv:1805.01902 [astro-ph.HE].

\bibitem{Smida:2015kga} 
  R.~Smida and R.~Engel,
   {\color{rossoCP3}   The ultra-high-energy cosmic rays image of Virgo A},
  PoS ICRC {\bf 2015}, 470 (2016)
  doi:10.22323/1.236.0470
  [arXiv:1509.09033 [astro-ph.HE]].

\bibitem{Anjos:2018mgr} 
 R.~C.~d.~Anjos, J. F. Soriano, L. A. Anchordoqui, T. C. Paul,
 D. F. Torres, J. F. Krizmanic, T. A. D. Paglione, R. J. Moncada,
 F. Sarazin, L. Wiencke, and A. V. Olinto,
 {\color{rossoCP3} Ultrahigh-energy cosmic ray composition from the
   distribution of arrival directions},
Phys.\ Rev.\ D {\bf 98}, 123018 (2018)
  doi:10.1103/PhysRevD.98.123018
  [arXiv:1810.04251 [astro-ph.HE]].

\bibitem{Biteau:2018paris}
J. Biteau {\it et al.}  [Telescope Array and Pierre Auger Collaborations],
 {\color{rossoCP3} Covering the sphere at ultra-high energies:
   full-sky cosmic-ray maps beyond the ankle and the flux
   suppression,}
To be published in Proceedings of Ultra High Energy Cosmic Rays 2018,
8 - 12 October 2018, Paris.

\bibitem{Abbasi:2018tqo} 
  R.~U.~Abbasi {\it et al.} [Telescope Array Collaboration],
   {\color{rossoCP3} Testing a reported correlation between arrival directions of ultra-high-energy cosmic rays and a flux pattern from nearby starburst galaxies using Telescope Array data},
  Astrophys.\ J.\  {\bf 867}, no. 2, L27 (2018)
  doi:10.3847/2041-8213/aaebf9
  [arXiv:1809.01573 [astro-ph.HE]].

\bibitem{AbuZayyad:2012ru}
  T.~Abu-Zayyad {\it et al.} [Telescope Array Collaboration],
   {\color{rossoCP3} The cosmic ray energy spectrum observed with the surface detector of the Telescope Array experiment},
  Astrophys.\ J.\  {\bf 768}, L1 (2013)
  doi:10.1088/2041-8205/768/1/L1
  [arXiv:1205.5067 [astro-ph.HE]].




\bibitem{Lemoine:2009pw} 
  M.~Lemoine and E.~Waxman,
    {\color{rossoCP3} Anisotropy vs chemical composition at ultra-high energies},
  JCAP {\bf 0911}, 009 (2009)
  doi:10.1088/1475-7516/2009/11/009
  [arXiv:0907.1354 [astro-ph.HE]].


\bibitem{Liu:2013ppa} 
  R.~Y.~Liu, A.~M.~Taylor, M.~Lemoine, X.~Y.~Wang and E.~Waxman,
    {\color{rossoCP3} Constraints on the source of ultra-high-energy cosmic rays using anisotropy versus chemical composition},
  Astrophys.\ J.\  {\bf 776}, 88 (2013)
  doi:10.1088/0004-637X/776/2/88
  [arXiv:1308.5699 [astro-ph.HE]].





\bibitem{Aab:2016vlz} 
  A.~Aab {\it et al.} [Pierre Auger Collaboration],
    {\color{rossoCP3} The Pierre Auger Observatory upgrade: Preliminary design report},
  arXiv:1604.03637 [astro-ph.IM].



\bibitem{Aab:2016eeq} 
  A.~Aab {\it et al.} [Pierre Auger Collaboration],
    {\color{rossoCP3} Measurement of the radiation energy in the radio signal of extensive air showers as a universal estimator of cosmic-ray energy},
  Phys.\ Rev.\ Lett.\  {\bf 116}, no. 24, 241101 (2016)
  doi:10.1103/PhysRevLett.116.241101
  [arXiv:1605.02564 [astro-ph.HE]].


\bibitem{Aab:2018ytv} 
  A.~Aab {\it et al.} [Pierre Auger Collaboration],
    {\color{rossoCP3} Observation of inclined EeV air showers with the radio detector of the Pierre Auger Observatory}, 
  arXiv:1806.05386 [astro-ph.IM].

\bibitem{Kido}
E. Kido [for the TA Collaboration],
  {\color{rossoCP3}  The TA$\times$4 experiment},
PoS ICRC {\bf 2017}, 386 (2017).








\bibitem{Hillas:1985is} 
  A.~M.~Hillas,
   {\color{rossoCP3}  The origin of ultrahigh-energy cosmic rays},
  Ann.\ Rev.\ Astron.\ Astrophys.\  {\bf 22}, 425 (1984).
  doi:10.1146/annurev.aa.22.090184.002233


\bibitem{Swann} W. F. G. Swann,
 {\color{rossoCP3} A mechanism of acquirement of cosmic ray energies by electrons},
Phys. Rev. {\bf 43}, 217 (1933).
doi:10.1103/PhysRev.43.217


\bibitem{deJager} 
O. C. de Jager,
 {\color{rossoCP3} Evidence for particle acceleration in a magnetized
   white dwarf from radio and gamma-ray observations}
Astrophys.\ J.\ Suppl.\  {\bf 90}, 775 (1994)
doi:10.1086/191902

\bibitem{Ikhsanov:2005qf} 
  N.~R.~Ikhsanov and P.~L.~Biermann,
 {\color{rossoCP3}  High-energy emission of fast rotating white dwarfs},
  Astron.\ Astrophys.\  {\bf 445}, 305 (2006)
  doi:10.1051/0004-6361:20053179
  [astro-ph/0509070].

\bibitem{Gunn:1969ej} 
  J.~E.~Gunn and J.~P.~Ostriker,
   {\color{rossoCP3}  Acceleration of high-energy cosmic rays by pulsars},
  Phys.\ Rev.\ Lett.\  {\bf 22}, 728 (1969).
  doi:10.1103/PhysRevLett.22.728



\bibitem{Blasi:2000xm} 
  P.~Blasi, R.~I.~Epstein and A.~V.~Olinto,
   {\color{rossoCP3}  Ultrahigh-energy cosmic rays from young neutron star winds},
  Astrophys.\ J.\  {\bf 533}, L123 (2000)
  doi:10.1086/312626
  [astro-ph/9912240].


\bibitem{Arons:2002yj} 
  J.~Arons,
   {\color{rossoCP3}  Magnetars in the metagalaxy: an origin for ultrahigh-energy cosmic rays in the nearby universe},
  Astrophys.\ J.\  {\bf 589}, 871 (2003)
  doi:10.1086/374776
  [astro-ph/0208444].


\bibitem{Fang:2012rx} 
  K.~Fang, K.~Kotera and A.~V.~Olinto,
   {\color{rossoCP3}  Newly-born pulsars as sources of ultrahigh energy cosmic rays},
  Astrophys.\ J.\  {\bf 750}, 118 (2012)
  doi:10.1088/0004-637X/750/2/118
  [arXiv:1201.5197 [astro-ph.HE]].


\bibitem{Fang:2013cba} 
  K.~Fang, K.~Kotera and A.~V.~Olinto,
   {\color{rossoCP3}  Ultrahigh energy cosmic ray nuclei from extragalactic pulsars and the effect of their Galactic counterparts},
  JCAP {\bf 1303}, 010 (2013)
  doi:10.1088/1475-7516/2013/03/010
  [arXiv:1302.4482 [astro-ph.HE]].





\bibitem{Blandford:1977ds} 
  R.~D.~Blandford and R.~L.~Znajek,
   {\color{rossoCP3}  Electromagnetic extractions of energy from Kerr black holes},
  Mon.\ Not.\ Roy.\ Astron.\ Soc.\  {\bf 179}, 433 (1977).


\bibitem{Znajek} R. L. Znajek
 {\color{rossoCP3} The electric and magnetic conductivity of a Kerr  hole},
 Mon.\ Not.\ Roy.\ Astron.\ Soc.\  {\bf 185}, 833 (1978).

\bibitem{Lovelace}
R. V. E. Lovelace,
 {\color{rossoCP3} Dynamo model of double radio sources}
Nature {\bf 262}, 649 (1976).



\bibitem{Fermi:1949ee} 
  E.~Fermi,
   {\color{rossoCP3}  On the origin of the cosmic radiation},
  Phys.\ Rev.\  {\bf 75}, 1169 (1949).
  doi:10.1103/PhysRev.75.1169

\bibitem{Fermi:1954ofk} 
  E.~Fermi,
  {\color{rossoCP3} Galactic magnetic fields and the origin of cosmic radiation},
  Astrophys.\ J.\  {\bf 119}, 1 (1954).
  doi:10.1086/145789




\bibitem{Jokipii:1971}
J. R. Jokipii,
 {\color{rossoCP3}  Propagation of cosmic rays in the solar wind},
Rev.\ Geophys.\  {\bf 9}, 27 (1971)
doi:10.1029/RG009i001p00027

\bibitem{Wenzel:1989}
K. P. Wenzel,
 {\color{rossoCP3}  Charged particle acceleration processes in the
   interplanetary medium}, 
Adv.\ Space Res.\  {\bf 9}, 179 (1989)
doi:10.1016/0273-1177(89)90112-9

\bibitem{Scott:1975}
J. S. Scott and R. A. Chevalier, 
{\color{rossoCP3} Cosmic-ray production in the Cassiopeia A supernova remnant}, 
Astrophys. J.  {\bf 197}  L5 (1975)
doi:10.1086/181763

\bibitem{Chevalier:1976}
R. A. Chevalier, J. W. Robertson, and J. S. Scott
{\color{rossoCP3} Cosmic ray acceleration and the radio evolution of
  Cassiopeia A}
Astrophys. J.  {\bf 207},  450 (1979)
doi:10.1086/154514

\bibitem{Chevalier:1978qk} 
  R.~A.~Chevalier, W.~R.~Oegerle and J.~S.~Scott,
   {\color{rossoCP3} Further studies of particle acceleration in Cassiopeia A},
  Astrophys.\ J.\  {\bf 222}, 527 (1978).
  doi:10.1086/156165

\bibitem{Cowsik:1984yya} 
  R.~Cowsik and S.~Sarkar,
  {\color{rossoCP3} The evolution of supernova remnants as radio sources},
  Mon.\ Not.\ Roy.\ Astron.\ Soc.\  {\bf 207}, 745 (1984)
  Erratum: [Mon.\ Not.\ Roy.\ Astron.\ Soc.\  {\bf 209}, 719 (1984)].
  doi:10.1093/mnras/209.4.719, 10.1093/mnras/207.4.745

\bibitem{Torres:2002af} 
  D.~F.~Torres, G.~E.~Romero, T.~M.~Dame, J.~A.~Combi and Y.~M.~Butt,
  {\color{rossoCP3}  Supernova remnants and gamma-ray sources},
  Phys.\ Rept.\  {\bf 382}, 303 (2003)
  doi:10.1016/S0370-1573(03)00201-1
  [astro-ph/0209565].

\bibitem{Blasi:2010gr} 
  P.~Blasi,
   {\color{rossoCP3} Cosmic ray acceleration in supernova remnants},
  doi:10.1142/9789814329033$_-$0061
  arXiv:1012.5005 [astro-ph.HE].

\bibitem{Jokipii:1985}
J. R. Jokipii and G. Morfill, 
 {\color{rossoCP3} On the origin of high-energy cosmic rays}
 Astrophys.\ J.\  {\bf 290}, L1 (1985)
doi:10.1086/184430

\bibitem{Jokipii:1987}
J. R. Jokipii and G. Morfill, 
 {\color{rossoCP3} Ultra-high-energy cosmic rays in a Galactic wind
   and its termination shock},
  Astrophys.\ J.\  {\bf 312}, 170 (1987)
doi:10.1086/164857


\bibitem{Bustard:2016swa} 
  C.~Bustard, E.~G.~Zweibel and C.~Cotter,
 {\color{rossoCP3}   Cosmic ray acceleration by a versatile family of galactic wind termination shocks},
  Astrophys.\ J.\  {\bf 835}, no. 1, 72 (2017)
  doi:10.3847/1538-4357/835/1/72
  [arXiv:1610.06565 [astro-ph.HE]].



\bibitem{Merten:2018qoa} 
  L.~Merten, C.~Bustard, E.~G.~Zweibel and J.~Becker Tjus,
   {\color{rossoCP3} The propagation of cosmic rays from the Galactic wind termination shock: Back to the Galaxy?},
  Astrophys.\ J.\  {\bf 859}, no. 1, 63 (2018)
  doi:10.3847/1538-4357/aabfdd
  [arXiv:1803.08376 [astro-ph.HE]].

\bibitem{Protheroe:1983}
R. J. Protheroe and D. Kazanas,
  {\color{rossoCP3} On the origin of relativistic particles and gamma-rays in quasars},
Astrophys.\ J.\  {\bf 265}, 620 (1983)
doi:10.1086/160707

\bibitem{Kazanas:1985ud} 
  D.~Kazanas and D.~C.~Ellison,
   {\color{rossoCP3} The central engine of quasars and AGNs: hadronic interactions of shock accelerated relativistic protons},
 Astrophys.\ J.\  {\bf 304}, 178 (1986)
doi:10.1086/164152

\bibitem{Protheroe:1992qs} 
  R.~J.~Protheroe and A.~P.~Szabo,
    {\color{rossoCP3} High-energy cosmic rays from active galactic nuclei},
  Phys.\ Rev.\ Lett.\  {\bf 69}, 2885 (1992).
  doi:10.1103/PhysRevLett.69.2885


\bibitem{Biermann:1987ep}  
  P.~L.~Biermann and P.~A.~Strittmatter,
    {\color{rossoCP3} Synchrotron emission from shock waves in active galactic nuclei},
  Astrophys.\ J.\  {\bf 322}, 643 (1987).
  doi:10.1086/165759
  



\bibitem{Rachen:1992pg} 
  J.~P.~Rachen and P.~L.~Biermann,
  {\color{rossoCP3}  Extragalactic ultrahigh-energy cosmic rays I: Contribution from hot spots in FR-II radio galaxies},
  Astron.\ Astrophys.\  {\bf 272}, 161 (1993)
  [astro-ph/9301010].


\bibitem{Romero:1995tn} 
  G.~E.~Romero, J.~A.~Combi, L.~A.~Anchordoqui and S.~E.~Perez Bergliaffa,
 {\color{rossoCP3} A possible source of extragalactic cosmic rays with arrival energies beyond the GZK cutoff},
  Astropart.\ Phys.\  {\bf 5}, 279 (1996)
  doi:10.1016/0927-6505(96)00029-1
  [gr-qc/9511031].


\bibitem{Blandford:1979za} 
  R.~D.~Blandford and A.~Konigl,
   {\color{rossoCP3} Relativistic jets as compact radio sources},
  Astrophys.\ J.\  {\bf 232}, 34 (1979).
  doi:10.1086/157262

\bibitem{Mannheim:1993jg} 
  K.~Mannheim,
   {\color{rossoCP3} The proton blazar},
  Astron.\ Astrophys.\  {\bf 269}, 67 (1993)
  [astro-ph/9302006].


\bibitem{Dermer:2008cy} 
  C.~D.~Dermer, S.~Razzaque, J.~D.~Finke and A.~Atoyan,
     {\color{rossoCP3} Ultrahigh energy cosmic rays from black hole jets of radio galaxies},
  New J.\ Phys.\  {\bf 11}, 065016 (2009)
  doi:10.1088/1367-2630/11/6/065016
  [arXiv:0811.1160 [astro-ph]].


\bibitem{Caprioli:2015zka} 
  D.~Caprioli,
   {\color{rossoCP3}"Espresso" acceleration of ultra-high-energy cosmic rays},
  Astrophys.\ J.\  {\bf 811}, no. 2, L38 (2015)
  doi:10.1088/2041-8205/811/2/L38
  [arXiv:1505.06739 [astro-ph.HE]].


\bibitem{Waxman:1995vg} 
  E.~Waxman,
  {\color{rossoCP3}   Cosmological gamma-ray bursts and the highest energy cosmic rays},
  Phys.\ Rev.\ Lett.\  {\bf 75}, 386 (1995)
  doi:10.1103/PhysRevLett.75.386
  [astro-ph/9505082].

\bibitem{Vietri:1995hs} 
  M.~Vietri,
   {\color{rossoCP3}  On the acceleration of ultrahigh-energy cosmic rays in gamma-ray bursts},
  Astrophys.\ J.\  {\bf 453}, 883 (1995)
  doi:10.1086/176448
  [astro-ph/9506081].


\bibitem{Anchordoqui:2018vji} 
  L.~A.~Anchordoqui,
   {\color{rossoCP3}  Acceleration of ultrahigh-energy cosmic rays in starburst superwinds},
  Phys.\ Rev.\ D {\bf 97}, no. 6, 063010 (2018)
  doi:10.1103/PhysRevD.97.063010
  [arXiv:1801.07170 [astro-ph.HE]].




\bibitem{Levinson:2001as} 
  A.~Levinson and E.~Waxman,
    {\color{rossoCP3}  Probing microquasars with TeV neutrinos},
  Phys.\ Rev.\ Lett.\  {\bf 87}, 171101 (2001)
  doi:10.1103/PhysRevLett.87.171101
  [hep-ph/0106102].


\bibitem{Aharonian:2005cx} 
  F.~A.~Aharonian, L.~A.~Anchordoqui, D.~Khangulyan and T.~Montaruli,
     {\color{rossoCP3}  Microquasar LS 5039: a TeV gamma-ray emitter and a potential TeV neutrino source},
  J.\ Phys.\ Conf.\ Ser.\  {\bf 39}, 408 (2006)
  doi:10.1088/1742-6596/39/1/106
  [astro-ph/0508658].



\bibitem{Norman:1995}
C. A. Norman, D. B. Melrose, and A. Achterberg,
{\color{rossoCP3} The origin of cosmic rays above $10^{18.5}~{\rm eV}$}
 Astrophys.\ J.\  {\bf 454}, 60 (1995).
doi:10.1086/176465

\bibitem{Kang:1996rp} 
  H.~Kang, J.~P.~Rachen and P.~L.~Biermann,
  {\color{rossoCP3} Contributions to the cosmic ray flux above the ankle: clusters of galaxies},
  Mon.\ Not.\ Roy.\ Astron.\ Soc.\  {\bf 286}, 257 (1997)
  doi:10.1093/mnras/286.2.257
  [astro-ph/9608071].



\bibitem{Ryu:2003cd} 
  D.~Ryu, H.~Kang, E.~Hallman and T.~W.~Jones,
 {\color{rossoCP3}  Cosmological shock waves and their role in the large scale structure of the universe},
  Astrophys.\ J.\  {\bf 593}, 599 (2003)
  doi:10.1086/376723
  [astro-ph/0305164].



\bibitem{yellowbook} M. Ahlers, L. A. Anchordoqui, J. K. Becker,
  T. K. Gaisser, F. Halzen, D. Hooper, S. R. Klein. P. M\'esz\'aros,
  S. Razzaque, and S. Sarkar,  {\color{rossoCP3}  Neutrinos on the rocks:
The IceCube yellow book}, FERMILAB-FN-0847-A, YITP-SB-10-01.

\bibitem{Ptitsyna:2008zs} 
  K.~V.~Ptitsyna and S.~V.~Troitsky,
 {\color{rossoCP3}  Physical conditions in potential sources of ultra-high-energy cosmic rays I: Updated Hillas plot and radiation-loss constraints},
  Phys.\ Usp.\  {\bf 53}, 691 (2010)
  doi:10.3367/UFNe.0180.201007c.0723
  [arXiv:0808.0367 [astro-ph]].






\bibitem{Chamel:2008ca} 
  N.~Chamel and P.~Haensel,
   {\color{rossoCP3} Physics of neutron star crusts},
  Living Rev.\ Rel.\  {\bf 11}, 10 (2008)
  doi:10.12942/lrr-2008-10
  [arXiv:0812.3955 [astro-ph]].


\bibitem{Ruderman:1975ju} 
  M.~A.~Ruderman and P.~G.~Sutherland,
  {\color{rossoCP3} Theory of pulsars: Polar caps, sparks, and coherent microwave radiation},
  Astrophys.\ J.\  {\bf 196}, 51 (1975).
  doi:10.1086/153393

\bibitem{Vigano:2013uia} 
  D.~Vigan\'o,
   {\color{rossoCP3} Magnetic fields in neutron stars},
  arXiv:1310.1243 [astro-ph.HE].

\bibitem{Goldreich:1969sb} 
  P.~Goldreich and W.~H.~Julian,
  {\color{rossoCP3} Pulsar electrodynamics},
  Astrophys.\ J.\  {\bf 157}, 869 (1969).
  doi:10.1086/150119




\bibitem{Berezinsky:1983}
V. S. Berezinsky,
  {\color{rossoCP3} Acceleration to ultra high energies in
    magnetospheres of young pulsars},
in Proceedings of the 18th International Cosmic Ray Conference {\bf 2},
275 (1983).



\bibitem{FaucherGiguere:2005ny} 
  C.~A.~Faucher-Giguere and V.~M.~Kaspi,
   {\color{rossoCP3}  Birth and evolution of isolated radio pulsars},
  Astrophys.\ J.\  {\bf 643}, 332 (2006)
  doi:10.1086/501516
  [astro-ph/0512585].




\bibitem{Haensel:1999mi} 
  P.~Haensel, J.~P.~Lasota and J.~L.~Zdunik,
  {\color{rossoCP3} On the minimum period of uniformly rotating meutron stars},
Astron. Astrophys. {\bf 344}, 151 (1999)  
[astro-ph/9901118].





\bibitem{Ochelkov} 
Y. P. Ochelkov and V. V. Usov,
{\color{rossoCP3} Curvature radiation of relativistic particles in the
  magnetosphere of pulsars}
 Astrophys.\ Space Sci.\  {\bf 69},  439 (1980)
doi:10.1007/BF00661929.



\bibitem{Kotera:2015pya} 
  K.~Kotera, E.~Amato and P.~Blasi,
   {\color{rossoCP3} The fate of ultrahigh energy nuclei in the immediate environment of young fast-rotating pulsars},
  JCAP {\bf 1508}, no. 08, 026 (2015)
  doi:10.1088/1475-7516/2015/08/026
  [arXiv:1503.07907 [astro-ph.HE]].

\bibitem{Spitkovsky:2006np} 
  A.~Spitkovsky,
   {\color{rossoCP3}  Time-dependent force-free pulsar magnetospheres: axisymmetric and oblique rotators},
  Astrophys.\ J.\  {\bf 648}, L51 (2006)
  doi:10.1086/507518
  [astro-ph/0603147].




\bibitem{Boldt:1999ge} 
  E.~Boldt and P.~Ghosh,
  {\color{rossoCP3} Cosmic rays from remnants of quasars?},
  Mon.\ Not.\ Roy.\ Astron.\ Soc.\  {\bf 307}, 491 (1999)
  doi:10.1046/j.1365-8711.1999.02600.x
  [astro-ph/9902342].


\bibitem{Boldt:2000dx} 
  E.~Boldt and M.~Loewenstein,
  {\color{rossoCP3} Cosmic ray generation by quasar remnants: Constraints and implications},
  Mon.\ Not.\ Roy.\ Astron.\ Soc.\  {\bf 316}, L29 (2000)
  doi:10.1046/j.1365-8711.2000.03768.x
  [astro-ph/0006221].




\bibitem{Neronov:2007mh} 
  A.~Y.~Neronov, D.~V.~Semikoz and I.~I.~Tkachev,
 {\color{rossoCP3} Ultra-high energy cosmic ray production in the polar cap regions of black hole magnetospheres},
  New J.\ Phys.\  {\bf 11}, 065015 (2009)
  doi:10.1088/1367-2630/11/6/065015
  [arXiv:0712.1737 [astro-ph]].


\bibitem{Moncada:2017hvq} 
  R.~J.~Moncada, R.~A.~Colon, J.~J.~Guerra, M.~J.~O'Dowd and L.~A.~Anchordoqui,
 {\color{rossoCP3}  Ultrahigh energy cosmic ray nuclei from remnants of dead quasars},
  JHEAp {\bf 13-14}, 32 (2017)
  doi:10.1016/j.jheap.2017.04.001
  [arXiv:1702.00053 [astro-ph.HE]].








\bibitem{Drury:1994fg} 
  L.~O.~Drury,
   {\color{rossoCP3}  Acceleration of cosmic rays},
  Contemp.\ Phys.\  {\bf 35}, 231 (1994).
  doi:10.1080/00107519408222090


\bibitem{Krymskii:1977}
G. F. Krymskii, 
  {\color{rossoCP3} A regular mechanism for the acceleration of
    charged particles on the front of a shock wave},
Akademiia Nauk SSSR Doklady {\bf 234}, 1306 (1977).



\bibitem{Axford:1977}
W. I. Axford, E. Leer, and G. Skadron
 {\color{rossoCP3}  The acceleration of cosmic rays by shock waves},
in Proceedings of the 15th International Cosmic Ray Conference {\bf 11},
132 (1977).



\bibitem{Bell:1978zc} 
  A.~R.~Bell,
   {\color{rossoCP3}  The acceleration of cosmic rays in shock fronts I},
  Mon.\ Not.\ Roy.\ Astron.\ Soc.\  {\bf 182}, 147 (1978).



\bibitem{Bell:1978fj} 
  A.~R.~Bell,
   {\color{rossoCP3}  The acceleration of cosmic rays in shock fronts II},
  Mon.\ Not.\ Roy.\ Astron.\ Soc.\  {\bf 182}, 443 (1978).



\bibitem{Blandford:1978ky} 
  R.~D.~Blandford and J.~P.~Ostriker,
    {\color{rossoCP3} Particle acceleration by astrophysical shocks},
  Astrophys.\ J.\  {\bf 221}, L29 (1978).
  doi:10.1086/182658


\bibitem{Lagage:1983zz} 
  P.~O.~Lagage and C.~J.~Cesarsky,
    {\color{rossoCP3} The maximum energy of cosmic rays accelerated by supernova shocks},
  Astron.\ Astrophys.\  {\bf 125}, 249 (1983).


\bibitem{Drury:1983zz} 
  L.~O.~Drury,
   {\color{rossoCP3}  An introduction to the theory of diffusive shock acceleration of energetic particles in tenuous plasmas},
  Rept.\ Prog.\ Phys.\  {\bf 46}, 973 (1983).
  doi:10.1088/0034-4885/46/8/002


\bibitem{Blandford:1987pw} 
  R.~Blandford and D.~Eichler,
   {\color{rossoCP3}  Particle acceleration at astrophysical shocks: A
     theory of cosmic ray 
origin},
  Phys.\ Rept.\  {\bf 154}, 1 (1987).
  doi:10.1016/0370-1573(87)90134-7


\bibitem{Protheroe:1998hp} 
  R.~J.~Protheroe,
   {\color{rossoCP3}  Acceleration and interaction of ultrahigh-energy cosmic rays},
  astro-ph/9812055.



\bibitem{Bell:2013vxa} 
  A.~R.~Bell,
 {\color{rossoCP3}    Cosmic ray acceleration},
  Astropart.\ Phys.\  {\bf 43}, 56 (2013).
  doi:10.1016/j.astropartphys.2012.05.022




\bibitem{Baerwald:2013pu} 
  P.~Baerwald, M.~Bustamante and W.~Winter,
 {\color{rossoCP3}  UHECR escape mechanisms for protons and neutrons from GRBs, and the cosmic ray-neutrino connection},
  Astrophys.\ J.\  {\bf 768}, 186 (2013)
  doi:10.1088/0004-637X/768/2/186
  [arXiv:1301.6163 [astro-ph.HE]].



\bibitem{Gaisser:1990vg} 
  T.~K.~Gaisser,
   {\color{rossoCP3}  Cosmic rays and particle physics},
  (Cambridge University Press, UK, 1990) 





\bibitem{Romero:2018mnb} 
  G.~E.~Romero, A.~L.~Müller and M.~Roth,
  {\color{rossoCP3}  Particle acceleration in the superwinds of starburst galaxies},
  arXiv:1801.06483 [astro-ph.HE].

\bibitem{Jokipii:1986} 
  J.~R.~Jokipii,
    {\color{rossoCP3} Rate of energy gain and maximum energy in
      diffusive shock acceleration} 
  Astrophys.\ J.\  {\bf 313}, 842 (1987).
  doi:10.1086/165022
 

\bibitem{Ferrand:2014laa} 
  G.~Ferrand, R.~J.~Danos, A.~Shalchi, S.~Safi-Harb, P.~Edmon and P.~Mendygral,
    {\color{rossoCP3} Cosmic ray acceleration at perpendicular shocks in supernova remnants},
  Astrophys.\ J.\  {\bf 792}, no. 2, 133 (2014)
  doi:10.1088/0004-637X/792/2/133
  [arXiv:1407.6728 [astro-ph.HE]].






\bibitem{Waxman:2005id} 
  E.~Waxman,
   {\color{rossoCP3}  Extra-galactic sources of high energy neutrinos},
  Phys.\ Scripta T {\bf 121}, 147 (2005)
  doi:10.1088/0031-8949/2005/T121/022
  [astro-ph/0502159].

\bibitem{Piran:2010yg} 
  T.~Piran,
   {\color{rossoCP3}   A new limit on the distances of nuclei UHECRs sources},
  arXiv:1005.3311 [astro-ph.HE].



\bibitem{Spruit:2013ud} 
  H.~C.~Spruit,
    {\color{rossoCP3}   Essential magnetohydrodynamics for astrophysics},
  arXiv:1301.5572 [astro-ph.IM].






\bibitem{Anchordoqui:2007tn} 
  L.~A.~Anchordoqui, D.~Hooper, S.~Sarkar and A.~M.~Taylor,
    {\color{rossoCP3}  High-energy neutrinos from astrophysical accelerators of cosmic ray nuclei},
  Astropart.\ Phys.\  {\bf 29}, 1 (2008)
  doi:10.1016/j.astropartphys.2007.10.006
  [astro-ph/0703001].


\bibitem{Wang:2007xj} 
  X.~Y.~Wang, S.~Razzaque and P.~M\'esz\'aros,
   {\color{rossoCP3} On the origin and survival of UHE cosmic-ray nuclei in GRBs and hypernovae},
  Astrophys.\ J.\  {\bf 677}, 432 (2008)
  doi:10.1086/529018
  [arXiv:0711.2065 [astro-ph]].


\bibitem{Murase:2008mr} 
  K.~Murase, K.~Ioka, S.~Nagataki and T.~Nakamura,
   {\color{rossoCP3} High-energy cosmic-ray nuclei from high- and low-luminosity gamma-ray bursts and implications for multi-messenger astronomy},
  Phys.\ Rev.\ D {\bf 78}, 023005 (2008)
  doi:10.1103/PhysRevD.78.023005
  [arXiv:0801.2861 [astro-ph]].


\bibitem{Globus:2014fka} 
  N.~Globus, D.~Allard, R.~Mochkovitch and E.~Parizot,
  {\color{rossoCP3}  UHECR acceleration at GRB internal shocks},
  Mon.\ Not.\ Roy.\ Astron.\ Soc.\  {\bf 451}, no. 1, 751 (2015)
  doi:10.1093/mnras/stv893
  [arXiv:1409.1271 [astro-ph.HE]].

\bibitem{Biehl:2017zlw} 
  D.~Biehl, D.~Boncioli, A.~Fedynitch and W.~Winter,
  {\color{rossoCP3}   Cosmic-ray and neutrino emission from gamma-ray bursts with a nuclear cascade},
  Astron.\ Astrophys.\  {\bf 611}, A101 (2018)
  doi:10.1051/0004-6361/201731337
  [arXiv:1705.08909 [astro-ph.HE]].


\bibitem{Zhang:2017moz} 
  B.~T.~Zhang, K.~Murase, S.~S.~Kimura, S.~Horiuchi and P.~M\'esz\'aros,
   {\color{rossoCP3} Low-luminosity gamma-ray bursts as the sources of ultrahigh-energy cosmic ray nuclei},
  Phys.\ Rev.\ D {\bf 97}, no. 8, 083010 (2018)
  doi:10.1103/PhysRevD.97.083010
  [arXiv:1712.09984 [astro-ph.HE]].


\bibitem{Blandford:1999hi} 
  R.~D.~Blandford,
    {\color{rossoCP3}  Acceleration of ultrahigh-energy cosmic rays},
  Phys.\ Scripta T {\bf 85}, 191 (2000)
  doi:10.1238/Physica.Topical.085a00191
  [astro-ph/9906026].

\bibitem{Dermer:2010iz} 
  C.~D.~Dermer and S.~Razzaque,
    {\color{rossoCP3} Acceleration of ultra-high energy cosmic rays in the colliding shells of blazars and GRBs: Constraints from the Fermi Gamma ray Space Telescope},
  Astrophys.\ J.\  {\bf 724}, 1366 (2010)
  doi:10.1088/0004-637X/724/2/1366
  [arXiv:1004.4249 [astro-ph.HE]].




\bibitem{Cannoni:2016hro} 
  M.~Cannoni,
     {\color{rossoCP3} Lorentz invariant relative velocity and relativistic binary collisions},
  Int.\ J.\ Mod.\ Phys.\ A {\bf 32}, no. 02n03, 1730002 (2017)
  doi:10.1142/S0217751X17300022
  [arXiv:1605.00569 [hep-ph]].





\bibitem{Stecker:1968uc} 
  F.~W.~Stecker,
    {\color{rossoCP3} Effect of photomeson production by the universal radiation field on high-energy cosmic rays},
  Phys.\ Rev.\ Lett.\  {\bf 21}, 1016 (1968).
  doi:10.1103/PhysRevLett.21.1016




\bibitem{Berezinsky:1987ed} 
  V.~S.~Berezinsky and S.~I.~Grigoreva,
   {\color{rossoCP3} The hump in the ultrahigh-energy cosmic ray spectrum},
  Sov.\ Phys.\ JETP {\bf 66}, 457 (1987)
  [Zh.\ Eksp.\ Teor.\ Fiz.\  {\bf 93}, 812 (1987)].


\bibitem{Berezinsky:1988wi} 
  V.~S.~Berezinsky and S.~I.~Grigor'eva,
   {\color{rossoCP3} A bump In the ultrahigh-energy cosmic ray spectrum},
  Astron.\ Astrophys.\  {\bf 199}, 1 (1988).

\bibitem{Fixsen:2009ug} 
  D.~J.~Fixsen,
 {\color{rossoCP3}  The temperature of the cosmic microwave background},
  Astrophys.\ J.\  {\bf 707}, 916 (2009)
  doi:10.1088/0004-637X/707/2/916
  [arXiv:0911.1955 [astro-ph.CO]].




\bibitem{Blumenthal:1970nn} 
  G.~R.~Blumenthal,
    {\color{rossoCP3} Energy loss of high-energy cosmic rays in pair-producing collisions with ambient photons},
  Phys.\ Rev.\ D {\bf 1}, 1596 (1970).
  doi:10.1103/PhysRevD.1.1596





\bibitem{Aharonian:1994nn} 
  F.~A.~Aharonian and J.~W.~Cronin,
   {\color{rossoCP3} Influence of the universal microwave background radiation on the extragalactic cosmic ray spectrum},
  Phys.\ Rev.\ D {\bf 50}, 1892 (1994).
  doi:10.1103/PhysRevD.50.1892

\bibitem{Armstrong:1971ns} 
  T.~A.~Armstrong {\it et al.},
   {\color{rossoCP3} Total hadronic cross-section of gamma rays in hydrogen in the energy range 0.265~GeV to 4.215~GeV},
  Phys.\ Rev.\ D {\bf 5}, 1640 (1972).
  doi:10.1103/PhysRevD.5.1640




\bibitem{Montanet:1994xu} 
   L.~Montanet {\it et al.} [Particle Data Group],
 {\color{rossoCP3}  Review of particle properties}
  Phys.\ Rev.\ D {\bf 50}, 1173 (1994). See p. 1335.
  doi:10.1103/PhysRevD.50.1173



\bibitem{Golyak:1992cz} 
  I.~Golyak,
  {\color{rossoCP3}  A connection of inelasticity with multiplicity distribution at high-energies},
  Mod.\ Phys.\ Lett.\ A {\bf 7}, 2401 (1992).
  doi:10.1142/S0217732392003839

\bibitem{Anchordoqui:1997gs} 
  L.~A.~Anchordoqui, M.~T.~Dova, L.~N.~Epele and J.~D.~Swain,
 {\color{rossoCP3} Opacity of the microwave background radiation to ultra-high-energy cosmic rays},
  Nucl.\ Phys.\ Proc.\ Suppl.\  {\bf 52B}, 249 (1997).
  doi:10.1016/S0920-5632(96)00898-5


\bibitem{Anchordoqui:1996ru} 
  L.~A.~Anchordoqui, M.~T.~Dova, L.~N.~Epele and J.~D.~Swain,
 {\color{rossoCP3} Effect of the 3-K background radiation on ultrahigh-energy cosmic rays},
  Phys.\ Rev.\ D {\bf 55}, 7356 (1997)
  doi:10.1103/PhysRevD.55.7356
  [hep-ph/9704387].






\bibitem{Abramowitz} M. Abramowitz, I. A. Stegun,  {\color{rossoCP3} Handbook of 
Mathematical Functions}, (Dover, New York, 1970).

\bibitem{Anchordoqui:1998ig} 
  L.~A.~Anchordoqui,
 {\color{rossoCP3} From 3K to $10^{20}~{\rm eV}$},
  PhD Thesis, UNLP 1998
  [astro-ph/9812445].





\bibitem{Chodorowski} M. J. Chodorowski, A. A. Zdziarski, and M. Sikora,
  {\color{rossoCP3} Reaction rate and energy-loss rate for photopair production by 
relativistic nuclei},
Astrophys. J. {\bf 400}, 181 (1992)
doi:10.1086/171984



\bibitem{Michalowski:1977eg} 
  S.~Michalowski, D.~Andrews, J.~Eickmeyer, T.~Gentile, N.~B.~Mistry, R.~Talman and K.~Ueno,
   {\color{rossoCP3} Experimental study of nuclear shadowing in photoproduction},
  Phys.\ Rev.\ Lett.\  {\bf 39}, 737 (1977).
  doi:10.1103/PhysRevLett.39.737




\bibitem{Hayward:1963zz} 
  E.~Hayward,
 {\color{rossoCP3}  Photodisintegration of light nuclei},
  Rev.\ Mod.\ Phys.\  {\bf 35}, 324 (1963).
  doi:10.1103/RevModPhys.35.324



\bibitem{Danos:1965yu} 
  M.~Danos and E.~G.~Fuller,
 {\color{rossoCP3}  Photonuclear reactions},
  Ann.\ Rev.\ Nucl.\ Part.\ Sci.\  {\bf 15}, 29 (1965).
  doi:10.1146/annurev.ns.15.120165.000333





\bibitem{Stecker:1998ib} 
  F.~W.~Stecker and M.~H.~Salamon,
   {\color{rossoCP3} Photodisintegration of ultrahigh-energy cosmic rays: A New determination},
  Astrophys.\ J.\  {\bf 512}, 521 (1999)
  doi:10.1086/306816
  [astro-ph/9808110].

\bibitem{Khan:2004nd} 
  E.~Khan, S.~Goriely, D.~Allard, E.~Parizot, T.~Suomijarvi, A.~J.~Koning, S.~Hilaire and M.~C.~Duijvestijn,
  {\color{rossoCP3} Photodisintegration of ultra-high-energy cosmic rays revisited},
  Astropart.\ Phys.\  {\bf 23}, 191 (2005)
  doi:10.1016/j.astropartphys.2004.12.007
  [astro-ph/0412109].


\bibitem{Boncioli:2016lkt} 
  D.~Boncioli, A.~Fedynitch and W.~Winter,
   {\color{rossoCP3} Nuclear physics meets the sources of the ultra-high-energy cosmic rays},
  Sci.\ Rep.\  {\bf 7}, no. 1, 4882 (2017)
  doi:10.1038/s41598-017-05120-7
  [arXiv:1607.07989 [astro-ph.HE]].



\bibitem{Karakula:1993he} 
  S.~Karakula and W.~Tkaczyk,
   {\color{rossoCP3} The formation of the cosmic ray energy spectrum by a photon field},
  Astropart.\ Phys.\  {\bf 1}, 229 (1993).
  doi:10.1016/0927-6505(93)90023-7



\bibitem{Anchordoqui:2006pe} 
  L.~A.~Anchordoqui, J.~F.~Beacom, H.~Goldberg, S.~Palomares-Ruiz and T.~J.~Weiler,
  {\color{rossoCP3} TeV $\gamma^-$ rays and neutrinos from photo-disintegration of nuclei in Cygnus OB2},
  Phys.\ Rev.\ D {\bf 75}, 063001 (2007)
  doi:10.1103/PhysRevD.75.063001
  [astro-ph/0611581].


\bibitem{Soriano:2018lly} 
  J.~F.~Soriano, L.~A.~Anchordoqui and D.~F.~Torres,
     {\color{rossoCP3} Photodisintegration of $^4$He on the cosmic microwave background is less severe than earlier thought},
 Phys. Rev. D, in press  [arXiv:1805.00409 [astro-ph.HE]].



\bibitem{Anchordoqui:2006pd} 
  L.~A.~Anchordoqui, J.~F.~Beacom, H.~Goldberg, S.~Palomares-Ruiz and T.~J.~Weiler,
   {\color{rossoCP3} TeV gamma-rays from photo-disintegration/de-excitation of cosmic-ray nuclei},
  Phys.\ Rev.\ Lett.\  {\bf 98}, 121101 (2007)
  doi:10.1103/PhysRevLett.98.121101
  [astro-ph/0611580].



\bibitem{Batista:2015mea} 
  R.~Alves Batista, D.~Boncioli, A.~di Matteo, A.~van Vliet and D.~Walz,
 {\color{rossoCP3}  Effects of uncertainties in simulations of extragalactic UHECR propagation, using CRPropa and SimProp},
  JCAP {\bf 1510}, no. 10, 063 (2015)
  doi:10.1088/1475-7516/2015/10/063
  [arXiv:1508.01824 [astro-ph.HE]].


\bibitem{Gilmore:2011ks} 
  R.~C.~Gilmore, R.~S.~Somerville, J.~R.~Primack and A.~Dominguez,
  {\color{rossoCP3} Semi-analytic modeling of the EBL and consequences for extragalactic gamma-ray spectra},
  Mon.\ Not.\ Roy.\ Astron.\ Soc.\  {\bf 422}, 3189 (2012)
  doi:10.1111/j.1365-2966.2012.20841.x
  [arXiv:1104.0671 [astro-ph.CO]].

\bibitem{Epele:1998ia} 
  L.~N.~Epele and E.~Roulet,
   {\color{rossoCP3} On the propagation of the highest energy cosmic ray nuclei},
  JHEP {\bf 9810}, 009 (1998)
  doi:10.1088/1126-6708/1998/10/009
  [astro-ph/9808104].


\bibitem{Hill:1983mk} 
  C.~T.~Hill and D.~N.~Schramm,
   {\color{rossoCP3} The ultrahigh-energy cosmic ray spectrum},
  Phys.\ Rev.\ D {\bf 31}, 564 (1985).
  doi:10.1103/PhysRevD.31.564

\bibitem{Stecker:1989ti} 
  F.~W.~Stecker,
    {\color{rossoCP3} Extragalactic radiation and the ultrahigh-energy cosmic ray spectrum},
  Nature {\bf 342}, 401 (1989).
  doi:10.1038/342401a0


\bibitem{Anchordoqui:1997rn} 
  L.~A.~Anchordoqui, M.~T.~Dova, L.~N.~Epele and J.~D.~Swain,
   {\color{rossoCP3} A depression before the bump in the highest energy cosmic ray spectrum},
  Phys.\ Rev.\ D {\bf 57}, 7103 (1998)
  doi:10.1103/PhysRevD.57.7103
  [astro-ph/9708082].



\bibitem{Allard:2008gj} 
  D.~Allard, N.~G.~Busca, G.~Decerprit, A.~V.~Olinto and E.~Parizot,
 {\color{rossoCP3}  Implications of the cosmic ray spectrum for the mass composition at the highest energies},
  JCAP {\bf 0810}, 033 (2008)
  doi:10.1088/1475-7516/2008/10/033
  [arXiv:0805.4779 [astro-ph]].



\bibitem{Allard:2011aa} 
  D.~Allard,
   {\color{rossoCP3} Extragalactic propagation of ultrahigh energy cosmic-rays},
  Astropart.\ Phys.\  {\bf 39-40}, 33 (2012)
  doi:10.1016/j.astropartphys.2011.10.011
  [arXiv:1111.3290 [astro-ph.HE]].


\bibitem{Kotera:2011cp} 
  K.~Kotera and A.~V.~Olinto,
   {\color{rossoCP3} The astrophysics of ultrahigh energy cosmic rays},
  Ann.\ Rev.\ Astron.\ Astrophys.\  {\bf 49}, 119 (2011)
  doi:10.1146/annurev-astro-081710-102620
  [arXiv:1101.4256 [astro-ph.HE]].

\bibitem{Waxman:1996zn} 
  E.~Waxman and J.~Miralda-Escude,
    {\color{rossoCP3} Images of bursting sources of high-energy cosmic
      rays I: Effects of magnetic fields},
  Astrophys.\ J.\  {\bf 472}, L89 (1996)
  doi:10.1086/310367
  [astro-ph/9607059].

\bibitem{Farrar:2012gm} 
  G.~R.~Farrar, R.~Jansson, I.~J.~Feain and B.~M.~Gaensler,
    {\color{rossoCP3} Galactic magnetic deflections and Centaurus A as a UHECR source},
  JCAP {\bf 1301}, 023 (2013)
  doi:10.1088/1475-7516/2013/01/023
  [arXiv:1211.7086 [astro-ph.HE]].





\bibitem{Farrar:2017lhm} 
  G.~R.~Farrar and M.~S.~Sutherland,
  {\color{rossoCP3} Deflections of UHECRs in the Galactic magnetic field},
  arXiv:1711.02730 [astro-ph.HE].



\bibitem{Fanaroff} B. L. Fannaroff and J. M. Riley,
 {\color{rossoCP3} The morphology of extragalactic radio sources of high and low luminosity},
Mon.\ Not.\ Roy.\ Astron.\ Soc.\  {\bf 167}, 31 (1974).

\bibitem{Blandford} R. D. Blandford and M. J. Rees,
 {\color{rossoCP3} A `twin-exhaust' model for double radio sources},
Mon.\ Not.\ Roy.\ Astron.\ Soc.\  {\bf 169}, 395 (1974).


\bibitem{Rosen:1999jm} 
  A.~Rosen, P.~A.~Hughes, G.~C.~Duncan and P.~E.~Hardee,
   {\color{rossoCP3} A comparison of the morphology and stability of relativistic and nonrelativistic jets},
  Astrophys.\ J.\  {\bf 516}, 729 (1999)
  doi:10.1086/307143
  [astro-ph/9901046].


\bibitem{Begelman:1984mw} 
  M.~C.~Begelman, R.~D.~Blandford and M.~J.~Rees,
    {\color{rossoCP3} Theory of extragalactic radio sources},
  Rev.\ Mod.\ Phys.\  {\bf 56}, 255 (1984).
  doi:10.1103/RevModPhys.56.255






\bibitem{Kolmogorov} 
A. N. Kolmogorov, 
   {\color{rossoCP3} The local structure of turbulence in
     incompressible 
viscous fluid for very large Reynolds numbers},
Compt. Rend. Acad. Sci. U.R.S.S. {\bf 30}, 201 (1941).


\bibitem{Anchordoqui:2001nt} 
  L.~A.~Anchordoqui, H.~Goldberg and T.~J.~Weiler,
    {\color{rossoCP3} An Auger test of the Cen A model of highest energy cosmic rays},
  Phys.\ Rev.\ Lett.\  {\bf 87}, 081101 (2001)
  doi:10.1103/PhysRevLett.87.081101
  [astro-ph/0103043].


\bibitem{Anchordoqui:2011ks} 
  L.~A.~Anchordoqui, H.~Goldberg and T.~J.~Weiler,
    {\color{rossoCP3} Update on tests of the Cen A neutron-emission model of highest energy cosmic rays},
  Phys.\ Rev.\ D {\bf 84}, 067301 (2011)
  doi:10.1103/PhysRevD.84.067301
  [arXiv:1103.0536 [astro-ph.HE]].



\bibitem{Anchordoqui:2001bs} 
  L.~A.~Anchordoqui and H.~Goldberg,
    {\color{rossoCP3} A lower bound on the local extragalactic magnetic field},
  Phys.\ Rev.\ D {\bf 65}, 021302 (2002)
  doi:10.1103/PhysRevD.65.021302
  [hep-ph/0106217].


\bibitem{Israel:1998ws} 
  F.~P.~Israel,
   {\color{rossoCP3} Centaurus A (NGC 5128)},
  Astron.\ Astrophys.\ Rev.\  {\bf 8}, 237 (1998)
  doi:10.1007/s001590050011
  [astro-ph/9811051].


\bibitem{Hardcastle:2003ye} 
  M.~J.~Hardcastle, D.~M.~Worrall, R.~P.~Kraft, W.~R.~Forman, C.~Jones and S.~S.~Murray,
  {\color{rossoCP3}  Radio and X-ray observations of the jet in Centaurus A},
  Astrophys.\ J.\  {\bf 593}, 169 (2003)
  doi:10.1086/376519
  [astro-ph/0304443].


\bibitem{Burns} J. O. Burns, E. D. Feigelson, and E. J. Schreier,
  {\color{rossoCP3}  The inner radio structure of Centaurus A: clues to the origin of the 
jet X-ray emission},
Astrophys. J. {\bf 273}, 128 (1983)
doi:10.1086/161353


\bibitem{Sreekumar:1999xw} 
  P.~Sreekumar, D.~L.~Bertsch, R.~C.~Hartman, P.~L.~Nolan and D.~J.~Thompson,
  {\color{rossoCP3} Gev emission from the nearby radio galaxy Centaurus A},
  Astropart.\ Phys.\  {\bf 11}, 221 (1999)
  doi:10.1016/S0927-6505(99)00054-7
  [astro-ph/9901277].

\bibitem{Grindlay} J. E. Grindlay, H. F. Helmken, R. H. Brown,
  J. Davis, and L. R. Allen, 
{\color{rossoCP3} Evidence for the detection of gamma rays from Centaurus A at 
$E_\gamma \geq 3 \times 10^{11}$~eV},
Astrophys. J. {\bf 197}, L9 (1975)
doi:10.1086/181764



\bibitem{Abdo:2009mg} 
  A.~A.~Abdo {\it et al.} [Fermi-LAT Collaboration],
 {\color{rossoCP3} Fermi Large Area Telescope bright gamma-ray source list},
  Astrophys.\ J.\ Suppl.\  {\bf 183}, 46 (2009)
  doi:10.1088/0067-0049/183/1/46
  [arXiv:0902.1340 [astro-ph.HE]].

\bibitem{Abdo:2009wu} 
  A.~A.~Abdo {\it et al.} [Fermi-LAT Collaboration],
 {\color{rossoCP3}  Bright AGN source list from the first three months of the Fermi Large Area Telescope all-sky survey},
  Astrophys.\ J.\  {\bf 700}, 597 (2009)
  doi:10.1088/0004-637X/700/1/597
  [arXiv:0902.1559 [astro-ph.HE]].



\bibitem{Aharonian:2009xn} 
  F.~Aharonian {\it et al.} [H.E.S.S. Collaboration],
 {\color{rossoCP3} Discovery of very high energy gamma-ray emission from Centaurus A with H.E.S.S.},
  Astrophys.\ J.\  {\bf 695}, L40 (2009)
  doi:10.1088/0004-637X/695/1/L40
  [arXiv:0903.1582 [astro-ph.CO]].

\bibitem{Falcone:2010fk} 
  A.~A.~Abdo {\it et al.} [Fermi Collaboration],
  {\color{rossoCP3} Fermi Large Area Telescope view of the core of the radio galaxy Centaurus A},
  Astrophys.\ J.\  {\bf 719}, 1433 (2010)
  doi:10.1088/0004-637X/719/2/1433
  [arXiv:1006.5463 [astro-ph.HE]].



\bibitem{Fermi-LAT:2010llz} 
  A.~A.~Abdo {\it et al.} [Fermi-LAT Collaboration],
 {\color{rossoCP3}  Fermi gamma-ray imaging of a radio galaxy},
  Science {\bf 328}, 725 (2010)
  doi:10.1126/science.1184656
  [arXiv:1006.3986 [astro-ph.HE]].


\bibitem{Abdalla:2018agf} 
  H.~Abdalla {\it et al.} [H. E. S. S. and Fermi-LAT Collaborations],
 {\color{rossoCP3}   The $\gamma$-ray spectrum of the core of Centaurus A as observed with H.E.S.S. and Fermi-LAT},
  arXiv:1807.07375 [astro-ph.HE].



\bibitem{Honda:2009xd} 
  M.~Honda,
   {\color{rossoCP3} Ultra-high energy cosmic-ray acceleration in the jet of Centaurus A},
  Astrophys.\ J.\  {\bf 706}, 1517 (2009)
  doi:10.1088/0004-637X/706/2/1517
  [arXiv:0911.0921 [astro-ph.HE]].

\bibitem{Junkes} N. Junkes, R. F. Haynes, J. I.
 Harnett, and  D. L.  Jauncey, 
  {\color{rossoCP3} Radio polarization surveys of Centaurus A (NGC 5128) I:The complete 
radio source at 6.3~cm},  
Astron. Astrophys. {\bf 269}, 29 (1993) [Erratum, {\it ibid} 
{\bf 274}, 1009 (1993)].



\bibitem{Rieger:2004jz} 
  F.~M.~Rieger and P.~Duffy,
  {\color{rossoCP3} Shear acceleration in relativistic astrophysical jets},
  Astrophys.\ J.\  {\bf 617}, 155 (2004)
  doi:10.1086/425167
  [astro-ph/0410269].


\bibitem{Rieger:2009pm} 
  F.~M.~Rieger and F.~A.~Aharonian,
 {\color{rossoCP3}   Cen A as TeV gamma-ray and possible UHE cosmic-ray source},
  Astron.\ Astrophys.\  {\bf 506}, L41 (2009)
  doi:10.1051/0004-6361/200912562
  [arXiv:0910.2327 [astro-ph.HE]].



\bibitem{Wykes:2013gba} 
  S.~Wykes {\it et al.},
  {\color{rossoCP3}   Mass entrainment and turbulence-driven acceleration of ultra-high energy cosmic rays in Centaurus A},
  Astron.\ Astrophys.\  {\bf 558}, A19 (2013)
  doi:10.1051/0004-6361/201321622
  [arXiv:1305.2761 [astro-ph.HE]].



\bibitem{Eilek:2014gya} 
  J.~A.~Eilek,
 {\color{rossoCP3}   The dynamic age of Centaurus A},
  New J.\ Phys.\  {\bf 16}, 045001 (2014)
  doi:10.1088/1367-2630/16/4/045001
  [arXiv:1402.4166 [astro-ph.GA]].


\bibitem{Farrar:2008ex} 
  G.~R.~Farrar and A.~Gruzinov,
 {\color{rossoCP3}  Giant AGN flares and cosmic ray bursts},
  Astrophys.\ J.\  {\bf 693}, 329 (2009)
  doi:10.1088/0004-637X/693/1/329
  [arXiv:0802.1074 [astro-ph]].



\bibitem{Farrar:2014yla} 
  G.~R.~Farrar and T.~Piran,
 {\color{rossoCP3} Tidal disruption jets as the source of ultra-high energy cosmic rays},
  arXiv:1411.0704 [astro-ph.HE].



\bibitem{AlvesBatista:2017shr} 
  R.~Alves Batista and J.~Silk,
  {\color{rossoCP3}  Ultrahigh-energy cosmic rays from tidally-ignited white dwarfs},
  Phys.\ Rev.\ D {\bf 96}, no. 10, 103003 (2017)
  doi:10.1103/PhysRevD.96.103003
  [arXiv:1702.06978 [astro-ph.HE]].

\bibitem{Zhang:2017hom} 
  B.~T.~Zhang, K.~Murase, F.~Oikonomou and Z.~Li,
 {\color{rossoCP3}   High-energy cosmic ray nuclei from tidal disruption events: origin, survival, and implications},
  Phys.\ Rev.\ D {\bf 96}, no. 6, 063007 (2017)
  Addendum: [Phys.\ Rev.\ D {\bf 96}, no. 6, 069902 (2017)]
  doi:10.1103/PhysRevD.96.063007, 10.1103/PhysRevD.96.069902
  [arXiv:1706.00391 [astro-ph.HE]].


\bibitem{Biehl:2017hnb} 
  D.~Biehl, D.~Boncioli, C.~Lunardini and W.~Winter,
   {\color{rossoCP3}  Tidally disrupted stars as a possible origin of both cosmic rays and neutrinos at the highest energies},
  Sci.\ Rep.\  {\bf 8}, no. 1, 10828 (2018)
  doi:10.1038/s41598-018-29022-4
  [arXiv:1711.03555 [astro-ph.HE]].




\bibitem{Heckman}
T. M. Heckman and T. A.  Thompson,
 {\color{rossoCP3}  A brief review of galactic winds}
asrXiv:1701.09062.



\bibitem{Veilleux:2005ia} 
  S.~Veilleux, G.~Cecil and J.~Bland-Hawthorn,
 {\color{rossoCP3}  Galactic winds},
  Ann.\ Rev.\ Astron.\ Astrophys.\  {\bf 43}, 769 (2005)
  doi:10.1146/annurev.astro.43.072103.150610
  [astro-ph/0504435].




\bibitem{Long:2014tsa} 
  K.~S.~Long, K.~D.~Kuntz, W.~P.~Blair, L.~Godfrey, P.~P.~Plucinsky, R.~Soria, C.~Stockdale and P.~F.~Winkler,
 {\color{rossoCP3}  A deep Chandra ACIS survey of M83},
  Astrophys.\ J.\ Suppl.\  {\bf 212}, 21 (2014)
  doi:10.1088/0067-0049/212/2/21
  [arXiv:1404.3218 [astro-ph.GA]].





\bibitem{MacFadyen:1998vz} 
  A.~MacFadyen and S.~E.~Woosley,
    {\color{rossoCP3} Collapsars: Gamma-ray bursts and explosions in ``failed supernovae''},
  Astrophys.\ J.\  {\bf 524}, 262 (1999)
  doi:10.1086/307790
  [astro-ph/9810274].




\bibitem{Dreyer:2009pj} 
  J.~Dreyer, J.~K.~Becker and W.~Rhode,
   {\color{rossoCP3} The starburst-GRB connection},
  arXiv:0909.0158 [astro-ph.HE].


\bibitem{Biermann:2016xzl} 
  P.~L.~Biermann {\it et al.},
   {\color{rossoCP3} The nature and origin of ultrahigh-energy cosmic ray particles},
  arXiv:1610.00944 [astro-ph.HE].


\bibitem{Chary:2001yx} 
  R.~Chary, E.~E.~Becklin and L.~Armus,
    {\color{rossoCP3} Are starburst galaxies the hosts of gamma-ray bursts ?},
  Astrophys.\ J.\  {\bf 566}, 229 (2002)
  doi:10.1086/337964
  [astro-ph/0110010].






\bibitem{Stanek:2006gc} 
  K.~Z.~Stanek {\it et al.},
    {\color{rossoCP3} Protecting life in the milky way: metals keep the GRBs away},
  Acta Astron.\  {\bf 56}, 333 (2006)
  [astro-ph/0604113].




\bibitem{Modjaz:2007st} 
  M.~Modjaz, L.~Kewley, R.~P.~Kirshner, K.~Z.~Stanek, P.~Challis, P.~M.~Garnavich, J.~E.~Greene and J.~L.~Prieto,
    {\color{rossoCP3} Measured metallicities at the sites of nearby broad-lined type Ic supernovae and implications for the SN-GRB connection},
  Astron.\ J.\  {\bf 135}, 1136 (2008)
  doi:10.1088/0004-6256/135/4/1136
  [astro-ph/0701246].


\bibitem{Jimenez:2013dka} 
  R.~Jimenez and T.~Piran,
    {\color{rossoCP3} Reconciling the gamma-ray burst rate and star formation histories},
  Astrophys.\ J.\  {\bf 773}, 126 (2013)
  doi:10.1088/0004-637X/773/2/126
  [arXiv:1303.4809 [astro-ph.HE]].

\bibitem{Wang:2007ya} 
  X.~Y.~Wang, S.~Razzaque, P.~Meszaros and Z.~G.~Dai,
  {\color{rossoCP3}  High-energy cosmic rays and neutrinos from semi-relativistic hypernovae},
  Phys.\ Rev.\ D {\bf 76}, 083009 (2007)
  doi:10.1103/PhysRevD.76.083009
  [arXiv:0705.0027 [astro-ph]].




\bibitem{Ptuskin:2010zn} 
  V.~S.~Ptuskin, V.~N.~Zirakashvili and E.~S.~Seo,
 {\color{rossoCP3}   Spectrum of Galactic cosmic rays accelerated in supernova remnants},
  Astrophys.\ J.\  {\bf 718}, 31 (2010)
  doi:10.1088/0004-637X/718/1/31
  [arXiv:1006.0034 [astro-ph.CO]].



\bibitem{Chevalier:1985pc} 
  R.~A.~Chevalier and A.~W.~Clegg,
    {\color{rossoCP3} Wind from a starburst galaxy nucleus},
  Nature {\bf 317}, 44 (1985).
  doi:10.1038/317044a0


\bibitem{Lacki:2013zsa} 
  B.~C.~Lacki,
   {\color{rossoCP3}  The Fermi Bubbles as starburst wind termination shocks},
  Mon.\ Not.\ Roy.\ Astron.\ Soc.\  {\bf 444}, L39 (2014)
  doi:10.1093/mnrasl/slu107
  [arXiv:1304.6137 [astro-ph.HE]].




\bibitem{Kroupa:2002ky} P.~Kroupa, 
{\color{rossoCP3} The Initial mass
    function of stars: Evidence for uniformity in variable systems},
  Science {\bf 295}, 82 (2002) doi:10.1126/science.1067524
  [astro-ph/0201098].




\bibitem{Strickland:2009we} 
  D.~K.~Strickland and T.~M.~Heckman,
    {\color{rossoCP3} Supernova feedback efficiency and mass loading in the starburst and galactic superwind exemplar M82},
  Astrophys.\ J.\  {\bf 697}, 2030 (2009)
  doi:10.1088/0004-637X/697/2/2030
  [arXiv:0903.4175 [astro-ph.CO]].



\bibitem{Heckman:2000sj} 
  T.~M.~Heckman, M.~D.~Lehnert, D.~K.~Strickland and L.~Armus,
    {\color{rossoCP3} Absorption-line probes of gas and dust in galactic superwinds},
  Astrophys.\ J.\ Suppl.\  {\bf 129}, 493 (2000)
  doi:10.1086/313421
  [astro-ph/0002526].



\bibitem{Hoopes:2006mz} 
  C.~G.~Hoopes {\it et al.},
    {\color{rossoCP3} The diverse properties of the most ultraviolet luminous galaxies discovered by the Galaxy Evolution Explorer},
  Astrophys.\ J.\ Suppl.\  {\bf 173}, 441 (2007)
  doi:10.1086/516644
  [astro-ph/0609415].


\bibitem{Beirao}
P.~Beir\~ao {\it et al},
  {\color{rossoCP3}  Spatially resolved Spitzer-IRS spectral maps of the superwind in M82},
Mon.\ Not.\ Roy.\ Astron.\ Soc.\  {\bf 451}, 2640 (2015). 
doi:10.1093/mnras/stv1101


\bibitem{Contursi:2012wa} 
  A.~Contursi {\it et al.},
    {\color{rossoCP3} Spectroscopic FIR mapping of the disk and galactic wind of M82 with Herschel-PACS},
  Astron.\ Astrophys.\  {\bf 549}, A118 (2013)
  doi:10.1051/0004-6361/201219214
  [arXiv:1210.3496 [astro-ph.GA]].




\bibitem{Heckman:1990fe} 
  T.~M.~Heckman, L.~Armus and G.~K.~Miley,
    {\color{rossoCP3} On the nature and implications of starburst-driven galactic superwinds},
  Astrophys.\ J.\ Suppl.\  {\bf 74}, 833 (1990).
  doi:10.1086/191522



\bibitem{Veilleux:2009rb} 
  S.~Veilleux, D.~S.~N.~Rupke and R.~Swaters,
   {\color{rossoCP3} Warm molecular hydrogen in the galactic wind of M82},
  Astrophys.\ J.\  {\bf 700}, L149 (2009)
  doi:10.1088/0004-637X/700/2/L149
  [arXiv:0907.1422 [astro-ph.CO]].


\bibitem{Leroy}
A. Leroy {\it et al.},
  {\color{rossoCP3} Themulti-phase cold fountain in M82 revealed by a
    wide, sensitive map of the molecular interstellar medium},
 Astrophys.\ J.\  {\bf 814}, 83 (2015)
 doi:10.1088/0004-637X/814/2/83.


\bibitem{Bolatto:2013aqa} 
  A.~D.~Bolatto {\it et al.},
    {\color{rossoCP3} The starburst-driven molecular wind in NGC 253 and the suppression of star formation},
  Nature {\bf 499}, 450 (2013)
  doi:10.1038/nature12351
  [arXiv:1307.6259 [astro-ph.CO]].


\bibitem{Lacki:2013sda} 
  B.~C.~Lacki,
   {\color{rossoCP3}  From 10K to 10 TK: Insights on the interaction between cosmic rays and gas in starbursts},
  Astrophys.\ Space Sci.\ Proc.\  {\bf 34}, 411 (2013)
  doi:10.1007/978-3-642-35410-6$_-$29
  [arXiv:1308.5241 [astro-ph.HE]].




\bibitem{Shukurov:2002hh} 
  A.~Shukurov, G.~R.~Sarson, A.~Nordlund, B.~Gudiksen and A.~Brandenburg,
    {\color{rossoCP3} The effects of spiral arms on the multi-phase ISM},
  Astrophys.\ Space Sci.\  {\bf 289}, 319 (2004)
  doi:10.1023/B:ASTR.0000014960.35780.2e
  [astro-ph/0212260].


\bibitem{Adebahr:2012ce} 
B.~Adebahr, M.~Krause, U.~Klein,
  M.~Wezgowiec, D.~J.~Bomans and R.-J.~Dettmar, 
{\color{rossoCP3} M82 - A radio continuum and polarisation study I: 
Data reduction and cosmic ray propagation}, 
Astron.\ Astrophys.\ {\bf 555}, A23
  (2013) doi:10.1051/0004-6361/201220226 [arXiv:1209.5552
  [astro-ph.GA]].

\bibitem{Beck}
R. Beck, C. L. Carilli, M. A. Holdaway, and U. Klein,
  {\color{rossoCP3}  Multifrequency observations of the radio continuum emission from NGC 253 I: Magnetic fields and rotation measures in the bar and halo},
Astron. Astrophys.  {\bf 292}, 409 (1994).


\bibitem{Heesen:2008cs} 
  V.~Heesen, R.~Beck, M.~Krause and R.-J.~Dettmar,
    {\color{rossoCP3}  Cosmic rays and the magnetic field in the
      nearby starburst galaxy NGC 253 I: The distribution and transport of cosmic rays},
  Astron.\ Astrophys.\  {\bf 494}, 563 (2009)
  doi:10.1051/0004-6361:200810543
  [arXiv:0812.0346 [astro-ph]].


\bibitem{Heesen:2009sg} 
  V.~Heesen, M.~Krause, R.~Beck and R.~J.~Dettmar,
    {\color{rossoCP3}  Cosmic rays and the magnetic field in the
      nearby starburst galaxy NGC 253 II: The magnetic field structure},
  Astron.\ Astrophys.\  {\bf 506}, 1123 (2009)
  doi:10.1051/0004-6361/200911698
  [arXiv:0908.2985 [astro-ph.CO]].


\bibitem{Heesen:2011kj} 
  V.~Heesen, R.~Beck, M.~Krause and R.~J.~Dettmar,
  {\color{rossoCP3}   Cosmic rays and the magnetic field in the nearby
    starburst galaxy NGC 253 III: Helical magnetic fields in the nuclear outflow},
  Astron.\ Astrophys.\  {\bf 535}, A79 (2011)
  doi:10.1051/0004-6361/201117618
  [arXiv:1109.0255 [astro-ph.CO]].




\bibitem{Krause:2014iza} 
  M.~Krause,
   {\color{rossoCP3}  Magnetic fields and halos in spiral galaxies},
  arXiv:1401.1317 [astro-ph.GA].



\bibitem{Thompson:2006is}
  T.~A.~Thompson, E.~Quataert, E.~Waxman, N.~Murray and C.~L.~Martin,
   {\color{rossoCP3} Magnetic fields in starburst galaxies and the origin of the fir-radio correlation},
  Astrophys.\ J.\  {\bf 645}, 186 (2006)
  [astro-ph/0601626].


\bibitem{Paglione:2012ma} 
  T.~A.~D.~Paglione and R.~D.~Abrahams,
  {\color{rossoCP3} Properties of nearby starburst galaxies based on their diffuse gamma-ray emission},
  Astrophys.\ J.\  {\bf 755}, 106 (2012)
  doi:10.1088/0004-637X/755/2/106
  [arXiv:1206.3530 [astro-ph.HE]].




\bibitem{Lacki:2013ry} 
  B.~C.~Lacki and R.~Beck,
    {\color{rossoCP3}  The equipartition magnetic field formula in starburst galaxies: Accounting for pionic secondaries and strong energy losses},
  Mon.\ Not.\ Roy.\ Astron.\ Soc.\  {\bf 430}, 3171 (2013)
  doi:10.1093/mnras/stt122
  [arXiv:1301.5391 [astro-ph.CO]].




\bibitem{DomingoSantamaria:2005qk} 
  E.~Domingo-Santamaria and D.~F.~Torres,
  {\color{rossoCP3}  High energy gamma-ray emission from the starburst nucleus of NGC 253},
  Astron.\ Astrophys.\  {\bf 444}, 403 (2005)
  doi:10.1051/0004-6361:20053613
  [astro-ph/0506240].




\bibitem{delPozo:2009mh} 
  E.~de Cea del Pozo, D.~F.~Torres and A.~Y.~R.~Marrero,
   {\color{rossoCP3}  Multi-messenger model for the starburst galaxy M82},
  Astrophys.\ J.\  {\bf 698}, 1054 (2009)
  doi:10.1088/0004-637X/698/2/1054
  [arXiv:0901.2688 [astro-ph.GA]].


\bibitem{Lacki:2013nda} 
  B.~C.~Lacki,
   {\color{rossoCP3} Sturm und drang: Supernova-driven turbulence, magnetic fields, and cosmic rays in the chaotic starburst interstellar medium},
  arXiv:1308.5232 [astro-ph.CO].


\bibitem{Thornley:2000ib} M.~D.~Thornley, N.~M.~Forster Schreiber,
  D.~Lutz, R.~Genzel, H.~W.~W.~Spoon, D.~Kunze and A.~Sternberg,
  {\color{rossoCP3} Massive star formation and evolution in starburst
    galaxies: mid-infrared spectroscopy with ISO-SWS}, Astrophys.\ J.\
  {\bf 539}, 641 (2000) doi:10.1086/309261 [astro-ph/0003334].





\bibitem{Torres:2004ui} 
  D.~F.~Torres,
   {\color{rossoCP3}  Theoretical modelling of the diffuse emission of gamma-rays from extreme regions of star formation: The Case of Arp 220},
  Astrophys.\ J.\  {\bf 617}, 966 (2004)
  doi:10.1086/425415
  [astro-ph/0407240].






\bibitem{Torres:2012xk} 
  D.~F.~Torres, A.~Cillis, B.~Lacki and Y.~Rephaeli,
   {\color{rossoCP3} Building up the spectrum of cosmic-rays in star-forming regions},
  Mon.\ Not.\ Roy.\ Astron.\ Soc.\  {\bf 423}, 822 (2012)
  doi:10.1111/j.1365-2966.2012.20920.x
  [arXiv:1203.2798 [astro-ph.HE]].



\bibitem{Meurer:2000yq} 
  G.~R.~Meurer,
   {\color{rossoCP3} Star clusters and the duration of starbursts},
  ASP Conf.\ Ser.\  {\bf 211}, 81 (2000)
  [astro-ph/0003161].




\bibitem{McQuinn:2009gc} 
  K.~B.~W.~McQuinn, E.~D.~Skillman, J.~M.~Cannon, J.~J.~Dalcanton, A.~Dolphin, D.~Stark and D.~Weisz,
   {\color{rossoCP3}  The true durations of starbursts: HST observations of three nearby dwarf starburst galaxies},
  Astrophys.\ J.\  {\bf 695}, 561 (2009)
  doi:10.1088/0004-637X/695/1/561
  [arXiv:0901.2361 [astro-ph.GA]].


\bibitem{McQuinn:2010kn} 
  K.~B.~W.~McQuinn {\it et al.},
    {\color{rossoCP3} The nature of starbursts  II: The duration of starbursts in dwarf galaxies},
  Astrophys.\ J.\  {\bf 724}, 49 (2010)
  doi:10.1088/0004-637X/724/1/49
  [arXiv:1009.2940 [astro-ph.CO]].






\bibitem{deGrijs:2001ec} 
  R.~de Grijs,
   {\color{rossoCP3}  Star formation time-scales in the nearby, prototype starburst galaxy M82},
Astron.\ Geophys.\  {\bf 42}, 14 (2001)  
[astro-ph/0106564].


\bibitem{deGrijs:2002qk} 
  R.~de Grijs, N.~Bastian and H.~J.~G.~L.~M.~Lamers,
    {\color{rossoCP3} Star cluster formation and disruption time-scales II: Evolution of the star cluster system in M82's fossil starburst},
  Mon.\ Not.\ Roy.\ Astron.\ Soc.\  {\bf 340}, 197 (2003)
  doi:10.1046/j.1365-8711.2003.06283.x
  [astro-ph/0211420].




\bibitem{Davidge:2010qg} 
  T.~J.~Davidge,
   {\color{rossoCP3}  Shaken, not stirred: The disrupted disk of the starburst galaxy NGC 253},
  Astrophys.\ J.\  {\bf 725}, 1342 (2010)
  doi:10.1088/0004-637X/725/1/1342
  [arXiv:1011.3006 [astro-ph.CO]].


\bibitem{Davidge}
T.~J.~Davidge
  {\color{rossoCP3} The compact star-forming complex at the heart of
    NGC 253},
Astrophys.\ J.\  {\bf 818}, 142 (2016)
doi:10.3847/0004-637X/818/2/142
[arXiv:1602.01400 [astro-ph.GA]].

\bibitem{Rieke:1980xt} 
  G.~H.~Rieke, M.~J.~Lebofsky, R.~I.~Thompson, F.~J.~Low and A.~T.~Tokunaga,
{\color{rossoCP3}  The nature of the nuclear sources in M82 and NGC 253},
  Astrophys.\ J.\  {\bf 238}, 24 (1980).
  doi:10.1086/157954



\bibitem{Vink:2002yx} 
  J.~Vink and J.~M.~Laming,
   {\color{rossoCP3} On the magnetic fields and particle acceleration in Cas A},
  Astrophys.\ J.\  {\bf 584}, 758 (2003)
  doi:10.1086/345832
  [astro-ph/0210669].


\bibitem{Yamazaki:2003xq} 
  R.~Yamazaki, T.~Yoshida, T.~Terasawa, A.~Bamba and K.~Koyama,
   {\color{rossoCP3} Constraints on the diffusive shock acceleration from the nonthermal X-ray thin shells in SN1006 NE rim},
  Astron.\ Astrophys.\  {\bf 416}, 595 (2004)
  doi:10.1051/0004-6361:20034212
  [astro-ph/0311345].



\bibitem{Volk:2004vi} 
  H.~J.~Volk, E.~G.~Berezhko and L.~T.~Ksenofontov,
   {\color{rossoCP3} Magnetic field amplification in Tycho and other shell-type supernova remnants},
  Astron.\ Astrophys.\  {\bf 433}, 229 (2005)
  doi:10.1051/0004-6361:20042015
  [astro-ph/0409453].


\bibitem{Lucek:2000} 
S. G. Lucek and A. R. Bell,
 {\color{rossoCP3}  Non-linear amplification of a magnetic field driven by cosmic ray streaming},
Mon.\ Not.\ Roy.\ Astron.\ Soc.\  {\bf 314},  65 (2000).
doi:10.1046/j.1365-8711.2000.03363.x

\bibitem{Bell:2004}
A. R. Bell,
 {\color{rossoCP3} Turbulent amplification of magnetic field and diffusive shock acceleration of cosmic rays},
Mon.\ Not.\ Roy.\ Astron.\ Soc.\  {\bf 353},  550 (2004)
doi:10.1111/j.1365-2966.2004.08097.x

\bibitem{Bell:2005}
A. R. Bell,
 {\color{rossoCP3} The interaction of cosmic rays and magnetized plasma},
Mon.\ Not.\ Roy.\ Astron.\ Soc.\  {\bf 358},  181 (2005)
doi:10.1111/j.1365-2966.2005.08774.x

\bibitem{Matthews:2017apu} 
  J.~H.~Matthews, A.~R.~Bell, K.~M.~Blundell and A.~T.~Araudo,
 {\color{rossoCP3} Amplification of perpendicular and parallel magnetic fields by cosmic ray currents},
  Mon.\ Not.\ Roy.\ Astron.\ Soc.\  {\bf 469}, no. 2, 1849 (2017)
  doi:10.1093/mnras/stx905
  [arXiv:1704.02985 [astro-ph.HE]].





\bibitem{Lipari}
S. L\'{\i}pari, Z. Tsvetanov, and F. Macchetto
 {\color{rossoCP3} Luminous ifrared galaxies II -- NGC 4945: A nearby obscured starburst/seyfert nucleus},
 Astrophys.\ J.\ Suppl.\ {\bf 111}, 369 (1997).


\bibitem{Marconi:2000xn} 
  A.~Marconi, E.~Oliva, P.~P.~van der Werf, R.~Maiolino, E.~J.~Schreier, F.~Macchetto and A.~F.~M.~Moorwood,
   {\color{rossoCP3} The elusive active nucleus of NGC 4945},
  Astron.\ Astrophys.\  {\bf 357}, 24 (2000)
  [astro-ph/0002244].

\bibitem{Levenson:2002fq} 
  N.~A.~Levenson, J.~H.~Krolik, P.~T.~Zycki, T.~M.~Heckman, K.~A.~Weaver, H.~Awaki and Y.~Terashima,
   {\color{rossoCP3} Extreme X-ray iron lines in active galactic nuclei},
  Astrophys.\ J.\  {\bf 573}, L81 (2002)
  doi:10.1086/342092
  [astro-ph/0206071].


\bibitem{Strickland:2003xk} 
  D.~K.~Strickland, T.~M.~Heckman, E.~J.~M.~Colbert, C.~G.~Hoopes and K.~A.~Weaver,
   {\color{rossoCP3} A high spatial resolution X-ray and H-alpha study of hot gas in the halos of star-forming disk galaxies I: Spatial and spectral properties of the diffuse X-ray emission},
  Astrophys.\ J.\ Suppl.\  {\bf 151}, 193 (2004)
  doi:10.1086/382214
  [astro-ph/0306592].


\bibitem{Levenson:2000qc} 
  N.~A.~Levenson, K.~A.~Weaver and T.~M.~Heckman,
   {\color{rossoCP3} The seyfert-starburst connection in X-rays I: the data},
  Astrophys.\ J.\ Suppl.\  {\bf 133}, 269 (2001)
  doi:10.1086/320355
  [astro-ph/0012035].


\bibitem{Levenson:2000qd} 
  N.~A.~Levenson, K.~A.~Weaver and T.~M.~Heckman,
   {\color{rossoCP3} The seyfert-starburst connection in X-rays 2: Results and implications},
  Astrophys.\ J.\  {\bf 550}, 230 (2001)
  doi:10.1086/319726
  [astro-ph/0012036].


\bibitem{Levenson:2003ez} 
  N.~A.~Levenson, K.~A.~Weaver, T.~M.~Heckman, H.~Awaki and Y.~Terashima,
   {\color{rossoCP3} Accretion and outflow in the AGN and starburst of NGC 5135},
  Astrophys.\ J.\  {\bf 602}, 135 (2004)
  doi:10.1086/380836
  [astro-ph/0310669].


\bibitem{Strickland:2004qk} 
  D.~K.~Strickland,
   {\color{rossoCP3} Winds from nuclear starbursts: Old truths and recent progress on superwinds},
  IAU Symp.\  {\bf 222}, 249 (2004)
  doi:10.1017/S1743921304002194
  [astro-ph/0404316].


\bibitem{Colbert:1995sz} 
  E.~J.~M.~Colbert, S.~A.~Baum, J.~F.~Gallimore, C.~P.~O'Dea, M.~D.~Lehnert, Z.~I.~Tsvetanov, J.~S.~Mulchaey and S.~Caganoff,
   {\color{rossoCP3} Large scale outflows in edge-on seyfert galaxies I: optical emission- line imaging and optical spectroscopy},
  Astrophys.\ J.\ Suppl.\  {\bf 105}, 75 (1996)
  doi:10.1086/192307
  [astro-ph/9512169].



\bibitem{Colbert:1997fa} 
  E.~J.~M.~Colbert, S.~A.~Baum, C.~P.~O'Dea and S.~Veilleux,
   {\color{rossoCP3} Large scale outflows in edge-on seyfert galaxies III: Kiloparsec-scale soft X-ray emission},
  Astrophys.\ J.\  {\bf 496}, 786 (1998)
  doi:10.1086/305417
  [astro-ph/9711137].

\bibitem{Krolik:1986}
J. H. Krolik and M. C. Begelman,
  {\color{rossoCP3} An X-ray heated wind in NGC 1068},
Astrophys. J. {\bf 308}, L55 (1986).


\bibitem{Soria:2002jk} 
  R.~Soria and K.~Wu,
   {\color{rossoCP3} X-ray sources in the starburst spiral galaxy M83: nuclear region and discrete source population},
  Astron.\ Astrophys.\  {\bf 384}, no. 1, 99 (2002)
  doi:10.1051/0004-6361:20020026
  [astro-ph/0201059].



\bibitem{Vogler:2005bg} 
  A.~Vogler, S.~C.~Madden, R.~Beck, A.~A.~Lundgren, M.~Sauvage, L.~Vigroux and M.~Ehle,
 {\color{rossoCP3} Dissecting the spiral galaxy M83: Mid-infrared emission and comparison with other tracers of star formation},
  Astron.\ Astrophys.\  {\bf 441}, 491 (2005)
  doi:10.1051/0004-6361:20042342
  [astro-ph/0508027].

\bibitem{Veilleux:1997nk} 
  S.~Veilleux and J.~Bland-Hawthorn,
  {\color{rossoCP3} Artillery shells over Circinus},
  Astrophys.\ J.\  {\bf 479}, L105 (1997)
  doi:10.1086/310588
  [astro-ph/9703040].

\bibitem{Elmouttie}
M. Elmouttie, B. Koribalski, S. Gordon,  K. Taylor, S. Houghton,
T. Lavezzi, R. Haynes, and K. Jones,
  {\color{rossoCP3} The kinematics of the ionized gas in the Circinus galaxy},
 Mon.\ Not.\ Roy.\ Astron.\ Soc.\  {\bf 297}, 49 (1998).


\bibitem{Luppino}
G. A. Luppino and J. L. Tonry
  {\color{rossoCP3} Infrared surface brightness fluctuations:
    $K'$-band observations of M32, M32, and Maffei 1},
Astrophys. J. {\bf 410}, 81 (1993).


\bibitem{Krismer}
M. Krismer, R. B. Tully, and I. Gioia,
 {\color{rossoCP3} IC 342/Maffei group of galaxies and distances for
   two of its members},
Astron. J. {\bf 110}, 1584 (1995).


\bibitem{Buta}
R. J. Buta and M. L. McCall,
  {\color{rossoCP3} The IC 342/Maffei group revealed} 
 Astrophys.\ J.\ Suppl.\  {\bf 124}, 33 (1999).

\bibitem{Tikhonov:2010in} 
  N.~A.~Tikhonov and O.~A.~Galazutdinova,
   {\color{rossoCP3} Distance to the galaxy IC 342},
  Astron.\ Lett.\  {\bf 36}, 167 (2010)
  doi:10.1134/S1063773710030023
  [arXiv:1003.0321 [astro-ph.CO]].

\bibitem{Schinnerer:2008pq} 
  E.~Schinnerer, T.~Boeker, D.~S.~Meier and D.~Calzetti,
    {\color{rossoCP3} Self-regulated fueling of galaxy centers: Evidence for star-formation feedback in IC342's nucleus},
  Astrophys.\ J.\  {\bf 684}, L21 (2008)
  doi:10.1086/592109
  [arXiv:0808.0793 [astro-ph]].


\bibitem{Bregman}
J. N. Bregman, C. V. Cox, and K. Tomisaka,
 {\color{rossoCP3} X-ray emission from the starburst galaxy IC 342},
 Astrophys.\ J.\  {\bf 415}, L79 (1993).


\bibitem{Taylor:2011ta} 
  A.~M.~Taylor, M.~Ahlers and F.~A.~Aharonian,
  {\color{rossoCP3} The need for a local source of UHE CR nuclei},
  Phys.\ Rev.\ D {\bf 84}, 105007 (2011)
  doi:10.1103/PhysRevD.84.105007
  [arXiv:1107.2055 [astro-ph.HE]].


\bibitem{Ahlers:2012az} 
  M.~Ahlers, L.~A.~Anchordoqui and A.~M.~Taylor,
  {\color{rossoCP3}  Ensemble fluctuations of the flux and nuclear composition of ultrahigh energy cosmic ray nuclei},
  Phys.\ Rev.\ D {\bf 87}, no. 2, 023004 (2013)
  doi:10.1103/PhysRevD.87.023004
  [arXiv:1209.5427 [astro-ph.HE]].


\bibitem{Ahlers:2013zxa}
  L.~A.~Anchordoqui, M.~Ahlers, A.~V.~Olinto, T.~C.~Paul and A.~M.~Taylor,
   {\color{rossoCP3}  Sensitivity of JEM-EUSO to ensemble fluctuations in the ultra-high energy cosmic ray flux},
  arXiv:1306.0910 [astro-ph.CO].

\bibitem{Robertson:2015uda} 
  B.~E.~Robertson, R.~S.~Ellis, S.~R.~Furlanetto and J.~S.~Dunlop,
   {\color{rossoCP3}  Cosmic reionization and early star-forming galaxies: a joint analysis of new constraints from Planck and the Hubble Space Telescope},
  Astrophys.\ J.\  {\bf 802}, no. 2, L19 (2015)
  doi:10.1088/2041-8205/802/2/L19
  [arXiv:1502.02024 [astro-ph.CO]].

\bibitem{Kampert:2012fi} 
  K.~H.~Kampert, J.~Kulbartz, L.~Maccione, N.~Nierstenhoefer, P.~Schiffer, G.~Sigl and A.~R.~van Vliet,
   {\color{rossoCP3}  CRPropa 2.0:  a public framework for propagating high energy nuclei, secondary gamma rays and neutrinos},
  Astropart.\ Phys.\  {\bf 42}, 41 (2013)
  doi:10.1016/j.astropartphys.2012.12.001
  [arXiv:1206.3132 [astro-ph.IM]].

\bibitem{Batista:2013gka} 
  R.~Alves Batista {\it et al.},
   {\color{rossoCP3}  CRPropa 3.0: a public framework for propagating UHE cosmic rays through Galactic and extragalactic space},
  arXiv:1307.2643 [astro-ph.IM].


\bibitem{Hooper:2008pm} 
  D.~Hooper, S.~Sarkar and A.~M.~Taylor,
   {\color{rossoCP3}  The intergalactic propagation of ultra-high energy cosmic ray nuclei: an analytic approach},
  Phys.\ Rev.\ D {\bf 77}, 103007 (2008)
  doi:10.1103/PhysRevD.77.103007
  [arXiv:0802.1538 [astro-ph]].

\bibitem{Ahlers:2010ty} 
  M.~Ahlers and A.~M.~Taylor,
   {\color{rossoCP3}  Analytic solutions of ultra-high energy cosmic ray nuclei revisited},
  Phys.\ Rev.\ D {\bf 82}, 123005 (2010)
  doi:10.1103/PhysRevD.82.123005
  [arXiv:1010.3019 [astro-ph.HE]].


\bibitem{Szabo:1994qx} 
  A.~P.~Szabo and R.~J.~Protheroe,
    {\color{rossoCP3}  Implications of particle acceleration in active galactic nuclei for cosmic rays and high-energy neutrino astronomy},
  Astropart.\ Phys.\  {\bf 2}, 375 (1994)
  doi:10.1016/0927-6505(94)90027-2
  [astro-ph/9405020].

\bibitem{Protheroe:1998pj} 
  R.~J.~Protheroe and T.~Stanev,
   {\color{rossoCP3} Cut-offs and pile-ups in shock acceleration spectra},
  Astropart.\ Phys.\  {\bf 10}, 185 (1999)
  doi:10.1016/S0927-6505(98)00055-3
  [astro-ph/9808129].

\bibitem{ThePierreAuger:2013eja} 
  A.~Aab {\it et al.} [Pierre Auger Collaboration],
 {\color{rossoCP3}  The Pierre Auger Observatory: Contributions to the 33rd International Cosmic Ray Conference (ICRC 2013)},
  arXiv:1307.5059 [astro-ph.HE].


\bibitem{Abreu:2013env} 
  P.~Abreu {\it et al.} [Pierre Auger Collaboration],
  {\color{rossoCP3} Interpretation of the Depths of Maximum of Extensive Air Showers Measured by the Pierre Auger Observatory},
  JCAP {\bf 1302}, 026 (2013)
  doi:10.1088/1475-7516/2013/02/026
  [arXiv:1301.6637 [astro-ph.HE]].

\bibitem{Pierog:2013ria} 
  T.~Pierog, I.~Karpenko, J.~M.~Katzy, E.~Yatsenko and K.~Werner,
    {\color{rossoCP3} EPOS LHC: Test of collective hadronization with data measured at the CERN Large Hadron Collider},
  Phys.\ Rev.\ C {\bf 92}, no. 3, 034906 (2015)
  doi:10.1103/PhysRevC.92.034906
  [arXiv:1306.0121 [hep-ph]].


\bibitem{Globus:2015xga} 
  N.~Globus, D.~Allard and E.~Parizot,
    {\color{rossoCP3} A complete model of the cosmic ray spectrum and composition across the Galactic to extragalactic transition},
  Phys.\ Rev.\ D {\bf 92}, no. 2, 021302 (2015)
  doi:10.1103/PhysRevD.92.021302
  [arXiv:1505.01377 [astro-ph.HE]].


\bibitem{Anchordoqui:2017acf} 
  L.~A.~Anchordoqui,
   {\color{rossoCP3} Unmasking the ultra-high-energy cosmic ray origin},
  PoS EPS {\bf -HEP2017}, 001 (2017)
  doi:10.22323/1.314.0001
  [arXiv:1707.09338 [astro-ph.HE]].


\bibitem{Supanitsky:2018jje} 
  A.~D.~Supanitsky, A.~Cobos and A.~Etchegoyen,
  {\color{rossoCP3}  Origin of the light cosmic ray component below the ankle},
  Phys.\ Rev.\ D {\bf 98}, no. 10, 103016 (2018)
  doi:10.1103/PhysRevD.98.103016
  [arXiv:1810.12367 [astro-ph.HE]].


\bibitem{Hasinger:2005sb} 
  G.~Hasinger, T.~Miyaji and M.~Schmidt,
   {\color{rossoCP3}  Luminosity-dependent evolution of soft X-ray selected AGN: New Chandra and XMM-Newton surveys},
  Astron.\ Astrophys.\  {\bf 441}, 417 (2005)
  doi:10.1051/0004-6361:20042134
  [astro-ph/0506118].


\bibitem{Stanev:2008un} 
  T.~Stanev,
   {\color{rossoCP3}  Ultra high energy cosmic rays and neutrinos after Auger},
  arXiv:0808.1045 [astro-ph].


\bibitem{Taylor:2015rla} 
  A.~M.~Taylor, M.~Ahlers and D.~Hooper,
   {\color{rossoCP3}  Indications of negative evolution for the sources of the highest energy cosmic rays},
  Phys.\ Rev.\ D {\bf 92}, no. 6, 063011 (2015)
  doi:10.1103/PhysRevD.92.063011
  [arXiv:1505.06090 [astro-ph.HE]].

\bibitem{Ajello:2013lka} 
  M.~Ajello {\it et al.},
    {\color{rossoCP3} The cosmic evolution of Fermi BL Lacertae objects},
  Astrophys.\ J.\  {\bf 780}, 73 (2014)
  doi:10.1088/0004-637X/780/1/73
  [arXiv:1310.0006 [astro-ph.CO]].



\bibitem{Waxman:1998yy} 
  E.~Waxman and J.~N.~Bahcall,
    {\color{rossoCP3} High-energy neutrinos from astrophysical sources: an upper bound},
  Phys.\ Rev.\ D {\bf 59}, 023002 (1999)
  doi:10.1103/PhysRevD.59.023002
  [hep-ph/9807282].

\bibitem{Ahlers:2005sn} 
  M.~Ahlers, L.~A.~Anchordoqui, H.~Goldberg, F.~Halzen, A.~Ringwald and T.~J.~Weiler,
    {\color{rossoCP3} Neutrinos as a diagnostic of cosmic ray galactic/extra-galactic transition},
  Phys.\ Rev.\ D {\bf 72}, 023001 (2005)
  doi:10.1103/PhysRevD.72.023001
  [astro-ph/0503229].

\bibitem{Gaisser:1997aw} 
  T.~K.~Gaisser,
    {\color{rossoCP3} Neutrino astronomy: Physics goals, detector parameters},
  astro-ph/9707283.


\bibitem{Waxman:1995dg} 
  E.~Waxman,
    {\color{rossoCP3} Cosmological origin for cosmic rays above $10^{19}~{\rm eV}$},
  Astrophys.\ J.\  {\bf 452}, L1 (1995)
  doi:10.1086/309715
  [astro-ph/9508037].



\bibitem{Ahlers:2018fkn} 
  M.~Ahlers and F.~Halzen,
    {\color{rossoCP3}  Opening a new window onto the universe with IceCube},
  Prog.\ Part.\ Nucl.\ Phys.\  {\bf 102}, 73 (2018)
doi:10.1016/j.ppnp.2018.05.001 
[arXiv:1805.11112 [astro-ph.HE]].


\bibitem{Anchordoqui:2003vc} 
  L.~A.~Anchordoqui, H.~Goldberg, F.~Halzen and T.~J.~Weiler,
    {\color{rossoCP3} Galactic point sources of TeV antineutrinos},
  Phys.\ Lett.\ B {\bf 593}, 42 (2004)
  doi:10.1016/j.physletb.2004.04.054
  [astro-ph/0311002].

\bibitem{Frichter:1997wh} G.~M.~Frichter, T.~K.~Gaisser and T.~Stanev,
  {\color{rossoCP3} Inelasticity in $p$-nucleus collisions and its
    application to high-energy cosmic ray cascades}, 
  Phys.\ Rev.\ D  {\bf 56}, 3135 (1997) doi:10.1103/PhysRevD.56.3135
  [astro-ph/9704061].


\bibitem{Learned:1994wg} 
  J.~G.~Learned and S.~Pakvasa,
   {\color{rossoCP3} Detecting tau-neutrino oscillations at PeV energies},
  Astropart.\ Phys.\  {\bf 3}, 267 (1995)
  doi:10.1016/0927-6505(94)00043-3
  [hep-ph/9405296, hep-ph/9408296].

\bibitem{Beresinsky:1969qj} 
  V.~S.~Berezinsky and G.~T.~Zatsepin,
   {\color{rossoCP3} Cosmic rays at ultrahigh-energies (neutrino?)},
  Phys.\ Lett.\  {\bf 28B}, 423 (1969).
  doi:10.1016/0370-2693(69)90341-4

\bibitem{Stecker:1978ah} 
  F.~W.~Stecker,
    {\color{rossoCP3} Diffuse fluxes of cosmic high-energy neutrinos},
  Astrophys.\ J.\  {\bf 228}, 919 (1979).
  doi:10.1086/156919


\bibitem{Hill:1983xs} 
  C.~T.~Hill and D.~N.~Schramm,
   {\color{rossoCP3}  Ultra-high-energy cosmic ray neutrinos},
  Phys.\ Lett.\ B {\bf 131}, 247 (1983)
  [Phys.\ Lett.\  {\bf 131B}, 247 (1983)].
  doi:10.1016/0370-2693(83)91130-9


\bibitem{Engel:2001hd} 
  R.~Engel, D.~Seckel and T.~Stanev,
    {\color{rossoCP3} Neutrinos from propagation of ultrahigh-energy protons},
  Phys.\ Rev.\ D {\bf 64}, 093010 (2001)
  doi:10.1103/PhysRevD.64.093010
  [astro-ph/0101216].




\bibitem{Fodor:2003ph} 
  Z.~Fodor, S.~D.~Katz, A.~Ringwald and H.~Tu,
    {\color{rossoCP3} Bounds on the cosmogenic neutrino flux},
  JCAP {\bf 0311}, 015 (2003)
  doi:10.1088/1475-7516/2003/11/015
  [hep-ph/0309171].

\bibitem{Hooper:2004jc} 
  D.~Hooper, A.~Taylor and S.~Sarkar,
    {\color{rossoCP3} The Impact of heavy nuclei on the cosmogenic neutrino flux},
  Astropart.\ Phys.\  {\bf 23}, 11 (2005)
  doi:10.1016/j.astropartphys.2004.11.002
  [astro-ph/0407618].



\bibitem{Ave:2004uj} 
  M.~Ave, N.~Busca, A.~V.~Olinto, A.~A.~Watson and T.~Yamamoto,
    {\color{rossoCP3} Cosmogenic neutrinos from ultra-high energy nuclei},
  Astropart.\ Phys.\  {\bf 23}, 19 (2005)
  doi:10.1016/j.astropartphys.2004.11.001
  [astro-ph/0409316].


\bibitem{Anchordoqui:2007fi} 
  L.~A.~Anchordoqui, H.~Goldberg, D.~Hooper, S.~Sarkar and A.~M.~Taylor,
   {\color{rossoCP3} Predictions for the cosmogenic neutrino flux in light of new data from the Pierre Auger Observatory},
  Phys.\ Rev.\ D {\bf 76}, 123008 (2007)
  doi:10.1103/PhysRevD.76.123008
  [arXiv:0709.0734 [astro-ph]].


\bibitem{Kotera:2010yn} 
  K.~Kotera, D.~Allard and A.~V.~Olinto,
    {\color{rossoCP3} Cosmogenic neutrinos: parameter space and detectabilty from PeV to ZeV},
  JCAP {\bf 1010}, 013 (2010)
  doi:10.1088/1475-7516/2010/10/013
  [arXiv:1009.1382 [astro-ph.HE]].


\bibitem{Ahlers:2012rz} 
  M.~Ahlers and F.~Halzen,
    {\color{rossoCP3} Minimal cosmogenic neutrinos},
  Phys.\ Rev.\ D {\bf 86}, 083010 (2012)
  doi:10.1103/PhysRevD.86.083010
  [arXiv:1208.4181 [astro-ph.HE]].






\bibitem{AlvesBatista:2018zui} 
  R.~Alves Batista, R.~M.~de Almeida, B.~Lago and K.~Kotera,
    {\color{rossoCP3} Cosmogenic photon and neutrino fluxes in the Auger era},
  arXiv:1806.10879 [astro-ph.HE].


\bibitem{Ahlers:2010fw} 
  M.~Ahlers, L.~A.~Anchordoqui, M.~C.~Gonzalez-Garcia, F.~Halzen and S.~Sarkar,
   {\color{rossoCP3} GZK neutrinos after the Fermi-LAT diffuse photon flux measurement},
  Astropart.\ Phys.\  {\bf 34}, 106 (2010)
  doi:10.1016/j.astropartphys.2010.06.003
  [arXiv:1005.2620 [astro-ph.HE]].


\bibitem{Gaisser:1994yf} 
  T.~K.~Gaisser, F.~Halzen and T.~Stanev,
  {\color{rossoCP3} Particle astrophysics with high-energy neutrinos},
  Phys.\ Rept.\  {\bf 258}, 173 (1995)
  Erratum: [Phys.\ Rept.\  {\bf 271}, 355 (1996)]
  doi:10.1016/0370-1573(95)00003-Y
  [hep-ph/9410384].

\bibitem{Learned:2000sw} 
  J.~G.~Learned and K.~Mannheim,
  {\color{rossoCP3} High-energy neutrino astrophysics},
  Ann.\ Rev.\ Nucl.\ Part.\ Sci.\  {\bf 50}, 679 (2000).
  doi:10.1146/annurev.nucl.50.1.679

\bibitem{Halzen:2002pg} 
  F.~Halzen and D.~Hooper,
   {\color{rossoCP3} High-energy neutrino astronomy: The Cosmic ray connection},
  Rept.\ Prog.\ Phys.\  {\bf 65}, 1025 (2002)
  doi:10.1088/0034-4885/65/7/201
  [astro-ph/0204527].

\bibitem{Anchordoqui:2009nf} 
  L.~A.~Anchordoqui and T.~Montaruli,
   {\color{rossoCP3} In search for extraterrestrial high energy neutrinos},
  Ann.\ Rev.\ Nucl.\ Part.\ Sci.\  {\bf 60}, 129 (2010)
  doi:10.1146/annurev.nucl.012809.104551
  [arXiv:0912.1035 [astro-ph.HE]].

\bibitem{Halzen:2007ip} 
  F.~Halzen,
   {\color{rossoCP3} Neutrino astrophysics experiments beneath the sea and ice},
  Science {\bf 315}, 66 (2007).
  doi:10.1126/science.1136504



\bibitem{Capelle:1998zz} 
  K.~S.~Capelle, J.~W.~Cronin, G.~Parente and E.~Zas,
   {\color{rossoCP3} On the detection of ultrahigh-energy neutrinos with the Auger Observatory},
  Astropart.\ Phys.\  {\bf 8}, 321 (1998)
  doi:10.1016/S0927-6505(97)00059-5
  [astro-ph/9801313].


\bibitem{Domokos:1997ve} 
  G.~Domokos and S.~Kovesi-Domokos,
    {\color{rossoCP3} Observation of UHE interactions neutrinos from outer space},
  AIP Conf.\ Proc.\  {\bf 433}, no. 1, 390 (1998)
  doi:10.1063/1.56127
  [hep-ph/9801362].


\bibitem{Domokos:1998hz} 
  G.~Domokos and S.~Kovesi-Domokos,
    {\color{rossoCP3} Observation of ultrahigh-energy neutrino interactions by orbiting detectors},
  hep-ph/9805221.


\bibitem{Bertou:2001vm} 
  X.~Bertou, P.~Billoir, O.~Deligny, C.~Lachaud and A.~Letessier-Selvon,
   {\color{rossoCP3} Tau neutrinos in the Auger Observatory: a new window to UHECR sources},
  Astropart.\ Phys.\  {\bf 17}, 183 (2002)
  doi:10.1016/S0927-6505(01)00147-5
  [astro-ph/0104452].



\bibitem{Feng:2001ue} 
  J.~L.~Feng, P.~Fisher, F.~Wilczek and T.~M.~Yu,
   {\color{rossoCP3} Observability of earth skimming ultrahigh-energy neutrinos},
  Phys.\ Rev.\ Lett.\  {\bf 88}, 161102 (2002)
  doi:10.1103/PhysRevLett.88.161102
  [hep-ph/0105067].

\bibitem{Fargion:2000iz} 
  D.~Fargion,
   {\color{rossoCP3} Discovering ultra high energy neutrinos by horizontal and upward tau air-showers: evidences in terrestrial gamma flashes?},
  Astrophys.\ J.\  {\bf 570}, 909 (2002)
  doi:10.1086/339772
  [astro-ph/0002453].


\bibitem{Abraham:2007rj} 
  J.~Abraham {\it et al.} [Pierre Auger Collaboration],
   {\color{rossoCP3} Upper limit on the diffuse flux of UHE tau neutrinos from the Pierre Auger Observatory},
  Phys.\ Rev.\ Lett.\  {\bf 100}, 211101 (2008)
  doi:10.1103/PhysRevLett.100.211101
  [arXiv:0712.1909 [astro-ph]].


\bibitem{Abraham:2009uy} 
  J.~Abraham {\it et al.} [Pierre Auger Collaboration],
    {\color{rossoCP3} Limit on the diffuse flux of ultra-high energy tau neutrinos with the surface detector of the Pierre Auger Observatory},
  Phys.\ Rev.\ D {\bf 79}, 102001 (2009)
  doi:10.1103/PhysRevD.79.102001
  [arXiv:0903.3385 [astro-ph.HE]].



\bibitem{Abreu:2011zze} 
  P.~Abreu {\it et al.} [Pierre Auger Collaboration],
    {\color{rossoCP3} A search for ultra-high-energy neutrinos in highly inclined events at the Pierre Auger Observatory},
  Phys.\ Rev.\ D {\bf 84}, 122005 (2011)
  Erratum: [Phys.\ Rev.\ D {\bf 84}, 029902 (2011)]
  doi:10.1103/PhysRevD.85.029902, 10.1103/PhysRevD.84.122005
  [arXiv:1202.1493 [astro-ph.HE]].


\bibitem{Abreu:2012zz} 
  P.~Abreu {\it et al.} [Pierre Auger Collaboration],
    {\color{rossoCP3} Search for point-like sources of ultra-high energy neutrinos at the Pierre Auger Observatory and improved limit on the diffuse flux of tau neutrinos},
  Astrophys.\ J.\  {\bf 755}, L4 (2012)
  doi:10.1088/2041-8205/755/1/L4
  [arXiv:1210.3143 [astro-ph.HE]].


\bibitem{Abreu:2013zbq} 
  P.~Abreu {\it et al.} [Pierre Auger Collaboration],
    {\color{rossoCP3} Ultra-high-energy neutrinos at the Pierre Auger Observatory},
  Adv.\ High Energy Phys.\  {\bf 2013}, 708680 (2013)
  doi:10.1155/2013/708680
  [arXiv:1304.1630 [astro-ph.HE]].



\bibitem{Aab:2015kma} 
  A.~Aab {\it et al.} [Pierre Auger Collaboration],
    {\color{rossoCP3} Improved limit to the diffuse flux of ultra-high-energy neutrinos from the Pierre Auger Observatory},
  Phys.\ Rev.\ D {\bf 91}, no. 9, 092008 (2015)
  doi:10.1103/PhysRevD.91.092008
  [arXiv:1504.05397 [astro-ph.HE]].




\bibitem{Anchordoqui:2002vb} 
  L.~A.~Anchordoqui, J.~L.~Feng, H.~Goldberg and A.~D.~Shapere,
   {\color{rossoCP3} Neutrino bounds on astrophysical sources and new physics},
  Phys.\ Rev.\ D {\bf 66}, 103002 (2002)
  doi:10.1103/PhysRevD.66.103002
  [hep-ph/0207139].

\bibitem{Feldman:1997qc} 
  G.~J.~Feldman and R.~D.~Cousins,
   {\color{rossoCP3} A unified approach to the classical statistical analysis of small signals},
  Phys.\ Rev.\ D {\bf 57}, 3873 (1998)
  doi:10.1103/PhysRevD.57.3873
  [physics/9711021 [physics.data-an]].


\bibitem{Gandhi:1998ri} 
  R.~Gandhi, C.~Quigg, M.~H.~Reno and I.~Sarcevic,
   {\color{rossoCP3} Neutrino interactions at ultrahigh-energies},
  Phys.\ Rev.\ D {\bf 58}, 093009 (1998)
  doi:10.1103/PhysRevD.58.093009
  [hep-ph/9807264].




\bibitem{Aartsen:2018vtx} 
  M.~G.~Aartsen {\it et al.} [IceCube Collaboration],
    {\color{rossoCP3} Differential limit on the extremely-high-energy cosmic neutrino flux in the presence of astrophysical background from nine years of IceCube data},
  arXiv:1807.01820 [astro-ph.HE].


\bibitem{Ahlers:2009rf} 
  M.~Ahlers, L.~A.~Anchordoqui and S.~Sarkar,
   {\color{rossoCP3}  Neutrino diagnostics of ultra-high energy cosmic ray protons},
  Phys.\ Rev.\ D {\bf 79}, 083009 (2009)
  doi:10.1103/PhysRevD.79.083009
  [arXiv:0902.3993 [astro-ph.HE]].


\bibitem{Aloisio:2015ega} 
  R.~Aloisio, D.~Boncioli, A.~di Matteo, A.~F.~Grillo, S.~Petrera and F.~Salamida,
  {\color{rossoCP3}   Cosmogenic neutrinos and ultra-high energy cosmic ray models},
  JCAP {\bf 1510}, no. 10, 006 (2015)
  doi:10.1088/1475-7516/2015/10/006
  [arXiv:1505.04020 [astro-ph.HE]].


\bibitem{Heinze:2015hhp} 
  J.~Heinze, D.~Boncioli, M.~Bustamante and W.~Winter,
    {\color{rossoCP3}  Cosmogenic neutrinos challenge the cosmic ray proton dip model},
  Astrophys.\ J.\  {\bf 825}, no. 2, 122 (2016)
  doi:10.3847/0004-637X/825/2/122
  [arXiv:1512.05988 [astro-ph.HE]].


\bibitem{Supanitsky:2016gke} 
  A.~D.~Supanitsky,
    {\color{rossoCP3}  Implications of gamma-ray observations on proton models of ultra-high-energy cosmic rays},
  Phys.\ Rev.\ D {\bf 94}, no. 6, 063002 (2016)
  doi:10.1103/PhysRevD.94.063002
  [arXiv:1607.00290 [astro-ph.HE]].


\bibitem{Aartsen:2016ngq} 
  M.~G.~Aartsen {\it et al.} [IceCube Collaboration],
    {\color{rossoCP3}  Constraints on ultra-high-energy cosmic-ray sources from a search for neutrinos above 10~PeV with IceCube},
  Phys.\ Rev.\ Lett.\  {\bf 117}, no. 24, 241101 (2016)
  Erratum: [Phys.\ Rev.\ Lett.\  {\bf 119}, no. 25, 259902 (2017)]
  doi:10.1103/PhysRevLett.117.241101, 10.1103/PhysRevLett.119.259902, 10.1103/PhysRevLett.117.241101, 10.1103/PhysRevLett.119.259902
  [arXiv:1607.05886 [astro-ph.HE]].




\bibitem{Fang:2013vla} 
  K.~Fang, K.~Kotera, K.~Murase and A.~V.~Olinto,
    {\color{rossoCP3} Testing the newborn pulsar origin of ultrahigh-energy cosmic rays with EeV neutrinos},
  Phys.\ Rev.\ D {\bf 90}, no. 10, 103005 (2014)
  [Phys.\ Rev.\ D {\bf 90}, 103005 (2014)]
  Erratum: [Phys.\ Rev.\ D {\bf 92}, no. 12, 129901 (2015)]
  doi:10.1103/PhysRevD.90.103005, 10.1103/PhysRevD.92.129901
  [arXiv:1311.2044 [astro-ph.HE]].




\bibitem{Fang:2015xhg} 
  K.~Fang, K.~Kotera, K.~Murase and A.~V.~Olinto,
    {\color{rossoCP3} IceCube constraints on fast-spinning pulsars as high-energy neutrino sources},
  JCAP {\bf 1604}, no. 04, 010 (2016)
  doi:10.1088/1475-7516/2016/04/010
  [arXiv:1511.08518 [astro-ph.HE]].




\bibitem{Loeb:2006tw} 
  A.~Loeb and E.~Waxman,
    {\color{rossoCP3} The cumulative background of high energy neutrinos from starburst galaxies},
  JCAP {\bf 0605}, 003 (2006)
  doi:10.1088/1475-7516/2006/05/003
  [astro-ph/0601695].

\bibitem{Aartsen:2014njl} 
  M.~G.~Aartsen {\it et al.} [IceCube Collaboration],
    {\color{rossoCP3} IceCube-Gen2: A vision for the future of neutrino astronomy in Antarctica},
  arXiv:1412.5106 [astro-ph.HE].






\bibitem{Ressell:1989rz} 
  M.~T.~Ressell and M.~S.~Turner,
   {\color{rossoCP3}  The grand unified photon spectrum: a coherent view of the diffuse extragalactic background radiation},
  Comments Astrophys.\  {\bf 14}, 323 (1990)
  [Bull.\ Am.\ Astron.\ Soc.\  {\bf 22}, 753 (1990)].

\bibitem{DeAngelis:2018lra} 
  A.~De Angelis and M.~Mallamaci,
 {\color{rossoCP3}  Gamma-ray astrophysics},
  arXiv:1805.05642 [astro-ph.HE].


\bibitem{Nikishov}
A. I. Nikishov,
  {\color{rossoCP3} Absorption of high-energy photons in the Universe} 
Sov.\ Phys.\ JETP  {\bf 14}, 393 (1962) 
[ZhETF {\bf 41} 549 (1962)].


\bibitem{Gould:1966pza} 
  R.~Gould and G.~Schreder,
   {\color{rossoCP3} Opacity of the Universe to high-energy photons},
  Phys.\ Rev.\ Lett.\  {\bf 16}, no. 6, 252 (1966).
  doi:10.1103/PhysRevLett.16.252


\bibitem{Gould:1967zza} 
  R.~J.~Gould and G.~P.~Schreder,
    {\color{rossoCP3} Opacity of the Universe to high-energy photons},
  Phys.\ Rev.\  {\bf 155}, 1408 (1967).
  doi:10.1103/PhysRev.155.1408

\bibitem{Stecker:1969gamma}
F. W. Stecker,
  {\color{rossoCP3} The cosmic gamma-ray spectrum from secondary-particle production in the metagalaxy},
 Astrophys.\ J.\  {\bf 157}, 507 (1969).
doi:10.1086/150091

\bibitem{Fazio:1970pr} 
  G.~G.~Fazio and F.~W.~Stecker,
   {\color{rossoCP3} Predicted high energy break in the isotropic gamma-ray spectrum: A test of cosmological origin},
  Nature {\bf 226}, 135 (1970).
  doi:10.1038/226135a0








\bibitem{Aartsen:2013bka}
  M.~G.~Aartsen {\it et al.} [IceCube Collaboration],
    {\color{rossoCP3}  First observation of PeV-energy neutrinos with IceCube},
  Phys.\ Rev.\ Lett.\  {\bf 111}, 021103 (2013)
  doi:10.1103/PhysRevLett.111.021103
  [arXiv:1304.5356 [astro-ph.HE]].

\bibitem{Schonert:2008is} 
  S.~Schonert, T.~K.~Gaisser, E.~Resconi and O.~Schulz,
   {\color{rossoCP3} Vetoing atmospheric neutrinos in a high energy neutrino telescope},
  Phys.\ Rev.\ D {\bf 79}, 043009 (2009)
  doi:10.1103/PhysRevD.79.043009
  [arXiv:0812.4308 [astro-ph]];

\bibitem{Gaisser:2014bja} 
  T.~K.~Gaisser, K.~Jero, A.~Karle and J.~van Santen,
   {\color{rossoCP3} Generalized self-veto probability for atmospheric neutrinos},
  Phys.\ Rev.\ D {\bf 90}, no. 2, 023009 (2014)
  doi:10.1103/PhysRevD.90.023009
  [arXiv:1405.0525 [astro-ph.HE]].





\bibitem{Aartsen:2013jdh} 
  M.~G.~Aartsen {\it et al.} [IceCube Collaboration],
    {\color{rossoCP3}  Evidence for high-energy extraterrestrial neutrinos at the IceCube detector},
  Science {\bf 342}, 1242856 (2013)
  doi:10.1126/science.1242856
  [arXiv:1311.5238 [astro-ph.HE]].


\bibitem{Aartsen:2014gkd} 
  M.~G.~Aartsen {\it et al.} [IceCube Collaboration],
    {\color{rossoCP3}  Observation of high-energy astrophysical neutrinos in three years of IceCube data},
  Phys.\ Rev.\ Lett.\  {\bf 113}, 101101 (2014)
  doi:10.1103/PhysRevLett.113.101101
  [arXiv:1405.5303 [astro-ph.HE]].


\bibitem{Aartsen:2014muf} 
  M.~G.~Aartsen {\it et al.} [IceCube Collaboration],
    {\color{rossoCP3}  Atmospheric and astrophysical neutrinos above 1 TeV interacting in IceCube},
  Phys.\ Rev.\ D {\bf 91}, no. 2, 022001 (2015)
  doi:10.1103/PhysRevD.91.022001
  [arXiv:1410.1749 [astro-ph.HE]].

\bibitem{Aartsen:2017mau} 
  M.~G.~Aartsen {\it et al.} [IceCube Collaboration],
  {\color{rossoCP3}   The IceCube neutrino observatory contributions to ICRC 2017 Part II: Properties of the atmospheric and astrophysical neutrino flux},
  arXiv:1710.01191 [astro-ph.HE].





\bibitem{Aartsen:2015knd} 
  M.~G.~Aartsen {\it et al.} [IceCube Collaboration],
    {\color{rossoCP3}  A combined maximum-likelihood analysis of the high-energy astrophysical neutrino flux measured with IceCube},
  Astrophys.\ J.\  {\bf 809}, no. 1, 98 (2015)
  doi:10.1088/0004-637X/809/1/98
  [arXiv:1507.03991 [astro-ph.HE]].


\bibitem{Aartsen:2015rwa} 
  M.~G.~Aartsen {\it et al.} [IceCube Collaboration],
    {\color{rossoCP3}  Evidence for astrophysical muon neutrinos from the Northern sky with IceCube},
  Phys.\ Rev.\ Lett.\  {\bf 115}, no. 8, 081102 (2015)
  doi:10.1103/PhysRevLett.115.081102
  [arXiv:1507.04005 [astro-ph.HE]].


\bibitem{Aartsen:2016xlq} 
  M.~G.~Aartsen {\it et al.} [IceCube Collaboration],
    {\color{rossoCP3}  Observation and characterization of a cosmic muon neutrino flux from the Northern hemisphere using six years of IceCube data},
  Astrophys.\ J.\  {\bf 833}, no. 1, 3 (2016)
  doi:10.3847/0004-637X/833/1/3
  [arXiv:1607.08006 [astro-ph.HE]].


\bibitem{Anchordoqui:2013dnh} 
  L.~A.~Anchordoqui {\it et al.},
    {\color{rossoCP3}  Cosmic neutrino pevatrons: A brand new pathway to astronomy, astrophysics, and particle physics},
  JHEAp {\bf 1-2}, 1 (2014)
  doi:10.1016/j.jheap.2014.01.001
  [arXiv:1312.6587 [astro-ph.HE]].


\bibitem{Ahlers:2015lln} 
  M.~Ahlers and F.~Halzen,
    {\color{rossoCP3}  High-energy cosmic neutrino puzzle: a review},
  Rept.\ Prog.\ Phys.\  {\bf 78}, no. 12, 126901 (2015).
  doi:10.1088/0034-4885/78/12/126901




\bibitem{Meszaros:2017fcs} 
  P.~Meszaros,
    {\color{rossoCP3}  Astrophysical sources of high energy neutrinos in the IceCube era},
  Ann.\ Rev.\ Nucl.\ Part.\ Sci.\  {\bf 67}, 45 (2017)
  doi:10.1146/annurev-nucl-101916-123304
  [arXiv:1708.03577 [astro-ph.HE]].



\bibitem{Aartsen:2017ujz} 
  M.~G.~Aartsen {\it et al.} [IceCube Collaboration],
    {\color{rossoCP3}  Constraints on Galactic neutrino emission with seven years of IceCube data},
  Astrophys.\ J.\  {\bf 849}, no. 1, 67 (2017)
  doi:10.3847/1538-4357/aa8dfb
  [arXiv:1707.03416 [astro-ph.HE]].


\bibitem{Anchordoqui:2013qsi} 
  L.~A.~Anchordoqui, H.~Goldberg, M.~H.~Lynch, A.~V.~Olinto, T.~C.~Paul and T.~J.~Weiler,
    {\color{rossoCP3}  Pinning down the cosmic ray source mechanism with new IceCube data},
  Phys.\ Rev.\ D {\bf 89}, no. 8, 083003 (2014)
  doi:10.1103/PhysRevD.89.083003
  [arXiv:1306.5021 [astro-ph.HE]].


\bibitem{Neronov:2013lza} 
  A.~Neronov, D.~V.~Semikoz and C.~Tchernin,
    {\color{rossoCP3}  PeV neutrinos from interactions of cosmic rays with the interstellar medium in the Galaxy},
  Phys.\ Rev.\ D {\bf 89}, no. 10, 103002 (2014)
  doi:10.1103/PhysRevD.89.103002
  [arXiv:1307.2158 [astro-ph.HE]].


\bibitem{Anchordoqui:2014rca} 
  L.~A.~Anchordoqui, H.~Goldberg, T.~C.~Paul, L.~H.~M.~da Silva and B.~J.~Vlcek,
    {\color{rossoCP3}  Estimating the contribution of Galactic sources to the diffuse neutrino flux},
  Phys.\ Rev.\ D {\bf 90}, no. 12, 123010 (2014)
  doi:10.1103/PhysRevD.90.123010
  [arXiv:1410.0348 [astro-ph.HE]].


\bibitem{Neronov:2015osa} 
  A.~Neronov and D.~V.~Semikoz,
    {\color{rossoCP3}  Evidence the Galactic contribution to the IceCube astrophysical neutrino flux},
  Astropart.\ Phys.\  {\bf 75}, 60 (2016)
  doi:10.1016/j.astropartphys.2015.11.002
  [arXiv:1509.03522 [astro-ph.HE]].



\bibitem{Neronov:2016bnp} 
  A.~Neronov and D.~Semikoz,
    {\color{rossoCP3}  Galactic and extragalactic contributions to the astrophysical muon neutrino signal},
  Phys.\ Rev.\ D {\bf 93}, no. 12, 123002 (2016)
  doi:10.1103/PhysRevD.93.123002
  [arXiv:1603.06733 [astro-ph.HE]].



\bibitem{Neronov:2018ibl} 
  A.~Neronov, M.~Kachelrieß and D.~V.~Semikoz,
    {\color{rossoCP3}  Multi-messenger gamma-ray counterpart of the IceCube neutrino signal},
  Phys.\ Rev.\ D {\bf 98}, no. 2, 023004 (2018)
  doi:10.1103/PhysRevD.98.023004
  [arXiv:1802.09983 [astro-ph.HE]].



\bibitem{Aartsen:2015dml} 
  M.~G.~Aartsen {\it et al.} [IceCube and Pierre Auger and Telescope Array Collaborations],
    {\color{rossoCP3}  Search for correlations between the arrival directions of IceCube neutrino events and ultrahigh-energy cosmic rays detected by the Pierre Auger Observatory and the Telescope Array},
  JCAP {\bf 1601}, no. 01, 037 (2016)
  doi:10.1088/1475-7516/2016/01/037
  [arXiv:1511.09408 [astro-ph.HE]].

\bibitem{Ahlers:2017wkk} 
  M.~Ahlers and F.~Halzen,
    {\color{rossoCP3}  IceCube: Neutrinos and multi-messenger astronomy},
  PTEP {\bf 2017}, no. 12, 12A105 (2017).
  doi:10.1093/ptep/ptx021


\bibitem{IceCube:2018cha} 
  M.~G.~Aartsen {\it et al.} [IceCube Collaboration],
  {\color{rossoCP3}  Neutrino emission from the direction of the blazar TXS 0506+056 prior to the IceCube-170922A alert},
  Science {\bf 361}, no. 6398, 147 (2018).
  doi:10.1126/science.aat2890
 [arXiv:1807.08794 [astro-ph.HE]].


\bibitem{Fermi-LAT:2011xmf} 
  M. Ackermann {\it et al.} [Fermi-LAT Collaboration],
    {\color{rossoCP3}  The second catalog of active galactic nuclei detected by the Fermi Large Area Telescope},
  Astrophys.\ J.\  {\bf 743}, 171 (2011)
  doi:10.1088/0004-637X/743/2/171
  [arXiv:1108.1420 [astro-ph.HE]].

\bibitem{Aartsen:2016lir} 
  M.~G.~Aartsen {\it et al.} [IceCube Collaboration],
    {\color{rossoCP3}  The contribution of Fermi-2LAC blazars to the diffuse TeV-PeV neutrino flux},
  Astrophys.\ J.\  {\bf 835}, no. 1, 45 (2017)
  doi:10.3847/1538-4357/835/1/45
  [arXiv:1611.03874 [astro-ph.HE]].




\bibitem{Mena:2014sja} 
  O.~Mena, S.~Palomares-Ruiz and A.~C.~Vincent,
     {\color{rossoCP3} Flavor composition of the high-energy neutrino events in IceCube},
  Phys.\ Rev.\ Lett.\  {\bf 113}, 091103 (2014)
  doi:10.1103/PhysRevLett.113.091103
  [arXiv:1404.0017 [astro-ph.HE]].

\bibitem{Chen:2014gxa} 
  C.~Y.~Chen, P.~S.~Bhupal Dev and A.~Soni,
   {\color{rossoCP3}   Two-component flux explanation for the high energy neutrino events at IceCube},
  Phys.\ Rev.\ D {\bf 92}, no. 7, 073001 (2015)
  doi:10.1103/PhysRevD.92.073001
  [arXiv:1411.5658 [hep-ph]].




\bibitem{Aartsen:2015ivb} 
  M.~G.~Aartsen {\it et al.} [IceCube Collaboration],
     {\color{rossoCP3} Flavor ratio of astrophysical neutrinos above 35~TeV in IceCube},
  Phys.\ Rev.\ Lett.\  {\bf 114}, no. 17, 171102 (2015)
  doi:10.1103/PhysRevLett.114.171102
  [arXiv:1502.03376 [astro-ph.HE]].


\bibitem{Palomares-Ruiz:2015mka} 
  S.~Palomares-Ruiz, A.~C.~Vincent and O.~Mena,
     {\color{rossoCP3} Spectral analysis of the high-energy IceCube neutrinos},
  Phys.\ Rev.\ D {\bf 91}, no. 10, 103008 (2015)
  doi:10.1103/PhysRevD.91.103008
  [arXiv:1502.02649 [astro-ph.HE]].


\bibitem{Vincent:2016nut} 
  A.~C.~Vincent, S.~Palomares-Ruiz and O.~Mena,
     {\color{rossoCP3} Analysis of the 4-year IceCube high-energy starting events},
  Phys.\ Rev.\ D {\bf 94}, no. 2, 023009 (2016)
  doi:10.1103/PhysRevD.94.023009
  [arXiv:1605.01556 [astro-ph.HE]].




\bibitem{Anchordoqui:2016ewn} 
  L.~A.~Anchordoqui, M.~M.~Block, L.~Durand, P.~Ha, J.~F.~Soriano and T.~J.~Weiler,
     {\color{rossoCP3} Evidence for a break in the spectrum of astrophysical neutrinos},
  Phys.\ Rev.\ D {\bf 95}, no. 8, 083009 (2017)
  doi:10.1103/PhysRevD.95.083009
  [arXiv:1611.07905 [astro-ph.HE]].

\bibitem{Muzio}
M. Muzio, G. R. Farrar, and M. Unger,
   {\color{rossoCP3} Detailed simulations of Fermi-LAT constraints on UHECR production scenarios}
  PoS ICRC {\bf 2017}, 557 (2017).


\bibitem{Kachelriess:2017tvs} 
  M.~Kachelriess, O.~Kalashev, S.~Ostapchenko and D.~V.~Semikoz,
   {\color{rossoCP3} Minimal model for extragalactic cosmic rays and neutrinos},
  Phys.\ Rev.\ D {\bf 96}, no. 8, 083006 (2017)
  doi:10.1103/PhysRevD.96.083006
  [arXiv:1704.06893 [astro-ph.HE]].


\bibitem{Abbott:2016blz} 
  B.~P.~Abbott {\it et al.} [LIGO Scientific and Virgo Collaborations],
  {\color{rossoCP3} Observation of gravitational waves from a binary black hole merger},
  Phys.\ Rev.\ Lett.\  {\bf 116}, no. 6, 061102 (2016)
  doi:10.1103/PhysRevLett.116.061102
  [arXiv:1602.03837 [gr-qc]].

\bibitem{TheLIGOScientific:2016wfe} 
  B.~P.~Abbott {\it et al.} [LIGO Scientific and Virgo Collaborations],
  {\color{rossoCP3} Properties of the binary black hole merger GW150914},
  Phys.\ Rev.\ Lett.\  {\bf 116}, no. 24, 241102 (2016)
  doi:10.1103/PhysRevLett.116.241102
  [arXiv:1602.03840 [gr-qc]].

\bibitem{Abbott:2016nmj} 
  B.~P.~Abbott {\it et al.} [LIGO Scientific and Virgo Collaborations],
  {\color{rossoCP3} GW151226: observation of gravitational waves from a 22-solar-mass binary black hole coalescence},
  Phys.\ Rev.\ Lett.\  {\bf 116}, no. 24, 241103 (2016)
  doi:10.1103/PhysRevLett.116.241103
  [arXiv:1606.04855 [gr-qc]].

\bibitem{TheLIGOScientific:2016pea} 
  B.~P.~Abbott {\it et al.} [LIGO Scientific and Virgo Collaborations],
  {\color{rossoCP3} Binary black hole mergers in the first Advanced LIGO observing run},
  Phys.\ Rev.\ X {\bf 6}, no. 4, 041015 (2016)
  doi:10.1103/PhysRevX.6.041015
  [arXiv:1606.04856 [gr-qc]].

\bibitem{Murase:2016etc} 
  K.~Murase, K.~Kashiyama, P.~Mészáros, I.~Shoemaker and N.~Senno,
  {\color{rossoCP3} Ultrafast outflows from black hole mergers with a minidisk},
  Astrophys.\ J.\  {\bf 822}, no. 1, L9 (2016)
  doi:10.3847/2041-8205/822/1/L9
  [arXiv:1602.06938 [astro-ph.HE]].


\bibitem{Kotera:2016dmp} 
  K.~Kotera and J.~Silk,
  {\color{rossoCP3} Ultra-high energy cosmic rays and black hole mergers},
  Astrophys.\ J.\  {\bf 823}, no. 2, L29 (2016)
  doi:10.3847/2041-8205/823/2/L29
  [arXiv:1602.06961 [astro-ph.HE]].


\bibitem{Anchordoqui:2016dcp} 
  L.~A.~Anchordoqui,
  {\color{rossoCP3} Neutrino lighthouse powered by Sagittarius A$^*$ disk dynamo},
  Phys.\ Rev.\ D {\bf 94}, 023010 (2016)
  doi:10.1103/PhysRevD.94.023010
  [arXiv:1606.01816 [astro-ph.HE]].


\bibitem{Aab:2016ras} 
  A.~Aab {\it et al.} [Pierre Auger Collaboration],
   {\color{rossoCP3} Ultra-high-energy neutrino follow-up of gravitational wave events GW150914 and GW151226 with the Pierre Auger Observatory},
  Phys.\ Rev.\ D {\bf 94}, no. 12, 122007 (2016)
  doi:10.1103/PhysRevD.94.122007
  [arXiv:1608.07378 [astro-ph.HE]].


\bibitem{TheLIGOScientific:2017qsa} 
  B.~P.~Abbott {\it et al.} [LIGO Scientific and Virgo Collaborations],
   {\color{rossoCP3} GW170817: Observation of gravitational waves from a binary neutron star inspiral},
  Phys.\ Rev.\ Lett.\  {\bf 119}, no. 16, 161101 (2017)
  doi:10.1103/PhysRevLett.119.161101
  [arXiv:1710.05832 [gr-qc]].


\bibitem{Monitor:2017mdv} 
  B.~P.~Abbott {\it et al.} [LIGO Scientific and Virgo and Fermi-GBM and INTEGRAL Collaborations],
   {\color{rossoCP3} Gravitational waves and gamma-rays from a binary neutron star merger: GW170817 and GRB 170817A},
  Astrophys.\ J.\  {\bf 848}, no. 2, L13 (2017)
  doi:10.3847/2041-8213/aa920c
  [arXiv:1710.05834 [astro-ph.HE]].

\bibitem{Coulter:2017wya} 
  D.~A.~Coulter {\it et al.},
   {\color{rossoCP3} Swope supernova survey 2017a (SSS17a), the optical counterpart to a gravitational wave source},
  Science
  [Science {\bf 358}, 1556 (2017)]
  doi:10.1126/science.aap9811
  [arXiv:1710.05452 [astro-ph.HE]].




\bibitem{Soares-Santos:2017lru} 
  M.~Soares-Santos {\it et al.} [DES and Dark Energy Camera GW-EM Collaborations],
   {\color{rossoCP3} The electromagnetic counterpart of the binary neutron star merger LIGO/Virgo GW170817 I: Discovery of the optical counterpart using the Dark Energy Camera},
  Astrophys.\ J.\  {\bf 848}, no. 2, L16 (2017)
  doi:10.3847/2041-8213/aa9059
  [arXiv:1710.05459 [astro-ph.HE]].


\bibitem{Valenti:2017ngx} 
  S.~Valenti {\it et al.},
   {\color{rossoCP3} The discovery of the electromagnetic counterpart of GW170817: kilonova AT 2017gfo/DLT17ck},
  Astrophys.\ J.\  {\bf 848}, no. 2, L24 (2017)
  doi:10.3847/2041-8213/aa8edf
  [arXiv:1710.05854 [astro-ph.HE]].


\bibitem{Arcavi:2017xiz} 
  I.~Arcavi {\it et al.},
   {\color{rossoCP3} Optical emission from a kilonova following a gravitational-wave-detected neutron-star merger},
  Nature {\bf 551}, 64 (2017)
  doi:10.1038/nature24291
  [arXiv:1710.05843 [astro-ph.HE]].


\bibitem{Lipunov:2017dwd} 
  V.~M.~Lipunov {\it et al.},
   {\color{rossoCP3} MASTER optical detection of the first LIGO/Virgo neutron star binary merger GW170817},
  Astrophys.\ J.\  {\bf 850}, no. 1, L1 (2017)
  doi:10.3847/2041-8213/aa92c0
  [arXiv:1710.05461 [astro-ph.HE]].








\bibitem{Tanvir:2017pws} 
  N.~R.~Tanvir {\it et al.},
   {\color{rossoCP3} The emergence of a Lanthanide-Rich kilonova following the merger of two neutron stars},
  Astrophys.\ J.\  {\bf 848}, no. 2, L27 (2017)
  doi:10.3847/2041-8213/aa90b6
  [arXiv:1710.05455 [astro-ph.HE]].




\bibitem{Pian:2017gtc} 
  E.~Pian {\it et al.},
   {\color{rossoCP3} Spectroscopic identification of r-process nucleosynthesis in a double neutron star merger},
  Nature {\bf 551}, 67 (2017)
  doi:10.1038/nature24298
  [arXiv:1710.05858 [astro-ph.HE]].


\bibitem{Troja:2017nqp} 
  E.~Troja {\it et al.},
   {\color{rossoCP3} The X-ray counterpart to the gravitational wave event GW 170817},
  Nature {\bf 551}, 71 (2017)
  [Nature {\bf 551}, 71 (2017)]
  doi:10.1038/nature24290
  [arXiv:1710.05433 [astro-ph.HE]].



\bibitem{Haggard:2017qne} 
  D.~Haggard, M.~Nynka, J.~J.~Ruan, V.~Kalogera, S.~Bradley Cenko, P.~Evans and J.~A.~Kennea,
   {\color{rossoCP3} A deep Chandra X-ray study of neutron star coalescence GW170817},
  Astrophys.\ J.\  {\bf 848}, no. 2, L25 (2017)
  doi:10.3847/2041-8213/aa8ede
  [arXiv:1710.05852 [astro-ph.HE]].



\bibitem{Hallinan:2017woc} 
  G.~Hallinan {\it et al.},
   {\color{rossoCP3} A radio counterpart to a neutron star merger},
  Science {\bf 358}, 1579 (2017)
  doi:10.1126/science.aap9855
  [arXiv:1710.05435 [astro-ph.HE]].



\bibitem{Kasliwal:2017ngb} 
  M.~M.~Kasliwal {\it et al.},
   {\color{rossoCP3} Illuminating gravitational waves: A concordant picture of photons from a neutron star merger},
  Science {\bf 358}, 1559 (2017)
  doi:10.1126/science.aap9455
  [arXiv:1710.05436 [astro-ph.HE]].

\bibitem{Li:1998bw} 
  L.~X.~Li and B.~Paczynski,
   {\color{rossoCP3} Transient events from neutron star mergers},
  Astrophys.\ J.\  {\bf 507}, L59 (1998)
  doi:10.1086/311680
  [astro-ph/9807272].


\bibitem{Metzger:2010sy} 
  B.~D.~Metzger {\it et al.},
   {\color{rossoCP3} Electromagnetic counterparts of compact object mergers powered by the radioactive decay of r-process nuclei},
  Mon.\ Not.\ Roy.\ Astron.\ Soc.\  {\bf 406}, 2650 (2010)
  doi:10.1111/j.1365-2966.2010.16864.x
  [arXiv:1001.5029 [astro-ph.HE]].

\bibitem{ANTARES:2017bia} 
  A.~Albert {\it et al.} [ANTARES and IceCube and Pierre Auger and LIGO Scientific and Virgo Collaborations],
   {\color{rossoCP3} Search for high-energy neutrinos from binary neutron star merger GW170817 with ANTARES, IceCube, and the Pierre Auger Observatory},
  Astrophys.\ J.\  {\bf 850}, no. 2, L35 (2017)
  doi:10.3847/2041-8213/aa9aed
  [arXiv:1710.05839 [astro-ph.HE]].


\bibitem{Biehl:2017qen} 
  D.~Biehl, J.~Heinze and W.~Winter,
    {\color{rossoCP3} Expected neutrino fluence from short gamma-ray burst 170817A and off-axis angle constraints},
  Mon.\ Not.\ Roy.\ Astron.\ Soc.\  {\bf 476}, no. 1, 1191 (2018)
  doi:10.1093/mnras/sty285
  [arXiv:1712.00449 [astro-ph.HE]].

\bibitem{Rodrigues:2018bjg} 
  X.~Rodrigues, D.~Biehl, D.~Boncioli and A.~M.~Taylor,
    {\color{rossoCP3} Binary neutron star merger remnants as sources of cosmic rays below the "ankle"},
  arXiv:1806.01624 [astro-ph.HE].







\bibitem{Capella:yb}
A.~Capella, U.~Sukhatme, C.~I.~Tan and J.~Tran Thanh Van,
  {\color{rossoCP3}  Dual parton model},
Phys.\ Rept.\  {\bf 236}, 225 (1994).

\bibitem{Predazzi:1998rp}
E.~Predazzi,
  {\color{rossoCP3}  Diffraction: past, present and future},
arXiv:hep-ph/9809454.



\bibitem{Cline:1973kv}
D.~Cline, F.~Halzen and J.~Luthe,
  {\color{rossoCP3}  High transverse momentum secondaries and rising total cross-sections in 
cosmic ray interactions},
Phys.\ Rev.\ Lett.\  {\bf 31}, 491 (1973).

\bibitem{Ellis:1973nb}
S.~D.~Ellis and M.~B.~Kislinger,
  {\color{rossoCP3}  Implications of parton model concepts for large transverse momentum 
production of hadrons},
Phys.\ Rev.\ D {\bf 9}, 2027 (1974).


\bibitem{Halzen:1974vh}
F.~Halzen,
  {\color{rossoCP3}  High transverse momentum secondaries In cosmic
    ray interactions up to $10,000,000~{\rm GeV}$},
Nucl.\ Phys.\ B {\bf 92}, 404 (1975).

\bibitem{Pancheri:sr}
G.~Pancheri and C.~Rubbia,
  {\color{rossoCP3}  Events of very high-energy density at the CERN
    S$p\bar p$S Collider},
Nucl.\ Phys.\ A {\bf 418}, 117C (1984).



\bibitem{Gaisser:1984pg}
T.~K.~Gaisser and F.~Halzen,
 {\color{rossoCP3} Soft hard scattering in the TeV range},
Phys.\ Rev.\ Lett.\  {\bf 54}, 1754 (1985).


\bibitem{DiasdeDeus:1984ip}
  J.~Dias de Deus,
   {\color{rossoCP3} Semihard physics at the SPS $p\bar p$ Colliders?},
  Nucl.\ Phys.\  B {\bf 252}, 369 (1985).






\bibitem{Pancheri:ix}
G.~Pancheri and Y.~Srivastava,
 {\color{rossoCP3} Jets in minimum bias physics},
Phys.\ Lett.\ B {\bf 159}, 69 (1985).

\bibitem{Pancheri:qg}
G.~Pancheri and Y.~N.~Srivastava,
 {\color{rossoCP3} Low $p_T$ jets and the rise with energy of the inelastic cross-section},
Phys.\ Lett.\ B {\bf 182}, 199 (1986).

\bibitem{Albajar:1988tt}
C.~Albajar {\it et al.}  [UA1 Collaboration],
 {\color{rossoCP3} Production of low transverse energy clusters in
   $p \bar p$ collisions 
at $\sqrt{s} = 0.2~{\rm TeV}$ to 0.9~TeV and their interpretation in terms of QCD
jets},
Nucl.\ Phys.\ B {\bf 309}, 405 (1988).


\bibitem{Gribov:rt}
V.~N.~Gribov and L.~N.~Lipatov,
 {\color{rossoCP3} $e^+ e^-$ pair annihilation and deep inelastic $ep$ scattering in perturbation theory},
Yad.\ Fiz.\  {\bf 15}, 1218 (1972) 
[Sov.\ J.\ Nucl.\ Phys.\  {\bf 15}, 675 (1972)].


\bibitem{Gribov:ri}
V.~N.~Gribov and L.~N.~Lipatov,
 {\color{rossoCP3} Deep inelastic $ep$ scattering in perturbation theory},
Yad.\ Fiz.\  {\bf 15}, 781 (1972)
[Sov.\ J.\ Nucl.\ Phys.\  {\bf 15}, 438 (1972)].



\bibitem{Dokshitzer:sg}
Y.~L.~Dokshitzer,
 {\color{rossoCP3} Calculation of the structure functions for deep inelastic 
scattering and $e^+e^-$ annihilation by perturbation theory in quantum 
chromodynamics},
Sov.\ Phys.\ JETP {\bf 46}, 641 (1977) 
[Zh.\ Eksp.\ Teor.\ Fiz.\  {\bf 73}, 1216 (1977)].

\bibitem{Altarelli:1977zs}
G.~Altarelli and G.~Parisi,
 {\color{rossoCP3} Asymptotic freedom in parton language},
Nucl.\ Phys.\ B {\bf 126}, 298 (1977).


\bibitem{Dittmar:2009ii}
  M.~Dittmar {\it et al.},
  {\color{rossoCP3} Parton distributions},
  arXiv:0901.2504 [hep-ph].




\bibitem{Engel:ac}
R.~Engel, 
 {\color{rossoCP3} Models of primary interactions},
Nucl.\ Phys.\ Proc.\ Suppl.\  {\bf 122}, 40 (2003).



\bibitem{Anchordoqui:2009eg}
  L.~Anchordoqui and F.~Halzen,
   {\color{rossoCP3} Lessons in particle physics},
  arXiv:0906.1271.


\bibitem{Kwiecinski:1990tb}
J.~Kwiecinski and A.~D.~Martin,
 {\color{rossoCP3} Semihard QCD expectations for $p\bar p$ scattering at CERN, 
Tevatron and SSC colliders},
Phys.\ Rev.\ D {\bf 43}, 1560 (1991).





\bibitem{DiasdeDeus:1987yw}
  J.~Dias de Deus and J.~Kwiecinski,
  {\color{rossoCP3} Semihard QCD: minijets and elastic scattering}, 
  Phys.\ Lett.\  B {\bf 196}, 537 (1987).

%
\bibitem{DiasDeDeus:1987bf}
  J.~Dias De Deus,
   {\color{rossoCP3}Geometric scaling, multiplicity distributions and cross-sections},
  Nucl.\ Phys.\  B {\bf 59}, 231 (1973).


\bibitem{Amaldi:1979kd}
  U.~Amaldi and K.~R.~Schubert,
   {\color{rossoCP3} Impact parameter interpretation of proton proton scattering from a critical
  review of all ISR data},
  Nucl.\ Phys.\  B {\bf 166}, 301 (1980).

\bibitem{Castaldi:1985ft}
  R.~Castaldi and G.~Sanguinetti,
 {\color{rossoCP3} Elastic scattering and total cross-section at very high-energies},
  Ann.\ Rev.\ Nucl.\ Part.\ Sci.\  {\bf 35}, 351 (1983).




\bibitem{Block:1984ru}
M.~M.~Block and R.~N.~Cahn,
 {\color{rossoCP3} High-energy $p \bar p$ and $pp$ forward elastic scattering and total
cross-sections},
Rev.\ Mod.\ Phys.\  {\bf 57}, 563 (1985).


\bibitem{Glauber:1970jm}
R.~J.~Glauber and G.~Matthiae,
 {\color{rossoCP3} High-energy scattering of protons by nuclei},
Nucl.\ Phys.\ B {\bf 21}, 135 (1970).



\bibitem{L'Heureux:jk}
P.~L'Heureux, B.~Margolis and P.~Valin,
 {\color{rossoCP3} Quark-gluon model for diffraction at high-energies},
Phys.\ Rev.\ D {\bf 32}, 1681 (1985).

 

\bibitem{Durand:prl} 
L. Durand and H. Pi, 
 {\color{rossoCP3} QCD and rising cross sections},
Phys. Rev. Lett. {\bf 58}, 303 (1987).



\bibitem{Durand:cr}
L.~Durand and H.~Pi,
 {\color{rossoCP3} High-energy nucleon nucleus scattering and cosmic ray cross-sections},
Phys.\ Rev.\ D {\bf 38}, 78 (1988).


 

\bibitem{Gaisser:1988ra}
T.~K.~Gaisser and T.~Stanev,
 {\color{rossoCP3} Minijets in minimum bias events},'
Phys.\ Lett.\ B {\bf 219}, 375 (1989).

\bibitem{Fletcher:1994bd}
R.~S.~Fletcher, T.~K.~Gaisser, P.~Lipari and T.~Stanev,
 {\color{rossoCP3} SIBYLL: An Event generator for simulation of high-energy cosmic 
ray cascades},
Phys.\ Rev.\ D {\bf 50}, 5710 (1994).

\bibitem{Kalmykov:te} 
N.~N.~Kalmykov, S.~S.~Ostapchenko and A.~I.~Pavlov,
 {\color{rossoCP3} Quark-gluon string model and EAS simulation problems at ultra-high 
energies},
Nucl.\ Phys.\ Proc.\ Suppl.\  {\bf 52B}, 17 (1997).

\bibitem{Alvarez-Muniz:2002ne}
J.~Alvarez-Muniz, R.~Engel, T.~K.~Gaisser, J.~A.~Ortiz and T.~Stanev,
 {\color{rossoCP3} Hybrid simulations of extensive air showers},
Phys.\ Rev.\ D {\bf 66}, 033011 (2002)
[arXiv:astro-ph/0205302].



\bibitem{Froissart:ux}
M.~Froissart,
 {\color{rossoCP3} Asymptotic behavior and subtractions in the Mandelstam representation},
Phys.\ Rev.\  {\bf 123}, 1053 (1961).




\bibitem{Ahn:2009wx} 
  E.~J.~Ahn, R.~Engel, T.~K.~Gaisser, P.~Lipari and T.~Stanev,
   {\color{rossoCP3} Cosmic ray interaction event generator SIBYLL 2.1},
  Phys.\ Rev.\ D {\bf 80}, 094003 (2009)
  doi:10.1103/PhysRevD.80.094003
  [arXiv:0906.4113 [hep-ph]].



\bibitem{Sjostrand:1987xj}
T.~Sjostrand,
 {\color{rossoCP3} Status of fragmentation models},
Int.\ J.\ Mod.\ Phys.\ A {\bf 3}, 751 (1988).








\bibitem{Engel:vf}
J.~Engel, T.~K.~Gaisser, T.~Stanev and P.~Lipari,
 {\color{rossoCP3} Nucleus-nucleus collisions and interpretation of cosmic ray cascades},
Phys.\ Rev.\ D {\bf 46}, 5013 (1992).




\bibitem{Belov:2006mb}
  K.~Belov  [HiRes Collaboration],
   {\color{rossoCP3} $p$-air cross-section measurement at $10^{18.5}$~eV},
  Nucl.\ Phys.\ Proc.\ Suppl.\  {\bf 151}, 197 (2006).



\bibitem{Block:1999ub}
M.~M.~Block, F.~Halzen and T.~Stanev,
  {\color{rossoCP3} Predicting proton air cross sections at $\sqrt{s}
    \approx 30~{\rm TeV}$, using accelerator and cosmic ray data},
Phys.\ Rev.\ Lett.\  {\bf 83}, 4926 (1999)
[arXiv:hep-ph/9908222].


\bibitem{Block:2007rq}
  M.~M.~Block,
   {\color{rossoCP3} Ultra-high energy predictions of proton-air cross sections from
  accelerator data},
  Phys.\ Rev.\  D {\bf 76}, 111503 (2007)
  [arXiv:0705.3037 [hep-ph]].



\bibitem{Aielli:2009ca}
  G.~Aielli {\it et al.}  [ARGO-YBJ Collaboration],
   {\color{rossoCP3} Proton-air cross section measurement with the ARGO-YBJ cosmic ray
  experiment},
  Phys.\ Rev.\  D {\bf 80}, 092004 (2009)
  [arXiv:0904.4198 [hep-ex]].


\bibitem{Aglietta:2009zza}
  M.~Aglietta {\it et al.},
   {\color{rossoCP3} Measurement of the proton-air Inelastic cross
     section at $\sqrt{s} \approx 2~{\rm TeV}$ from
  the EAS-TOP experiment},
  Phys.\ Rev.\  D {\bf 79}, 032004 (2009).



\bibitem{Mielke:1994un}
  H.~H.~Mielke, M.~Foeller, J.~Engler and J.~Knapp,
  {\color{rossoCP3} Cosmic ray hadron flux at sea level up to 15~TeV},
  J.\ Phys.\ G {\bf 20}, 637 (1994).



\bibitem{Honda:1992kv}
  M.~Honda {\it et al.},
   {\color{rossoCP3} Inelastic cross-section for $p$-air collisions from air shower experiment and
  total cross-section for $p p$ collisions at SSC energy},
  Phys.\ Rev.\ Lett.\  {\bf 70}, 525 (1993).



\bibitem{Dyakonov:1990cd}
  M.~N.~Dyakonov {\it et al.},
   {\color{rossoCP3} Parameters of hadron interactions at $E_0 > 10^{17}$~eV on EAS development
  fluctuation data},
in 
Proceedings of 21st International Cosmic Ray Conference (Adelaide, Australia) {\bf 9}, 252 (1990).


\bibitem{Knurenko:1999cr}
  S.~P.~Knurenko, V.~R.~Sleptsova, I.~E.~Sleptsov, N.~N.~Kalmykov and S.~S.~Ostapchenko,
   {\color{rossoCP3} Longitudinal EAS development at $E_0 = 10^{18}$~eV to $3 \times 10^{19}$~eV and  the
  QGSJET model},
in Proceedings of 26th International Cosmic Ray Conference (Salt Lake City, Utah) {\bf 1}, 372 (1999).


\bibitem{Nam:1975xk}
  R.~A.~Nam, S.~I.~Nikolsky, V.~P.~Pavlyuchenko, A.~P.~Chubenko and V.~I.~Yakovlev,
    {\color{rossoCP3} Investigation of nucleon-nuclei of air cross-section at energy greater than
  10~TeV},
in Proceedings of 14th International Cosmic Ray Conference (Munich, Germany) {\bf 7}, 2258 (1975).


\bibitem{Baltrusaitis:1984ka}
  R.~M.~Baltrusaitis {\it et al.},
  {\color{rossoCP3} Total proton proton cross-section at $\sqrt{s} = 30$~TeV},
  Phys.\ Rev.\ Lett.\  {\bf 52}, 1380 (1984).



\bibitem{Collaboration:2012wt} 
  P.~Abreu {\it et al.} [Pierre Auger Collaboration],
   {\color{rossoCP3} Measurement of the proton-air cross-section at $\sqrt{s}=57$ TeV with the Pierre Auger Observatory},
  Phys.\ Rev.\ Lett.\  {\bf 109}, 062002 (2012)
  doi:10.1103/PhysRevLett.109.062002
  [arXiv:1208.1520 [hep-ex]].




\bibitem{Abbasi:2015fdr} 
  R.~U.~Abbasi {\it et al.} [Telescope Array Collaboration],
   {\color{rossoCP3} Measurement of the proton-air cross section with Telescope Array’s Middle Drum detector and surface array in hybrid mode},
  Phys.\ Rev.\ D {\bf 92}, no. 3, 032007 (2015)
  doi:10.1103/PhysRevD.92.032007
  [arXiv:1505.01860 [astro-ph.HE]].





\bibitem{dEnterria:2011twh} 
  D.~d'Enterria, R.~Engel, T.~Pierog, S.~Ostapchenko and K.~Werner,
 {\color{rossoCP3}  Constraints from the first LHC data on hadronic event generators for ultra-high energy cosmic-ray physics},
  Astropart.\ Phys.\  {\bf 35}, 98 (2011)
  doi:10.1016/j.astropartphys.2011.05.002
  [arXiv:1101.5596 [astro-ph.HE]].



\bibitem{Barbosa:2003dc} 
  H.~M.~J.~Barbosa, F.~Catalani, J.~A.~Chinellato and C.~Dobrigkeit,
   {\color{rossoCP3}  Determination of the calorimetric energy in extensive air showers},
  Astropart.\ Phys.\  {\bf 22}, 159 (2004)
  doi:10.1016/j.astropartphys.2004.06.007
  [astro-ph/0310234].





\bibitem{Mccomb:tp}
T.~J.~L.~Mccomb, R.~J.~Protheroe and K.~E.~Turver,
 {\color{rossoCP3}  Photoproduction in Large cosmic ray showers},
J.\ Phys.\ G {\bf 5}, 1613 (1979).



\bibitem{Rossi:1941zza} 
  B.~Rossi and K.~Greisen,
  {\color{rossoCP3} Cosmic-ray theory},
  Rev.\ Mod.\ Phys.\  {\bf 13}, 240 (1941).
  doi:10.1103/RevModPhys.13.240



\bibitem{Bethe:1934za}
H.~Bethe and W.~Heitler,
  {\color{rossoCP3} On the stopping of fast particles and on the creation of positive
electrons},
Proc.\ Roy.\ Soc.\ Lond.\ A {\bf 146}, 83 (1934).






\bibitem{Rossi:book}
B. Rossi, 
 {\color{rossoCP3} High Energy Particles} 
(Prentice--Hall, Inc., Englewood Cliffs, NY, 1952).




\bibitem{Weast} R. C. Weast, 
 {\color{rossoCP3} CRC Handbook of Chemistry and Physics}, 
(CRC Press, Boca Raton, FL, USA, 1981). 





\bibitem{Tsai:1973py}
Y.~S.~Tsai,
 {\color{rossoCP3} Pair production and bremsstrahlung of charged leptons},
Rev.\ Mod.\ Phys.\  {\bf 46}, 815 (1974)
[Erratum Rev.\ Mod.\ Phys.\  {\bf 49}, 421 (1977)].



\bibitem{Stanev:au}
T.~Stanev, C.~Vankov, R.~E.~Streitmatter, R.~W.~Ellsworth and T.~Bowen,
  {\color{rossoCP3} Development of ultra-high-energy electromagnetic cascades in water and lead
including the Landau-Pomeranchuk-Migdal effect},
Phys.\ Rev.\ D {\bf 25}, 1291 (1982).


\bibitem{Cillis:1998hf}
A.~N.~Cillis, H.~Fanchiotti, C.~A.~Garcia Canal and S.~J.~Sciutto,
 {\color{rossoCP3} Influence of the LPM effect and dielectric suppression on particle air
showers},
Phys.\ Rev.\ D {\bf 59}, 113012 (1999)
[arXiv:astro-ph/9809334].







\bibitem{Vankov:2002cb}
H.~P.~Vankov, N.~Inoue and K.~Shinozaki,
 {\color{rossoCP3} Ultra-high energy gamma rays in geomagnetic field and atmosphere},
Phys.\ Rev.\ D {\bf 67}, 043002 (2003)
[arXiv:astro-ph/0211051].




\bibitem{Cillis:2000xc}
A.~N.~Cillis and S.~J.~Sciutto,
 {\color{rossoCP3} Extended air showers and muon interactions},
Phys.\ Rev.\ D {\bf 64}, 013010 (2001)
[arXiv:astro-ph/0010488].



\bibitem{Anchordoqui:2003gm} 
  L.~Anchordoqui and H.~Goldberg,
   {\color{rossoCP3} Footprints of superGZK cosmic rays in the Pilliga State Forest},
  Phys.\ Lett.\ B {\bf 583}, 213 (2004)
  doi:10.1016/j.physletb.2003.12.072
  [hep-ph/0310054].














\bibitem{Heitler} W. Heitler. 
   {\color{rossoCP3} The Quantum Theory of Radiation}, 2nd. Edition, (Oxford
University Press, London, 1944). 


\bibitem{Linsley:P3} 
J. Linsley,
  {\color{rossoCP3} Structure of large air showers  at depth 834 g/cm$^2$: Applications},
in Proceedings of 15th International Cosmic Ray Conference (Plovdiv, Bulgaria) {\bf 12}, 89 (1977). 


\bibitem{Matthews:2005sd}
  J.~Matthews,
    {\color{rossoCP3} A Heitler model of extensive air showers},
  Astropart.\ Phys.\  {\bf 22}, 387 (2005).


\bibitem{Ulrich:2009zq}
  R.~Ulrich, J.~Blumer, R.~Engel, F.~Schussler and M.~Unger,
    {\color{rossoCP3} On the measurement of the proton-air cross section using air shower data},
  New J.\ Phys.\  {\bf 11}, 065018 (2009)
  [arXiv:0903.0404 [astro-ph.HE]].



\bibitem{AbuZayyad:1999xa} 
  T.~Abu-Zayyad {\it et al.} [HiRes and MIA Collaborations],
  {\color{rossoCP3} Evidence for changing of cosmic ray composition between $10^{17}$~eV and $10^{18}$eV from multicomponent measurements},
  Phys.\ Rev.\ Lett.\  {\bf 84}, 4276 (2000)
  doi:10.1103/PhysRevLett.84.4276
  [astro-ph/9911144].





\bibitem{Aab:2014pza} 
  A.~Aab {\it et al.} [Pierre Auger Collaboration],
  {\color{rossoCP3} Muons in air showers at the Pierre Auger Observatory: Mean number in highly inclined events},
  Phys.\ Rev.\ D {\bf 91}, no. 3, 032003 (2015)
  Erratum: [Phys.\ Rev.\ D {\bf 91}, no. 5, 059901 (2015)]
  doi:10.1103/PhysRevD.91.059901, 10.1103/PhysRevD.91.032003
  [arXiv:1408.1421 [astro-ph.HE]].

\bibitem{Abbasi:2018fkz} 
  R.~U.~Abbasi {\it et al.} [Telescope Array Collaboration],
  {\color{rossoCP3}  Study of muons from ultrahigh energy cosmic ray air showers measured with the Telescope Array experiment},
  Phys.\ Rev.\ D {\bf 98}, no. 2, 022002 (2018)
  doi:10.1103/PhysRevD.98.022002
  [arXiv:1804.03877 [astro-ph.HE]].



\bibitem{Farrar:2013sfa} 
  G.~R.~Farrar and J.~D.~Allen,
  {\color{rossoCP3} A new physical phenomenon in ultrahigh energy collisions},
  EPJ Web Conf.\  {\bf 53}, 07007 (2013)
  doi:10.1051/epjconf/20135307007
  [arXiv:1307.2322 [hep-ph]].




 





\bibitem{Anchordoqui:2016oxy} 
  L.~A.~Anchordoqui, H.~Goldberg and T.~J.~Weiler,
 {\color{rossoCP3} Strange fireball as an explanation of the muon excess in Auger data},
  Phys.\ Rev.\ D {\bf 95}, no. 6, 063005 (2017)
  doi:10.1103/PhysRevD.95.063005
  [arXiv:1612.07328 [hep-ph]].


\bibitem{Tomar:2017mgc} 
  G.~Tomar,
 {\color{rossoCP3}  Lorentz invariance violation as an explanation of the muon excess in Auger data},
  Phys.\ Rev.\ D {\bf 95}, no. 9, 095035 (2017)
  doi:10.1103/PhysRevD.95.095035
  [arXiv:1701.05890 [hep-ph]].

\bibitem{Soriano:2017bvs} 
  J.~F.~Soriano, L.~A.~Anchordoqui, T.~C.~Paul and T.~J.~Weiler,
   {\color{rossoCP3} Probing QCD approach to thermal equilibrium with ultrahigh energy cosmic rays},
  PoS ICRC {\bf 2017}, 342 (2018)
  doi:10.22323/1.301.0342
  [arXiv:1811.07728 [hep-ph]].


\bibitem{Knapp:2002vs}
J.~Knapp, D.~Heck, S.~J.~Sciutto, M.~T.~Dova and M.~Risse,
   {\color{rossoCP3} Extensive air shower simulations at the highest energies},
Astropart.\ Phys.\  {\bf 19}, 77 (2003)
[arXiv:astro-ph/0206414].


\bibitem{Ulrich:2010rg} 
  R.~Ulrich, R.~Engel and M.~Unger,
    {\color{rossoCP3} Hadronic multiparticle production at ultra-high energies and extensive air showers},
  Phys.\ Rev.\ D {\bf 83}, 054026 (2011)
  doi:10.1103/PhysRevD.83.054026
  [arXiv:1010.4310 [hep-ph]].







\bibitem{Lemaitre:1931zzb} 
  G.~Lemaitre,
    {\color{rossoCP3} Republication of: The beginning of the world from the point of view of quantum theory},
  Nature {\bf 127}, 706 (1931)
  [Gen.\ Rel.\ Grav.\  {\bf 43}, 2929 (2011)].
  doi:10.1007/s10714-011-1214-6, 10.1038/127706b0


\bibitem{Hill:1982iq}
C.~T.~Hill,
 {\color{rossoCP3} Monopolonium},
Nucl.\ Phys.\ B {\bf 224}, 469 (1983).



\bibitem{Chudnovsky:1986hc} 
  E.~M.~Chudnovsky, G.~B.~Field, D.~N.~Spergel and A.~Vilenkin,
  {\color{rossoCP3}  Superconducting cosmic strings},
  Phys.\ Rev.\ D {\bf 34}, 944 (1986).
  doi:10.1103/PhysRevD.34.944



\bibitem{Hill:1986mn}
C.~T.~Hill, D.~N.~Schramm and T.~P.~Walker,
 {\color{rossoCP3}  Ultra-high-energy cosmic rays from superconducting cosmic strings},
Phys.\ Rev.\ D {\bf 36}, 1007 (1987).


\bibitem{Bhattacharjee:1989vu} 
  P.~Bhattacharjee,
    {\color{rossoCP3} Cosmic strings and ultra-high-energy cosmic rays},
  Phys.\ Rev.\ D {\bf 40}, 3968 (1989).
  doi:10.1103/PhysRevD.40.3968



\bibitem{Bhattacharjee:1990js} 
  P.~Bhattacharjee and N.~C.~Rana,
    {\color{rossoCP3} Ultra-high-energy particle flux from cosmic strings},
  Phys.\ Lett.\ B {\bf 246}, 365 (1990).
  doi:10.1016/0370-2693(90)90615-D





\bibitem{Bhattacharjee:1991zm}
P.~Bhattacharjee, C.~T.~Hill and D.~N.~Schramm,
 {\color{rossoCP3}  Grand unified theories, topological defects and ultra-high-energy cosmic  rays},
Phys.\ Rev.\ Lett.\  {\bf 69}, 567 (1992).




\bibitem{Bhattacharjee:1994pk} 
  P.~Bhattacharjee and G.~Sigl,
    {\color{rossoCP3} Monopole annihilation and highest energy cosmic rays},
  Phys.\ Rev.\ D {\bf 51}, 4079 (1995)
  doi:10.1103/PhysRevD.51.4079
  [astro-ph/9412053].



\bibitem{Berezinsky:1997kd}
V.~Berezinsky, X.~Martin and A.~Vilenkin,
 {\color{rossoCP3}  High energy particles from monopoles connected by strings},
Phys.\ Rev.\ D {\bf 56}, 2024 (1997)
[arXiv:astro-ph/9703077].


\bibitem{Berezinsky:1997td}
V.~Berezinsky and A.~Vilenkin,
 {\color{rossoCP3} Cosmic necklaces and ultrahigh energy cosmic rays},
Phys.\ Rev.\ Lett.\  {\bf 79}, 5202 (1997)
[arXiv:astro-ph/9704257].



\bibitem{Bhattacharjee:1998qc} 
  P.~Bhattacharjee and G.~Sigl,
   {\color{rossoCP3} Origin and propagation of extremely high-energy cosmic rays},
  Phys.\ Rept.\  {\bf 327}, 109 (2000)
  doi:10.1016/S0370-1573(99)00101-5
  [astro-ph/9811011].

\bibitem{Kuzmin:1997cm}
V.~A.~Kuzmin and V.~A.~Rubakov,
 {\color{rossoCP3} Ultra-high-energy cosmic rays: A window on postinflationary reheating  
epoch of the universe?},
Phys.\ Atom.\ Nucl.\  {\bf 61}, 1028 (1998)
[Yad.\ Fiz.\  {\bf 61}, 1122 (1998)]
[arXiv:astro-ph/9709187].



\bibitem{Kuzmin:1998uv} 
  V.~Kuzmin and I.~Tkachev,
  {\color{rossoCP3} Ultrahigh-energy cosmic rays, superheavy long living particles, and matter creation after inflation},
  JETP Lett.\  {\bf 68}, 271 (1998)
  [Pisma Zh.\ Eksp.\ Teor.\ Fiz.\  {\bf 68}, 255 (1998)]
  doi:10.1134/1.567858
  [hep-ph/9802304].



\bibitem{Dubovsky:1998pu} 
  S.~L.~Dubovsky and P.~G.~Tinyakov,
  {\color{rossoCP3} Galactic anisotropy as signature of CDM related ultrahigh-energy cosmic rays},
  JETP Lett.\  {\bf 68}, 107 (1998)
  doi:10.1134/1.567830
  [hep-ph/9802382].





\bibitem{Berezinsky:1998ed} 
  V.~Berezinsky and M.~Kachelriess,
  {\color{rossoCP3} Limiting SUSY QCD spectrum and its application for decays of superheavy particles},
  Phys.\ Lett.\ B {\bf 434}, 61 (1998)
  doi:10.1016/S0370-2693(98)00728-X
  [hep-ph/9803500].


\bibitem{Birkel:1998nx} 
  M.~Birkel and S.~Sarkar,
  {\color{rossoCP3} Extremely high-energy cosmic rays from relic particle decays},
  Astropart.\ Phys.\  {\bf 9}, 297 (1998)
  doi:10.1016/S0927-6505(98)00028-0
  [hep-ph/9804285].



\bibitem{Sarkar:2001se} 
  S.~Sarkar and R.~Toldra,
  {\color{rossoCP3} The high-energy cosmic ray spectrum from relic particle decay},
  Nucl.\ Phys.\ B {\bf 621}, 495 (2002)
  doi:10.1016/S0550-3213(01)00565-X
  [hep-ph/0108098].


\bibitem{Kuzmin:1999zk} 
  V.~A.~Kuzmin and I.~I.~Tkachev,
    {\color{rossoCP3} Ultrahigh-energy cosmic rays and inflation relics},
  Phys.\ Rept.\  {\bf 320}, 199 (1999)
  doi:10.1016/S0370-1573(99)00064-2
  [hep-ph/9903542].




\bibitem{Hamaguchi:1998wm}
K.~Hamaguchi, Y.~Nomura and T.~Yanagida,
  {\color{rossoCP3} Superheavy dark matter with discrete gauge symmetries},
Phys.\ Rev.\ D {\bf 58}, 103503 (1998)
[arXiv:hep-ph/9805346].


\bibitem{Hamaguchi:1998nj}
K.~Hamaguchi, Y.~Nomura and T.~Yanagida,
  {\color{rossoCP3} Long lived superheavy dark matter with discrete gauge symmetries},
Phys.\ Rev.\ D {\bf 59}, 063507 (1999)
[arXiv:hep-ph/9809426].


\bibitem{Hamaguchi:1999cv}
K.~Hamaguchi, K.~I.~Izawa, Y.~Nomura and T.~Yanagida,
  {\color{rossoCP3} Long-lived superheavy particles in dynamical supersymmetry-breaking  models in supergravity},
Phys.\ Rev.\ D {\bf 60}, 125009 (1999)
[arXiv:hep-ph/9903207].


\bibitem{Ellis:1990iu}
J.~R.~Ellis, J.~L.~Lopez and D.~V.~Nanopoulos,
  {\color{rossoCP3} Confinement of fractional charges yields integer charged relics in string models},
Phys.\ Lett.\ B {\bf 247}, 257 (1990).

\bibitem{Benakli:1998ut}
K.~Benakli, J.~R.~Ellis and D.~V.~Nanopoulos,
  {\color{rossoCP3} Natural candidates for superheavy dark matter in string and M theory},
Phys.\ Rev.\ D {\bf 59}, 047301 (1999)
[arXiv:hep-ph/9803333].

\bibitem{Berezinsky:1997hy}
V.~Berezinsky, M.~Kachelriess and A.~Vilenkin,
  {\color{rossoCP3} Ultra-high-energy cosmic rays without GZK cutoff},
Phys.\ Rev.\ Lett.\  {\bf 79}, 4302 (1997)
[arXiv:astro-ph/9708217].




\bibitem{Blasi:2001hr} 
  P.~Blasi, R.~Dick and E.~W.~Kolb,
  {\color{rossoCP3} Ultrahigh-energy cosmic rays from annihilation of superheavy dark matter},
  Astropart.\ Phys.\  {\bf 18}, 57 (2002)
  doi:10.1016/S0927-6505(02)00113-5
  [astro-ph/0105232].


\bibitem{Coriano:2001rt}
C.~Coriano and A.~E.~Faraggi,
  {\color{rossoCP3} SUSY QCD and high energy cosmic rays I: Fragmentation functions of  SUSY QCD},
Phys.\ Rev.\ D {\bf 65}, 075001 (2002)
[arXiv:hep-ph/0106326].


\bibitem{Barbot:2002gt}
C.~Barbot and M.~Drees,
  {\color{rossoCP3} Detailed analysis of the decay spectrum of a super-heavy $X$ particle},
Astropart.\ Phys.\  {\bf 20}, 5 (2003)
[arXiv:hep-ph/0211406].

\bibitem{Barbot:2003cj}
C.~Barbot,
  {\color{rossoCP3} Decay of super-heavy particles: User guide of the SHdecay program},
Comput.\ Phys.\ Commun.\  {\bf 157}, 63 (2004)
[arXiv:hep-ph/0306303].



\bibitem{Aharonian:1992qf}
F.~A.~Aharonian, P.~Bhattacharjee and D.~N.~Schramm,
  {\color{rossoCP3} Photon/proton ratio as a diagnostic tool for topological defects  as the sources of extremely high-energy cosmic rays},
Phys.\ Rev.\ D {\bf 46}, 4188 (1992).





\bibitem{Sigl:1998vz}
G.~Sigl, S.~Lee, P.~Bhattacharjee and S.~Yoshida,
  {\color{rossoCP3} Probing grand unified theories with cosmic ray, gamma-ray and
neutrino  astrophysics},
Phys.\ Rev.\ D {\bf 59}, 043504 (1999)
[arXiv:hep-ph/9809242].


\bibitem{Sigl:1995kk}
G.~Sigl, K.~Jedamzik, D.~N.~Schramm and V.~S.~Berezinsky,
  {\color{rossoCP3} Helium photodisintegration and nucleosynthesis: Implications for topological defects, high-energy cosmic rays, and massive black holes},
Phys.\ Rev.\ D {\bf 52}, 6682 (1995)
[arXiv:astro-ph/9503094].



\bibitem{Sigl:1996gm}
G.~Sigl, S.~Lee, D.~N.~Schramm and P.~Coppi,
  {\color{rossoCP3} Cosmological neutrino signatures for grand unification scale physics},
Phys.\ Lett.\ B {\bf 392}, 129 (1997)
[arXiv:astro-ph/9610221].




\bibitem{Protheroe:1996pd}
R.~J.~Protheroe and T.~Stanev,
  {\color{rossoCP3} Limits on models of the ultrahigh energy cosmic rays based on  topological defects},
Phys.\ Rev.\ Lett.\  {\bf 77}, 3708 (1996)
[Erratum-ibid.\  {\bf 78}, 3420 (1997)]
[arXiv:astro-ph/9605036].





\bibitem{Protheroe:1996zg}
R.~J.~Protheroe and P.~A.~Johnson,
  {\color{rossoCP3} Are topological defects responsible for the 300-EeV cosmic rays?},
Nucl.\ Phys.\ Proc.\ Suppl.\  {\bf 48}, 485 (1996)
[arXiv:astro-ph/9605006].

\bibitem{Sreekumar:1997un} 
  P.~Sreekumar {\it et al.} [EGRET Collaboration],
    {\color{rossoCP3} EGRET observations of the extragalactic gamma-ray emission},
  Astrophys.\ J.\  {\bf 494}, 523 (1998)
  doi:10.1086/305222
  [astro-ph/9709257].

\bibitem{Berezinsky:1998ft} 
  V.~Berezinsky, P.~Blasi and A.~Vilenkin,
    {\color{rossoCP3} Ultra-high-energy gamma-rays as signature of topological defects},
  Phys.\ Rev.\ D {\bf 58}, 103515 (1998)
  doi:10.1103/PhysRevD.58.103515
  [astro-ph/9803271].


\bibitem{Abdo:2010nz} 
  A.~A.~Abdo {\it et al.} [Fermi-LAT Collaboration],
    {\color{rossoCP3} The spectrum of the isotropic diffuse gamma-ray emission derived from first-year Fermi Large Area Telescope data},
  Phys.\ Rev.\ Lett.\  {\bf 104}, 101101 (2010)
  doi:10.1103/PhysRevLett.104.101101
  [arXiv:1002.3603 [astro-ph.HE]].

\bibitem{Berezinsky:2010xa} 
  V.~Berezinsky, A.~Gazizov, M.~Kachelriess and S.~Ostapchenko,
    {\color{rossoCP3} Restricting UHECRs and cosmogenic neutrinos with Fermi-LAT},
  Phys.\ Lett.\ B {\bf 695}, 13 (2011)
  doi:10.1016/j.physletb.2010.11.019
  [arXiv:1003.1496 [astro-ph.HE]].


\bibitem{Ackermann:2014usa} 
  M.~Ackermann {\it et al.} [Fermi-LAT Collaboration],
 {\color{rossoCP3}  The spectrum of isotropic diffuse gamma-ray emission between 100 MeV and 820 GeV},
  Astrophys.\ J.\  {\bf 799}, 86 (2015)
  doi:10.1088/0004-637X/799/1/86
  [arXiv:1410.3696 [astro-ph.HE]].

\bibitem{Berezinsky:2016feh} 
  V.~Berezinsky and O.~Kalashev,
  {\color{rossoCP3}  High energy electromagnetic cascades in extragalactic space: physics and features},
  Phys.\ Rev.\ D {\bf 94}, no. 2, 023007 (2016)
  doi:10.1103/PhysRevD.94.023007
  [arXiv:1603.03989 [astro-ph.HE]].



\bibitem{Weiler:1982qy} 
  T.~J.~Weiler,
   {\color{rossoCP3} Resonant absorption of cosmic ray neutrinos by the relic neutrino background},
  Phys.\ Rev.\ Lett.\  {\bf 49}, 234 (1982).
  doi:10.1103/PhysRevLett.49.234


\bibitem{Weiler:1997sh} 
  T.~J.~Weiler,
    {\color{rossoCP3} Cosmic ray neutrino annihilation on relic neutrinos revisited: A Mechanism for generating air showers above the Greisen-Zatsepin-Kuzmin cutoff},
  Astropart.\ Phys.\  {\bf 11}, 303 (1999)
  doi:10.1016/S0927-6505(98)00068-1
  [hep-ph/9710431].


\bibitem{Fargion:1997ft} 
  D.~Fargion, B.~Mele and A.~Salis,
    {\color{rossoCP3} Ultra-high-energy neutrino scattering onto relic light neutrinos in galactic halo as a possible source of highest energy extragalactic cosmic rays},
  Astrophys.\ J.\  {\bf 517}, 725 (1999)
  doi:10.1086/307203
  [astro-ph/9710029].


\bibitem{Aloisio:2015lva} 
  R.~Aloisio, S.~Matarrese and A.~V.~Olinto,
    {\color{rossoCP3} Super heavy dark matter in light of BICEP2, Planck and ultra high energy cosmic ray observations},
  JCAP {\bf 1508}, no. 08, 024 (2015)
  doi:10.1088/1475-7516/2015/08/024
  [arXiv:1504.01319 [astro-ph.HE]].


\bibitem{Christiansen:2010zi} 
  J.~L.~Christiansen, E.~Albin, T.~Fletcher, J.~Goldman, I.~P.~W.~Teng, M.~Foley and G.~F.~Smoot,
    {\color{rossoCP3} Search for cosmic strings in the COSMOS survey},
  Phys.\ Rev.\ D {\bf 83}, 122004 (2011)
  doi:10.1103/PhysRevD.83.122004
  [arXiv:1008.0426 [astro-ph.CO]].

\bibitem{vanHaasteren:2011ni} 
  R.~van Haasteren {\it et al.},
    {\color{rossoCP3} Placing limits on the stochastic gravitational-wave background using European Pulsar Timing Array data},
  Mon.\ Not.\ Roy.\ Astron.\ Soc.\  {\bf 414}, no. 4, 3117 (2011)
  Erratum: [Mon.\ Not.\ Roy.\ Astron.\ Soc.\  {\bf 425}, no. 2, 1597 (2012)]
  doi:10.1111/j.1365-2966.2011.18613.x, 10.1111/j.1365-2966.2012.20916.x
  [arXiv:1103.0576 [astro-ph.CO]].

\bibitem{Damour:2004kw} 
  T.~Damour and A.~Vilenkin,
    {\color{rossoCP3} Gravitational radiation from cosmic (super)strings: Bursts, stochastic background, and observational windows},
  Phys.\ Rev.\ D {\bf 71}, 063510 (2005)
  doi:10.1103/PhysRevD.71.063510
  [hep-th/0410222].

\bibitem{Olmez:2010bi} 
  S.~Olmez, V.~Mandic and X.~Siemens,
    {\color{rossoCP3} Gravitational-wave stochastic background from kinks and cusps on cosmic ctrings},
  Phys.\ Rev.\ D {\bf 81}, 104028 (2010)
  doi:10.1103/PhysRevD.81.104028
  [arXiv:1004.0890 [astro-ph.CO]].


\bibitem{Berezinsky:2011cp} 
  V.~Berezinsky, E.~Sabancilar and A.~Vilenkin,
    {\color{rossoCP3} Extremely high energy neutrinos from cosmic strings},
  Phys.\ Rev.\ D {\bf 84}, 085006 (2011)
  doi:10.1103/PhysRevD.84.085006
  [arXiv:1108.2509 [astro-ph.CO]].


\bibitem{Boyle:2018tzc} 
  L.~Boyle, K.~Finn and N.~Turok,
   {\color{rossoCP3}{$CPT$ symmetric universe}},
  arXiv:1803.08928 [hep-ph].

\bibitem{Boyle:2018rgh} 
  L.~Boyle, K.~Finn and N.~Turok,
   {\color{rossoCP3}{The Big Bang, CPT, and neutrino dark matter}},
  arXiv:1803.08930 [hep-ph].


\bibitem{Aartsen:2018mxl} 
  M.~G.~Aartsen {\it et al.} [IceCube Collaboration],
   {\color{rossoCP3}{ Search for neutrinos from decaying dark matter with IceCube}},
  arXiv:1804.03848 [astro-ph.HE].



\bibitem{Gorham:2008dv} 
  P.~W.~Gorham {\it et al.} [ANITA Collaboration],
   {\color{rossoCP3}{The Antarctic Impulsive Transient Antenna ultra-high energy neutrino detector design, performance, and sensitivity for 2006-2007 balloon flight}},
  Astropart.\ Phys.\  {\bf 32}, 10 (2009)
  doi:10.1016/j.astropartphys.2009.05.003
  [arXiv:0812.1920 [astro-ph]].


\bibitem{Allison:2018cxu} 
  P.~W.~Gorham {\it et al.} [ANITA Collaboration],
   {\color{rossoCP3}{Constraints on the diffuse high-energy neutrino flux from the third flight of ANITA}},
  Phys.\ Rev.\ D {\bf 98}, no. 2, 022001 (2018)
  doi:10.1103/PhysRevD.98.022001
  [arXiv:1803.02719 [astro-ph.HE]].



\bibitem{Gorham:2016zah} 
  P.~W.~Gorham {\it et al.},
  {\color{rossoCP3}{Characteristics of four upward-pointing cosmic-ray-like events observed with ANITA}},
  Phys.\ Rev.\ Lett.\  {\bf 117}, no. 7, 071101 (2016)
  doi:10.1103/PhysRevLett.117.071101
  [arXiv:1603.05218 [astro-ph.HE]].

\bibitem{Gorham:2018ydl} 
  P.~W.~Gorham {\it et al.} [ANITA Collaboration],
  {\color{rossoCP3}{Observation of an unusual upward-going cosmic-ray-like event in the third flight of ANITA}},
  arXiv:1803.05088 [astro-ph.HE].



\bibitem{Anchordoqui:2018ucj} 
  L.~A.~Anchordoqui, V.~Barger, J.~G.~Learned, D.~Marfatia and T.~J.~Weiler,
  {\color{rossoCP3}{ Upgoing ANITA events as evidence of the CPT symmetric universe}},
  LHEP {\bf 1}, no. 1, 13 (2018)
  doi:10.31526/LHEP.1.2018.03
  [arXiv:1803.11554 [hep-ph]].



\bibitem{Patterson}
C. Patterson, G. Tilton, and M.  Inghram,
 {\color{rossoCP3}Age of the Earth},
Science {\bf 121}, 69 (1955)
doi:10.1126/science.121.3134.69


\bibitem{Bienayme:2014kva} 
  O.~Bienaymé {\it et al.},
   {\color{rossoCP3} Weighing the local dark matter with RAVE red clump stars},
  Astron.\ Astrophys.\  {\bf 571}, A92 (2014)
  doi:10.1051/0004-6361/201424478
  [arXiv:1406.6896 [astro-ph.GA]].



\bibitem{Piffl:2014mfa} 
  T.~Piffl {\it et al.},
   {\color{rossoCP3} Constraining the Galaxy's dark halo with RAVE stars},
  Mon.\ Not.\ Roy.\ Astron.\ Soc.\  {\bf 445}, no. 3, 3133 (2014)
  doi:10.1093/mnras/stu1948
  [arXiv:1406.4130 [astro-ph.GA]].


\bibitem{McKee:2015hwa} 
  C.~F.~McKee, A.~Parravano and D.~J.~Hollenbach,
   {\color{rossoCP3} Stars, gas, and dark matter in the solar neighborhood},
  Astrophys.\ J.\  {\bf 814}, no. 1, 13 (2015)
  doi:10.1088/0004-637X, 10.1088/0004-637X/814/1/13
  [arXiv:1509.05334 [astro-ph.GA]].




\bibitem{Sivertsson:2017rkp} 
  S.~Sivertsson, H.~Silverwood, J.~I.~Read, G.~Bertone and P.~Steger,
   {\color{rossoCP3} The local dark matter density from SDSS-SEGUE G-dwarfs},
  Mon.\ Not.\ Roy.\ Astron.\ Soc.\ 
  doi:10.1093/mnras/sty977
  [arXiv:1708.07836 [astro-ph.GA]].



\bibitem{Neufeld:2018slx} 
  D.~A.~Neufeld, G.~R.~Farrar and C.~F.~McKee,
   {\color{rossoCP3} Dark matter that interacts with baryons: density distribution within the Earth and new constraints on the interaction cross-section},
  arXiv:1805.08794 [astro-ph.CO].





  \bibitem{Adams:2017fjh} 
  J.~H.~Adams {\it et al.},
   {\color{rossoCP3}{ White paper on EUSO-SPB2 }},
  arXiv:1703.04513 [astro-ph.HE].


\bibitem{Redei:1966} 
L.~B.~R\'edei, 
{\color{rossoCP3}{Possible experimental test of the existence of a universal length}},
Phys.\ Rev.\ {\bf 145}, 999  (1966)  
doi:10.1103/PhysRev.145.999

\bibitem{Redei:1967zz} 
L.~B.~R\'edei, 
{\color{rossoCP3}{Validity of  Special Relativity at small distances and the velocity
      dependence of the muon lifetime}}, 
   Phys.\ Rev.\ {\bf 162}, 1299  (1967).  
doi:10.1103/PhysRev.162.1299

\bibitem{Anchordoqui:1995im} 
  L.~Anchordoqui, M.~T.~Dova, D.~Gomez Dumm and P.~Lacentre,
    {\color{rossoCP3} Possible test of local Lorentz invariance from tau decays},
  Z.\ Phys.\ C {\bf 73}, 465 (1997)
  doi:10.1007/s002880050336
  [gr-qc/9512015].


\bibitem{Coleman:1998ti} 
  S.~R.~Coleman and S.~L.~Glashow,
   {\color{rossoCP3} High-energy tests of Lorentz invariance},
  Phys.\ Rev.\ D {\bf 59}, 116008 (1999)
  doi:10.1103/PhysRevD.59.116008
  [hep-ph/9812418].


\bibitem{Coleman:1998en} 
  S.~R.~Coleman and S.~L.~Glashow,
    {\color{rossoCP3} Evading the GZK cosmic ray cutoff},
  hep-ph/9808446.


\bibitem{Aloisio:2000cm} 
  R.~Aloisio, P.~Blasi, P.~L.~Ghia and A.~F.~Grillo,
  {\color{rossoCP3} Probing the structure of space-time with cosmic rays},
  Phys.\ Rev.\ D {\bf 62}, 053010 (2000)
  doi:10.1103/PhysRevD.62.053010
  [astro-ph/0001258].

\bibitem{Jankiewicz:2003sm} 
  M.~Jankiewicz, R.~V.~Buniy, T.~W.~Kephart and T.~J.~Weiler,
   {\color{rossoCP3} Space-time foam and cosmic ray interactions},
  Astropart.\ Phys.\  {\bf 21}, 651 (2004)
  doi:10.1016/j.astropartphys.2004.04.008
  [hep-ph/0312221].




\bibitem{Galaverni:2007tq} 
  M.~Galaverni and G.~Sigl,
   {\color{rossoCP3} Lorentz violation in the photon sector and ultrahigh-energy cosmic rays},
  Phys.\ Rev.\ Lett.\  {\bf 100}, 021102 (2008)
  doi:10.1103/PhysRevLett.100.021102
  [arXiv:0708.1737 [astro-ph]].



\bibitem{Galaverni:2008yj} 
  M.~Galaverni and G.~Sigl,
   {\color{rossoCP3} Lorentz violation and ultrahigh-energy photons},
  Phys.\ Rev.\ D {\bf 78}, 063003 (2008)
  doi:10.1103/PhysRevD.78.063003
  [arXiv:0807.1210 [astro-ph]].


\bibitem{Mattingly:2009jf} 
  D.~M.~Mattingly, L.~Maccione, M.~Galaverni, S.~Liberati and G.~Sigl,
   {\color{rossoCP3}  Possible cosmogenic neutrino constraints on Planck-scale Lorentz violation},
  JCAP {\bf 1002}, 007 (2010)
  doi:10.1088/1475-7516/2010/02/007
  [arXiv:0911.0521 [hep-ph]].

\bibitem{Bietenholz:2008ni} 
  W.~Bietenholz,
    {\color{rossoCP3}  Cosmic rays and the search for a Lorentz invariance violation},
  Phys.\ Rept.\  {\bf 505}, 145 (2011)
  doi:10.1016/j.physrep.2011.04.002
  [arXiv:0806.3713 [hep-ph]].


\bibitem{Stecker:2017gdy} 
  F.~W.~Stecker,
  {\color{rossoCP3}   Testing Lorentz symmetry using high energy astrophysics observations},
  Symmetry {\bf 9}, no. 10, 201 (2017)
  doi:10.3390/sym9100201
  [arXiv:1708.05672 [astro-ph.HE]].


\bibitem{Scully:2008jp} 
  S.~T.~Scully and F.~W.~Stecker,
  {\color{rossoCP3} Lorentz invariance violation and the observed spectrum of ultra-high energy cosmic rays},
  Astropart.\ Phys.\  {\bf 31}, 220 (2009)
  doi:10.1016/j.astropartphys.2009.01.002
  [arXiv:0811.2230 [astro-ph]].

\bibitem{Stecker:2009hj} 
  F.~W.~Stecker and S.~T.~Scully,
   {\color{rossoCP3} Searching for new physics with ultra-high energy cosmic rays},
  New J.\ Phys.\  {\bf 11}, 085003 (2009)
  doi:10.1088/1367-2630/11/8/085003
  [arXiv:0906.1735 [astro-ph.HE]].






\bibitem{Saveliev:2011vw} 
  A.~Saveliev, L.~Maccione and G.~Sigl,
   {\color{rossoCP3} Lorentz invariance violation and chemical composition of ultra-high-energy cosmic rays}
  JCAP {\bf 1103}, 046 (2011)
  doi:10.1088/1475-7516/2011/03/046
  [arXiv:1101.2903 [astro-ph.HE]].

\bibitem{Anchordoqui:2017pmf} 
  L.~A.~Anchordoqui and J.~F.~Soriano,
  {\color{rossoCP3} New test of Lorentz symmetry using ultra-high-energy cosmic rays},
  Phys.\ Rev.\ D {\bf 97}, no. 4, 043010 (2018)
  doi:10.1103/PhysRevD.97.043010
  [arXiv:1710.00750 [hep-ph]].


\bibitem{Boncioli:2015cqa} 
  D.~Boncioli {\it et al.},
   {\color{rossoCP3} Future prospects of testing Lorentz invariance with UHECRs},
  PoS ICRC {\bf 2015}, 521 (2016)
  doi:10.22323/1.236.0521
  [arXiv:1509.01046 [astro-ph.HE]].


\bibitem{Pavlidou:2018yux} 
  V.~Pavlidou and T.~Tomaras,
   {\color{rossoCP3} What do the highest-energy cosmic-ray data suggest about possible new physics around 50 TeV?},
  arXiv:1802.04806 [astro-ph.HE].

\bibitem{Kusenko:2001gj} 
  A.~Kusenko and T.~J.~Weiler,
   {\color{rossoCP3} Neutrino cross-sections at high-energies and the future observations of ultrahigh-energy cosmic rays},
  Phys.\ Rev.\ Lett.\  {\bf 88}, 161101 (2002)
  doi:10.1103/PhysRevLett.88.161101
  [hep-ph/0106071].


\bibitem{Anchordoqui:2001cg} 
  L.~A.~Anchordoqui, J.~L.~Feng, H.~Goldberg and A.~D.~Shapere,
   {\color{rossoCP3} Black holes from cosmic rays: Probes of extra dimensions and new limits on TeV scale gravity},
  Phys.\ Rev.\ D {\bf 65}, 124027 (2002)
  doi:10.1103/PhysRevD.65.124027
  [hep-ph/0112247].


\bibitem{Anchordoqui:2005is} 
  L.~Anchordoqui and F.~Halzen,
   {\color{rossoCP3} IceHEP high energy physics at the south pole},
  Annals Phys.\  {\bf 321}, 2660 (2006)
  doi:10.1016/j.aop.2005.11.015
  [hep-ph/0510389].

\bibitem{Anchordoqui:2005ey} 
  L.~Anchordoqui, T.~Han, D.~Hooper and S.~Sarkar,
    {\color{rossoCP3} Exotic neutrino interactions at the Pierre Auger Observatory},
  Astropart.\ Phys.\  {\bf 25}, 14 (2006)
  doi:10.1016/j.astropartphys.2005.10.006
  [hep-ph/0508312].

\bibitem{Anchordoqui:2010hq} 
  L.~A.~Anchordoqui, H.~Goldberg, D.~Gora, T.~Paul, M.~Roth, S.~Sarkar and L.~L.~Winders,
   {\color{rossoCP3} Using cosmic neutrinos to search for non-perturbative physics at the Pierre Auger Observatory},
  Phys.\ Rev.\ D {\bf 82}, 043001 (2010)
  doi:10.1103/PhysRevD.82.043001
  [arXiv:1004.3190 [hep-ph]].

\bibitem{PalomaresRuiz:2005xw} 
  S.~Palomares-Ruiz, A.~Irimia and T.~J.~Weiler,
   {\color{rossoCP3} Acceptances for space-based and ground-based
     fluorescence detectors, and inference of the neutrino-nucleon
     cross-section above $10^{19}~{\rm eV}$},
  Phys.\ Rev.\ D {\bf 73}, 083003 (2006)
  doi:10.1103/PhysRevD.73.083003
  [astro-ph/0512231].



\bibitem{Olinto:2017xbi} 
  A.~V.~Olinto {\it et al.},
   {\color{rossoCP3}  POEMMA: Probe Of Extreme Multi-Messenger Astrophysics},
  PoS ICRC {\bf 2017}, 542 (2017)
  [arXiv:1708.07599 [astro-ph.IM]].

\bibitem{Stecker:2004wt} 
  F.~W.~Stecker, J.~F.~Krizmanic, L.~M.~Barbier, E.~Loh, J.~W.~Mitchell, P.~Sokolsky and R.~E.~Streitmatter,
   {\color{rossoCP3}  Observing the ultra-high-energy universe with OWL eyes},
  Nucl.\ Phys.\ Proc.\ Suppl.\  {\bf 136C}, 433 (2004)
  doi:10.1016/j.nuclphysbps.2004.10.027
  [astro-ph/0408162].

\bibitem{Neronov:2016zou} 
  A.~Neronov, D.~V.~Semikoz, L.~A.~Anchordoqui, J.~Adams and A.~V.~Olinto,
   {\color{rossoCP3}  Sensitivity of a proposed space-based Cherenkov astrophysical-neutrino telescope},
  Phys.\ Rev.\ D {\bf 95}, no. 2, 023004 (2017)
  doi:10.1103/PhysRevD.95.023004
  [arXiv:1606.03629 [astro-ph.IM]].



\bibitem{Cowan:2010js} 
  G.~Cowan, K.~Cranmer, E.~Gross and O.~Vitells,
    {\color{rossoCP3} Asymptotic formulae for likelihood-based tests of new physics},
  Eur.\ Phys.\ J.\ C {\bf 71}, 1554 (2011)
  Erratum: [Eur.\ Phys.\ J.\ C {\bf 73}, 2501 (2013)]
  doi:10.1140/epjc/s10052-011-1554-0, 10.1140/epjc/s10052-013-2501-z
  [arXiv:1007.1727 [physics.data-an]].

\bibitem{Cousins:2018tiz} 
  R.~D.~Cousins,
     {\color{rossoCP3} Lectures on statistics in theory: Prelude to statistics in practice},
  arXiv:1807.05996 [physics.data-an].




\bibitem{Wilks:1938dza} 
  S.~S.~Wilks,
    {\color{rossoCP3} The large-sample distribution of the likelihood ratio for testing composite hypotheses},
  Annals Math.\ Statist.\  {\bf 9},  60 (1938).
  doi:10.1214/aoms/1177732360

\bibitem{Deligny:2018blo} 
  O.~Deligny,
   {\color{rossoCP3} Measurements and implications of cosmic ray anisotropies from TeV to trans-EeV energies},
  Astropart.\ Phys.\  {\bf 104}, 13 (2019)
  doi:10.1016/j.astropartphys.2018.08.005
  [arXiv:1808.03940 [astro-ph.HE]].



\end{thebibliography}
\end{document}